\documentclass[12pt]{article}

\usepackage{amssymb}
\usepackage{cite}

\newcommand{\be}{\begin{equation}}
\newcommand{\ee}{\end{equation}}

\newcommand{\ra}{\rightarrow}
\newcommand{\prt}{\partial}
\newcommand{\al}{\alpha}
\newcommand{\bt}{\beta}
\newcommand{\dlt}{\delta}
\newcommand{\Dlt}{\Delta}
\newcommand{\Om}{\Omega}
\newcommand{\om}{\omega}
\newcommand{\gm}{\gamma}
\newcommand{\sgm}{\sigma}
\newcommand{\Gm}{\Gamma}
\newcommand{\lbd}{\lambda}
\newcommand{\Lbd}{\Lambda}
\newcommand{\dgr}{\dagger}
\newcommand{\ep}{\varepsilon}
\newcommand{\vp}{\varphi}
\newcommand{\vt}{\vartheta}
\newcommand{\bB}{{\bf B}}
\newcommand{\bE}{{\bf E}}

\newcommand{\bP}{{\bf P}}
\newcommand{\bJ}{{\bf J}}
\newcommand{\bS}{{\bf S}}

\newcommand{\bj}{{\bf j}}
\newcommand{\br}{{\bf r}}
\newcommand{\bk}{{\bf k}}
\newcommand{\bv}{{\bf v}}
\newcommand{\bp}{{\bf p}}
\newcommand{\bd}{{\bf d}}

\newcommand{\cF}{{\cal F}}

\font\tenmsb=msbm10 scaled\magstep 1
   \font\sevenmsb=msbm7 scaled \magstep 1
   \font\faivemsb=msbm5 scaled \magstep 1
\newfam\msbfam
      \textfont\msbfam=\tenmsb
      \scriptfont\msbfam=\sevenmsb
      \scriptscriptfont\msbfam=\faivemsb

\begin{document}

{\bf TUTORIAL}

\vskip 5mm

\begin{center}

{\Large{\bf Theory of cold atoms: Bose-Einstein statistics} \\ [5mm]
V.I. Yukalov} \\ [3mm]

{\it Bogolubov Laboratory of Theoretical Physics, \\
Joint Institute for Nuclear Research, \\
Dubna 141980, Russia}

\end{center}

\vskip 2cm

\begin{abstract}
This Tutorial is the continuation of the previous tutorial part, published
in Laser Phys. {\bf 23}, 062001 (2013), where the basic mathematical techniques
required for an accurate description of cold atoms for both types of
quantum statistics are expounded. In the present part, the specifics of the
correct theoretical description of atoms obeying Bose-Einstein statistics
are explained, including trapped Bose atoms. In the theory of systems
exhibiting the phenomenon of Bose-Einstein condensation, there exists a
number of delicate mathematical points, whose misunderstanding often
results in principally wrong conclusions. This is why the consideration
in the present Tutorial is sufficiently detailed in order that the reader
could clearly understand the underlying mathematics and would avoid
confusions.
\end{abstract}

{\parindent=0pt
\vskip 1cm
{\bf Keywords}: Bose-Einstein condensate, superfluidity, trapped atoms, condensation
phase transition, thermodynamic stability


}

\newpage

\section{Introduction}

In the previous Tutorial \cite{Yukalov_1}, the basic mathematical notions
that are necessary for an accurate description of quantum systems composed
of atoms or molecules are elucidated. That consideration has been done for
the general case of atoms satisfying either Bose-Einstein or Fermi-Dirac
statistics. The present tutorial is devoted to the consideration of specifics
related to atomic systems of Bose-Einstein statistics. These systems exhibit
the phase transition of Bose-Einstein condensation and the related
superfluid transition. The theoretical description of Bose-condensed systems
confronts several delicate points that are often misunderstood in literature,
resulting in wrong conclusions. Among such misconceptions, it is possible
to mention the following.

One assumes that the ideal Bose gas is a realistic object and can be a
good approximation for Bose-condensed systems with weak interactions.
However, this is not exactly correct, since the ideal Bose gas and an
interacting Bose gas pertain to different classes of universality. Moreover,
the uniform ideal Bose-condensed gas is unstable, being a pathological system
with infinite compressibility. The ideal Bose-condensed gas can be
stabilized inside a trap. But not each trap can stabilize the Bose-condensed 
gas. For instance, a three-dimensional harmonic trap does stabilize the ideal
Bose-condensed gas, although the lower dimensional harmonic traps cannot
make it stable.

Sometimes, in the literature it is possible to meet the statement that
Bose-Einstein condensation does not necessarily require gauge symmetry
breaking. But this is wrong, since there exist strict mathematical theorems
proving that gauge symmetry breaking is a necessary and sufficient condition
for Bose-Einstein condensation. As a consequence, a mean-field approximation
of the Hartree-Fock type, where gauge symmetry is not broken, is not
applicable to Bose-condensed systems, but the Hartree-Fock-Bogolubov
approximation, with the broken gauge symmetry, is compulsory.

Neglecting the broken gauge symmetry leads to the appearance of
thermodynamically anomalous particle fluctuations, when using the grand
canonical ensemble, which one sometimes calls "grand canonical catastrophe".
However, there is no any catastrophe when correctly using the grand ensemble.
Catastrophes arise only because of incorrect calculations.

Similarly, thermodynamically anomalous particle fluctuations in a
Bose-condensed system formally occur in the Bogolubov approximation in any
ensemble, whether grand canonical or canonical. This, however, is also caused
by the incorrect use of the second-order, with respect to field operators,
Bogolubov approximation to calculating the fourth-order operator products.
Being in the frame of the applicability of the approximation order always
leads to thermodynamically normal fluctuations.

There are discussions stating that canonical and grand canonical ensembles
are not equivalent for Bose-condensed systems. This is again is due to an
incorrect use of the ensembles, for example, when one forgets to break
gauge symmetry under Bose-Einstein condensation. Generally, it is necessary
to employ representative ensembles, where all necessary restrictions, uniquely
characterising the system, are taken into account. Then such representative
ensembles are always equivalent in thermodynamic limit. 

A very widespread mistake is the neglect of the so-called anomalous averages, 
when describing Bose-condensed systems, while retaining the normal averages 
corresponding to uncondensed atoms. The fact that such a neglect is principally 
inadmissible is evident, if one remembers that Bose-Einstein condensation is 
necessarily accompanied by gauge symmetry breaking, which is responsible for 
the appearance of the condensate fraction as well as of the anomalous averages. 
Therefore it should be logically clear that omitting one of the basic 
characteristics of the Bose-condensed system, while keeping the other, is 
principally wrong. Moreover, by direct calculations, it is straightforward to 
check that the anomalous averages, by the order of magnitude, are always larger, 
depending on temperature, than either the normal averages, representing the 
fraction of normal atoms, or than the condensate fraction. Thus omitting the 
larger terms, while keeping the smaller ones is mathematically wrong. Such a 
contradiction results in the system instability and in the incorrect description 
of the Bose-Einstein condensation transition.

One sometimes involves the Dirac representation of field operators, by
defining a phase operator. This representation, unfortunately, is
mathematically incorrect and can lead to wrong results.

Because of a number of misconceptions in the theoretical study of
Bose-condensed systems, it looks important to give a thorough analysis
of the mathematics providing the tools for a correct description of systems
with Bose-Einstein condensate.

In the following parts, the knowledge of the previous Tutorial \cite{Yukalov_1}
is assumed, because of which many definitions are not repeated here. Since
the present paper is a Tutorial, but not a review article, only the most
important, from the point of view of the author, material is discussed. Also,
no extensive citations of numerous literature are given, but only the
publications directly related to the concrete points under consideration
are cited.

Throughout the paper, the system of units is employed where the Planck
constant $\hbar \ra 1$ and the Boltzmann constant $k_B \ra 1$.

\section{Basic notions}

In this part, the main notions and specific characteristics of atomic systems, 
satisfying the Bose-Einstein statistics, are described, giving the general 
understanding of the peculiarities accompanying the phenomenon of Bose-Einstein 
condensation.

\subsection{Phenomenon of superfluidity}

Probably, the most physically important property of Bose systems is the
occurrence of superfluidity. Therefore, let us start the consideration from
the problem of how this property could be quantified.

The occurrence of superfluidity implies the appearance of superfluid fraction
of atoms
\be
\label{1}
n_s \equiv \frac{N_s}{N} \; ,
\ee
where $N_s$ is the average number of superfluid atoms among the total number
of atoms $N$. This number $N_s$ represents the average number of atoms, but
is not related to particular atoms, since the latter are not distinguishable.

The system of atoms, without internal degrees of freedom, is characterized 
by the Hamiltonian
\be
\label{2}
H = \int \psi^\dgr(\br)\left ( -\; \frac{\nabla^2}{2m} + U - \mu\right )
\psi(\br)\; d\br + \frac{1}{2}
\int \psi^\dgr(\br)\psi^\dgr(\br')\Phi(\br-\br')\psi(\br')\psi(\br)\;
d\br d\br' \; ,
\ee
in which the field operators satisfy Bose commutation relations and,
generally, depend on time $t$ that is not shown explicitly. Here $U = U(\br)$
is an external potential.

The motion of the system is associated with the momentum operator
\be
\label{3}
\hat{\bf P} = \int \psi^\dgr(\br)(-i\nabla)\psi(\br)\; d\br \; ,
\ee
defining the observable quantity, the average system momentum
\be
\label{4}
\langle \hat{\bf P} \rangle \; =\; {\rm Tr}\hat\rho\; \hat{\bf P} \; ,
\ee
where the statistical averaging is done with respect to a statistical
operator $\hat{\rho}$. For an equilibrium system, the latter is
\be
\label{5}
 \hat\rho = \frac{e^{-\bt H}}{{\rm Tr}e^{-\bt H}} \; ,
\ee
with $\beta = 1/T$ being the inverse temperature.

The property of superfluidity is related to the nontrivial response of the
system with respect to a boost with velocity $\bv$. Under this boost, the
field operator, in the laboratory frame, satisfies the Galilean
transformation
\be
\label{6}
\psi_v(\br,t) = \psi(\br-\bv t,t)\; \exp\left \{ i\left ( m\bv \cdot\br \;
- \; \frac{mv^2}{2}\; t\right )\right \} \; .
\ee
Then the Hamiltonian for the moving system becomes
\be
\label{7}
H_v =  H + \int \psi^\dgr(\br) \left ( - i\bv \cdot \nabla +
\frac{mv^2}{2}\right ) \psi(\br)\; d\br \; ,
\ee
and the momentum operator is
\be
\label{8}
\hat{\bf P}_v = \hat{\bf P} + m\bv N \; .
\ee
The average momentum reads as
$$
\langle \hat{\bf P}_v \rangle_v \; \equiv \; {\rm Tr}\hat\rho_v \hat{\bf P}_v \; ,
$$
with the statistical operator
$$
\hat\rho_v \equiv \frac{e^{-\bt H_v}}{{\rm Tr}e^{-\bt H_v}} \; .
$$

To find the number of particles nontrivially responding to the velocity boost,
\be
\label{9}
N_s \equiv \frac{1}{3m}\; \lim_{v\ra 0} \; \frac{\prt}{\prt\bv} \;
\cdot \langle \hat{\bf P}_v \rangle_v \; ,
\ee
one needs the derivatives
$$
\frac{\prt}{\prt\bv}\; \cdot \hat{\bf P} \equiv \sum_{\al=1}^3
\frac{\prt P^\al}{\prt v^\al} \; , \qquad
\frac{\prt}{\prt\bv} \; e^{-\bt H_v} = - \;
\frac{\bt}{2} \left [  e^{-\bt H_v}, \; \frac{\prt H_v}{\prt\bv}
\right ]_+ \; .
$$
The statistical operator, for small $v \ra 0$, varies as
$$
\hat\rho_v \simeq \hat\rho + \bt \left ( \hat\rho \langle \hat\bP \rangle -
\; \frac{1}{2} \left [ \hat\rho, \; \hat\bP \right ]_+ \right )
\cdot \bv \; .
$$

The average momentum of the moving system is
$$
\langle \hat\bP_v \rangle_v \; = \; \langle \hat\bP \rangle_v \; + \; m\bv N \; .
$$
For small $v$, we have
$$
\langle \hat\bP \rangle_v \; \simeq \; \langle \hat\bP \rangle + \bt \left [
\langle \hat\bP \rangle \langle \hat\bP \cdot \bv \rangle -
\langle \hat\bP \left ( \hat\bP \cdot \bv \right ) \rangle \right ]
$$
and
$$
\frac{\prt}{\prt\bv}\; \cdot \; \langle \hat{\bf P} \rangle_v \; \simeq \;
-\bt {\rm var}(\hat{\bf P}) \; ,
$$
where
$$
{\rm var}(\hat\bP) \equiv\; \langle \hat\bP^2 \rangle - \langle \hat\bP \rangle^2 \; .
$$

In this way, the average numbers of superfluid and normal atoms, respectively, are
$$
N_s = N -\; \frac{\bt}{3m}\; {\rm var}(\hat{\bf P}) \; , \qquad
N_n \equiv \frac{\bt}{3m}\; {\rm var}(\hat{\bf P}) \; ,
$$
so that the superfluid fraction is
\be
\label{10}
n_s = 1 -\; \frac{\bt}{3mN}\; {\rm var}(\hat{\bf P}) \; .
\ee

In equilibrium
$$
\langle \hat{\bf P} \rangle \; =\; 0 \; , \qquad {\rm var}(\hat{\bf P}) = \;
\langle \hat{\bf P}^2 \rangle \; .
$$
Hence, we need to find the quantity
$$
\langle \hat{\bf P}^2 \rangle \; = \; - \int \langle \psi^\dgr(\br_1)\nabla_1\psi(\br_1)
\psi^\dgr(\br_2)\nabla_2\psi(\br_2) \rangle \; d\br_1 d\br_2 \; ,
$$
which can also be written as
$$
\langle \hat{\bf P}^2 \rangle\; = \; \int \lim_{\br_3\ra\br_1} \;
\lim_{\br_4\ra\br_2} \; \nabla_3\cdot\nabla_4 \left [
\rho(\br_4,\br_3)\dlt(\br_1-\br_2) +
\rho_2(\br_1,\br_4,\br_3,\br_2)\right ] \; d\br_1 d\br_2 \; ,
$$
where
$$
\rho(\br,\br')\equiv \langle \psi^\dgr(\br')\psi(\br) \rangle
$$
and $\rho_2$ is the second-order density matrix (see \cite{Yukalov_1}).

In the case of a uniform system, the field operators can be expanded over
the plane waves,
$$
\psi(\br) = \sum_k a_k\vp_k(\br) \; , \qquad
\vp_k(\br) = \frac{1}{\sqrt{V}}\; e^{i\bk\cdot\br} \; ,
$$
which yields
$$
\langle \hat{\bf P}^2 \rangle \; = \; \sum_{kp} (\bk\cdot\bp)
\langle \hat n_k \hat n_p \rangle
\qquad (\hat n_k \equiv a_k^\dgr a_k) \; .
$$

Introducing the notation for the dissipated heat
\be
\label{11}
Q \equiv \frac{{\rm var}(\hat\bP)}{2mN} \; ,
\ee
that for an equilibrium system reads as
$$
Q = \frac{\langle \hat\bP^2 \rangle}{2mN} \; ,
$$
we get the superfluid fraction
\be
\label{12}
n_s = 1 \; - \; \frac{2Q}{3T} \; .
\ee

Whether the considered system exhibits superfluidity or not depends on
particular system properties yielding the nonzero dissipated heat. The
existence of the latter essentially depends on the character of pair
atomic correlations. Note that the Bose-Einstein statistics has not been
used in deriving the above formulas. Hence, in principle, the effect of
superfluidity is not prohibited for Fermi systems.

\subsection{Moment of inertia}

The existence of a nontrivial response to a velocity boost should be
connected with a nontrivial response to a rotation boost that also
produces a velocity boost with the angular velocity vector $\vec\om$.

The occurrence of quantum effects, such as superfluidity, happens when
a quantum system behaves differently from a classical one. In
{\it classical mechanics}, the angular momentum is
$$
{\bf L} = m \int (\br\times\bv ) \rho(\br)\; d\br \; ,
$$
where the linear velocity $\bv=\vec\om\times\br$ and $\rho(\br)$ is the
density distribution. The classical angular momentum takes the form
$$
{\bf L} = m \int [ r^2\vec\om - (\br\cdot\vec\om)\br ]
\rho(\br) \; d\br \; .
$$
Directing the axis $z$ along the rotation axis, so that
$$
{\bf e}_z = \frac{\vec\om}{\om} \qquad ( \om\equiv|\vec\om|) \; ,
$$
gives the $z$-component of the angular momentum
$$
L_z = m\om \int \left ( r^2 - z^2 \right ) \rho(\br) \; d\br \; ,
$$
which can also be written as
$$
L_z = m\om \int (\br \times {\bf e}_z )^2 \rho(\br) \; d\br \; .
$$

The classical moment of inertia with respect to the rotation axis is defined
by the derivative
$$
I_0 \equiv \lim_{\om\ra 0} \; \frac{\prt L_z}{\prt\om} \; .
$$
Under a slow rotation, when $\omega \ra 0$, the density $\rho(\br)$ does
not change. Then the classical moment of inertia is
$$
I_0 = m \int (\br \times {\bf e}_z )^2 \rho(\br) \; d\br \; .
$$

As examples, let us consider the case where $\rho(\br)=\rho=const$. Then,
for a sphere of radius $R$, the classical moment of inertia is
$I_0 = 2 m N R^2/ 5$. For a cylinder of radius $R$, we have
$I_0 = m N R^2/2$.

In {\it quantum theory}, the local angular momentum operator is
$\hat{\bf L}(\br)\equiv -i\br\times\vec\nabla$. The total angular momentum
operator reads as
\be
\label{13}
\hat{\bf L} \equiv \int \psi^\dgr(\br) \hat{\bf L}(\br)
\psi(\br) \; d\br \; .
\ee

The axial component of the angular momentum is
\be
\label{14}
\langle \hat L_z \rangle_\om \; = \; \int \langle \psi^\dgr(\br)
\hat L_z(\br) \psi(\br) \rangle_\om \; d\br \; ,
\ee
where
\be
\label{15}
\hat L_z(\br) = - i{\bf e}_z\cdot(\br\times\nabla) = - i\; \frac{\prt}{\prt\vp} \; ,
\ee
with the angle $\varphi$ in the cylindrical system of coordinates. The
statistical averages $\langle \ldots \rangle_\om$ in the rotating frame
are taken with the Hamiltonian
\be
\label{16}
H_\om\equiv H-\vec\om\cdot\hat{\bf L} \; .
\ee
Then the average of an operator $\hat A$ is
$$
\langle \hat A \rangle_\om={\rm Tr}\hat\rho_\om\hat A \; ,\qquad
\hat\rho_\om \equiv \frac{e^{-\bt H_\om}}{{\rm Tr}e^{-\bt H_\om}} \; .
$$
Thus the moment of inertia with respect to the rotation axis is
\be
\label{17}
I \equiv \lim_{\om\ra 0} \;
\frac{\prt}{\prt\om} \langle \hat L_z \rangle_\om \; .
\ee

Nontrivial response to the rotation boost arises, when the quantum moment
of inertia differs from the classical one. This defines the superfluid
fraction
\be
\label{18}
n_s \equiv 1 \; - \; \frac{I}{I_0} \; .
\ee
Invoking the formulas of parametric variation \cite{Yukalov_1}, we have
$$
\frac{\prt H_\om}{\prt\om} = -\hat L_z \; , \qquad
\frac{\prt}{\prt\om} \langle \hat L_z \rangle_\om = \bt {\rm var}(\hat L_z) \; .
$$

In an equilibrium non-rotating system,
$$
\lim_{\om\ra 0} \langle \hat L_z \rangle_\om = 0 \; .
$$
Then the moment of inertia (\ref{17}) becomes
$$
I = \lim_{\om\ra 0} \; \frac{\prt}{\prt\om} \langle \hat L_z \rangle_\om =
\bt \langle \hat L_z^2 \rangle \; .
$$
Here in the right-hand side, the average $\langle \ldots \rangle$ is taken
with respect to the Hamiltonian $H$ of a system at rest, which leads to the
expression
$$
\langle \hat L_z^2 \rangle \; = \; \int \langle \psi^\dgr(\br_1) \hat L_z(\br_1)
\psi(\br_1) \psi^\dgr(\br_2) \hat L_z(\br_2) \psi(\br_2) \rangle \;
d\br_1 d\br_2 \; .
$$
In terms of density matrices, this can be written as
$$
\langle L_z^2 \rangle \; = \; - \int \lim_{\br_3\ra\br_1} \;
\lim_{\br_4\ra\br_2} \hat L_z(\br_3) \hat L_z(\br_4)
\left [ \rho(\br_4,\br_3) \dlt(\br_1-\br_2) +
\rho_2(\br_1,\br_4,\br_3,\br_2) \right ] \; d\br_1 \; d\br_2 \; .
$$
In this way, we come to the superfluid fraction
\be
\label{19}
n_s =  1 \; - \; \frac{\langle \hat L_z^2 \rangle}{I_0T} \; .
\ee

Note again that the Bose-Einstein statistics is not used in the derivation
of the above formulas. Hence superfluidity is not prohibited for Fermi
systems. The importance of pair correlations is also evident.

The response to the rotation boost is related to the response to the velocity
boost, so that the previous definition (\ref{12}) is equivalent to definition
(\ref{19}). To show this, let us take a cylindrical annulus of radius $R$,
width $\dlt\ll R$, height $L$, and volume $V\cong 2\pi RL\dlt$. Then the
classical moment of inertia is $I_0\cong mNR^2$. When $R\ra\infty$, then the
elementary length, in the direction ${\bf e}_l$ perpendicular to the radius,
is $dl=Rd\vp$, so that
$$
\hat L_z(\br) \simeq - i R \; \frac{\prt}{\prt l} \; ,
$$
and we get
$$
\langle \hat L_z^2 \rangle \; \simeq \; R^2 \int \lim_{\br_3\ra\br_1} \;
\lim_{\br_4\ra\br_2} \; \frac{\prt^2}{\prt l_3\prt l_4} \left [
\rho(\br_4,\br_3) \dlt(\br_1-\br_2) +
\rho_2(\br_1,\br_4,\br_3,\br_2) \right ] \; d\br_1 \; d\br_2 \; .
$$
Therefore
$$
\langle \hat L_z^2 \rangle \; \simeq \; R^2 \langle \hat P_l^2 \rangle \;  ,
$$
where $\hat P_l \equiv{\bf e}_l\cdot\hat\bP$. As a result, the superfluid
fraction (\ref{19}) transforms into
\be
\label{20}
n_s \simeq 1 \; - \; \frac{ \langle \hat P_l^2 \rangle}{mNT} \; .
\ee
The same formula follows from the previous definition (\ref{12}), if
the velocity boost $\bv$ is taken along the direction ${\bf e}_l$, when
$\bv=v{\bf e}_l$.

\subsection{Bose-Einstein condensation}

Bose systems exhibit the phase transition of Bose-Einstein condensation.
Before discussing the relation between Bose-Einstein condensation and
superfluidity, let us study the peculiarities of the Bose condensation
transition.

Historically, Bose-Einstein condensation was first considered for an
ideal uniform Bose gas (see, e.g., \cite{Mayer_2,Ter_3}). The
noninteracting gas is described by the Hamiltonian
\be
\label{21}
H = \int \psi^\dgr(\br)\left ( - \; \frac{\nabla^2}{2m} \; -\;
\mu \right ) \psi(\br)\; d\br \; .
\ee
Assuming spinless particles and expanding the field operator over the
plane waves,
$$
\psi(\br) = \sum_k a_k \vp_k(\br) \; , \qquad
\vp_k(\br) = \frac{1}{\sqrt{V}} \; e^{i\bk\cdot\br} \; ,
$$
we get
\be
\label{22}
H = \sum_k \om_k a_k^\dgr a_k \; , \qquad \om_k = \frac{k^2}{2m}\; - \; \mu \; .
\ee
The chemical potential $\mu$ is defined from the normalization to the
total number of particles
\be
\label{23}
N=\sum_k n_k \; ,
\ee
with the momentum distribution
$$
n_k =\; \langle a_k^\dgr a_k \rangle = \left ( e^{\bt\om_k} - 1 \right )^{-1} \; .
$$
The latter, by definition, is non-negative, $n_k\geq 0$, which requires that
$\om_k\geq 0$, hence $\mu \leq 0$. The dangerous point is where
$\min_k\om_k = 0$. If the minimum occurs at $k = 0$, then $\mu = 0$.

The behaviour of $\mu$ with varying temperature is described by the
derivatives
$$
\frac{\prt\mu}{\prt T} + \frac{\prt\mu}{\prt N}
\frac{\prt N}{\prt T} = 0 \; , \qquad
\frac{\prt\mu}{\prt T} = -\; \frac{\prt N/\prt T}{\prt N/\prt\mu} \; ,
$$
$$
\frac{\prt\mu}{\prt T} = -\;
\frac{\sum_k \bt\om_k n_k(1+n_k)}{\sum_k n_k(1+n_k)} \; , \qquad
\frac{\prt\mu}{\prt T} \leq 0 \; .
$$
This shows that $\mu$ increases with decreasing $T$, so that at a critical
temperature $T_c$ it touches zero, $\mu=0$. Since $\mu$ cannot be positive,
ir remains zero everywhere below the critical temperature, $T\leq T_c$.
To satisfy the normalization condition (\ref{23}), in the sum for $N$, it
is necessary to separate the term $N_0$ corresponding to the number of
condensed atoms. Then normalization (\ref{23}) changes to
\be
\label{24}
N=N_0+N_1 \; , \qquad N_1 \equiv \sum_{k>0} n_k \; .
\ee
Therefore the total density is the sum of the density of condensed atoms
$\rho_0$ and of the density of uncondensed atoms $\rho_1$,
\be
\label{25}
\rho =\rho_0 +\rho_1 \; , \qquad \rho_0 \equiv \frac{N_0}{V} \; ,
\qquad \rho_1 \equiv \frac{N_1}{V} \; .
\ee
Respectively, the fractions of condensed and uncondensed atoms satisfy
the normalization condition
\be
\label{26}
n_0 + n_1 = 1 \; , \qquad n_0 \equiv \frac{N_0}{N_1}\; , \qquad
n_1 \equiv \frac{N_1}{N} \; .
\ee

At the critical temperature $T_c$, one has $N_1=N$, $N_0=0$, $\rho_0=0$,
and $\rho_1=\rho$. For a macroscopically large system with volume $V$,
it is admissible to replace the summation over momenta by the related
integration, which in a three-dimensional space reads as
$$
\sum_k \longrightarrow \; V \int\frac{d\bk}{(2\pi)^3} \; .
$$
Then the density of uncondensed atoms, coinciding at the critical
temperature with the total atomic density, is
$$
\rho = \int n_k \; \frac{d\bk}{(2\pi)^3} \qquad (T=T_c) \; .
$$
This yields
\be
\label{27}
\rho = \frac{4\pi}{(2\pi)^3} \int_0^\infty
\frac{k^2dk}{\exp(\bt k^2/2m)-1} =
\frac{\pi}{4}\left ( \frac{2m T_c}{\pi^2}\right )^{3/2}\;
\Gm\left (\frac{3}{2}\right ) \zeta\left (\frac{3}{2}\right ) \; .
\ee
Here $\Gm(x)$ is the gamma function
$$
\Gm(x) \equiv \int_0^\infty  t^{x-1} e^{-t}\; dt \qquad ({\rm Re}\; x>0)
$$
and $\zeta(x)$ is the Riemann zeta function
$$
\zeta(x) = \frac{1}{\Gm(x)} \int_0^\infty\frac{t^{x-1}}{e^t-1}\; dt =
\sum_{n=1}^\infty \frac{1}{n^x} \qquad ({\rm Re}\; x>1) \; .
$$
In particular, $\Gm(3/2) = \sqrt{\pi}/2$ and $\zeta(3/2) = 2.61238$.
This gives the critical temperature
\be
\label{28}
T_c = \frac{2\pi}{m} \left [ \frac{\rho}{\zeta(3/2)} \right ]^{2/3} \; .
\ee
The density of condensed atoms plays the role of the order parameter. Since
this density, at the transition temperature, continuously deviates from zero,
Bose-Einstein condensation is a second-order phase transition \cite{Yukalov_1}.

Bose-Einstein condensation is characterized by the arising
{\it macroscopic number of particles} in a lowest single-particle state:
\be
\label{29}
N_0 \equiv \lim_{k\ra 0} n_k\; , \qquad N_0 \sim N \; .
\ee
From the expression for $n_k$, it follows that the appearance of a
macroscopic number of particles in a single-particle quantum state can
occur when the single-particle spectrum $\om_k \equiv \om(k)$ touches zero
at a point labelled by a momentum $k_0$:
\be
\label{30}
\om(k_0) \equiv \min_k \om(k) = 0 \; .
\ee
In general, $k$ is a set of quantum numbers labelling single-particle
quantum states $\vp_k$. When $\bk$ is a momentum and $k_0=0$, then
$\mu =k_0^2/2m = 0$.

Let us consider whether Bose-Einstein condensation can happen in other
dimensions. In one-dimensional real space, $d=1$, the existence of a
critical temperature $T_c$ would require that at this temperature
$$
\rho = 2 \int_0^\infty n_k\; \frac{dk}{2\pi} =
\sqrt{\frac{mT_c}{2\pi}}\; \zeta\left ( \frac{1}{2}\right ) \; ,
$$
where the value $\Gm(1/2) = \sqrt{\pi}$ is taken into account.

The representations for the Riemann $\zeta$ function, used above, assumed
that its argument satisfies the condition ${\rm Re}\; x>1$. To define the
function $\zeta(x)$ for $x<1$, it is possible to resort to an analytical
continuation of $\zeta(x)$ to the entire complex plane, except $x=1$. This
analytical continuation is
$$
\zeta(x) = \frac{1}{1-2^{1-x}}\; \sum_{n=0}^\infty \;
\frac{1}{2^{n+1}} \; \sum_{m=0}^n \; \frac{(-1)^mC_n^m}{(1+m)^x} \; ,
$$
where the binomial coefficients are the combinations of $m$ from $n$,
$$
C_n^m \equiv \left ( ^n_m\right ) \equiv \frac{n!}{(n-m)!\;m!} \; .
$$
From this analytical continuation, it follows that $\zeta(x)<0$ for $x<1$.
For instance, $\zeta(1/2)=-1.46035$. Since $\rho>0$, but $\zeta(1/2)<0$,
the temperature $T_c$ does not exist. In one-dimensional space, there is
no Bose condensation at any temperature.

Consider now the general case of a $d$-dimensional space. For macroscopically
large systems, the summation over momenta is replaced as
$$
\sum_k \; \ra \; V \int \; \frac{d\bk}{(2\pi)^d} \; .
$$
In spherical coordinates,
$$
d\bk \longrightarrow \; \frac{2\pi^{d/2}}{\Gm(d/2)}\; k^{d-1}\; dk \; ,
$$
with $k\equiv |\bk|$.

The system density is given by the integral
\be
\label{31}
\rho = \frac{2}{(4\pi)^{d/2}\Gm(d/2)} \; \int_0^\infty n_k k^{d-1} \; dk \; .
\ee
At $T=T_c$, when $\mu=0$, we get
\be
\label{32}
\rho =\left ( \frac{mT_c}{2\pi}\right )^{d/2} \; \zeta\left (\frac{d}{2} \right ) \; ,
\ee
which results in the critical temperature
\be
\label{33}
T_c = \frac{2\pi}{m}\left [ \frac{\rho}{\zeta(d/2)} \right ]^{2/d} \; .
\ee

We know that for $d = 1$ there is no condensation at any temperature. For $d=2$,
we have $\Gm(1)=1$ and the $\zeta$ function behaves as
$$
\zeta(x) \simeq \gm + \frac{1}{x-1} \qquad (x\ra 1) \; ,
$$
where the Euler-Mascheroni constant $\gm=0.577216$. Therefore
$\zeta(x)\ra\infty$ when $x\ra 1+0$. This means that $T_c=0$ for $d=2$.
Thus in a two-dimensional space, Bose-Einstein condensation can happen only
at $T=0$.

Finite condensation temperatures, for a uniform ideal Bose gas, exist only 
for spatial dimensions $d > 2$.

\subsection{Condensation conditions}

Bose-Einstein condensation can occur not only in an ideal Bose gas, but
also in interacting systems. What would be general conditions for the
possibility of this phenomenon?

A reasonable expected condition of Bose-Einstein condensation is a low
temperature, when the temperature wavelength
$$
\lbd_T  \equiv \sqrt{ \frac{2\pi}{mT} }
$$
becomes larger than the mean interatomic distance $a=\rho^{-1/3}$, where
$\rho\equiv N/V$ is the average system density. This condition, meaning
that quantum correlations between atoms become noticeable, can be written as
\be
\label{34}
\frac{a}{\lbd_T} < 1 \; , \qquad \rho\lbd_T^3 >1 \; .
\ee
For example, in the case of a three-dimensional ideal Bose gas, considered
in the previous section, $\rho\lbd_T^3(T_c) = \zeta(3/2) = 2.612$.

Since Bose-Einstein condensation implies a macroscopic occupation of a
lowest single-particle energy level, while atomic interactions tend to
disperse the particles over higher levels, one should expect that other
conditions, favourable for Bose condensation, are connected with the
weak influence of atomic interactions. Such a condition is the
low-density condition, when the mean interatomic distance $a$ is larger
than the effective radius of atomic interactions $r_0$, that is
\be
\label{35}
\frac{r_0}{a} \ll 1 \; , \qquad \rho r_0^3 \ll 1 \; .
\ee
This condition should facilitate the appearance of condensate, but it
is not compulsory.

The other condition that should facilitate the occurrence of Bose
condensation is the condition of weak interactions, when the atomic
scattering length $a_s$ is shorter than the mean distance $a$,
\be
\label{36}
\frac{|a_s|}{a} \ll 1 \; , \qquad \rho |a_s|^3 \ll 1 \; .
\ee

Let us emphasize that conditions (\ref{35}) and (\ref{36}) make the
condensation easier. However, if they do not hold, it does not necessarily
mean that the condensation is impossible. Just it may happen at a lower
temperature and the condensate fraction will be smaller.

The mathematical formalization of condensation conditions can be done in
several ways. The first evident condition is the {\it Bose-Einstein
condition} requiring that the number of condensed particles
$N_0 = \langle a_0^\dgr a_0 \rangle$ be macroscopic. This means that in 
thermodynamic limit it should be:
\be
\label{37}
\lim_{N\ra\infty} \; \frac{N_0}{N} > 0 \; .
\ee
Note that for finite systems $N_0\sim N$ and the phase transition becomes
a crossover \cite{Yukalov_1}.

A general condition of Bose condensation was formulated by Penrose and
Onsager \cite{Penrose_4}. They clarified what has to be understood as
single-particle states. The latter are defined as the eigenstates of the
eigenproblem
\be
\label{38}
\int \rho(\br,\br')\vp_k(\br')\; d\br' = n_k\vp_k(\br)
\ee
for the single-particle density matrix
$\rho(\br,\br')\equiv \langle\psi^\dgr(\br')\psi(\br)\rangle$. Generally,
$k$ is a set of quantum numbers labelling the single-particle states.
The field operators can be expanded over the basis of these states,
$$
\psi(\br) = \sum_k a_k\vp_k(\br) \; .
$$
The density matrix is representable as the expansion
\be
\label{39}
\rho(\br,\br') = \sum_k n_k \vp_k(\br)\vp_k^*(\br') \; ,
\ee
where $n_k=\langle a_k^\dgr a_k \rangle$. The eigenstates $\vp_k(\br)$ of the
single-particle density matrix, allowing for the diagonal expansion (\ref{39}),
are called {\it natural orbitals} \cite{Coleman_5}.

The {\it Penrose-Onsager condition} tells us that Bose condensation occurs, when
the largest eigenvalue of the density matrix is macroscopic, that is when
\be
\label{40}
N_0\equiv\sup_k n_k \sim N \; .
\ee
This largest eigenvalue is identified with the number of condensed atoms.
Then expansion (\ref{39}) can be written in the form
\be
\label{41}
\rho(\br,\br') = N_0 \vp_0(\br)\vp_0^*(\br') +
\sum_{k\neq 0} n_k\vp_k(\br)\vp_k^*(\br') \; .
\ee

The {\it Yang criterion of condensation} \cite{Yang_6}
\be
\label{42}
\lim_{|\br-\br'|\ra\infty}\; \rho(\br,\br') = \rho_0 > 0
\ee
requires the existence of the {\it off-diagonal long-range order}, with
the limit $\rho_0$ being the condensate density. This is a good criterion 
for uniform systems, when
$$
\vp_0(\br) = \frac{1}{\sqrt{V}} \; , \qquad
\rho(\br,\br')\longrightarrow \frac{N_0}{V} = \rho_0 \; .
$$
But the Yang criterion is not suitable for finite and for nonuniform systems.
This is because for a confined system $\vp_k(\br)\ra 0$ as $\br\ra\infty$.
Hence the limit $\rho(\br,\br')\ra 0$ for $|\br-\br'|\ra\infty$.

Quantification of the order, arising under the Bose-Einstein condensation,
can be conveniently done with the help of the {\it order indices}
\cite{Coleman_5,Yukalov_7}. For this purpose, we can consider
$$
\hat\rho_n = \left[ \rho_n(\br_1,\ldots,\br_n,\br_1',\ldots,\br_n')\right ]
$$
as a matrix with the elements
$$
\rho_n(\br_1,\ldots,\br_n,\br_1',\ldots,\br_n')\; = \;
\langle \psi^\dgr(\br_n')\ldots\psi^\dgr(\br_1')\psi(\br_1)\ldots\psi(\br_n) \rangle \; .
$$

The order index of the matrix $\hat\rho_n$ is defined as
\be
\label{43}
\om(\hat\rho_n) \equiv \frac{\ln||\hat\rho_n||}{\ln|{\rm Tr}\hat\rho_n|} \; .
\ee
Here $||\cdot||$ means an operator norm. For a semipositive matrix,
$||\hat\rho_n||\leq |{\rm Tr}\hat\rho_n|$. Then $\om(\hat\rho_n)\leq 1$.
For a large number of atoms $N\gg 1$, one has
$\ln{\rm Tr}\hat\rho_n \simeq n\ln N$. Keeping in mind the Hermitian norm
gives $||\hat\rho_1||=N_0$ and $||\hat\rho_n||\sim N_0^n$, which is correct
if there exists $N_0\geq 1$.

In the case of a finite system,
\be
\label{44}
\om(\hat\rho_n) = \frac{\ln N_0}{\ln N}\; .
\ee
This tells us that $N_0 = N^{\om(\hat\rho_n)}$.

With the help of the order indices, it is straightforward to classify
different types of order. Thus, when $\om(\hat\rho_n)<0$, there is no
order. If $\om(\hat\rho_n)=0$, there can be short-range order. When
$0<\om(\hat\rho_n)<1$, there occurs mid-range order. And for
$\om(\hat\rho_n)=1$, long-range order is present. This classification
suits for any system.

Bogolubov \cite{Bogolubov_8,Bogolubov_9} connected Bose condensation
with global gauge symmetry breaking. He noticed that in the field-operator
expansion
$$
\psi(\br) =\sum_k a_k\vp_k(\br) \; ,
$$
one can separate out the condensate term getting
$$
\psi(\br) = a_0\vp_0(\br) +\psi_1(\br)\; , \qquad
\psi_1(\br) \equiv \sum_{k\neq 0} a_k\vp_k(\br) \; .
$$
In thermodynamic limit, the first operator term in $\psi(\br)$ can be
replaced by a function,
$$
a_0\vp_0(\br)\longrightarrow \eta(\br) \; .
$$
Equivalently, it is possible to employ the {\it Bogolubov shift}
\be
\label{45}
\psi(\br) =\eta(\br) +\psi_1(\br) \; ,
\ee
where $\eta(\br)$ plays the role of a condensate wave function normalized
to the number of condensed atoms
$$
N_0 = \int |\eta(\br)|^2\; d\br \; .
$$
The condition of quantum-number conservation requires that
\be
\label{46}
\langle \psi_1(\br) \rangle=0 \; .
\ee
The condensate wave function
\be
\label{47}
\eta(\br) = \langle \psi(\br) \rangle = \langle \eta(\br)+\psi_1(\br) \rangle
\ee
is zero in the normal thermodynamic phase and becomes nonzero after Bose
condensation, when $N_0>0$. That is, the condensate wave function can be
treated as an order parameter.

The Bogolubov shift (\ref{45}) leads to global gauge symmetry breaking.
However this does not contradict to the experimental setup, where the
average number of atoms is conserved \cite{Yukalov_10}.

\subsection{Uncondensed gas}

The properties of the ideal Bose gas above the condensation temperature
have been described in many publications
\cite{Kubo_11,Ter_12,Ziff_13,Huang_14,Yukalov_15,Bogolubov_16}.
For completeness, we give here a brief account of the results related to
the ideal Bose gas, which can be useful in the following.

An important notion in characterizing Bose gas is the
{\it Bose-Einstein integral function}
\be
\label{48}
g_n(z) \equiv \frac{1}{\Gm(n)} \; \int_0^\infty
\frac{zu^{n-1}}{e^u-z}\; du \; .
\ee
This function is finite for $n\geq 0$, while diverges for $n<0$, if $z<1$.
It is such that $g_n(z)\geq 0$, when $z\leq 1$. And
$\Gm(n) \simeq 1/n$, if $n\ra 0$.

In applications, $z$ denotes the {\it fugacity}
\be
\label{49}
z\equiv e^{\bt\mu} \; .
\ee
And the dimensionless variable $u \equiv \bt k^2/2m$ is used. Then, since
$\om_k =k^2/2m-\mu$, one has $e^{\bt\om_k}=e^u/z$.

Differentiating and integrating by parts yields the relation
$$
\frac{\prt}{\prt z}\; g_n(z) = \frac{1}{z}\; g_{n-1}(z) \; ,
$$
valid for all $n$.

If $|z|\leq 1$, then the Bose-Einstein function can be written as
$$
g_n(z) = \sum_{j=1}^\infty \; \frac{z^j}{j^n} \qquad (|z|\leq 1, \; n>0) \; .
$$
For ${\rm Re}\;n>1$, it is called the polylogarithmic function. And
$$
g_n(1)=\zeta(n) \quad (n>1) \; .
$$
In low orders, one has
$$
g_0(z) = \frac{z}{z-1} \; , \qquad g_1(z) = -\ln(1-z)  \qquad (|z|<1) \; .
$$

When approaching the condensation temperature from above, $\mu \ra 0$ and
$z\sim 1$. Then the quantity
$$
\al\equiv-\ln z=-\bt\mu
$$
tends to zero. The derivative with respect to fugacity can be transformed
into
$$
\frac{\prt}{\prt\al}\; g_n\left ( e^{-\al}\right ) = -
 g_{n-1}\left ( e^{-\al}\right ) \; .
$$

There exists the {\it Robinson representation}
\be
\label{50}
g_n(e^{-\al}) = \Gm(1-n) \al^{n-1} +
\sum_{j=0}^\infty \; \frac{(-1)^j}{j!}\; \zeta(n-j)\al^j
\ee
that is valid for all $n$ on the complex plane, except integers 
$n = 1,2,3,\ldots $, since the values of $\Gm(0)$, $\Gm(-1)$, $\Gm(-2)$, 
etc. diverge. This representation is especially convenient when $\alpha \ra 0$.

For integer $n$, one has
$$
g_n \left ( e^{-\al} \right ) \; = \; (-1)^{n-1} \; \al^{n-1}
\left ( \sum_{j=1}^{n-1} \; \frac{1}{j} \; - \; \ln \al \right ) \;
+ \; \sum_{j=0\; (\neq n-1)}^\infty \;
\frac{(-1)^j}{j!}\; \zeta(n-j) \al^j \qquad
(n=1,2,3,\ldots) \; ,
$$
with
$$
\lim_{n\ra 1} \; \sum_{j=1}^{n-1} \; \frac{1}{j} \; = \; 0 \; .
$$
From here $g_n(1)=\zeta(n)$ for $n>1$.

Using the Robinson representation and the values
$$
\Gm\left ( -\; \frac{3}{2}\right ) =\frac{4\sqrt{\pi}}{3}\; , \qquad
\Gm\left ( -\; \frac{1}{2}\right ) = -2\sqrt{\pi}\; , \qquad
\Gm\left (\frac{1}{2}\right ) = \sqrt{\pi} \; ,
$$
$$
\zeta\left ( \frac{1}{2}\right ) = -1.46035\; , \qquad
\zeta\left ( \frac{3}{2}\right ) = 2.61238 \; , \qquad
\zeta\left ( \frac{5}{2}\right ) = 1.34149 \; ,
$$
for $z \ra 1$, one gets 
$$
g_{5/2}(z) \simeq \Gm\left ( -\; \frac{3}{2}\right )\; \al^{3/2} +
\zeta\left (\frac{5}{2}\right ) \; ,  \qquad
g_{3/2}(z) \simeq \Gm\left ( -\; \frac{1}{2}\right )\; \al^{1/2} +
\zeta\left (\frac{3}{2}\right ) \; ,
$$
$$
g_{1/2}(z) \simeq \Gm\left (\frac{1}{2}\right )\; \al^{-1/2} +
\zeta\left (\frac{1}{2}\right ) \; .
$$

Above the condensation temperature, $T\geq T_c$, when $n_0=0$, one has
$$
N = \frac{2V}{\sqrt{\pi}\;\lbd_T^3} \;
\int_0^\infty \frac{z\sqrt{u}\; du}{e^u-z} = \frac{V}{\lbd_T^3}\;
g_{3/2}(z) \; ,
$$
which defines $\mu$. This can be rewritten in the compact form
\be
\label{51}
\rho\lbd_T^3=g_{3/2}(z) \; .
\ee

At high temperatures $T\gg T_c$, where $z\ll 1$, we get $g_n(z)\simeq z$
and $z\simeq \rho\lbd_T^3$. Then $\mu\simeq T\ln(\rho\lbd_T^3)$. In the
limit $T\ra\infty$, one has $\rho\lbd_T^3\ra 0$, hence $\mu\ra-\infty$.

Close to the condensation temperature, when $T\ra T_c+0$, with $z\ra 1$,
and $\al\ra+0$, the use of the Robinson representation for $g_{3/2}(z)$
gives
\be
\label{52}
\rho\lbd_T^3\simeq -2\sqrt{\pi}\;\al^{1/2} +\zeta\left (\frac{3}{2}\right ) \; ,
\ee
where $\lbd_T= \sqrt{2\pi/mT}$. Taking into account that
$$
\rho\lbd_T^3(T_c) = \zeta\left (\frac{3}{2} \right )\; ,
$$
it is also possible to write
\be
\label{53}
\rho\lbd_T^3 \simeq \zeta\left (\frac{3}{2} \right ) \left (\frac{T_c}{T}
\right )^{3/2} \; ,
\ee
Also, since
$$
\al^{1/2} \simeq \frac{\zeta(3/2)}{2\sqrt{\pi}}\left [ 1 -
\left (\frac{T_c}{T} \right )^{3/2} \right ]
$$
and $\al=-\bt\mu$, we find
\be
\label{54}
\mu\simeq -T \frac{\zeta^2(3/2)}{4\pi}\left [ 1 -
\left (\frac{T_c}{T} \right )^{3/2} \right ]^2 \; .
\ee
This shows that $\mu\ra-0$ as $T\ra T_c+0$.

The grand potential $\Om = - PV$ defines the pressure 
\be
\label{55}
P = -T \int \ln\left ( 1  - e^{-\bt\om_k}\right ) \;
\frac{d\bk}{(2\pi)^3} = \frac{T}{\lbd_T^3}\; g_{5/2}(z) \; .
\ee
For high temperatures, we find the virial expansion,
\be
\label{56}
P = \rho T \sum_{n=0}^\infty a_n\left ( \rho\lbd_T^3\right )^n
\ee
that is useful for $\rho\lbd_T^3\ll 1$. The low virial coefficients are
$$
a_0 = 1 \; , \qquad a_1 = -\; \frac{1}{4\sqrt{2}}\; , \qquad
a_2 = -\left (\frac{2}{9\sqrt{3}}\; - \; \frac{1}{8}\right ) \; ,
$$
$$
a_3 = -\left (\frac{3}{32} +\frac{5}{32\sqrt{2}}\; - \;
\frac{1}{2\sqrt{6}}\right ) \; .
$$
At very high temperatures $(T\gg T_c)$ one has $P\simeq\rho T$.

It is straightforward to derive other thermodynamic quantities above $T_c$,
such as energy $E$, entropy $S$, and specific heat $C_V$, for which
we find
$$
\frac{E}{N} = \frac{3T}{2\rho\lbd_T^3}\; g_{5/2}(z) \; , \qquad
\frac{S}{N} = \frac{5}{2\rho\lbd_T^3}\; g_{5/2}(z)- \ln z \; ,
$$
\be
\label{57}
PV= \frac{2}{3}\; E \; , \qquad
C_V = \frac{15}{4\rho\lbd_T^3}\; g_{5/2}(z) - \;
\frac{9\rho\lbd_T^3}{4g_{1/2}(z)} \; .
\ee

At $T=T_c$, when $z=1$, then $g_{1/2}(1)$ diverges and the specific heat
becomes
$$
C_V(T_c) = \frac{15\zeta(5/2)}{4\zeta(3/2)}\cong 1.925 \; .
$$

For all temperatures $T\geq T_c$, a good approximation for the specific heat,
following from the expansion of $g_n(e^{-\al})$ and using $\al$ above $T_c$, is
\be
\label{58}
C_V \simeq 1.496 + 0.341\left (\frac{T_c}{T}\right )^{3/2} +
0.089 \left (\frac{T_c}{T}\right )^3 \; .
\ee
This expression is exact at $T_c$ and is accurate within an error of $1\%$
for all $T\geq T_c$, with the limit $\lim_{T\ra\infty}\; C_V =3/2$.

The {\it isothermal compressibility} is
\be
\label{59}
\kappa_T = \frac{1}{\rho N}\left ( \frac{\prt N}{\prt\mu} \right )_{TV}
= \frac{1}{\rho^2}\left ( \frac{\prt^2 P}{\prt\mu^2}\right )_{TV}  =
\frac{\bt}{\rho^2\lbd_T^3}\; g_{1/2}(z) \; .
\ee
At high temperature $T\gg T_c$, when $z\ll 1$,
\be
\label{60}
\kappa_T \simeq \frac{1}{\rho T} \qquad (T \gg T_c) \; .
\ee
While in the vicinity of the condensation point, where $z\sim 1$ and
$g_{1/2}(e^{-\al})\simeq \sqrt{\pi}\;\al^{-1/2}$, the compressibility is
\be
\label{61}
\kappa_T \simeq \frac{0.921}{\rho T}\left [ 1 -
\left (\frac{T_c}{T}\right )^{3/2}\right ]^{-1} \; ,
\ee
which diverges at $T_c$.

We should not expect that the normal gas above the condensation temperature
could exhibit superfluidity. But it is instructive to demonstrate this
explicitly. To this end, we need to find $\langle\hat\bP^2 \rangle$.
Applying to the average
$$
\langle \hat n_k\hat n_p \rangle\; = \; \langle a_k^\dgr a_p^\dgr a_k a_p \rangle +
\dlt_{kp}\langle a_k^\dgr a_k \rangle
$$
the Wick theorem gives
$$
\langle \hat n_k\hat n_p \rangle\; = \dlt_{kp} n_k(1+n_k) + n_k n_p \; .
$$
In equilibrium,
$$
\langle \hat\bP \rangle =\sum_k \bk n_k=0 \; .
$$
And we obtain
$$
\langle \hat\bP^2 \rangle \; = \sum_k k^2 n_k (1+n_k) =
\frac{8\sqrt{\pi}}{\lbd_T^5}\; V \int_0^\infty \frac{z u^{3/2} e^u}{(e^u-z)^2}\; du \;.
$$
Then we use the integrals
$$
\int_0^\infty \frac{u^{3/2} e^u}{(e^u-z)^2}\; du = -
\lim_{\lbd\ra 1}\; \frac{d}{d\lbd}
\int_0^\infty \frac{\sqrt{u}\; du}{e^{\lbd u}-z} \; ,
$$
$$
\int_0^\infty \frac{\sqrt{u}\; du}{e^{\lbd u}-z} =
\frac{1}{\lbd^{3/2}}\; \int_0^\infty \frac{\sqrt{u}\; du}{e^u-z} \; ,
$$
$$
\int_0^\infty \frac{u^{3/2}e^u}{(e^u-z)^2}\; du =
\frac{3}{2}\; \int_0^\infty \frac{\sqrt{u}\; du}{e^u-z} \; .
$$
As a result, we come to the value
$$
\langle \hat\bP^2 \rangle \; = \frac{6\pi}{\lbd_T^5}\; g_{3/2}(z)\; V = 3mTN \; .
$$
Therefore
$$
n_s =  1  -\; \frac{\langle \hat\bP^2 \rangle}{3mTN} = 0 \; ,
$$
which means that there is no superfluidity above $T_c$.

\subsection{Condensed gas}

Below the condensation temperature, the chemical potential of the ideal
Bose gas $\mu\ra -0$ and the fugacity $z\ra 1$. The total number of atoms
is the sum $N=N_0+N_1$ of the number of condensed atoms
$$
N_0 =\left ( e^{-\bt\mu} -1 \right )^{-1} = \frac{z}{1-z}  \qquad (\mu\ra -0)
$$
and the number of uncondensed atoms
$$
N_1  =\frac{V}{\lbd_T^3}\; g_{3/2}(1) \; .
$$
The chemical potential, for large $N_0$, tends to zero as
$$
\mu\ra - T\ln\left (1 + \frac{1}{N_0}\right )\ra  -0 \; .
$$

The fractions of condensed and uncondensed atoms, respectively, are
\be
\label{62}
n_0 = 1 - \left (\frac{T}{T_c}\right )^{3/2} \; , \qquad
n_1 =\left (\frac{T}{T_c}\right )^{3/2} \; .
\ee

Taking into account that $g_n(1)=\zeta(n)$ for $n > 1$, we have
\be
\label{63}
\rho\lbd^3_T =\zeta\left ( \frac{3}{2}\right )
\left ( \frac{T_c}{T}\right )^{3/2} \; .
\ee
For the pressure, energy, entropy, and specific heat, we find
$$
P = \frac{\zeta(5/2)}{\zeta(3/2)} \; \rho T_c
\left ( \frac{T}{T_c} \right )^{5/2} \; , \qquad
\frac{E}{N} = \frac{3\zeta(5/2)}{2\zeta(3/2)} \; T_c
\left ( \frac{T}{T_c} \right )^{5/2} \; ,
$$
\be
\label{64}
\frac{S}{N} = \frac{5\zeta(5/2)}{2\zeta(3/2)}
\left ( \frac{T}{T_c} \right )^{3/2} \; , \qquad
C_V = \frac{15\zeta(5/2)}{4\zeta(3/2)}
\left ( \frac{T}{T_c} \right )^{3/2} \; .
\ee
At the critical temperature, the specific heat displays the jump in the value,
$$
C_V(T_c-0) - C_V(T_c+0)=1.925 \; ,
$$
and in the derivative,
$$
\left (\frac{\prt C_V}{\prt T} \right )_{T_c-0}  -
\left (\frac{\prt C_V}{\prt T} \right )_{T_c+0}  =
\frac{27\zeta^2(3/2)}{16\pi T_c} = \frac{3.666}{T_c} \; .
$$

To find the superfluid fraction, we need to calculate
$ \langle \hat\bP^2 \rangle\ = 3mT N_1$. As a result, we obtain
\be
\label{65}
n_s = 1 - n_1 = 1 -\left (\frac{T}{T_c}\right )^{3/2} \; .
\ee
It turns out that for an ideal Bose gas, $n_s=n_0$. That is, the
Bose-Einstein condensate of an ideal gas is totally superfluid.
However, the Landau criterion of superfluidity does not hold, since
$$
\min_k\; \frac{\om_k}{k} = 0 \; ,
$$
while this limit should be positive.

Moreover, the compressibility exhibits anomalous behaviour. For the latter,
we have
$$
\kappa_T = \frac{1}{\rho N} \; \frac{\prt N}{\prt \mu} = \frac{1}{\rho N}
\left ( \frac{\prt N_0}{\prt\mu} + \frac{\prt N_1}{\prt\mu}\right ) \; ,
$$
with the relations
$$
{\rm var}(\hat N_0) = T \; \frac{\prt N_0}{\prt\mu} \; , \qquad
{\rm var}(\hat N_1) = T \; \frac{\prt N_1}{\prt\mu} \; .
$$
Therefore the compressibility can be written as
$$
\kappa_T = \frac{1}{\rho TN} \left [ {\rm var}(\hat N_0) + {\rm var}(\hat N_1)
\right ] \; .
$$
For the derivatives of the numbers of particles, we have
$$
\frac{\prt N_0}{\prt\mu} =\bt N_0(1+N_0) \; , \qquad
\frac{\prt N_1}{\prt\mu} = \frac{\bt N}{\rho\lbd_T^3}\; g_{1/2}(z)
\quad (z\ra 1) \; .
$$

Here we confront the problem, since the value $g_{1/2}(1)$ diverges.
Strictly speaking, this divergence is caused by thermodynamic limit. In
order to avoid the divergence, we have to consider a finite system possessing
a minimal energy $k_{min}^2/2m$. For a system in a box, it is admissible
to take the minimal wave vector
\be
\label{66}
k_{min} = \frac{\pi}{L} \; , \qquad  (L^3=V) \; .
\ee
Then we can define \cite{Yukalov_17} the modified Bose-Einstein functions,
where the integration starts not from zero, but from a minimal value
\be
\label{67}
u_{min} =\frac{\bt k_{min}^2}{2m}  = \frac{\pi^2 \bt}{2m L^2} =
\frac{\pi \lbd_T^2}{4 L^2}  \; .
\ee
In particular, the modified function $g_{1/2}(1)$ becomes
$$
g_{1/2}(1) = \frac{1}{\sqrt{\pi}}\;
\int_{u_{min}}^\infty \; \frac{u^{-1/2}}{e^u-1}\; du \; .
$$
This yields
$$
g_{1/2}(1) = \frac{1}{\sqrt{\pi}}\; \int_{u_{min}}^\infty \;
\frac{du}{u^{3/2}} = \frac{2}{\sqrt{\pi u_{min}}} \; .
$$
In view of the minimal value (\ref{67}), one has
$$
g_{1/2}(1) = \frac{4L}{\pi \lbd_T} \; .
$$
Taking into account
$$
\frac{\prt N_1}{\prt\mu} = \frac{m^2 T}{\pi^3}\; V^{4/3} \; ,
$$
we get
$$
{\rm var}(\hat N_1) = \frac{m^2T^2}{\pi^3}\; V^{4/3} \; .
$$
In this way, the compressibility is
\be
\label{68}
\kappa_T = \frac{{\rm var}(\hat N_1)}{\rho T N} =
\frac{m^2 T}{\pi^3 \rho^2}\; V^{1/3} \; .
\ee

If one assumes the occurrence of Bose condensation, so that $N_0\propto N$,
but without gauge symmetry breaking, then ${\rm var}(\hat N_0)\propto N^2$.
These fluctuations are, of course, thermodynamically anomalous, and such
a system cannot be stable \cite{Yukalov_1}, since $\kappa_T \propto N$.
This is what one names the "grand canonical catastrophe". However, here
there is no any catastrophe, but there is just a simple mistake of
describing a Bose-condensed system without gauge symmetry breaking.
Breaking the latter makes the condensate fluctuations effectively zero.
Using the Bogolubov shift implies that ${\rm var}(\hat N_0)=0$.

Nevertheless, even breaking gauge symmetry, one has
${\rm var}(\hat N_1)\propto N^{4/3}$, which means thermodynamically
anomalous fluctuations, as far as the compressibility
$\kappa_T \propto {\rm var}(\hat N)/N$ diverges as $\kappa_T\propto N^{1/3}$.
This tells us that the condensed ideal Bose gas is a pathological object
that cannot exist in nature.

It is worth emphasizing that it is not the grand canonical ensemble that
is guilty for the divergence of ${\rm var}(\hat N_1)$, but  the same
happens in the canonical ensemble, where also
${\rm var}(\hat N_1)\propto N^{4/3}$. On the other hand, making the formal
analytic continuation $g_{1/2}(1)\ra\zeta(1/2)$, with $\zeta(1/2)=-1.46035$,
leads to a negative $\kappa_T$, that is, again to instability.

In this way, the sole conclusion is that the ideal uniform Bose-condensed
gas is an unstable object. Fortunately, any real gas possesses atomic
interactions, although they can be weak, but not exactly zero. And, as will
be shown in the following sections, atomic interactions can stabilize the
Bose-condensed gas.

\subsection{Superfluidity and condensation}

As we have seen above, in the ideal Bose gas, the condensate and superfluid
fractions are identical. Generally, this does not need to be always so.
Moreover, the phenomena of condensation and superfluidity can exist
separately from each other. To illustrate this, let us consider a simple case,
where the particle distribution has the form $n_k=(e^{\bt\om_k}-1)^{-1}$,
with an effective spectrum
\be
\label{69}
\om_k = Ak^n - \mu \; .
\ee
A $d$-dimensional space is assumed. The spectrum can differ from $k^2/2m$
because of interactions or external fields. The superfluid fraction in a
$d$-dimensional space is given by the formula
\be
\label{70}
n_s = 1 - \; \frac{\langle \hat\bP^2 \rangle}{mTNd} \; .
\ee

For any $\om_k$, we have
$$
\frac{\prt n_k}{\prt\om_k} = -\bt n_k(1+n_k) \; ,
\qquad \langle \hat\bP^2 \rangle \; = \; - TV \int k^2 \;
\frac{\prt n_k}{\prt\om_k} \; \frac{d\bk}{(2\pi)^d} \; .
$$

When the system is uniform and isotropic, the replacement
$$
\frac{d\bk}{(2\pi)^d} \; \ra \;
\frac{2k^{d-1}dk}{(4\pi)^{d/2}\Gm(d/2)}
$$
is valid, which gives
$$
\langle \hat\bP^2 \rangle \; = \; TV \int \frac{n_k}{k^{d-1}} \;
\frac{\prt}{\prt k} \left ( \frac{k^{d+1}}{\om_k'}
\right ) \; \frac{d\bk}{(2\pi)^d} \; ,
$$
where $\om_k' \equiv \prt\om_k/\prt k$.

Under the effective spectrum (\ref{69}),
$$
\frac{\langle \hat\bP^2 \rangle}{V} = \frac{d+2-n}{n\bt A} \;
\int n_k k^{2-n} \; \frac{d\bk}{(2\pi)^d} \; .
$$
For any $d$, if $n\ra 2$ and $A\ra 1/2m$, then
$\langle \hat\bP^2 \rangle \ra mTNd$, and $n_s\ra 0$.

But generally, for the given effective spectrum, we have
$$
\frac{\langle \hat\bP^2 \rangle}{V} =
\frac{2(d+2-n)\Gm((d+2-n)/n)}{(4\pi)^{d/2}\Gm(d/2)n^2(\bt A)^{(d+2)/n}}
\; g_{(d+2-n)/n}(z) \; ,
$$
with the fugacity $z$ and, respectively, the chemical potential
$\mu=\mu(\rho,T)$ defined by the equation
$$
\rho =
\frac{2\Gm(d/n)g_{d/n}(z)}{(4\pi)^{d/2}\Gm(d/2)n (\bt A)^{d/n}} \; .
$$

Bose-Einstein condensation can happen at the temperature
\be
\label{71}
T_c = A \left [
\frac{(4\pi)^{d/2}\Gm(d/2)n\rho}{2\Gm(d/n)\zeta(d/n)}
\right ]^{n/d} \; .
\ee
Since $\Gm(x) > 0$ for $x>0$, the condensation temperature $T_c$ does
not exist for $d<n$. It is zero, $T_c=0$, for $d=n$, and $T_c>0$
when $d>n$. The condensate fraction is
\be
\label{72}
n_0 = 1 - \left ( \frac{T}{T_c}\right )^{d/n} \qquad (T\leq T_c) \; .
\ee
In the presence of the condensate, the superfluid fraction reads as
\be
\label{73}
n_s = 1 \; - B \zeta(\frac{d+2-n}{n}) T^{(d+2-n)/n} \; ,
\ee
where
$$
B \equiv \frac{2(d+2-n)\Gm((d+2-n)/n)}
{(4\pi)^{d/2}\Gm(d/2)n^2\rho m A^{(d+2)/n}d} \; .
$$
As is seen, in general, $n_s$ does not coincide with $n_0$. The
superfluid fraction $n_s$ coincides with the condensate fraction $n_0$ only
for $n=2$ and $A=1/2m$.

But, in principle, superfluidity can exist without condensation. Then the
superfluid fraction is given by the expression
\be
\label{74}
n_s = 1 \; - B g_{(d+2-n)/n}(z) T^{(d+2-n)/n} \; .
\ee

Thus, Bose-Einstein condensation can exist without superfluidity. And
superfluidity can occur without Bose-Einstein condensation. The possibility
for the occurrence of superfluidity without Bose condensation and vice
versa can be understood remembering that these phenomena are due to
different causes. Bose-Einstein condensation happens if and only if global
gauge symmetry is broken. While superfluidity does not require broken gauge
symmetry, while requires the existence of  strong pair correlations.
Strictly speaking, Bose-Einstein condensation is neither a necessary nor
sufficient condition for superfluidity.

Treating the condensate, $n_0$, and superfluid, $n_s$, fractions as order
parameters, generally, it is admissible to have four possibilities
corresponding to four different phases: \\

(i) uncondensed non-superfluid phase,
$$
n_0=0, \quad n_s=0 \; ,
$$

(ii) Bose-condensed non-superfluid phase,
$$
n_0>0, \quad n_s=0 \; ,
$$

(iii) uncondensed superfluid,
$$
n_0=0, \quad n_s>0 \; ,
$$

(iv) Bose-condensed superfluid,
$$
n_0>0, \quad n_s>0 \; .
$$

\subsection{Bogolubov shift}

It has been mentioned above that Bose-Einstein condensation implies global
gauge symmetry breaking that can be realized by means of the Bogolubov
shift for the field operator. In the present section, the physical meaning
of this shift is explained in more details.

For the beginning, let us keep in mind a uniform system, as has been done
by Bogolubov \cite{Bogolubov_8,Bogolubov_9,Bogolubov_18,Bogolubov_19}.
For a uniform system, the field operator can be expanded over the plane
waves,
$$
\psi(\br) = \sum_k a_k \vp_k(\br) \; , \qquad
\vp_k(\br) = \frac{1}{\sqrt{V}}\; e^{i\bk\cdot\br} \; , \qquad
\vp_0(\br) = \frac{1}{\sqrt{V}} \; .
$$
The operators in the momentum representation satisfy the commutation
relations
$$
[a_k,a_p]=0 \; , \quad [a_k,a_p^\dgr]=\dlt_{kp} \; .
$$

In the field-operator expansion, it is possible to separate out the
zero-momentum term, getting
\be
\label{75}
\psi(\br) = \psi_0 + \psi_1(\br) \; ,
\ee
where
$$
\psi_0 \equiv \frac{a_0}{\sqrt{V}} \; , \qquad \psi_1(\br) \equiv
\sum_{k\neq 0} a_k\vp_k(\br) \; .
$$
The zero-momentum term is assumed to correspond to Bose-condensed atoms,
while the second term, to uncondensed atoms.

The number-of-particle operators of condensed and uncondensed atoms,
respectively, are
\be
\label{76}
\hat N_0 =\psi_0^\dgr \psi_0 V = a_0^\dgr a_0 \; , \qquad
\hat N_1 = \int \psi_1^\dgr(\br)\psi_1(\br)\; d\br = \sum_{k\neq 0}
a_k^\dgr a_k \; .
\ee

The field operators of condensed and uncondensed atoms are orthogonal
to each other,
$$
\int \psi_0^\dgr \psi_1(\br)\; d\br = 0 \; ,
$$
since
$$
\sum_{k\neq 0} a_k \; \int \vp_0^*(\br)\vp_k(\br)\; d\br =
\sum_{k\neq 0} a_k\dlt_{k0} = 0 \; .
$$
Therefore the total number-of-particle operator is the sum
$$
\hat N= \hat N_0 + \hat N_1 \; .
$$

For the operators of condensed atoms, we have the commutation relations
$$
[a_0,a_0^\dgr]=1 \; , \quad [\psi_0,\psi_0^\dgr]=\frac{1}{V} \; .
$$
Bose-Einstein condensation implies a microscopic occupation of the lowest
single-particle state, $N_0 = \langle a_0^\dgr a_0 \rangle \sim N$. Therefore
$$
\frac{[a_0,a_0^\dgr]}{\langle a_0^\dgr a_0 \rangle} = \frac{1}{N_0} \ll 1 \; .
$$
This shows that the commutator $[a_0,a_0^\dgr] = 1$ is negligible as
compared to the number of atoms $N$. On the contrary
$n_k = \langle a_k^\dgr a_k \rangle \sim 1$, or even much smaller, $n_k \ll 1$. 
Because of this the expression
$$
\frac{[a_k,a_k^\dgr]}{\langle a_k^\dgr a_k \rangle} = \frac{1}{n_k}
$$
is of order $1$ or even much larger than $1$.

The commutator of uncondensed atoms
$$
[\psi_1(\br),\; \psi_1^\dgr(\br')]  =\sum_{k\neq 0}
\vp_k(\br)\vp_k^*(\br') \; ,
$$
in thermodynamic limit, is close to the commutator
$$
[\psi(\br),\; \psi^\dgr(\br')]  = [\psi_0,\; \psi_0^\dgr] +
[\psi_1(\br),\; \psi_1^\dgr(\br')] = \frac{1}{V}  +
\sum_{k\neq 0} \vp_k(\br)\vp_k^*(\br') = \dlt(\br-\br') \; .
$$

The above properties demonstrate that, in thermodynamic limit, the operator
$a_0$ can be treated as a number, together with the operator $\psi_0$. Thus,
in thermodynamic limit, the following replacements are admissible:
$\psi_0\ra\eta$ and $|\eta|^2=\rho_0$, with $a_0 \ra \sqrt{N_0}$. Summarizing,
for the field operator $\psi(\br)$, one can use the {\it Bogolubov shift}
\be
\label{77}
\psi(\br) \ra \eta(\br) + \psi_1(\br) \; .
\ee
The first term is called the {\it condensate wave function}. And the second
term corresponds to the field operator of uncondensed atoms. This shift
is valid for non-uniform systems as well. In the case of a uniform system,
$\eta(\br)$ becomes a constant $\eta$.

The Bogolubov shift breaks global gauge symmetry. Another method of gauge
symmetry breaking is by introducing infinitesimal sources \cite{Yukalov_1}.
The method of gauge symmetry breaking by means of infinitesimal sources is
equivalent to the Bogolubov shift. This is illustrated below by a simple
model with the Hamiltonian
$$
H = \sum_k \om_k a_k^\dgr a_k \; ,
$$
in which $\om_k\geq 0$ is arbitrary, and $k$ is a multi-index labelling
quantum microstates. It is easy to see that
$$
\langle a_k^\dgr a_k \rangle \; = \; \left ( e^{\bt\om_k} - 1\right )^{-1} \; ,
\qquad \langle a_0^\dgr a_0 \rangle \; = \; \left ( e^{\bt\om_0} - 1\right )^{-1} \; ,
$$
where $\om_0 \equiv {\rm inf}_k \; \om_k$. The criterion for the condensate
existence is
$$
\lim_{N\ra\infty} \; \frac{\langle a_0^\dgr a_0 \rangle}{N} \; > \; 0 \; .
$$
This means that $\om_0\ra 0$ as $N\ra\infty$. Therefore
$$
\langle a_0^\dgr a_0 \rangle \; \simeq \;
\frac{T}{\om_0} \qquad (\om_0\ra 0) \; .
$$

To break the gauge symmetry by an infinitesimal source, one adds to the
system Hamiltonian a small term, thus defining
$$
H_\ep \equiv H - \ep\; \sqrt{N_0} \left ( a_0^\dgr + a_0 \right ) \; .
$$
The latter Hamiltonian is not gauge-invariant with respect to $a_0$. It
can be diagonalized by the canonical transformation
$$
a_0 = b_0 + \frac{\ep}{\om_0}\; \sqrt{N_0} \; .
$$
In terms of the new field operators, the Hamiltonian is diagonal,
$$
H_\ep = \om_0 b_0^\dgr b_0 +
\sum_{k\neq 0} \om_k a_k^\dgr a_k \; - \;
\frac{\ep^2}{\om_0}\; N_0 \; .
$$
This new Hamiltonian is gauge-invariant with respect to $b_0$ and $a_k$,
where $k\neq 0$.

Denoting by $\langle \ldots \rangle_\ep$ the statistical averages with
respect to $H_\ep$, we find
$$
\langle b_0 \rangle_\ep \; = \; \langle b_0 b_0 \rangle_\ep\; = \; 0 \; , \qquad
\langle b_0^\dgr b_0 \rangle_\ep\; = \;
\left ( e^{\bt\om_0} - 1\right )^{-1} \; , \qquad
\langle b_0^\dgr b_0 \rangle_\ep\; = \; \langle a_0^\dgr a_0 \rangle \; .
$$
and
$$
\langle a_k \rangle_\ep \; = \; 0 \quad (k \neq 0)\; , \qquad
\langle a_0 \rangle_\ep \; = \; \frac{\ep}{\om_0}\; \sqrt{N_0} \; , \qquad
\langle a_0^\dgr a_0 \rangle_\ep \; = \; \langle b_0^\dgr b_0 \rangle_\ep +
\frac{\ep^2}{\om_0^2}\;N_0 \; .
$$
Using these equations gives the inequalities
$$
\langle b_0^\dgr b_0 \rangle_\ep\;\leq \;\; \langle a_0^\dgr a_0 \rangle_\ep \; ,
\qquad \langle a_0^\dgr a_0 \rangle \;\; \leq \;\; \langle a_0^\dgr a_0 \rangle_\ep \; .
$$
The latter yield the inequality
$$
\lim_{N\ra\infty} \; \frac{ \langle a_0^\dgr a_0 \rangle}{N} \; \leq \;
\lim_{\ep\ra 0} \; \lim_{N\ra\infty} \;
\frac{\langle a_0^\dgr a_0 \rangle_\ep}{N} \; .
$$
By definition, $\langle a_0^\dgr a_0 \rangle_\ep\; = \; N_0$. From the
above equations, we have
$$
N_0 \simeq \frac{T}{\om_0} + \frac{\ep^2}{\om_0^2}\; N_0 \; ,
$$
which results in the expression
$$
\om_0 \simeq \frac{T}{2N_0} \left [ 1 + \sqrt{1+(2\bt\ep N_0)^2}\right ] \; .
$$
Under $\ep\neq 0$ and large $N_0$, we get $\om_0\simeq\ep$. In this way,
we come to the equation $a_0\simeq b_0+\sqrt{N_0}$, which is a particular 
kind of the Bogolubov shift.

In addition, we have
$$
\langle a_0 \rangle_\ep \; \simeq\sqrt{N_0} \; , \qquad
\langle a_0^\dgr a_0 \rangle_\ep\; \simeq | \langle a_0 \rangle_\ep|^2 \; .
$$
Summarizing, we obtain the equation
$$
\lim_{\ep\ra 0}\; \lim_{N\ra\infty} \;
\frac{\langle a_0^\dgr a_0 \rangle_\ep}{N} =
\lim_{\ep\ra 0}\; \lim_{N\ra\infty} \; \frac{|\langle a_0 \rangle_\ep|^2}{N} \; .
$$
The left-hand side of this equation defines the fraction of condensed atoms,
while the right-hand side demonstrates gauge symmetry breaking. Therefore
gauge symmetry breaking is a sufficient condition for the existence of
Bose-Einstein condensation.

Also we have the inequality
$$
\lim_{N\ra\infty} \; \frac{\langle a_0^\dgr a_0 \rangle}{N} \; \leq
\lim_{\ep\ra 0}\; \lim_{N\ra\infty} \; \frac{|\langle a_0 \rangle_\ep|^2}{N} \; ,
$$
which tells us that Bose-Einstein condensation necessarily implies spontaneous
gauge symmetry breaking.

These findings can be formulated in the statement: {\it Spontaneous gauge
symmetry breaking is the necessary and sufficient condition for Bose-Einstein
condensation}. General theorems confirming this statement have been proven
by Bogolubov \cite{Bogolubov_18,Bogolubov_19}, Ginibre \cite{Ginibre_20},
Roepstorff \cite{Roepstorff_21}, and Lieb et al. \cite{Lieb_22}. A detailed
discussion can be found in review articles \cite{Yukalov_10,Yukalov_23,Yukalov_24}.

The Bogolubov shift of the field operator, in general, is applicable to
arbitrary Bose systems, whether uniform or nonuniform, equilibrium or not.
In the general case, the {\it Bogolubov shift} is
\be
\label{78}
\psi(\br,t) \ra \eta(\br,t)+\psi_1(\br,t) \; .
\ee
The orthogonality condition remains:
\be
\label{79}
\int\eta^*(\br,t)\psi_1(\br,t)\; d\br = 0 \; .
\ee
And the quantum-number conservation condition is also valid:
\be
\label{80}
\langle \psi_1(\br,t) \rangle =0 \; .
\ee
More mathematical details, related to the Bogolubov shift, will be considered
in the following sections.

\subsection{Bogolubov approximation}

The first microscopic approach to the description of weakly interacting
Bose-condensed gas was developed by Bogolubov
\cite{Bogolubov_8,Bogolubov_9,Bogolubov_18,Bogolubov_19}. This is what is
now named the Bogolubov approximation.

The main assumption of this approximation is that the system is almost
completely Bose-condensed, so that the number of condensed atoms is very
close to the total number of atoms in the system,
\be
\label{81}
\frac{N-N_0}{N} \ll 1 \; .
\ee
Physically this can happen when atomic interactions are very weak and
the system temperature is very low, which can be represented in the form
of the inequalities
\be
\label{82}
\rho|a_s|^3\ll 1 \; , \qquad T\ll T_c \; ,
\ee
where $a_s$ is scattering length and $T_c$, critical temperature.

The condition of weak interactions can also be written in another way,
noticing that the potential energy per particle is $\rho\Phi_0/2$, where
\be
\label{83}
\Phi_0 \equiv \int \Phi(\br)\; d\br \; ,
\ee
while the kinetic energy per particle is $\rho^{2/3}/2m$. The condition
of weak interactions can be understood as the smallness of the potential
energy as compared to the kinetic energy, which gives
$$
|m\Phi_0 \rho^{1/3}| \ll 1 \; .
$$

A system of spinless Bose atoms is characterized by the Hamiltonian
\be
\label{84}
H= \int \psi^\dgr(\br)\left ( -\; \frac{\nabla^2}{2m} + U - \mu
\right ) \psi(\br)\; d\br + \frac{1}{2} \int \psi^\dgr(\br)\psi^\dgr(\br')
\Phi(\br-\br')\psi(\br')\psi(\br)\; d\br d\br' \; .
\ee
Here $U = U(\br)$ is an external potential and the pair interaction
potential is symmetric, so that $\Phi(-\br)=\Phi(\br)$.

For a uniform system, $U(\br)=0$. The interaction potential is assumed
to possess the Fourier transform,
\be
\label{85}
\Phi(\br) = \frac{1}{V} \; \sum_k \Phi_k e^{i\bk\cdot\br} \; , \qquad
\Phi_k = \int \Phi(\br) e^{-i\bk\cdot\br} \; d\br \; .
\ee
The transform $\Phi_k$ is symmetric, such that $\Phi_{-k}=\Phi_k$,
since $\Phi(-\br)=\Phi(\br)$.

Expanding the field operator over the plane waves,
$$
\psi(\br) = \sum_k a_k\vp_k(\br) \; , \qquad
a_k =\int \vp_k^*(\br)\psi(\br)\; d\br \; ,
$$
one separates out the condensate term:
$$
\psi(\br) = \frac{a_0}{\sqrt{V}} + \psi_1(\br) \; .
$$

The form $a_0^\dgr a_0=\hat N_0$ is the number-operator of condensed
atoms. Generally, $a_0a_0=\hat N_0e^{i\al}$, with $\alpha$ being a
real number. It is always possible to choose $\al=0$ by means of
a gradient transformation.

Substituting the above field operator into the initial Hamiltonian
yields the Hamiltonian
\be
\label{86}
H =\sum_{n=0}^4 H^{(n)} \; ,
\ee
containing five terms classified according to the number of the field
operators $\psi_1(\br)$ in the products. The zero-order term is
$$
H^{(0)} = \frac{\hat N_0^2}{2V}\; \Phi_0 - \mu\hat N_0 \; .
$$
The first-order term $H^{(1)}=0$, since $\int\psi_1(\br)\; d\br=0$.
The second-order term leads to 
$$
H^{(2)} = \sum_{k\neq 0} \left [ \frac{k^2}{2m} + \frac{\hat N_0}{V}
\left (\Phi_0 +\Phi_k\right ) - \mu\right ]\; a_k^\dgr a_k +
\frac{\hat N_0}{2V}\; \sum_{k\neq 0} \Phi_k\left ( a_k^\dgr a_{-k}^\dgr
+ a_{-k} a_k\right ) \; .
$$
And the third-order term yields
$$
H^{(3)} = \frac{1}{V}\; {\sum_{k,p}} ' \; \Phi_k\left ( a_0^\dgr a_p^\dgr
a_{k+p}a_{-k} + a_{-k}^\dgr a_{k+p}^\dgr a_p a_0\right ) \; .
$$
Here the prime means $\bk\neq 0$, $\bp\neq 0$, and $\bk+\bp\neq 0$. The
fourth-order term gives
$$
H^{(4)} = \frac{1}{2V} \; \sum_k {\sum_{k,p}}' \; \Phi_k a_p^\dgr
a_q^\dgr a_{k+p} a_{q-k} \; ,
$$
with $\bp \neq 0$, ${\bf q}\neq 0$, $\bk+\bp\neq 0$, and $\bk-{\bf q}\neq 0$.
In the expressions above, the symmetry $\Phi_{-k}=\Phi_k$ is used.

The main steps of the Bogolubov approximation are as follows. First, in
view of the assumption that almost all atoms are Bose-condensed, the terms
$H^{(3)}$ and $H^{(4)}$ in the Hamiltonian are considered as small
perturbations and are omitted. This gives
$$
H = \frac{\hat N_0^2}{2V} \; \Phi_0 - \mu\hat N_0 +
\sum_{k\neq 0} \left [ \frac{k^2}{2m} + \frac{\hat N_0}{V} \;
(\Phi_0 + \Phi_k) - \mu\right ] \; a^\dgr_k a_k +
\frac{\hat N_0}{2V}\; \sum_{k\neq 0} \Phi_k\left ( a_k^\dgr a_{-k}^\dgr
+ a_{-k} a_k\right ) \; .
$$

At the second step, one replaces in the Hamiltonian the operator $\hat N_0$
by the number $N_0$, keeping in mind that $a_0$ becomes a number in
thermodynamic limit. Then
$$
H = \frac{N_0^2}{2V}\; \Phi_0 - \mu N_0 + \sum_{k\neq 0}
\om_k a_k^\dgr a_k + \frac{N_0}{2V} \; \sum_{k\neq 0} \Phi_k\left (
a_k^\dgr a_{-k}^\dgr + a_{-k}a_k\right ) \; ,
$$
where
$$
\om_k \equiv \frac{k^2}{2m} + \frac{N_0}{V} \; (\Phi_0 + \Phi_k) -\mu \; .
$$

At the third step, according to assumption (\ref{81}), one replaces
$N_0$ by $N$, getting the Hamiltonian
\be
\label{87}
H = \frac{N^2}{2V}\; \Phi_0 + \sum_{k\neq 0} \om_k a_k^\dgr a_k -\mu N +
\frac{1}{2}\; \sum_{k\neq 0} \Dlt_k \left ( a_k^\dgr a_{-k}^\dgr +
a_{-k}a_k\right )\; ,
\ee
with the notation
\be
\label{88}
\Dlt_k \equiv \rho\Phi_k \; , \qquad
\om_k = \frac{k^2}{2m} +\rho (\Phi_0 + \Phi_k) -\mu \; .
\ee

It is important to pay attention to the condition that the studied system 
be stable, since we remember that the ideal Bose-condensed gas is unstable.
To this end, we consider the compressibility
$$
\kappa_T = - \; \frac{1}{V} \left (
\frac{\prt P}{\prt V}\right )^{-1}_{TN} = \frac{1}{V} \left (
\frac{\prt^2  F}{\prt V^2}\right )^{-1}_{TN} \; .
$$
At low temperature $T \ra 0$, we have $F \ra E$. Then, under very weak
interactions, we find
$$
E \simeq \frac{N^2}{2V}\; \Phi_0 \; , \qquad
\kappa_T \simeq \frac{1}{\rho^2\Phi_0} \; .
$$
This tells us that the system is stable when
\be
\label{89}
\Phi_0 >0 \qquad (0 \leq \kappa_T < \infty) \; .
\ee
By the way, from here it is clear why the ideal Bose-condensed gas becomes
unstable, since for the latter $\Phi_0 \ra 0$, hence $\kappa_T \ra \infty$.
But interactions can stabilize the Bose-condensed gas, provided the
stability condition (\ref{89}) holds true.

Hamiltonian (\ref{87}) can be diagonalized by means of the Bogolubov
canonical transformation
$$
a_k = u_k b_k + v_{-k}^* b_{-k}^\dgr \; , \qquad
a_k^\dgr = u_k^* b_k^\dgr + v_{-k} b_{-k} \; ,
$$
\be
\label{90}
b_k = u_k^* a_k - v_{k}^* a_{-k}^\dgr \; , \qquad
b_k^\dgr = u_k a_k^\dgr - v_{k} a_{-k} \; ,
\ee
where the quantities $u_k$ and $v_k$ are called the Bogolubov
coefficient functions and the operators $b_k$ correspond to collective
excitations, called {\it bogolons}.

The new operators $b_k$ obey the Bose-Einstein statistics, with the
commutation relations $[b_k,b_p]=0$ and $[b_k,b_p^\dgr]=\dlt_{kp}$.
The commutator $[a_k,a_p]=0$ gives $u_kv_k^*-v_{-k}^*u_{-k}=0$. The
commutator $[a_k,a_p^\dgr]=\dlt_{kp}$ gives $|u_k|^2-|v_{-k}|^2=1$.
We also use the interaction-potential symmetry $\Phi_{-k}=\Phi_k=\Phi_k^*$,
which results in the properties $\om_{-k}=\om_k=\om_k^*$. By the
gradient transformations for the operators $a_k$ and $b_k$, it is
admissible to take the Bogolubov coefficient functions to be symmetric
and real, such that $u_k^*=u_{-k}=u_k$ and $v_k^*=v_{-k}=v_k$.

The diagonalization yields the equations
$$
2\om_k u_k v_k +\left ( u_k^2 + v_k^2\right )\Dlt_k = 0 \; ,
$$
$$
u_k^2 = \frac{\sqrt{\ep_k^2+\Dlt_k^2}+\ep_k}{2\ep_k} \; , \qquad
v_k^2 = \frac{\sqrt{\ep_k^2+\Dlt_k^2}-\ep_k}{2\ep_k} \; .
$$
Here $\ep_k$ is the {\it Bogolubov spectrum}
\be
\label{91}
\ep_k =\sqrt{\om_k^2 -\Dlt_k^2} \; .
\ee
The equations for the coefficient functions can be represented as
$$
u_k^2 -v_k^2 =1 \; , \qquad u_k v_k = -\; \frac{\Dlt_k}{2\ep_k} \; ,
\qquad u_k^2+v_k^2 = \frac{\sqrt{\ep_k^2+\Dlt_k^2}}{\ep_k} \; .
$$
Keeping in mind the spectrum (\ref{91}) gives
$$
\Dlt_k^2=\om_k^2 -\ep_k^2 \; , \qquad
u_k^2  = \frac{|\om_k|+\ep_k}{2\ep_k}\; , \qquad
v_k^2 = \frac{|\om_k|-\ep_k}{2\ep_k} \; .
$$
Finally, we obtain the diagonal Bogolubov Hamiltonian
\be
\label{92}
H_B = E_0 + \sum_{k\neq 0} \ep_k b_k^\dgr b_k - \mu N \; ,
\ee
in which the first term plays the role of the ground-state energy
$$
E_0 = \frac{1}{2}\; N\rho\Phi_0 - \sum_{k\neq 0} \ep_k v_k^2 \; .
$$
The latter can also be written as
\be
\label{93}
E_0 = \frac{1}{2}\; N\rho\Phi_0 - \; \frac{1}{2}\; \sum_{k\neq 0}
\left (|\om_k| -\ep_k\right ) \; .
\ee

The reason of separating the terms with $k=0$ is the existence of Bose
condensate. Otherwise, these terms would be absent. And the existence
of the condensate requires the validity of the {\it condition for
condensate existence}:
\be
\label{94}
\lim_{k\ra 0} \ep_k = 0 \; , \qquad \ep_k \geq 0 \; .
\ee

In the limit of zero momentum,
$$
\lim_{k\ra 0}\left (\om_k^2 - \Dlt_k^2\right ) = 0 \; , \qquad
\lim_{k\ra 0} \om_k = 2\rho\Phi_0 - \mu \; , \qquad
\lim_{k\ra 0}\Dlt_k = \rho\Phi_0 \; .
$$
Then from condition (\ref{94}) we find the chemical potential
\be
\label{95}
\mu=\rho\Phi_0 \; .
\ee
With
$$
\om_k = \frac{k^2}{2m} + \rho\Phi_k \; ,
$$
we get the Bogolubov spectrum
\be
\label{96}
\ep_k = \sqrt{\frac{\Dlt_k}{m}\; k^2 + \left (\frac{k^2}{2m}\right )^2} \; .
\ee
In the long-wave limit, the spectrum is of phonon type, $\ep \simeq ck$,
with the sound velocity
\be
\label{97}
c\equiv \lim_{k\ra 0} \; \sqrt{\frac{\Dlt_k}{m}} =\sqrt{\frac{\rho}{m} \;
\Phi_0} \; ,
\ee
where $\Phi_0 = \int \Phi(\br)\; d\br > 0$.

The terms under the square root of $\ep_k$ are approximately equal
at $k_0=2mc$. For long waves, when $k\ll k_0$, the spectrum is
$$
\ep_k \simeq ck \left [ \sqrt{\frac{\Phi_k}{\Phi_0}} + \frac{1}{2}
\left (\frac{k}{k_0}\right )^2 \; - \; \frac{1}{8}\left (
\frac{k}{k_0}\right )^4\right ] \; ,
$$
restoring the phonon form $\ep_k\simeq ck$ as $k\ra 0$.

For short waves, when $k\gg k_0$, one has
$$
\ep_k \simeq \frac{k^2}{2m} + \Dlt_k  -\; \frac{m\Dlt_k^2}{k^2} \; .
$$
And the spectrum is of the single-particle type
$$
\ep_k \simeq \frac{k^2}{2m} \qquad (k\ra\infty) \; .
$$

Important quantities, defining thermodynamic characteristics, are the
{\it normal average}
\be
\label{98}
n_k\equiv \langle a_k^\dgr a_k \rangle \; ,
\ee
which is the momentum distribution of atoms, and the
{\it anomalous average}
\be
\label{99}
\sgm_k\equiv \langle a_ka_{-k} \rangle \; ,
\ee
for which $|\sgm_k|$ describes the distribution of pair-correlated atoms.

Averaging with the Bogolubov Hamiltonian $H_B$, we have
$$
n_k =\left ( u_k^2 + v_k^2\right ) \pi_k + v_k^2 \; , \qquad
\sgm_k = u_k v_k (1 + 2\pi_k) \; ,
$$
with the bogolon distribution
$$
\pi_k\equiv\; \langle b_k^\dgr b_k \rangle \; = \left ( e^{\bt\ep_k}-1\right )^{-1} \; .
$$
Invoking the coefficient functions $u_k$ and $v_k$ yields the normal and
anomalous averages
\be
\label{100}
n_k = \frac{\om_k}{2\ep_k}\; (1 +2\pi_k) -\; \frac{1}{2}\; , \qquad
\sgm_k = -\; \frac{\Dlt_k}{2\ep_k}\; (1 +2\pi_k) \; .
\ee
Here we can use the equality
$$
1+2\pi_k = {\rm coth}\left(\frac{\ep_k}{2T}\right ) \; .
$$

Recall that, according to the main assumption of the Bogolubov approximation,
it is applicable when the system is almost completely Bose-condensed. This
implies very low temperatures, close to zero, and asymptotically weak
interactions, when the condensate depletion is very small. Under these
conditions, we can expect that the influence of uncondensed atoms is weak,
because of which in some estimates it is admissible to replace the condensate
density $\rho_0$ by the average density $\rho$.

\subsection{Dilute gas}

A system is dilute, when atomic interactions are short-range, such that
the effective interaction radius $r_0$ is much shorter than the mean
interatomic distance, which can be expressed as
\be
\label{101}
\rho r_0^3 \ll 1 \; .
\ee
Then it is possible to assume that the physical processes are mostly
influenced by small momenta, where kinetic energy is much less than
potential energy, $k^2/2m \ll \rho\Phi_0/2$, hence where
$k\ll \sqrt{m\rho\Phi_0}$. In other words, when $k\ll k_0$, where
$k_0\equiv 2mc=2\sqrt{m\rho\Phi_0}$.

For a dilute system, the exact form of the interaction potential
$\Phi(\br)$ is not of great importance, and it is admissible to use
the local effective potential.
\be
\label{102}
\Phi(\br) =\Phi_0\dlt(\br) \; , \qquad
\Phi_0 = 4\pi\; \frac{a_s}{m} \; .
\ee
With this potential, $\Phi_k=\Phi_0$ and, resorting to the Bogolubov
approximation of the previous section, we have
$$
\Dlt_k =\rho\Phi_0 \; , \qquad \mu=\rho\Phi_0 \; , \qquad
\om_k  =\frac{k^2}{2m} + \rho\Phi_0 \; .
$$
The Bogolubov sound velocity is
\be
\label{103}
c \equiv \sqrt{\frac{\rho}{m}\;\Phi_0} = \frac{\sqrt{4\pi\rho a_s}}{m} \; .
\ee
And the Bogolubov spectrum takes the form
\be
\label{104}
\ep_k=\sqrt{\om_k^2-m^2c^4} = \sqrt{(ck)^2+\left (\frac{k^2}{2m}\right )^2} \; .
\ee
The wave vector, where $ck$ equals $k^2/2m$, is $k_0=2mc$.

Thus, in the Bogolubov approximation, one has
$$
\Dlt_k=\mu=mc^2\; , \qquad \om_k = \frac{k^2}{2m} + mc^2 \; , \qquad
\om_k^2 = \ep_k^2 + m^2 c^4 \; .
$$

The normal and anomalous averages become
\be
\label{105}
n_k = \frac{\sqrt{\ep_k^2+m^2c^4}}{2\ep_k}\; {\rm coth}\left (
\frac{\ep_k}{2T}\right ) -\; \frac{1}{2}\; , \qquad
\sgm_k = -\; \frac{mc^2}{2\ep_k}\; {\rm coth}\left (
\frac{\ep_k}{2T}\right ) \; .
\ee

Note that under the asymptotically weak interactions, when $\Phi_0\ra 0$,
hence $c\ra 0$, the anomalous average tends to zero, $\sgm_k\ra 0$.
Because of this, in the Bogolubov approximation, that itself is valid
only for weak interactions, it is tempting to omit the anomalous averages.
This, however, is not quite correct, as is shown below.

Let us consider expressions (\ref{105}) in the region of small excitation
energies
$$
\ep_k\ll mc^2 \; , \qquad \ep_k\ll T \; .
$$
In this region, the bogolon distribution is
$$
\pi_k \simeq \frac{T}{\ep_k} \; - \; \frac{1}{2} + \frac{\ep_k}{12T} \;
- \; \frac{\ep_k^3}{720T^3} \; .
$$
And we find the expansions for the normal average
$$
n_k \simeq \frac{Tmc^2}{\ep_k^2} +\frac{mc^2}{12T} + \frac{T}{2mc^2}
 \; - \; \frac{1}{2} +\left ( \frac{mc^2}{3T}\; - \; \frac{T}{mc^2}\; -
\; \frac{m^3c^6}{90T^3}\right ) \frac{\ep_k^2}{8m^2c^4}
$$
and for the anomalous average
$$
\sgm_k \simeq -\; \frac{Tmc^2}{\ep_k^2} \; - \; \frac{mc^2}{12T} +
\frac{mc^2\ep_k^2}{720T^3} \; .
$$
From these expressions, it is evident that these averages are of the same
order of magnitude. In particular,
$$
\lim_{k\ra 0}\; \left ( |\sgm_k|- n_k\right ) = \frac{1}{2}\left ( 1 -\;
\frac{T}{mc^2}\right ) \; .
$$

Under large excitation energies, when
$$
\ep_k\gg mc^2 \; , \qquad \ep_k\gg T \; ,
$$
we get the normal average
$$
n_k \simeq \left ( \frac{mc^2}{2\ep_k}\right )^2 - \left (
\frac{mc^2}{2\ep_k}\right )^4 + e^{-\bt\ep_k}
$$
and the anomalous average
$$
\sgm_k \simeq -\; \frac{mc^2}{2\ep_k}\left ( 1 + 2e^{-\bt\ep_k}
\right ) \; .
$$
Here $|\sgm_k|$ is much larger than $n_k\ll|\sgm_k|$.

In this way, at low temperatures $T\ll T_c$, where the Bogolubov
approximation is assumed to be valid, the anomalous average is as
important as the normal one. And omitting $\sgm_k$ is not admissible.

The relation between the normal and anomalous averages can be derived
from the Bogolubov inequality (see \cite{Yukalov_1})
$$
|\langle \hat A\hat B\rangle|^2 \leq \left ( \langle \hat A\hat A^+ \rangle
\langle \hat B^+\hat B \rangle \right ) \; ,
$$
which gives
$$
|\langle a_k a_{-k} \rangle|^2 \leq \left ( \langle a_k a_k^\dgr \rangle
\langle a_{-k}^\dgr a_{-k} \rangle \right ) \; .
$$
Taking into account that $n_k = n_{-k}$ and $\sgm_k^*=\sgm_k$, we get
$$
\sgm_k^2\leq n_k(1+ n_k) \; .
$$
Instead of this inequality, by invoking expressions (\ref{105}), we can
get the equality
$$
n_k(1+n_k)-\sgm_k^2 = \left [ 2{\rm sinh}\left (
\frac{\ep_k}{2T}\right )\right ]^{-2} \; .
$$
Formally setting here $T=0$ gives $\sgm_k^2 = n_k(1+ n_k)$. This again
shows that the normal and anomalous averages are closely related and
the latter is not negligible as compared to the former.

The normal average defines the density of normal, uncondensed, atoms
\be
\label{106}
\rho_1 \equiv \frac{N_1}{V} = \int n_k \; \frac{d\bk}{(2\pi)^3}\; .
\ee
With the replacement $x = \ep_k/mc^2$, this yields
$$
\rho_1 = \frac{\sqrt{2}(mc)^3}{(2\pi)^2} \; \int_0^\infty \;
\left ( \sqrt{1+x^2} -1 \right )^{1/2} \left [ {\rm coth}\left (
\frac{mc^2}{2T}\; x\right ) - \; \frac{x}{\sqrt{1+x^2}}\right ]\; dx \; .
$$
Adding and subtracting $1$ in the square brackets and using the integral
$$
\int_0^\infty \left ( \sqrt{1+x^2}-1\right )^{1/2} \left ( 1 -\;
\frac{x}{\sqrt{1+x^2}}\right ) \; dx = \frac{2\sqrt{2}}{3}
$$
results in the expression
\be
\label{107}
\rho_1 =\frac{(mc)^3}{3\pi^2}\left \{ 1 + \frac{3\sqrt{2}}{4}\;
\int_0^\infty \left (\sqrt{1+x^2} - 1\right )^{1/2} \left [
{\rm coth}\left (\frac{mc^2}{2T}\; x\right ) - 1 \right ]\; dx
\right \} \; .
\ee

To analyze the behaviour of the normal density at low temperatures, when
$$
\frac{T}{mc^2} \ll 1 \; ,
$$
we accomplish the following steps. Employing the expansion
$$
\sqrt{2}\; \left (\sqrt{1+x^2} - 1\right )^{1/2} \simeq
x\; - \; \frac{1}{8}\; x^3 + \frac{7}{128}\; x^5 \; - \;
\frac{33}{1024}\; x^7 \; ,
$$
we calculate the integral
$$
\int_0^\infty x^{n-1} \left [ {\rm coth}(px) -1\right ]\; dx =
\frac{\Gm(n)\zeta(n)}{2^{n-1}p^n} \; ,
$$
in which ${\rm Re}\; n > 1$ and the zeta function of the even argument
$$
\zeta(2n) = \frac{(2\pi)^{2n}}{2(2n)!} \; |B_{2n}|
$$
is expressed through the Bernoulli numbers $B_n$. These are defined as
$$
B_{2n+1}=0 \qquad (n=1,2,\ldots) \; , \qquad
B_0=1\; , \qquad B_1=-\frac{1}{2} \; , \qquad B_2=\frac{1}{6} \; ,
$$
$$
B_4 = -\;\frac{1}{30}\; , \qquad B_6 =\frac{1}{42}\; , \qquad
B_8 = -\; \frac{1}{30} \; , \qquad B_{10} = \frac{5}{66} \; ,
$$
and so on. Then we have the integral
$$
\int_0^\infty \; x^{2n-1}\left [ {\rm coth}(px) -1\right ]\; dx =
\frac{\pi^{2n}|B_{2n}|}{2n\; p^{2n}} \; .
$$
Thus, we find the low-temperature behaviour of the normal density
$$
\rho_1 \simeq \frac{(mc)^3}{3\pi^2} \left [ 1 + \frac{\pi^2}{4}
\left (\frac{T}{mc^2}\right )^2 \; - \; \frac{\pi^4}{80}
\left (\frac{T}{mc^2}\right )^4 + \frac{\pi^6}{96}
\left (\frac{T}{mc^2}\right )^6 \; - \; \frac{33\pi^8}{1280}
\left (\frac{T}{mc^2}\right )^8\right ] \; .
$$

For the normal fraction $n_1=\rho_1/\rho$, we obtain
\be
\label{108}
n_1 \simeq \frac{8}{3\sqrt{\pi}} \; \sqrt{\rho a_s^3}\left [ 1 +
\frac{\pi^2}{4}\left ( \frac{mT}{4\pi\rho a_s}\right )^2 \right ] \; .
\ee
Therefore the fraction of condensed atoms $n_0=1-n_1$ becomes
\be
\label{109}
n_0 =  1 - \; \frac{8}{3\sqrt{\pi}} \; \sqrt{\rho a_s^3}\left [ 1 +
\frac{\pi^2}{4}\left ( \frac{T}{mc^2}\right )^2 \right ] \; .
\ee
The existence of atomic interactions and temperature depletes the condensate.

In the opposite limit of weak interactions, but finite temperature, when
$$
\frac{mc^2}{T_c} \ll 1 \; \qquad
T_c \equiv \frac{2\pi}{m} \left [ \frac{\rho}{\zeta(3/2)}\right ]^{2/3} \; ,
$$
the normal density (\ref{107}) reduces to
$$
\rho_1 \simeq \rho \left (\frac{T}{T_c}\right )^{3/2} +
\frac{(mc)^3}{3\pi^2} \; .
$$
This gives the condensate fraction
\be
\label{110}
n_0 \simeq 1  - \left (\frac{T}{T_c}\right )^{3/2} \; -\;
\frac{(mc)^3}{3\pi^2\rho} \; .
\ee

Recall that the limit $c\ra 0$ is not well defined, since the ideal gas
is unstable everywhere for $T\leq T_c$, as has been explained above.
This instability is easy to understand calculating the compressibility
that, in the Bogolubov approximation, is
$$
\kappa_T = \frac{1}{\rho m c^2} \; .
$$
As is evident, the compressibility diverges when $c\ra 0$, which means
instability.

Let us stress again the importance of the anomalous average. Sometimes,
appealing to the fact that the anomalous average tends to zero in the
limit of asymptotically weak interactions, one assumes that it can be
neglected, while retaining the normal average. In doing this, one mentions 
Popov who allegedly advocated such an omission. This, however, is wrong, 
since Popov, as can be easily inferred form his book \cite{Popov_25}, has 
never suggested such an incorrect trick. As is shown in the present section, 
the anomalous average can be important even under weak interactions. 
Moreover, the omission of the anomalous average results in the divergence 
of the compressibility, thus, implying instability of the system 
(see details in Ref. \cite{Yukalov_26}). However, those quantities that do 
not require the use of the anomalous averages can be estimated in the 
Bogolubov approximation, providing reasonable evaluations for asymptotically 
weak interactions and low temperatures. But, strictly speaking, the omission 
of the anomalous average is justified only under the simultaneous omission 
of the normal average defining the density $\rho_1$ of uncondensed atoms. 
That is, under the Bogolubov approximation, both the normal as well as the 
anomalous averages are assumed to be negligible as compared to the condensate 
fraction.

\subsection{Particle energy}

In the Bogolubov approximation, it is straightforward to define the energy
and the related characteristics of the dilute system. Thus the internal energy
$E =\; \langle H_B \rangle + \mu N$ yields
$$
E = E_0 + V \int \ep_k \pi_k \; \frac{d\bk}{(2\pi)^3} \; ,
$$
with the ground-state energy
\be
\label{111}
\frac{E_0}{N} = \frac{1}{2}\; \rho \Phi_0 + \frac{1}{2\rho} \;
\int (\ep_k -\om_k)\; \frac{d\bk}{(2\pi)^3} \; .
\ee
Using the integral
$$
\int \ep_k \pi_k \frac{d\bk}{(2\pi)^3} = \frac{1}{(2\pi)^2} \;
\int_0^\infty \; \ep_k \left [ {\rm coth}\left (
\frac{\ep_k}{2T}\right ) - 1\right ] k^2 \; dk \; ,
$$
the internal energy can be written as
\be
\label{112}
\frac{E}{N} = \frac{E_0}{N} + \frac{\sqrt{2}(mc)^5}{(2\pi)^2 m\rho}\;
\int_0^\infty \; x^2 \; \frac{(\sqrt{1+x^2}-1)^{1/2}}{\sqrt{1+x^2}}
\left [ {\rm coth}\left ( \frac{mc^2}{2T}\; x\right ) - 1\right ] dx\; .
\ee

The grand potential $\Om = -T \ln{\rm Tr}\; e^{-\bt H_B}$ takes the form
\be
\label{113}
\Om = E_0 - \mu N + TV \int \ln\left ( 1 - e^{-\bt\ep_k}\right )
\frac{d\bk}{(2\pi)^3} \; .
\ee

The free energy $F = \Om + \mu N$ becomes
\be
\label{114}
F = E_0 + \frac{\sqrt{2}(mc)^3}{2\pi^2\rho}\; NT \;
\int_0^\infty \; x\; \frac{(\sqrt{1+x^2}-1)^{1/2}}{\sqrt{1+x^2}}\;
\ln\left ( 1  - e^{-\bt mc^2 x}\right )\; dx \; .
\ee
The same expression for $F$ can also be found from the relation between
$F$ and $E$.

The average kinetic energy per particle, defined as
\be
\label{115}
\overline K \equiv \frac{1}{\rho} \; \int \; \frac{k^2}{2m} \; n_k \;
\frac{d\bk}{(2\pi)^3} \; ,
\ee
takes the form
$$
\overline K = K_0  +  \frac{1}{4m\rho} \; \int k^2 \; \frac{\om_k}{\ep_k}
\left [ {\rm coth}\left ( \frac{\ep_k}{2T} \right ) -1 \right ]
\frac{d\bk}{(2\pi)^3} \; ,
$$
where
\be
\label{116}
K_0 \equiv \frac{1}{4m\rho} \; \int k^2 \left (  \frac{\om_k}{\ep_k} - 1
\right ) \frac{d\bk}{(2\pi)^3} \; .
\ee
So that expression (\ref{115}) transforms into
\be
\label{117}
\overline K = K_0 + \frac{\sqrt{2}(mc)^5}{(2\pi)^2m\rho} \;
\int_0^\infty \; \left (\sqrt{1+x^2}-1\right )^{3/2}
\left [ {\rm coth}\left ( \frac{mc^2}{2T}\; x \right ) -1 \right ] dx\; .
\ee

Considering the internal energy, free energy, and kinetic energy at low
temperatures, where $T/mc^2 \ll 1$, we use the integral
$$
\int_0^\infty \; x^n \ln\left ( 1 - e^{-ax}\right ) dx =
- \left ( \frac{2}{a}\right )^{n+1}\;
\frac{\pi^{n+2}|B_{n+2}|}{(n+1)(n+2)} \; .
$$
As a result, we obtain the low-temperature expressions
$$
\frac{E}{N} \simeq \frac{E_0}{N} + \frac{\pi^2(mc)^5}{30m\rho}
\left ( \frac{T}{mc^2}\right )^4 \; ,
$$
$$
\frac{F}{N} \simeq \frac{E_0}{N} \; - \; \frac{\pi^2(mc)^5}{90m\rho}
\left ( \frac{T}{mc^2}\right )^4 \; ,
$$
\be
\label{118}
\overline K \simeq K_0 + \frac{\pi^2(mc)^5}{60m\rho} \left (
\frac{T}{mc^2}\right )^4 \; .
\ee

In the opposite limit of weak interactions, such that $mc^2/T_c \ll 1$,
where $T_c$ is the ideal-gas critical temperature, we find
$$
\frac{E}{N} \simeq \frac{E_0}{N} + \frac{3\zeta(5/2)}{2\zeta(3/2)} \;
T_c \left ( \frac{T}{T_c}\right )^{5/2} - \; \frac{1}{2}\; mc^2 \left (
\frac{T}{T_c}\right )^{3/2} \; ,
$$
$$
\frac{F}{N} \simeq \frac{E_0}{N} \; -\; \frac{\zeta(5/2)}{\zeta(3/2)}\;
T_c \left ( \frac{T}{T_c}\right )^{5/2} + mc^2 \left (
\frac{T}{T_c}\right )^{3/2} \; ,
$$
\be
\label{119}
\overline K \simeq K_0 +  \frac{3\zeta(5/2)}{2\zeta(3/2)}\;
T_c \left ( \frac{T}{T_c}\right )^{5/2} - \; \frac{3}{2}\; mc^2
\left ( \frac{T}{T_c}\right )^{3/2} \; .
\ee

In the limit $c \ra 0$, we return to the expressions for the ideal gas.
However, since this limit is not well defined, the following corrections,
proportional to $c^2$, should be considered only as approximate estimates.

\subsection{Regularization procedure}

Accomplishing the calculations of the previous section, we meet divergences
in the expressions for $E_0$ and $K_0$. Physically, this is connected with
the used assumption of the dilute gas, when the exact interaction potential
is replaced by its local form (\ref{102}). There are several ways to avoid
this divergence by regularizing the divergent integrals.

First of all, it is possible to take not the local, but a more realistic
interaction potential leading to the momentum dependence of the Fourier
transform $\Phi_k$. Then all equations become essentially more complicated.

The second way could be by means of the cutoff regularization, when the upper 
limit in the divergent integrals is cut by a cutoff momentum. For example, 
one can take as such a limit $k_0=2mc$. But the disadvantage of this method 
is the arising cutoff dependence.

The other way is to resort to an analytic continuation with respect to some
parameters, extending the divergent integrals into the region, where they
become convergent, and then return to the considered region of interest.

A convenient method is the {\it dimensional regularization}, whose idea is
as follows \cite{Andersen_27,Kleinert_28}. Let the integral $\int f_kd\bk$
be divergent in the $3$-dimensional space. Consider this integral in a
$d$-dimensional space, where it is convergent. Then make an analytic
continuation to $d\ra 3$:
\be
\label{120}
\int f_k \; \frac{d\bk}{(2\pi)^3} = \lim_{d\ra 3} \int  f_k \;
\frac{d\bk}{(2\pi)^d} \; .
\ee

When the function $f_k$ depends only on $k\equiv|\bk|$, then
$$
d\bk \ra \frac{2\pi^{d/2}}{\Gm(d/2)} \; k^{d-1}\; dk \; .
$$
And the $d$-dimensional integral reads as
$$
 \int  f_k \; \frac{d\bk}{(2\pi)^d} = \frac{2}{(4\pi)^{d/2}\Gm(d/2)} \;
\int_0^\infty \; f_k k^{d-1} \; dk \; .
$$
In this case, the dimensional regularization (\ref{120}) becomes
$$
 \int  f_k \; \frac{d\bk}{(2\pi)^3} = \frac{1}{2\pi^2} \lim_{d\ra 3} \;
\int_0^\infty f_k k^{d-1}\; dk \; .
$$

In the process of the regularization, one often meets the integrals, that
are divergent in the $3$-dimensional space, but are expressed through the
Euler beta functions in the $d$-dimensional space. The often met such
integrals are of the type
$$
I_d(m,n) \equiv \int_0^\infty
\frac{x^{m+d-n-1}}{(1+x^2)^{n/2}} \; dx = \; \frac{1}{2}\; B(\mu,\nu)\; ,
$$
where the Euler beta function is
$$
B(\mu,\nu) \equiv \int_0^\infty \;
\frac{x^{\mu-1}}{(1+x)^{\mu+\nu}} \; dx \; ,
$$
with
$$
\mu= \frac{m+d-n}{2} \; , \qquad \nu = \frac{2n-m-d}{2} \; .
$$
The integral for $B(\mu,\nu)$ converges for $\mu>0$, $\nu>0$. Therefore,
the integral for $I_d(n,m)$ converges for
$$
n < m+d < 2n \; , \qquad n-m < d < 2n-m \; .
$$

The Euler beta function is represented through the gamma functions as
$$
B(\mu,\nu) =\frac{\Gm(\mu)\Gm(\nu)}{\Gm(\mu+\nu)} = B(\nu,\mu)\; .
$$
And the gamma function $\Gm(z)$ can be analytically continued to the whole
plane of complex $z$. Thus we come to the integral
$$
I_d(m,n) = \frac{\Gm\left ( \frac{m+d-n}{2}\right ) \Gm\left (
\frac{2n-m-d}{2}\right )}{2\Gm(n/2)} \; ,
$$
where $m$, $n$, and $d$ are any complex numbers.

In many cases, the relations
$$
\Gm(z)\Gm(1-z) = \frac{\pi}{\sin(\pi z)} \; , \qquad
\Gm(-z)\Gm(z) = -\; \frac{\pi}{z\sin(\pi z)}
$$
are useful.

For example, let us calculate the ground-state energy (\ref{111}),
in which we meet a divergent integral. This integral, in $d$ dimensions,
has the form
$$
\int \; \frac{\ep_k-\om_k}{2\rho} \; \frac{d\bk}{(2\pi)^d}  =
\frac{(mc)^{d+2}}{\pi^{d/2}\Gm(d/2)m\rho}\; \int_0^\infty \left (
2x\sqrt{1+x^2}-1 -2x^2\right ) x^{d-1}\; dx \; .
$$
Employing the dimensional regularization, we analytically continue the
result of the integration to $d=3$, using the values
$$
\Gm\left ( -\; \frac{1}{2}\right ) = -2\sqrt{\pi} \; , \qquad
\Gm\left ( -\; \frac{3}{2}\right ) = \frac{4\sqrt{\pi}}{3} \; , \qquad
\Gm\left ( -\; \frac{5}{2}\right ) = -\; \frac{8\sqrt{\pi}}{15} \; ,
\qquad I_3(0,-1) = \frac{2}{15} \; .
$$
This gives
$$
\int \; \frac{\ep_k-\om_k}{2\rho} \; \frac{d\bk}{(2\pi)^3}  =
\frac{8(mc)^5}{15\pi^2m\rho} \; .
$$
And we get the ground-state energy
$$
\frac{E_0}{N} = \frac{\rho\Phi_0}{2} + \frac{8(mc)^5}{15\pi^2m\rho} \; .
$$
Taking into account that in the Bogolubov approximation for a dilute gas,
one has
$$
\Dlt=mc^2 = \rho\Phi_0 \; , \qquad mc = \sqrt{m\rho\Phi_0} =
2\sqrt{\pi\rho a_s}\; ,
$$
we obtain
\be
\label{121}
\frac{E_0}{N} = 2\pi \; \frac{\rho a_s}{m} \left ( 1 +
\frac{128}{15\sqrt{\pi}}\; \sqrt{\rho a_s^3}\right ) \; .
\ee

\vskip 2mm

If $\int f_kd\bk$, by its physical definition, is strictly sign-defined,
say positive (non-negative), then the dimensional regularization for
$\int f_kd\bk \geq 0$ can be understood as the limiting procedure
\be
\label{122}
\int f_k\; \frac{d\bk}{(2\pi)^3} = \lim_{d\ra 3} \left | \int
f_k \; \frac{d\bk}{(2\pi)^d} \right | \; .
\ee

As an example, let us consider the kinetic energy at zero temperature,
$K_0\geq 0$. This can be written as
$$
K_0 = \lim_{d\ra 3} \left | \int \frac{k^2}{4m\rho} \left (
\frac{\om_k}{\ep_k} \; - \; 1 \right )  \frac{d\bk}{(2\pi)^d} \right | \; .
$$
Calculating the integral
$$
\int \; \frac{k^2}{4m\rho} \left ( \frac{\om_k}{\ep_k} \; - \; 1 \right )
 \frac{d\bk}{(2\pi)^d} = \frac{2(mc)^{d+2}}{\pi^{d/2}\Gm(d/2)m\rho} \;
\int_0^\infty \; x^{d+1}\left ( \frac{1+2x^2}{2x\sqrt{1+x^2}}\; -\; 1
\right )\; dx \; ,
$$
and using the values
$$
I_3(2,1) = \frac{2}{3}\; , \qquad I_3(4,1) = -\; \frac{8}{15}\; ,
$$
we come to the expression
\be
\label{123}
K_0 = \frac{4(mc)^5}{5\pi^2m\rho}\; = \frac{128\sqrt{\pi}\rho a_s}{5m} \;
\sqrt{\rho a_s^3} \; .
\ee

As is mentioned above, it would be possible to use the cutoff regularization.
However this would lead to the cutoff dependence, so that choosing a
particular cutoff would give an expression that could be different from
(\ref{123}). Thus, if we cut the integral for the average kinetic energy
by a finite value $k_0= 2mc$, we get the value of $K_0$ that is eight times
smaller than (\ref{123}). Therefore the dimensional regularization,
containing no free parameters, seems to be preferable.

\subsection{Particle fluctuations}

The system properties essentially depend on particle fluctuations that
influence many observable quantities \cite{Yukalov_1}. First of all, these
fluctuations show whether the considered system is stable.

Investigating particle fluctuations, it is necessary to use correct
calculations, otherwise it is easy to come to wrong unphysical conclusions.
Unfortunately, there exist numerous publications claiming that particle
fluctuations in Bose-condensed systems are thermodynamically anomalous.
This is certainly principally wrong, since thermodynamically anomalous
particle fluctuations are connected with a divergent compressibility, which
makes the system unstable \cite{Yukalov_1}. If particle fluctuations in
Bose-condensed systems would be thermodynamically anomalous, there could
exist neither equilibrium Bose-condensed systems, nor equilibrium superfluid 
helium. In the present section, we explain the standard mistake made by 
those who claim the occurrence of thermodynamically anomalous fluctuations 
and we show the correct way of calculating these fluctuations. The correct 
calculations yield thermodynamically normal fluctuations, as it has to be
\cite{Yukalov_29,Yukalov_30}.

Particle fluctuations are quantified by the variance of the particle-number
operator,
\be
\label{124}
{\rm var}(\hat N)= \langle \hat N^2 \rangle- \langle \hat N \rangle^2 \; .
\ee
In the presence of Bose-Einstein condensate, the particle-number operator
is $\hat N=\hat N_0 +\hat N_1$. Therefore the variance reads as
$$
{\rm var}(\hat N)={\rm var}(\hat N_0)+{\rm var}(\hat N_1) +
2{\rm cov}(\hat N_0,\hat N_1) \; ,
$$
with the covariance
$$
{\rm cov}(\hat N_0,\hat N_1) \equiv \frac{1}{2} \langle \hat N_0\hat N_1 +
\hat N_1 \hat N_0 \rangle- \langle\hat N_0 \rangle \langle\hat N_1 \rangle \; .
$$
By their definition, $[\hat N_0,\hat N_1]=0$, hence the covariance
$$
{\rm cov}(\hat N_0,\hat N_1) = \langle \hat N_0\hat N_1 \rangle-
\langle \hat N_0 \rangle \langle \hat N_1 \rangle
$$
acquires the form of a correlation function.

In the Bogolubov theory, $\langle \hat N_0\hat N_1 \rangle=
\langle \hat N_0 \rangle \langle \hat N_1 \rangle$, because of which
${\rm cov}(\hat N_0,\hat N_1)=0$. Thus variance (\ref{124}) takes the form
\be
\label{125}
{\rm var}(\hat N) ={\rm var}(\hat N_0)+{\rm var}(\hat N_1) \; .
\ee
According to the Bogolubov approximation, the operator $\hat N_0$ is
replaced by the number $N_0$. Hence $\langle\hat N_0^2\rangle =
\langle \hat N_0 \rangle^2$. Consequently, ${\rm var}(\hat N_0)=
\langle\hat N_0^2\rangle-\langle\hat N_0\rangle^2=0$. In this way,
$$
{\rm var}(\hat N) = {\rm var}(\hat N_1)= \langle \hat N_1^2 \rangle -
\langle \hat N_1 \rangle^2 \; .
$$

The variance of the particle-number operator of uncondensed atoms can be
written as
$$
\langle \hat N_1^2 \rangle \; = \;
\sum_{k,p\neq 0} \langle a_k^\dgr a_k a_p^\dgr a_p \rangle \; .
$$
Here, one substitutes the Bogolubov transformation
$a_k=u_kb_k+v_{-k}^*b_{-k}^\dgr$. As a result, one comes to the
four-operator terms containing the products of four operators
$b_i\equiv b_{k_i}$. By employing the Bogolubov Hamiltonian $H_B$,
one has
$$
\langle b_1^\dgr b_2b_3b_4 \rangle \; = \;
\langle b_1^\dgr b_2^\dgr b_3^\dgr b_4\rangle \; = \; 0 \; ,
$$
and also
$$
\langle b_1^\dgr b_2^\dgr b_3 b_4 \rangle \; = \;
\langle b_1^\dgr b_4\rangle \langle  b_2^\dgr b_3 \rangle +
\langle b_1^\dgr b_3 \rangle \langle b_2^\dgr b_4 \rangle  \; ,
$$
with
$$
\langle b_k^\dgr b_p \rangle=\dlt_{kp} \langle b_k^\dgr b_k \rangle \; .
$$

In that way, one gets
$$
{\rm var}(\hat N_1) = \sum_{k\neq 0} \left \{ \left [ \left (
u_k^2 + v_k^2\right )^2 + 4u_k^2 v_k^2 \right ] \pi_k (1 +\pi_k) +
2u_k^2 v_k^2 \right \} \; .
$$
Here, one substitutes the expressions corresponding to the dilute gas,
$$
u^2_k =\frac{\sqrt{\ep_k^2+m^2c^4}+\ep_k}{2\ep_k} \; , \qquad
v^2_k =\frac{\sqrt{\ep_k^2+m^2c^4}-\ep_k}{2\ep_k} \; ,
$$
$$
u_k^2-v_k^2=1 \; , \qquad u_kv_k = -\; \frac{mc^2}{2\ep_k}\; , \qquad
u_k^2+v_k^2 = \frac{\sqrt{\ep_k^2+m^2c^4}}{\ep_k} \; ,
$$
where $\ep_k$ is the Bogolubov spectrum (\ref{104}). This gives
$$
{\rm var}(\hat N_1) = \frac{N}{\rho} \; \int \left [
\frac{m^2c^4}{2\ep_k^2} + \left (1 + \frac{2m^2 c^4}{\ep_k^2}\right )\;
\pi_k(1+\pi_k)\right ] \; \frac{d\bk}{(2\pi)^3} \; .
$$
The main contribution to the above integral comes form the infrared region,
where $k\ra 0$, $\ep_k\simeq ck$, and $\pi_k\simeq T/\ep_k$. As a result,
there occurs the infrared divergence
$$
\int \frac{\pi_k}{\ep_k^2} \; k^2\; dk \propto \ln k_{min}\; , \qquad
\int \frac{\pi_k^2}{\ep_k^2} \; k^2\; dk \propto \frac{1}{k_{min}} \; ,
$$
where $k_{min} \sim \pi/L \propto 1/N^{1/3}$. Then the anomalous behaviour
of the variance ${\rm var}(\hat N_1) \propto N^{4/3}$ leads to the
divergent compressibility $\kappa_T \propto {\rm var}(\hat N)/N \propto N^{1/3}$,
which implies instability.

The arising instability, caused by thermodynamically anomalous particle
fluctuations, has nothing to do with physics, but is the result of incorrect
calculations. The mistake is in considering the fourth-order operator terms
$\langle a_k^\dgr a_k a_p^\dgr a_p \rangle$ invoking the Bogolubov
approximation that is a second-order approximation with respect to such
operator products. Going outside of the region of applicability of the
approximation leads to wrong results.

To correctly calculate particle fluctuations, it is necessary to be able
to separate out the terms appropriate for the chosen approximation and to
omit the terms of higher orders, which have been also omitted in the used
approximation.

The correct way of studying particle fluctuations is as follows. It is
possible to present the number-operator variance as
\be
\label{126}
{\rm var}(\hat N) = N \left \{ 1 + \rho \int [g(\br)-1]\; d\br\right \} \; ,
\ee
involving the pair correlation function
\be
\label{127}
g(\br_{12}) =\frac{1}{\rho^2} \langle \psi^\dgr(\br_1)\psi^\dgr(\br_2)
\psi(\br_2)\psi(\br_1) \rangle \; ,
\ee
with $\br_{12}\equiv \br_1 -\br_2$.

We substitute here the Bogolubov shift $\psi(\br) \ra \eta+\psi_1(\br)$. And
we use the following property: {\it If a set of operators satisfies the Wick
theorem, then their linear combinations also satisfy this theorem}. In our
case the operators $b_k$ do satisfy the Wick theorem because of the
structure of the Bogolubov Hamiltonian $H_B$. Hence the operators $a_k$,
as well as $\psi_1(\br)$, also satisfy this theorem. This gives
$$
\langle \psi_1^\dgr(\br_1) \psi_1(\br_1) \psi_1(\br_2) \rangle \; = \;
\langle \psi_1^\dgr(\br_1)\psi_1(\br_1) \rangle \langle \psi_1(\br_2) \rangle +
\langle \psi_1^\dgr(\br_1)\psi_1(\br_2)\rangle \langle \psi_1(\br_1) \rangle +
$$
$$
+ \langle \psi_1^\dgr(\br_1) \rangle \langle \psi_1(\br_1) \psi_1(\br_2) \rangle
\; = \; 0 \; ,
$$
since $ \langle \psi_1(\br_1) \rangle =0$ and $\langle a_k \rangle=0$
for $k\neq 0$.

In the real space, the normal average is the first-order density matrix
\be
\label{128}
\rho_1(\br_1,\br_2)\equiv \langle \psi_1^\dgr(\br_2)\psi_1(\br_1) \rangle \; ,
\ee
and the anomalous average is
\be
\label{129}
\sgm_1(\br_1,\br_2)\equiv \langle \psi_1(\br_2)\psi_1(\br_1) \rangle \; .
\ee
Using this gives
$$
\langle \psi_1^\dgr(\br_1) \psi_1^\dgr(\br_2)\psi_1(\br_2)\psi_1(\br_1) \rangle \; =
\; \rho_1^2 +|\rho_1(\br_1,\br_2)|^2 + |\sgm_1(\br_1,\br_2)|^2 \; .
$$
Thus, we obtain the pair correlation function (\ref{127}) in the form
\be
\label{130}
g(\br_{12}) = 1 +\frac{2\rho_0}{\rho^2}\; {\rm Re}\left [
\rho_1(\br_1,\br_2) +\sgm_1(\br_1,\br_2)\right ] + \frac{1}{\rho^2}
\left [ |\rho_1(\br_1,\br_2)|^2 + |\sgm_1(\br_1,\br_2)|^2 \right ] \; .
\ee

Fourier transforming averages (\ref{128}) and (\ref{129}), we have
$$
\rho_1(\br_1,\br_2) = \int n_k \; e^{i\bk\cdot\br_{12}}\;
\frac{d\bk}{(2\pi)^3}\; , \qquad
\sgm_1(\br_1,\br_2) = \int \sgm_k \; e^{i\bk\cdot\br_{12}}\;
\frac{d\bk}{(2\pi)^3}\; .
$$
Notice that $\rho_1(\br_1,\br_2)$ and $\sgm_1(\br_1,\br_2)$ are real due
to the symmetry properties of $n_k$ and $\sgm_k$. Using the equality
$$
\int \dlt(\bk) (n_k +\sgm_k) \; d\bk \equiv \lim_{k\ra 0} (n_k +\sgm_k) \; ,
$$
we find
$$
\int [g(\br)-1]\; d\br = \frac{2\rho_0}{\rho^2}\; \lim_{k\ra 0}
(n_k +\sgm_k) + \frac{1}{\rho^2} \; \int \left ( n_k^2 +\sgm_k^2\right )
\; \frac{d\bk}{(2\pi)^3} \; .
$$

Now let us remember that in the Bogolubov approximation, in addition to
the replacement $\rho_0\ra \rho$, the operator terms have been neglected
having the order higher then two with respect to the operators $a_k$.
But here the term, containing $n_k^2 +\sgm_k^2$, is of fourth order with
respect to these operators $a_k$. Hence, there is no any reason of taking
account of such terms that, in the Bogolubov second-order approximation
have to be neglected. Thus we come to the variance
$$
{\rm var}(\hat N) = N\left [ 1 + 2\lim_{k\ra 0} (n_k +\sgm_k)\right ] \; .
$$

Employing the expressions of $n_k$ and $\sgm_k$ gives
$$
\lim_{k\ra 0} (n_k +\sgm_k) = \frac{1}{2}\left (
\frac{T}{mc^2}\; - \; 1\right ) \; .
$$
Therefore we obtain the variance
\be
\label{131}
{\rm var}(\hat N) =\frac{TN}{mc^2} \; ,
\ee
corresponding to thermodynamically normal fluctuations. Respectively, the
compressibility
\be
\label{132}
\kappa_T = \frac{{\rm var}(\hat N)}{\rho TN} = \frac{1}{\rho mc^2}
\ee
does not diverge, as it has to be for an equilibrium stable system.

From here, we again see that in the case of the ideal gas, when $\Phi_0\ra 0$
and $c\ra 0$, the compressibility diverges, $\kappa_T\ra\infty$, signifying
the instability of the ideal Bose-condensed gas. But interactions do
stabilize the system.

Also, let us recall the general exact relations (see \cite{Yukalov_1})
for the central structural factor $S(0)$ at $\bk=0$,
$$
S(0) = \frac{{\rm var}(\hat N)}{N} = \rho T \kappa_T \; ,
$$
the sound velocity $s$,
$$
s^2 = \frac{1}{m} \left ( \frac{\prt P}{\prt\rho} \right )_T =
\frac{1}{m\rho\kappa_T} \; ,
$$
and the isothermal compressibility,
$$
\kappa_T = \frac{S(0)}{\rho T} = \frac{1}{m\rho s^2} \; .
$$
These relations clearly demonstrate that in an unstable system, with a
divergent compressibility, there would occur senseless values of an infinite
structure factor and zero sound velocity.

\subsection{Superfluid fraction}

The general formula for the superfluid fraction in a three-dimensional
space is
\be
\label{133}
n_s =  1 -\; \frac{\bt}{3mN}\; {\rm var}(\hat\bP) \; .
\ee
In equilibrium, $\langle \hat\bP \rangle=0$, hence
${\rm var}(\hat\bP) =\; \langle \hat\bP^2 \rangle$.

The Bogolubov shift for a uniform system reads as
$$
\psi(\br) \ra \hat\psi(\br) = \eta+\psi_1(\br) \; , \quad \rho_0=\eta^2 \; .
$$
Using this, we have
$$
\langle \hat\psi^\dgr(\br_2)\hat\psi(\br_1) \rangle \; = \rho_0 +\rho_1(\br_1,\br_2) \; ,
$$
where
$$
\rho_1(\br_1,\br_2) =\; \langle \psi_1^\dgr(\br_2)\psi_1(\br_1) \rangle \; =
\int n_k \; e^{i\bk\cdot\br_{12}} \; \frac{d\bk}{(2\pi)^3} \; .
$$
Therefore
$$
\int \lim_{\br_3,\br_4\ra\br_1} \nabla_3\cdot \nabla_4 \;
\rho_1(\br_4,\br_3) \; d\br_1 =  V \int k^2 n_k\; \frac{d\bk}{(2\pi)^3} \; .
$$
In the Bogolubov approximation, the Wick theorem is valid for the operators
$\psi_1(\br)$. This makes it possible to represent the second-order density
matrix as
$$
\rho_2(\br_3,\br_4,\br_2,\br_1) =\;
\langle \psi^\dgr(\br_1)\psi^\dgr(\br_2)\psi(\br_3)\psi(\br_4) \rangle \; =
$$
$$
= \rho_0^2 + \rho_0 \left [ \rho_1(\br_3,\br_1)+ \rho_1(\br_4,\br_1) +
\rho_1(\br_3,\br_2) + \rho_1(\br_4,\br_2) + \right.
$$
$$
\left. + \sgm_1(\br_3,\br_4) + \sgm_1^*(\br_1,\br_2) \right ]
+ \rho_1(\br_3,\br_1)\rho_1(\br_4,\br_2) +
\rho_1(\br_4,\br_1)\rho_1(\br_3,\br_2) +
\sgm_1^*(\br_1,\br_2)\sgm_1(\br_3,\br_4) \; .
$$

For the anomalous average
$$
\sgm_1(\br_1,\br_2)=\langle \psi_1(\br_2)\psi_1(\br_1) \rangle \; =
$$
$$
= \int \sgm_k \; e^{i\bk\cdot\br_{12}}\; \frac{d\bk}{(2\pi)^3} \; ,
$$
we meet the integrals of the type
$$
\int \dlt(\bk) f(\bk)\sgm_k\; d\bk \equiv f(0)\lim_{k\ra 0} \sgm_k \; ,
$$
$$
\int \dlt(\bk) k^2 \sgm_k\; d\bk = 0 \; .
$$

Using the above expressions, we have
$$
\int \; \lim_{\br_3\ra\br_1,\br_4\ra\br_2} \; \nabla_3\cdot\nabla_4\;
\rho_2(\br_3,\br_4,\br_2,\br_1)\; d\br_1 d\br_2 = - V \int k^2
\left (n_k^2 -\sgm_k^2\right ) \; \frac{d\bk}{(2\pi)^3} \; .
$$
So that finally, we get
$$
\langle \hat\bP^2 \rangle \; = V \int k^2 \left ( n_k + n_k^2 -\sgm_k^2\right ) \;
\frac{d\bk}{(2\pi)^3} \; .
$$
The same result can be obtained from the expression
$$
\langle \hat\bP^2 \rangle \; =\sum_{kp} (\bk\cdot\bp) \langle \hat n_k\hat n_p \rangle \; ,
$$
using the Wick theorem for
$$
\langle \hat n_k \hat n_p \rangle \; = n_k n_p +\dlt_{kp} n_k(1 + n_k) +
\dlt_{-kp}\sgm_k^2
$$
and the sum $\sum_k \bk n_k = 0$.

In this way, we find
\be
\label{134}
n_s =  1 \; - \frac{2Q}{3T} \; , \qquad
Q \equiv \frac{1}{\rho}\; \int \; \frac{k^2}{2m} \left (
n_k + n_k^2 -\sgm_k^2\right ) \; \frac{d\bk}{(2\pi)^3} \; .
\ee
It is necessary to stress that here the term containing $n_k^2 -\sgm_k^2$
cannot be omitted because, although each of the quantities $n_k^2$ and
$\sgm_k^2$ are of the fourth order with respect to the operators $a_k$, but
the difference $n_k^2 -\sgm_k^2$ is effectively of second order. This is
explained in section 2.10 showing that in the most important long-wave
region the values $n_k$ and $|\sgm_k|$ are close to each other, exactly
coinciding for the asymptotically small excitation energy. Therefore, the
main terms in $n_k^2$ and $\sgm_k^2$ cancel each other and their difference
$n_k^2 -\sgm_k^2$ plays the role of a term of the lower order than each
of the components.

Such a cancellation is crucial for the calculation of the dissipated heat, 
where the dangerous terms cancel each other, removing divergences that would 
exist for each of the terms separately. The sum in the integrand of $Q$ reads as
$$
n_k+n_k^2 -\sgm_k^2 =\frac{1/2}{{\rm cosh}(\bt\ep_k)-1} \; =
- T\; \frac{\prt\pi_k}{\prt\ep_k} \; .
$$
Here we also can use the equality ${\rm cosh}x -1=2{\rm sinh}^2 (x/2)$.

The normal density, that is, the density of uncondensed atoms, is
$$
\rho_n = \frac{2Q}{3T}\; \rho \; = 
-\; \frac{1}{3m} \; \int k^2 \; \frac{\prt\pi_k}{\prt\ep_k} \; \frac{d\bk}{(2\pi)^3}\; .
$$
In the Bogolubov approximation, the dissipated heat becomes
$$
Q = \frac{1}{8\pi^2m\rho} \; \int_0^\infty \;
\frac{k^4\;dk}{{\rm cosh}(\bt\ep_k)-1} \; .
$$
Using the change of the variable $x=\ep_k/mc^2$, we get
$$
\frac{Q}{mc^2} =\frac{\sqrt{2}(mc)^3}{(2\pi)^2\rho} \; \int_0^\infty \;
\frac{(\sqrt{1+x^2}-1)^{3/2}x\;dx}{\sqrt{1+x^2}[{\rm cosh}(\bt mc^2 x) -1]} \; .
$$

Let us study the low-temperature behaviour of the superfluid fraction, when
$T/mc^2 \ll 1$. Then we use the expansion
$$
\frac{\sqrt{2}(\sqrt{1+x^2}-1)^{3/2}}{\sqrt{1+x^2}} \simeq \frac{1}{2}\;
x^3 \; - \; \frac{7}{16}\; x^5 + \frac{99}{256}\; x^7
$$
and calculate the integral
$$
\int_0^\infty\; \frac{x^\mu\; dx}{{\rm cosh}x+\cos\al} =
\frac{2\Gm(1+\mu)}{\sin\al} \; \sum_{k=1}^\infty (-1)^{k-1}
\frac{\sin(k\al)}{k^{1+\mu}} \; ,
$$
in which ${\rm Re}\;\mu > -1$ and $|\al|<\pi$. The result can be extended to
$\al\ra\pi$, yielding
$$
\lim_{\al\ra \pi} \; \sum_{k=1}^\infty \;
\frac{(-1)^{k-1}\sin(k\al)}{k^{1+\mu}\sin\al} = \zeta(\mu) \; .
$$

Thus we get the integral
$$
\int_0^\infty \; \frac{x^\mu\; dx}{{\rm cosh}x-1} = 2\Gm(1+\mu)\zeta(\mu) \; .
$$
For even powers, this reduces to
$$
\int_0^\infty \; \frac{x^{2n}\; dx}{{\rm cosh}x-1} = (2\pi)^{2n}|B_{2n}| \; ,
$$
with $B_n$ being the Bernoulli numbers. In this way, we find the expansion for
the dissipated heat
$$
\frac{Q}{mc^2} \simeq \frac{(mc)^3}{\rho} \left [ \frac{\pi^2}{15} \left (
\frac{T}{mc^2}\right )^5 \; - \; \frac{\pi^4}{6}\left ( \frac{T}{mc^2}
\right )^7 + \frac{33\pi^6}{40} \left (
\frac{T}{mc^2}\right )^9\right ] \; .
$$
Then the low-temperature behaviour of the superfluid fraction reads as
\be
\label{135}
n_s \simeq 1 \; - \; \frac{2\pi^2(mc)^3}{45\rho}\left [
\left ( \frac{T}{mc^2}\right )^4 \; - \; \frac{5\pi^2}{2}
\left ( \frac{T}{mc^2}\right )^6 + \frac{99\pi^4}{8}
\left ( \frac{T}{mc^2}\right )^8 \right ]\; .
\ee

In the limit of zero $c$, we return to the expressions for the ideal gas,
$$
Q \simeq \frac{3}{2} \; T \left (\frac{T}{T_c}\right )^{3/2} \; , \quad
n_s \simeq 1 - \left (\frac{T}{T_c}\right )^{3/2} \; .
$$
The fraction $n_s$ coincides with $n_0$ only in the limit of the ideal
gas, when $c=0$. But they are different for any finite $c$. As has been
mentioned above, the limit $c=0$ corresponds to an unstable system.
Corrections to the zero-interaction case cannot be derived by the direct
expansion of $Q$ in powers of $c$, since this leads to divergent terms.

\subsection{Bogolubov theorem}

An important property of Bose systems follows from the investigation of
the response induced by local gauge transformations
\be
\label{136}
\hat U_\al \equiv \exp \left\{ i \int \al(\br) \hat n(\br) \; d\br
\right \} \; ,
\ee
where $\al^*(\br)=\al(\br)$ is a real function and
$\hat n(\br)\equiv \psi^\dgr(\br)\psi(\br)$. In particular, this property
imposes restrictions specifying the behavior of the spectrum of
collective excitations. The mathematical formulation of this property
composes the Bogolubov theorem \cite{Bogolubov_16,Bogolubov_19}.

For any type of statistics, there exist the commutation relations
$$
[\psi(\br),\hat n(\br')] = \psi(\br)\dlt(\br-\br') \; , \qquad
\left [ \psi(\br), \;
\int \al(\br')\hat n(\br')\; d\br'\right ] = \al(\br) \psi(\br) \; ,
$$
from which one has
$$
\psi(\br) \hat U_\al = e^{i\al(\br)} \hat U_\al \psi(\br) \; .
$$
The local gauge transformation is unitary, $\hat U_\al^+\hat U_\al=1$.

Let us introduce the transformed field operator
$$
\tilde\psi(\br) \equiv
\hat U_\al^+ \psi(\br) \hat U_\al = e^{i\al(\br)} \psi(\br) \; ,
$$
with the inverse transformation
$$
\psi(\br) = \hat U_\al \tilde\psi(\br) \hat U_\al^+ = e^{-i\al(\br)}
\tilde\psi(\br) \; .
$$
Respectively, for a Hamiltonian $H \equiv H[\psi]$, we define the
transformed Hamiltonian
$$
\tilde H \equiv H[\tilde\psi] \; = \hat U_\al^+ H[\psi] \hat U_\al \; ,
$$
and, similarly, for an operator of observable $\hat A \equiv \hat A[\psi]$,
the transformed operator
$$
\tilde A \equiv \hat A[\tilde\psi] \; = \hat U_\al^+\hat A[\psi]\hat U_\al \; .
$$

It is straightforward to see that for the averages
$$
\langle \hat A \rangle_H \; \equiv \;
\frac{{\rm Tr} e^{-\bt H}\hat A}{{\rm Tr}e^{-\bt H}} \; ,
$$
there is the equality
\be
\label{137}
\langle \hat A[\tilde\psi] \rangle_{\tilde H} \; =\; \langle \hat A[\psi] \rangle_H \; .
\ee

Owing to this equality, we have
$$
\langle \prod_i \tilde\psi^\dgr(\br_i) \;
\prod_j \tilde\psi(\br_j)\rangle_{\tilde H} \; = \;
\langle \prod_i \psi^\dgr(\br_i) \; \prod_j \psi(\br_j) \rangle_H \; .
$$
From direct calculations it follows that
$$
\langle \prod_i \tilde\psi^\dgr(\br_i) \;
\prod_j \tilde\psi(\br_j) \rangle_{\tilde H} \; = \;
\langle \prod_i \psi^\dgr(\br_i) \; \prod_j \psi(\br_j) \rangle_{\tilde H}
 \; \exp\left\{ - i \left [ \sum_i \al(\br_i) \; - \;
\sum_j \al(\br_j) \right ] \right \} \; .
$$
Using equation (\ref{137}), we get
$$
\langle \prod_i \psi^\dgr(\br_i) \; \prod_j \psi(\br_j) \rangle_{\tilde H} \; = \;
\langle \prod_i \psi^\dgr(\br_i) \; \prod_j \psi_j(\br_j) \rangle_H \;
\exp\left\{ i \left [ \sum_i \al(\br_i) \; - \;
\sum_j \al(\br_j) \right ] \right \} \; .
$$

Let us consider the infinitesimal transformation, where $\al(\br)\ra 0$. Then
$$
\tilde\psi(\br)\simeq[1+i\al(\br)]\psi(\br) \; .
$$ 
And let us define the increment of an operator
\be
\label{138}
\dlt \langle \hat A \rangle\; \equiv \; \langle \hat A[\psi] \rangle_{\tilde H} -
\langle \hat A[\psi]\rangle_H \; .
\ee
We shall be interested in the increment
$$
\dlt \langle \prod_i \psi^\dgr(\br_i)\; \prod_j\psi(\br_j) \rangle \; = \;
i\left [ \sum_i \al(\br_i) \; - \; \sum_j \al(\br_j)\right ] \; .
$$

Consider a uniform system, with the field operators in the momentum space
$$
a_k \equiv \int \vp_k^*(\br)\psi(\br)\; d\br \; , \qquad
\vp_k(\br) = \frac{e^{i\bk\cdot\br}}{\sqrt{V}} \; ,
$$
whose transformed operators are
$$
\tilde a_k \equiv \int \vp_k^*(\br)\tilde\psi(\br)\; d\br \; .
$$
Let us set $\al(\br) = 2\cos({\bf q}\cdot\br)\dlt\al$, with $\dlt\al \ra 0$.
Then
$$
\tilde a_k = a_k + i(a_{k+q} + a_{k-q} ) \dlt \al \; , \qquad
\dlt \langle a_k \rangle \; = \; - i \langle a_{k-q} + a_{k+q} \rangle \dlt \al \; .
$$

For a Bose-condensed system, gauge symmetry is broken, because of which
$$
\langle a_k \rangle_H \; = \; \sqrt{N_0}\; \dlt_{k0} \; , \qquad
\dlt \langle a_k \rangle \; = \; -i (\dlt_{kq} + \dlt_{-kq})\sqrt{N_0}\; \dlt \al\; ,
$$
$$
\dlt \langle a_k^\dgr a_p \rangle \; = \; i \langle \left ( a_{k-q}^\dgr + a_{k+q}^\dgr
\right ) a_p - a_k^\dgr ( a_{p-q} + a_{p+q} ) \rangle_H \dlt \al \; .
$$
With the notation
$$
\langle a_k^\dgr a_p \rangle_H \; = \; \dlt_{kp} n_k \; , \qquad
n_k \; \equiv \; \langle a_k^\dgr a_k \rangle_H \; ,
$$
we get
$$
\dlt \langle a_k^\dgr a_p \rangle \; = \; i ( \dlt_{k,p+q} + \dlt_{k,p-q})
(n_p - n_k) \dlt \al \; .
$$
Setting here ${\bf q}=\bk$ gives
$$
\dlt \langle a_k \rangle \; = \; -i\sqrt{N_0}\; \dlt\al \; , \qquad
\dlt\langle a_k^\dgr - a_{-k} \rangle \; = \; 2i\; \sqrt{N_0}\; \dlt\al \; .
$$

The transformed Hamiltonian can be written as $H[\tilde\psi]= H[\psi]+\dlt H$,
with
$$
\dlt H = -\; \frac{i}{2m} \; \int \psi^\dgr(\br) \left\{ \left [
\nabla^2 \al(\br)\right ] + 2\left [\vec\nabla\al(\br)\right ] \cdot
\vec\nabla \right \} \psi(\br) \; d\br \; ,
$$
where only the terms linear in $\al$ are left, while containing $\al^2$ are
omitted. Passing to the momentum space, we have $\dlt H = H_q \dlt\al$,
with
$$
H_q = -\; \frac{i}{2m}\; \sum_p
\left ( q^2 + 2{\bf p}\cdot {\bf q}\right ) \left ( a_p^\dgr a_{p+q}
- a_{p+q}^\dgr a_p \right ) \; .
$$
The latter form is clearly Hermitian, $H_q^+ = H_q$, and the related
increment is
$$
\dlt \langle H_k \rangle \; = \; -2N\; \frac{k^2}{m}\; \dlt \al \; .
$$

Now let us use the Kubo formula in the Bogolubov-Tyablikov representation,
as is explained in Ref. \cite{Yukalov_1}. Then, using the notation for the
linear response function $\chi(\hat A, \hat B) = \chi(\hat A, \hat B, 0)$,
defined in Ref. \cite{Yukalov_1}, we obtain
$$
\dlt \langle a_k \rangle= \chi(a_k,H_k)\dlt\al \; ,
$$
$$
\dlt \langle a_k^\dgr - a_{-k} \rangle \; = \; \chi\left ( a_k^\dgr-a_{-k},H_k
\right )\dlt\al \; ,
$$
$$
\dlt \langle H_k \rangle \; = \; \chi(H_k,H_k) \dlt\al \; .
$$
For the related response functions, we find
$$
|\chi(a_k,H_k)|^2 = N_0 \; , \qquad
\left | \chi\left ( a_k^\dgr-a_{-k},H_k\right )
\right |^2 = 4N_0 \; , \qquad \chi(H_k,H_k) = - 2N\;
\frac{k^2}{m} \; .
$$

The expression $|\chi(\hat A,\hat B)|$ is a bilinear form, for which the
Bogolubov inequality
$$
| \chi(\hat A,\hat B)|^2 \leq
| \chi(\hat A,\hat A^+)\; \chi(\hat B^+,\hat B)|
$$
is valid (see \cite{Yukalov_1}). In the present case, we have
$$
| \chi(a_k,H_k)|^2 \leq \left | \chi\left ( a_k,a_k^\dgr\right )
\chi\left ( H_k^+,H_k \right ) \right | \; ,
$$
$$
\left | \chi\left ( a_k^\dgr-a_{-k},H_k\right ) \right |^2 \leq
\left | \chi\left ( a_k^\dgr - a_{-k},a_k-a_{-k}^\dgr\right )
\chi\left ( H_k^+,H_k \right ) \right | \; .
$$

Finally, the Bogolubov theorem takes the form
\be
\label{139}
\left | \chi \left ( a_k,a_k^\dgr\right ) \right | \geq
\frac{mn_0}{2k^2} \; , \qquad
\left | \chi \left ( a_k^\dgr-a_{-k},a_k-a_{-k}^\dgr\right )
\right | \geq \frac{2mn_0}{k^2} \; ,
\ee
where $n_0\equiv N_0/N$ and
$$
\chi \left ( a_k^\dgr-a_{-k},a_k-a_{-k}^\dgr\right ) =
\chi \left ( a_k^\dgr,a_k\right ) - \chi \left ( a_{-k},a_k\right )
+ \chi \left ( a_{-k},a_{-k}^\dgr \right ) +
\chi \left ( a_k^\dgr,a_{-k}^\dgr\right ) \; .
$$

Another convenient form of the Bogolubov theorem, which is often used,
follows from the above inequalities, after introducing the matrix
$[G_{\al\bt}(\bk)]$ composed of the Green functions
$$
G_{11}(\bk) \equiv \chi \left ( a_k,a_k^\dgr\right ) \; , \qquad
G_{12}(\bk) \equiv \chi \left ( a_k,a_{-k}\right ) \; ,
$$
$$
G_{21}(\bk) \equiv \chi \left ( a_k^\dgr,a_{-k}^\dgr\right ) \; ,
\qquad G_{22}(\bk) \equiv \chi \left ( a_k^\dgr,a_k\right ) \; ,
$$
such that
$$
\chi\left ( a_k^\dgr - a_{-k}, a_k - a_{-k}^\dgr\right ) =
2 G_{11}(\bk) - 2 G_{12}(\bk) \; .
$$
The Green functions enjoy the symmetry properties
$$
G_{11}(\bk) = G_{22}(\bk) \; , \qquad
G_{12}(\bk)=G_{21}(\bk) \; , \qquad
G_{\al\bt}(-\bk) = G_{\al\bt}(\bk) \; .
$$

Then the Bogolubov theorem acquires the form
\be
\label{140}
|G_{11}(\bk)| \geq \frac{mn_0}{2k^2}\; , \qquad
| G_{11}(\bk) - G_{12}(\bk)| \geq \frac{mn_0}{k^2} \; .
\ee
As will be shown below, this theorem prescribes that the spectrum
of collective excitations in a Bose-condensed system has to be gapless.

\section{Main features}

In the theoretical description of Bose-condensed systems there exist several 
delicate points that have to be taken into account for developing a correct 
theory. Otherwise, it is easy to come to wrong conclusions having no physical 
meaning.  

\subsection{Nonuniqueness of vacuum}

As has been explained in the first part of the Tutorial \cite{Yukalov_1}, 
the field operators $\psi(\br)$ and $\psi^\dgr(\br)$ are defined on the 
Fock space ${\cal F}(\psi)$ generated by these field operators from a vacuum 
state $|0 \rangle$, such that $\psi(\br)|0 \rangle = 0$. Any function 
$\vp\in{\cal F}(\psi)$ can be constructed from the vacuum by the rule
$$
\vp = \sum_{n=0}^\infty \; \frac{1}{\sqrt{n!}} \;
\int f_n(\br_1,\ldots,\br_n) \prod_{i=1}^n \psi^\dgr(\br_i)\; d\br_i |\; 0 \rangle \; ,
$$ 
where $f_n(\br_1,\ldots,\br_n)$ is a symmetrized function of its arguments. 

When accomplishing the Bogolubov shift $\psi(\br) \ra \eta(\br) + \psi_1(\br)$,
with a nonoperator function $\eta(\br)$ that is not identically zero, one passes 
from the field operators $\psi(\br)$ to the field operators $\psi_1(\br)$, both
satisfying the Bose commutation relations. The transformation between these field
operators can be described by the transformation operator 
\be
\label{141}
\hat C \equiv \exp \left\{ \int \left [ \eta^*(\br)\psi(\br) -
\eta(\br)\psi^\dgr(\br)\right ] \; d\br \right \} \; ,
\ee
whose inverse is
\be
\label{142}
\hat C^{-1} =  \exp \left\{ - \int \left [ \eta^*(\br)\psi(\br) -
\eta(\br)\psi^\dgr(\br)\right ] \; d\br \right \} \; .
\ee
The transformation operator can be written as 
$$
\hat C = \sum_{n=0}^\infty \; \frac{1}{n!} \left \{ \int \left [
\eta^*(\br)\psi(\br) - \eta(\br)\psi^\dgr(\br)\right ] \; d\br \right \}^n \; .
$$
Using the commutation relation
$$
[\psi(\br),\; \hat C] = -\eta(\br) \hat C \; , 
$$
the transformation between the field operators can be represented in the form 
\be
\label{143}
\psi(\br) = \hat C\psi_1(\br)\hat C^{-1} \; , \qquad
\psi_1(\br) = \hat C^{-1}\psi(\br)\hat C \; .
\ee
However, the state $|0 \rangle$ is not a vacuum for $\psi_1(\br)$, since
\be
\label{144}
\psi_1(\br)|0 \rangle \; = -\eta(\br)|0 \rangle \; \neq 0 \; ,
\ee
when $\eta(\br)\not\equiv 0$. In turn, $|0>_1$ is not a vacuum for $\psi(\br)$,
as far as
\be
\label{145}
\psi(\br)|0 \rangle_1 = \eta(\br)|0 \rangle_1 \neq 0 \; ,
\ee
if $\eta(\br)\not\equiv 0$.

The vacuum for $\psi_1(\br)$ is the state
\be
\label{146}
|0 \rangle_1 = \hat C^{-1}|0 \rangle \; , 
\ee
since $\psi_1(\br)|0 \rangle_1 = 0$. Over this vacuum, one can generate 
the Fock space ${\cal F}(\psi_1)$ by means of the field operator 
$\psi_1^\dgr(\br)$. But the Fock spaces ${\cal F}(\psi)$ and ${\cal F}(\psi_1)$ 
are different. Moreover, the spaces ${\cal F}(\psi)$ and ${\cal F}(\psi_1)$ 
are {\it orthogonal} in thermodynamic limit. 

To show their orthogonality, we resort to the Baker-Hausdorff formula for the
operators $\hat A$ and $\hat B$, for which $[\hat A,\hat B]$ commutes with both 
$\hat A$ and $\hat B$,
$$
\left [ \hat A,\; [\hat A,\hat B]\right ] =
\left [ \hat B,\; [\hat A,\hat B]\right ] = 0 \; .
$$
Then we have
$$
e^{\hat A} e^{\hat B} =\exp \left ( \hat A + \hat B + \frac{1}{2}\;
[\hat A,\hat B] \right ) \; , \qquad
e^{\hat A+ \hat B} = e^{\hat A} e^{\hat B} \exp \left (-\;
\frac{1}{2}\; [\hat A,\hat B] \right ) \; .
$$
Operator (\ref{142}) reads as
$$
\hat C^{-1} = \exp\left\{ \int \eta(\br)\psi^\dgr(\br)\; d\br\right \}\;
\exp\left \{ - \int \eta^*(\br)\psi(\br)\; d\br \right \} \;
\exp\left \{ -\; \frac{1}{2}\; \int |\eta(\br)|^2\; d\br \right \} \; .
$$
The action of this operator on the vacuum $|0 \rangle$ is
$$
\hat C^{-1}|0 \rangle \; = \exp\left \{ -\; \frac{1}{2}\; \int |\eta(\br)|^2\;
d\br \right \} \; \exp \left \{  \int \eta(\br)\psi^\dgr(\br)\; d\br \right \}|0 \rangle \; .
$$

By definition (see \cite{Yukalov_1}) a {\it coherent state} $|\eta \rangle$ 
is the eigenstate of the destruction operator,
$$
\psi(\br)|\eta \rangle =\eta(\br)|\eta \rangle \; ,
$$
with the eigenvalue $\eta(\br) = \langle \eta|\psi(\br)|\eta \rangle$ called 
the coherent field. The coherent state can be represented as
$$
|\eta \rangle\; = \eta_0  \exp \left \{ \int \eta(\br)\psi^\dgr(\br)\; d\br
\right \}|0 \rangle \; , \qquad
|\eta_0| = \exp \left \{ -\; \frac{1}{2}\; \int |\eta(\br)|^2 \; d\br \right \} \; .
$$
This shows that the vacuum of the Fock space ${\cal F}(\psi_1)$ is the coherent 
state
\be
\label{147}
|0 \rangle_1 = \hat C^{-1}|0 \rangle \; = |\eta \rangle \; .
\ee

Considering the scalar product $\langle 0|0 \rangle_1 =\; \langle 0|\eta \rangle$,
we use the equality 
$$
\langle 0| \exp \left \{ \int \eta(\br)\psi^\dgr(\br) \; d\br \right \}|0 \rangle \; = 1 \; ,
$$
which gives
$$
\langle 0|\eta \rangle \; = \exp \left \{ -\; \frac{1}{2}\; \int |\eta(\br)|^2 \; d\br
\right \} \; .
$$

The number of condensed particles is given by the integral over the condensate 
density, 
$$
N_0 = \int \rho_0(\br)d\br \; , \qquad \rho_0(\br) = |\eta(\br)|^2 \; .
$$
Therefore
$$
\langle0 |\eta \rangle \; = \exp\left ( -\; \frac{1}{2}\; N_0\right ) \; .
$$
Bose-condensation implies that $N_0\propto N$, because of which we see that 
the vacua $|0\rangle$ and $|\eta \rangle$ are asymptotically orthogonal:
\be
\label{148}
\langle 0|\eta \rangle \; \simeq 0 \qquad (N\ra \infty) \; . 
\ee
And all states generated from these vacua are orthogonal to each other,
except $|\eta \rangle$ that is a coherent state in ${\cal F}(\psi)$, and it 
is the vacuum $|\eta \rangle = |0 \rangle_1$ in ${\cal F}(\psi_1)$.

The sole intersection between the spaces, containing continuum number of states,
is the intersection of zero measure. To eliminate the influence of this 
intersection, it is possible to impose the orthogonality condition
\be
\label{149}
\int \eta^*(\br)\psi_1(\br) \; d\br = 0 \; , \qquad
\int \eta(\br)\psi_1^\dgr(\br) \; d\br = 0  \; .
\ee

In this way, we come to the conclusion that the spaces ${\cal F}(\psi)$ and 
${\cal F}(\psi_1)$ are asymptotically orthogonal with each other. The spaces 
have their own vacua that are orthogonal to each other. Thus, a given 
statistical system may possess several vacua and different orthogonal Fock 
spaces. 

It is also worth noting that the transformation operator $\hat C$ is not 
defined in one space. The operator $\hat C$ transforms ${\cal F}(\psi_1)$ 
into ${\cal F}(\psi)$. And the inverse operator $\hat C^{-1}$ transforms 
${\cal F}(\psi)$ into ${\cal F}(\psi_1)$. Therefore $\hat C$ is nonunitary, 
since $\hat C^+$ is not defined on the same space.

Strictly speaking, the Bogolubov shift has to be understood as a replacement
$\psi(\br) \longrightarrow \hat\psi(\br)$ of the field operator $\psi(\br)$
by the field operator $\hat \psi(\br)$, where
\be
\label{150}
\hat\psi(\br) \equiv \eta(\br ) + \psi_1(\br ) \; ,
\ee
since these operators are defined on different Fock spaces. The operator
$\psi(\br)$ is defined on ${\cal F}(\psi)$, but $\hat\psi(\br)$ is defined 
on ${\cal F}(\psi_1)$. The left-hand and right-hand sides of the replacement 
are defined on different spaces, orthogonal to each other. The same 
conclusions are valid irrespectively to whether the gauge symmetry has been 
broken by the Bogolubov shift or by infinitesimal sources. More details can 
be found in \cite{Yukalov_31}.

\subsection{Nonequivalent representations}

The Bogolubov shift can be considered as a canonical transformation between
the field operators. Although such transformations for operators, enjoying 
the same commutation relations, are usually written as equalities, they 
actually are not equalities, but rather are replacements, since the left-hand 
and right-hand sides of the equalities correspond to operators defined on 
different Fock spaces. The operators connected by such transformations 
demonstrate {\it unitary nonequivalent representations of canonical 
commutation relations} or {\it nonequivalent operator representations}.
The nonequivalent operator representations occur in many commutation relations,
not necessarily involving a symmetry breaking.

As an example, let us consider the field operators in the momentum 
representation. Suppose the field operators $a_k$ are defined on the Fock space 
${\cal F}(a_k)$ generated by $a_k^\dgr$ and possessing a vacuum $|0 \rangle_a$, 
such that $a_k|0 \rangle_a = 0$. The Bogolubov canonical transformation
\be
\label{151}
a_k = u_k b_k + v^*_{-k} b_{-k}^\dgr \; , \qquad
b_k = u_k^* a_k - v_k^* a_{-k}^\dgr \; , \qquad |u_k|^2-|v_k|^2=1 \; ,
\ee
provides such an example of nonequivalent representations of canonical commutation 
relations. Note that the coefficient functions here can be chosen as real and 
symmetric, such that $u_k^* = u_{-k} = u_k$ and $v_k^* = v_{-k} = v_k$. 

Let us introduce the transformation operators
\be
\label{152}
\hat D \equiv \exp\left \{ \frac{1}{2}\; \sum_k \gm_k \left ( a_k a_{-k}
- a^\dgr_{-k} a_k ^\dgr \right ) \right \} \; , \qquad
\hat D^{-1} \equiv \exp\left \{-\; \frac{1}{2}\; \sum_k \gm_k \left (
a_k a_{-k} - a^\dgr_{-k} a_k ^\dgr \right ) \right \} \; ,
\ee
in which we use the notation
$$
u_k ={\rm cosh}\; \gm_k \; , \qquad v_k ={\rm sinh}\; \gm_k \; ,
\qquad \gm_k =\ln(u_k + v_k) \; . 
$$
Because of the commutator
$$
\left [ a_k, \; \frac{1}{2}\sum_p \gm_p \left ( a_p a_{-p} -
a^\dgr_{-p} a_p^\dgr \right )\right ] = - \gm_k a^\dgr_{-k} \; ,
$$
we have the transformations
\be
\label{153}
a_k = \hat D b_k\hat D^{-1} \; , \qquad b_k = \hat D^{-1} a_k \hat D\; .
\ee

The state $|0 \rangle_a$, that is a vacuum for $a_k$, is not a vacuum for $b_k$,
since
$$
b_k|0 \rangle_a \; = -v_k^* a_{-k}^\dgr|0 \rangle_a \neq 0 \; . 
$$
The vacuum for $b_k$, for which $b_k |0 \rangle_b \; = 0$, is 
\be
\label{154}
|0 \rangle_b \; =\hat D^{-1}|0 \rangle_a \; .
\ee
This state $|0 \rangle_b$ is not a vacuum for $a_k$, since
$$
a_k|0 \rangle_b \; = -v_k^* b_{-k}^\dgr|0 \rangle_b \; \neq 0 \; .
$$
The operators $b_k^\dgr$ generate the Fock space ${\cal F}(b_k)$. 

The commutator of the terms in the exponential (\ref{152}) is 
$\sum_k \gm_k^2 (1/2 + a_k^\dgr a_k)$, which is not proportional to the 
unity operator. Therefore the Baker-Hausdorff formula does not simplifies,
as in the previous section. However, this commutator gives a constant when
acting on the vacuum $|0\rangle_a$. Keeping this in mind, it is possible 
to use the approximate Baker-Hausdorff formula for $\hat D^{-1}$ acting 
on $|0>_a$, omitting higher-order commutators \cite{Umezawa_32}. This gives 
$$
\hat D^{-1}|0 \rangle_a \; \approx \exp\left ( \frac{1}{2}\; \sum_k \gm_k a_{-k}^\dgr
a_k^\dgr \right ) \exp\left (-\; \frac{1}{2}\; \sum_k \gm_k a_k
a_{-k} \right ) \exp\left \{ -\;\frac{1}{4}\; \sum_k \gm_k^2 (1+ 2
a_k^\dgr a_k) \right \} |0 \rangle_a \; .
$$
In view of the equality
$$
\exp\left ( -\; \frac{1}{2}\; \sum_k \gm_k^2 a_k^\dgr a_k \right )
|0 \rangle_a \; = |0 \rangle_a \; ,
$$
it follows
$$
|0 \rangle_b\; = \hat D^{-1}|0 \rangle_a = \exp\left ( - \; \frac{1}{4} \; \sum_k
\gm_k^2 \right ) \; \exp\left (\frac{1}{2}\; \sum_k \gm_k a_{-k}^\dgr
a_k^\dgr \right ) |0 \rangle_a \; .
$$
For a macroscopic system
$$
\sum_k \gm_k^2 =  V \int \gm_k^2 \; \frac{d\bk}{(2\pi)^3} \; .
$$
Taking into account that
$$
_a\langle 0| \exp \left (\frac{1}{2} \; \sum_k \gm_k a_{-k}^\dgr
a_k^\dgr \right ) |0 \rangle_a \; = 1 \; , 
$$
we find
$$
_a\langle 0|0 \rangle_b \; = \exp \left \{ -\;\frac{V}{4}\; \int \gm_k^2 \;
\frac{d\bk}{(2\pi)^3} \right \} \; .
$$
Since the integral in the exponential is positive, we get
\be
\label{155}
_a\langle 0|0 \rangle_b \; \simeq 0 \qquad (N\ra \infty) \; .
\ee
That is, the vacua are asymptotically orthogonal, hence the spaces 
${\cal F}(a_k)$ and ${\cal F}(b_k)$ are also asymptotically orthogonal. 

Note that the transformation operator $\hat D$ is not unitary, because 
$\hat D^+$ is not defined on the same space. Thus, we have a unitary 
nonequivalent operator representations, where $\hat D$ transforms 
${\cal F}(b_k)$ into ${\cal F}(a_k)$ and $\hat D^{-1}$ transforms 
${\cal F}(a_k)$ into ${\cal F}(b_k)$. The unitary nonequivalent 
representation of canonical commutation relations (\ref{151}), strictly 
speaking, should be understood as the replacement
$a_k \longrightarrow u_k b_k + v^*_{-k} b^\dgr_{-k}$. 

Calculating the averages of operators, it is important to keep in mind the 
spaces these operators are defined on.

\subsection{Phase operator}

One often defines a phase operator by resorting to the Dirac representation
$$
\psi(\br) \models e^{i\hat\vp(\br)} \; \sqrt{\hat n(\br)} \; ,
$$
with $\hat\vp(\br)$ being the phase operator and 
$\hat n(\br) \equiv \psi^\dgr(\br)\psi(\br)$. This representation requires
that $\hat\vp(\br)$ be Hermitian, $\hat\vp^+(\br) =\hat\vp(\br)$. Then we have
$$
\psi^\dgr(\br) \models \sqrt{\hat n(\br)}\; e^{-i\hat\vp(\br)} \; , \qquad
e^{i\hat\vp(\br)} \models \psi(\br) \left [ \hat n(\br)\right ]^{-1/2} \; .
$$
Since $[\psi(\br),\hat n(\br')]=\psi(\br)\dlt(\br-\br')$,  one has
$$
\left [ e^{i\hat\vp(\br)},\; \hat n(\br')\right ] \models
e^{i\hat\vp(\br)}\; \dlt(\br-\br') \; ,
$$
from where one gets the commutation relation
$$
\left [ \hat n(\br),\; \hat\vp(\br')\right ]  = i \dlt(\br-\br') \; . 
$$
But then one comes to the commutator
$$
[\hat N,\hat\vp(\br)] \models i \; , \qquad \hat N \equiv \int \hat n(\br)\; d\br  
$$
valid for any $\br$. This implies that $\hat\vp(\br)$ does not depend on $\br$, 
which is certainly strange. 

An operator equality is valid in the weak sense, when it holds for all matrix 
elements with respect to the functions from the considered space. For the number 
basis $\{|n \rangle\}$, such that 
$$
\hat N|n>=n|n> \; , \qquad
(n - n') < n|\hat\vp(\br)|n'>\; = i\dlt_{nn'} \; ,
$$
setting $n = n'$, we obtain the equality $i \models 0$, which does not have sense.  
The conclusion is that such a representation does not exist.

Another often used incorrect representation is
$$
\psi(\br) \models \sqrt{\rho_0(\br)}\; e^{i\hat\vp(\br)} \; ,
$$
with $\rho_0(\br)$ being the density of condensed particles. But then
$$
\hat n(\br) =\psi^\dgr(\br)\psi(\br) \models \rho_0(\br) \; , \qquad
\hat N = \int \rho_0(\br)\; d\br \models N_0 \; .
$$
From the commutator $[\psi(\br),\hat N] = \psi(\br)$, and for $\hat N = N_0$, 
one has $\psi(\br) \models 0$. The conclusion is again that this 
representation does not exist. The phase operator cannot be introduced by the
considered relations. 

This does not contradict the possibility of introducing a phase for 
non-operator functions. For instance, the coherent state, given by the 
eigenproblem $\psi(\br)|\eta \rangle = \eta(\br)|\eta \rangle$, defines the
coherent field, for which the Madelung form 
$$
\eta(\br) = \sqrt{\rho_0(\br)}e^{iS(\br)}, 
$$
where $\rho_0(\br) = |\eta(\br)|^2$, can be employed. 

The ways of correctly introducing phase operators have been discussed in 
the review articles \cite{Carruthers_33,Lynch_34,Kochetov_35,Pegg_36}.

\subsection{Noncommutativity of transformations}

As is explained above, canonical transformations can connect operators 
acting on different Fock spaces. Therefore it is necessary to be careful 
accomplishing practical calculations. The general rule is to consistently
work with one Fock space, accomplishing calculations after canonical
transformations.   

To illustrate what kind of problems can arise, let us pose the question: 
Is there a difference in the result of the following procedures?

\vskip 2mm

(i) Suppose we, first, realize the Bogolubov shift, that is, we pass from
the Fock space ${\cal F}(\psi)$ to the Fock space ${\cal F}(\psi_1)$. Then 
we accomplish some actions with the operators $\psi_1(\br)$ on the space 
${\cal F}(\psi_1)$.

\vskip 2mm

(ii) First, we accomplish the equivalent actions with $\psi(\br)$ on the 
space ${\cal F}(\psi)$. And then, we realize the Bogolubov shift, hence the 
transformation from the space ${\cal F}(\psi)$ to the space ${\cal F}(\psi_1)$.

\vskip 2mm

The general answer to this question is "no" - these actions are not commutative.

\vskip 2mm

All transformations with operators must be accomplished in the same
space. As an illustration, let us consider the Bogolubov shift
$\psi(\br)\longrightarrow\hat\psi(\br) = \eta(\br)+\psi_1(\br)$ and the 
Hartree-Fock-Bogolubov (HFB) approximation. For simplicity, we assume a 
uniform system, when $\eta(\br)=\eta=\sqrt{\rho_0}$. 

Let us be interested in the pair correlation function
$$
g(\br_{12}) = \frac{1}{\rho^2} \langle \hat\psi^\dgr(\br_1)
\hat\psi^\dgr(\br_2) \hat\psi(\br_2)
\hat\psi(\br_1) \rangle_{ {\cal F}(\psi_1)} \; ,
$$
where $\br_{12}\equiv\br_1-\br_2$. In what follows, for short, we write
$\langle \ldots \rangle_{ {\cal F}(\psi_1)} = \langle \ldots \rangle$.

Realizing, first, the Bogolubov shift and then the HFB approximation 
for $\psi_1(\br)$, we have
$$
\langle \psi_1^\dgr(\br_1) \psi_1(\br_1) \psi_1(\br_2) \rangle \; =
\rho_1 \langle \psi_1(\br_2) \rangle + \rho_1(\br_2,\br_1) \langle \psi_1(\br_1) \rangle +
\sgm_1(\br_1,\br_2) \langle \psi_1^\dgr(\br_1) \rangle \; ,
$$
where the normal and anomalous averages, respectively, are
$$
\rho_1(\br_1,\br_2)\equiv \langle \psi_1^\dgr(\br_2)\psi_1(\br_1) \rangle \; ,
\quad \rho_1=\rho_1(\br,\br) \; , \quad
\sgm_1(\br_1,\br_2)\equiv \langle \psi_1(\br_2)\psi_1(\br_1) \rangle \; ,
\quad \sgm_1=\sgm_1(\br,\br) \; .
$$
In view of the conservation of quantum numbers, $\langle \psi_1(\br) \rangle = 0$, 
the HFB approximation gives
$$
\langle \psi_1^\dgr(\br_1) \psi_1(\br) \psi_1(\br_2) \rangle \; = 0 \; .
$$
For the four-operator term, the HFB approximation yields 
$$
\langle \psi_1^\dgr(\br_1) \psi_1^\dgr(\br_2) \psi_1(\br_2) \psi_1(\br_1) \rangle \; =
\rho_1^2 + |\rho_1(\br_1,\br_2)|^2  + |\sgm_1(\br_1,\br_2)|^2 \; .
$$
In this way, we find the pair correlation function
$$
g(\br_{12})  = 1 + \frac{2\rho_0}{\rho^2} \; {\rm Re}\; \left [
\rho_1(\br_1,\br_2) + \sgm_1(\br_1,\br_2)\right ] + \frac{1}{\rho^2}
\left\{ |\rho_1(\br_1,\br_2)|^2 + |\sgm_1(\br_1,\br_2)|^2 \right \} \; .
$$
This function satisfies the correct limit
$$
\lim_{r_{12}\ra\infty} g(\br_{12}) = 1 \qquad (r_{12} \equiv |\br_{12}|) \; .
$$
 
Now let us, first, use the HFB approximation for $\psi(\br)$ and then, 
make the Bogolubov shift. This gives
$$
g(\br_{12}) \models 1 + \frac{2\rho_0^2}{\rho^2} + \frac{2\rho_0}{\rho^2}\;
{\rm Re}\; \left [\rho_1(\br_1,\br_2) + \sgm_1(\br_1,\br_2)\right ]
+ \frac{1}{\rho^2} \left\{ |\rho_1(\br_1,\br_2)|^2
+ |\sgm_1(\br_1,\br_2)|^2 \right \} \; .
$$
The latter correlation function yields the limit
$$
\lim_{r_{12}\ra\infty} g(\br_{12}) \models 1 + \frac{2\rho_0^2}{\rho^2} \; ,
$$
which is not correct when $\rho_0 \neq 0$. 

Thus the correct order of actions is when, first, one accomplishes the 
Bogolubov shift and then invokes the HFB approximation for $\psi_1(\br)$.
The origin of the noncommutativity of these actions is that the function 
$\eta$ does not have the same commutation relations as $\psi_1(\br)$.

\subsection{Action functional}

The evolution equations can be derived in two ways, either by employing
the Heisenberg equations of motion or by varying an effective action 
functional. These two ways are completely equivalent 
\cite{Yukalov_24,Yukalov_37}. The variation of an action functional is
called the principle of minimal action \cite{Kleinert_28}.

For a system with Bose-Einstein condensate, we break the gauge symmetry by 
accomplishing the Bogolubov shift, thus, passing from the field operator 
$\psi(\br)$, defined on the Fock space ${\cal F}(\psi)$, to the operator 
$\hat\psi(\br)$, acting on the space ${\cal F}(\psi_1)$,
\be
\label{156}
\hat\psi(\br)=\eta(\br)+\psi_1(\br) \; .
\ee
Thus we need to consider the energy Hamiltonian
\be
\label{157}
\hat H = \int \hat\psi^\dgr(\br) \left ( - \; \frac{\nabla^2}{2m} +
U\right ) \hat\psi(\br) \; d\br + \frac{1}{2} \; \int
\hat\psi^\dgr(\br) \hat\psi^\dgr(\br') \Phi(\br-\br')
\hat\psi(\br') \hat\psi(\br)\; d\br d\br' \; ,
\ee
in which $U = U(\br,t)$ is an external potential, and the interaction 
potential $\Phi(-\br) = \Phi(\br)$ is assumed to be integrable.

The Lagrangian is 
\be
\label{158}
\hat L[\hat\psi] = \int \hat\psi^\dgr(\br)\; i\;
\frac{\prt}{\prt t}\; \hat\psi(\br) \; d\br - \hat H \; ,
\ee
whose time integral gives the action functional. The evolution equations
are derived from the conditional extremum of the action functional. 
Additional constraints, that are to be taken into account, include the 
normalization condition for the condensate function
\be
\label{159}
N_0 =\int \rho_0(\br)\; d\br \; , \qquad
\rho_0\equiv |\eta(\br)|^2 
\ee
and the normalization condition for the number of uncondensed atoms
\be
\label{160}
N_1= \langle \hat N_1 \rangle \; , \qquad 
\hat N_1 \equiv \int \psi_1^\dgr(\br) \psi_1(\br) \; d\br \; .
\ee
The total number of particles $N$ is given by the average
$$
N = \langle \hat N \rangle \; , \qquad 
\hat N = \int \hat\psi^\dgr(\br) \hat\psi(\br)\; d\br \; ,
$$ 
which can be written as
$$
N = \int\rho(\br)\; d\br \; , \qquad
\rho(\br) \equiv \; \langle \hat\psi^\dgr(\br) \hat\psi(\br) \rangle \; .
$$
All averages are calculated in the space ${\cal F}(\psi_1)$. Because 
of the orthogonality condition (\ref{149}), the operator of the total 
number of atoms is the sum $\hat N = N_0 +\hat N_1$ .

Since the gauge symmetry is broken, it could happen that 
$\langle \psi_1(\br)\rangle$ could be nonzero, which would imply that
quantum numbers, such as spin and momentum, are not conserved and which 
would contradict the quantum-number conservation condition 
$\langle \psi_1(\br) \rangle = 0$. Therefore it is necessary to impose 
one more restriction 
\be
\label{161}
\langle \hat\Lbd \rangle = 0 \; , \qquad  
\hat\Lbd \equiv \int \left [\lbd(\br)\psi_1^\dgr(\br) +
\lbd^*(\br)\psi_1(\br)\right ] \; d\br \; ,
\ee
guaranteeing the quantum-number conservation condition (\ref{46}). 

Thus the effective action functional, taking into account the above
constraints, reads as 
\be
\label{162}
A[\eta,\psi_1]  =\int \left (\hat L[\hat\psi] + \mu_0 N_0 +
\mu_1\hat N_1 + \hat\Lbd\right ) \; dt \ ,
\ee
where $\mu_0,\; \mu_1$, and $\lbd(\br)$ are Lagrange multipliers.

Introducing the grand Hamiltonian $H[\eta,\psi_1]$ as 
\be
\label{163}
H[\eta,\psi_1] \equiv \hat H - \mu_0 N_0 -\mu_1\hat N_1 -\hat\Lbd 
\ee
and the effective Lagrangian
$$
L[\eta,\psi_1]  =\int \left [ \eta^*(\br)\; i\; \frac{\prt}{\prt t}\;
\eta(\br) + \psi_1^\dgr(\br)  i\; \frac{\prt}{\prt t}\; \psi_1(\br)
\right ]\; d\br - H[\eta,\psi_1] \; ,
$$
we get the effective action functional 
\be
\label{164}
A[\eta,\psi_1]  = \int L [\eta,\psi_1] \; dt \; .
\ee

The equations of motion for the condensate function and the field operator 
of uncondensed atoms are given by the extrema of the effective action 
functional:
\be
\label{165}
 \left \langle \frac{\dlt A[\eta,\psi_1]}{\dlt\eta^*(\br,t)} \right \rangle = 0 \; , 
\qquad \frac{\dlt A[\eta,\psi_1]}{\dlt\psi_1^\dgr(\br,t)} = 0 \; .
\ee
The effective action functional correctly represents the considered 
statistical system only when it takes into account all constraints uniquely 
defining this system \cite{Yukalov_30,Yukalov_38}.

\subsection{Grand Hamiltonian}

Substituting into the grand Hamiltonian (\ref{163}) the Bogolubov shift 
(\ref{156}) leads to the sum 
\be
\label{166}
H[\eta,\psi_1] = \sum_{n=0}^4 H^{(n)} 
\ee
of five terms $H^{(n)}$ arranged according to the number of factors 
$\psi_1(\br)$ or $\psi_1^\dgr(\br)$ in the operator products entering the
corresponding parts. Thus, the zero-order term 
\be
\label{167}
H^{(0)} = \int \eta^*(\br) \left ( -\; \frac{\nabla^2}{2m} + U
- \mu_0\right )\eta(\br)\; d\br + \frac{1}{2} \; \int \Phi(\br-\br')
|\eta(\br)|^2 |\eta(\br')|^2\; d\br d\br' 
\ee
contains no operators of uncondensed atoms. Here $\mu_0$ is the condensate 
chemical potential guaranteeing the normalization condition (\ref{159}).

For the quantum-number conservation condition, it is necessary and sufficient 
that there would be no terms linear in $\psi_1(\br)$, which requires the 
Lagrange multiplier
$$
\lbd(\br) = \left ( -\; \frac{\nabla^2}{2m} + U \right )
\eta(\br) + \int \Phi(\br-\br') |\eta(\br')|^2 \eta(\br)\; d\br' \; .
$$
Then the first-order term is zero, $H^{(1)} = 0$. If the Hamiltonian would
contain linear in $\psi_1(\br)$ terms, the average $\langle \psi_1(\br) \rangle$ 
would be nonzero \cite{Yukalov_38}.

The second-order term is
$$
H^{(2)} = \int \psi_1^\dgr(\br) \left ( -\;\frac{\nabla^2}{2m} + U -
\mu_1 \right ) \psi_1(\br)\; d\br + \int \Phi(\br-\br')
\left [ |\eta(\br)|^2 \psi_1^\dgr(\br')\psi_1(\br') + \right.
$$
\be
\label{168}
\left.
+ \eta^*(\br)\eta(\br')\psi_1^\dgr(\br')\psi_1(\br) +
\frac{1}{2}\; \eta^*(\br)\eta^*(\br')\psi_1(\br')\psi_1(\br) +
\frac{1}{2}\; \eta(\br)\eta(\br')\psi_1^\dgr(\br) \psi_1^\dgr(\br')
\right ]\; d\br d\br' \; .
\ee
The third-order term reads as
\be
\label{169}
H^{(3)} = \int \Phi(\br-\br') \left [ \eta^*(\br) \psi_1^\dgr(\br')
\psi_1(\br') \psi_1(\br) + \psi_1^\dgr(\br) \psi_1^\dgr(\br')
\psi_1(\br') \eta(\br) \right ]\; d\br d\br' \; .
\ee
And the fourth-order term is
\be
\label{170}
H^{(4)} =  \frac{1}{2}\; \int \psi_1^\dgr(\br) \psi_1^\dgr(\br')
\Phi(\br-\br') \psi_1(\br') \psi_1(\br) \; d\br d\br' \; .
\ee

The variation of action (\ref{165}) is equivalent to the equations involving 
the variation of the grand Hamiltonian. The condensate function is described
by the equation
\be
\label{171}
i\; \frac{\prt}{\prt t}\; \eta(\br,t) =
\left \langle \frac{\dlt H[\eta,\psi_1]}{\dlt\eta^*(\br,t)} \right \rangle \; , 
\ee
while the equation for the field operator of uncondensed atoms is
\be
\label{172}
i\; \frac{\prt}{\prt t}\; \psi_1(\br,t) =
\frac{\dlt H[\eta,\psi_1]}{\dlt\psi_1^\dgr(\br,t)} \; .
\ee
As has been mentioned above, the variational equation (\ref{172}) is 
equivalent to the Heisenberg equation of motion, since
\be
\label{173} 
\frac{\dlt H[\eta,\psi_1]}{\dlt\psi_1^\dgr(\br,t)} =
\left [ \psi_1(\br,t),\; H[\eta,\; \psi_1]\right ] \; .
\ee

Equation (\ref{171}) results in the condensate-function equation 
\be
\label{174}
i\; \frac{\prt}{\prt t}\; \eta(\br,t) = \left ( -\; \frac{\nabla^2}{2m}
+ U -\mu_0\right ) \eta(\br) + \int \Phi(\br-\br') \left [ |\eta(\br')|^2
\eta(\br) + \langle \hat X(\br,\br') \rangle \right ] \; d\br' \; ,
\ee
in which $\eta(\br) \equiv \eta(\br,t)$, $\psi_1(\br) \equiv \psi_1(\br,t)$,
with the notation
$$
\hat X(\br,\br') \equiv \psi_1^\dgr(\br') \psi_1(\br') \eta(\br) +
\psi_1^\dgr(\br') \eta(\br') \psi_1(\br) + \eta^*(\br') \psi_1(\br')
\psi_1(\br) + \psi_1^\dgr(\br') \psi_1(\br') \psi_1(\br) \; .
$$
The field operator of uncondensed atoms satisfies the equation 
$$
i\; \frac{\prt}{\prt t}\; \psi_1(\br) = \left ( -\; \frac{\nabla^2}{2m}
+ U -\mu_1 \right ) \psi_1(\br) +
$$
\be
\label{175}
+ \int \Phi(\br-\br') \left [ |\eta(\br')|^2  \psi_1(\br) + \eta^*(\br')
\eta(\br) \psi_1(\br') + \eta(\br') \eta(\br) \psi_1^\dgr(\br') +
\hat X(\br,\br') \right ]\; d\br' \; .
\ee

Recall that, after accomplishing the Bogolubov shift, all operators are
defined on the Fock space ${\cal F}(\psi_1)$.

\subsection{Conservation laws}

Correctly defining the operators on the Fock space ${\cal F}(\psi_1)$, with
broken gauge symmetry, preserves all conservation laws corresponding to the
relations between the operators of local observables. 

Let us derive the continuity equations. For this purpose, we define the 
density of current for condensed atoms, 
\be
\label{176}
\bj_0(\br) \equiv -\; \frac{i}{2m}\left [ \eta^*(\br) \;
\nabla\eta(\br) -\eta(\br)\; \nabla\eta^*(\br)\right ] \; ,
\ee
and the density of current for uncondensed atoms, 
\be
\label{177}
\hat\bj_1(\br) \equiv -\; \frac{i}{2m}\left \{ \psi_1^\dgr(\br)\;
\nabla\psi_1(\br) -\left [\nabla\psi_1^\dgr(\br)\right ]\psi_1(\br)
\right \} \; .
\ee
Also, we introduce the operator source 
$$
\hat\Gm(\br) \equiv \int \Phi(\br-\br') \left [ \hat \Xi(\br,\br')
+ \hat\Xi^+(\br,\br')\right ] \; d\br' \; ,
$$
in which
$$
\hat\Xi(\br,\br') = i \eta^*(\br)\left [ \psi_1^\dgr(\br')
\eta(\br') + \eta^*(\br') \psi_1(\br') +
\psi_1^\dgr(\br') \psi_1(\br') \right ] \; \psi_1(\br) \; .
$$

Using these notations, we find the equation connecting the density 
and current of condensate,
\be
\label{178}
\frac{\prt}{\prt t}\; \rho_0(\br) +\nabla \cdot\bj_0(\br) =
- \hat\Gm(\br)  \; , 
\ee
and the operator relation between the operator density of uncondensed 
atoms and their density of current 
\be
\label{179}
\frac{\prt}{\prt t} \; \hat n_1(\br) + \nabla \cdot\hat\bj_1(\br) = \hat\Gm(\br) \; ,
\ee
where $\hat n_1(\br) \equiv \psi_1^\dgr(\br)\psi_1(\br)$. 

For the total operator density of atoms and the total density of current  
$$
\hat n(\br)\equiv \rho_0(\br) +\hat n_1(\br) \; , \qquad
\hat\bj(\br)\equiv \bj_0(\br) +\hat\bj_1(\br) \; ,
$$
we obtain the continuity equation
\be
\label{180}
\frac{\prt}{\prt t}\;  \hat n(\br) + \nabla \cdot\hat\bj(\br)= 0 \; .
\ee

Similarly, it is straightforward to derive other relations between the 
operators of local observables characterizing operator conservation laws.

\subsection{Condensate function}

The condensate function $\eta({\bf r}, t)$ is the solution to equation (\ref{174}).
Accomplishing there the averaging over the space ${\cal F}(\psi_1)$, we have
$$
\langle \hat X(\br,\br') \rangle \; = \rho_1(\br')\eta(\br) + \rho_1(\br,\br')\eta (\br')
+ \sgm_1(\br,\br')\eta^*(\br') + \langle \psi_1^\dgr(\br')\psi_1(\br') \psi_1(\br) \rangle \; ,
$$
where
$$
\rho(\br)=\rho_0(\br)+\rho_1(\br) \; , \qquad
\rho_0(\br) \equiv |\eta(\br)|^2 \; , \qquad \rho_1(\br)\equiv \;
\langle \psi_1^\dgr(\br)\psi_1(\br) \rangle \; .
$$
Then equation (\ref{174}) becomes
$$
i\; \frac{\prt}{\prt t}\; \eta(\br) = \left ( -\; \frac{\nabla^2}{2m} +
U - \mu_0\right ) \eta(\br) +
$$
\be
\label{181}
+ \int \Phi(\br-\br') \left [ \rho(\br')\eta(\br) + \rho_1(\br,\br')
\eta(\br') + \sgm_1(\br,\br')\eta^*(\br') + \langle \psi_1^\dgr(\br')\psi_1(\br')
\psi_1(\br) \rangle \right ]\; d\br'\; .
\ee

In equilibrium, the function $\eta(\br)$ does not depend on time, which gives
the eigenproblem
$$
\mu_0\eta(\br) = \left [ -\; \frac{\nabla^2}{2m} + U \right ] \eta(\br) +
$$
\be
\label{182}
+ \int \Phi(\br-\br') \left [ \rho(\br')\eta(\br) + \rho_1(\br,\br')
\eta(\br') + \sgm_1(\br,\br')\eta^*(\br') + \langle \psi_1^\dgr(\br')\psi_1(\br')
\psi_1(\br) \rangle \right ]\; d\br'\; ,
\ee
where $\mu_0$ plays the role of an eigenvalue. Generally, this eigenproblem can
define the whole spectrum of eigenvalues. The condition of equilibrium implies
that $\mu_0$ is given by the lowest eigenvalue. 

Note that if we would take $\eta(\br,t)\propto e^{-i\lbd t}$, this would be 
just a redefinition of $\mu_0$. So, it is sufficient to accept that in 
equilibrium the condensate function $\eta(\br)$ does not depend on time. 

In the case of a uniform system, where $U = 0$, we have 
$\eta(\br) = \eta = \sqrt{\rho_0}$ and $\rho(\br) = \rho$. The quantities
$\rho_1(\br,\br')$ and $\sgm_1(\br,\br')$ depend on $\br-\br'$. Then equation
(\ref{182}) reduces to 
\be
\label{183}
\mu_0\eta = \rho \Phi_0 \eta + \int \Phi(\br) \left [ \rho_1(\br,0)\eta
+ \sgm_1(\br,0)\eta^* + \langle \psi_1^\dgr(0)\psi_1(0)\psi_1(\br) \rangle \right ] \; d\br \; .
\ee

When we transform these general equations by resorting to the 
Hartree-Fock-Bogolubov (HFB) approximation, we take into account that in this 
approximation 
$$
\psi_1^\dgr(\br') \psi_1(\br') \psi_1(\br) = \rho_1(\br')\psi_1(\br)
+ \rho_1(\br,\br') \psi_1(\br') + \sgm_1(\br,\br') \psi_1^\dgr(\br')\; .
$$
Hence
$$
\langle \psi_1^\dgr(\br') \psi_1(\br') \psi_1(\br)\rangle \; = \; 0 \; .
$$
The four-operator term takes the form
$$
\psi_1^\dgr(\br) \psi_1^\dgr(\br') \psi_1(\br') \psi_1(\br) =
\rho_1(\br)\psi_1^\dgr(\br') \psi_1(\br') +
\rho_1(\br')\psi_1^\dgr(\br)\psi_1(\br) +
\rho_1(\br,\br')\psi_1^\dgr(\br)  \psi_1(\br') +
$$
$$
+ \rho_1(\br',\br)\psi_1^\dgr(\br')\psi_1(\br) +
\sgm_1(\br,\br')\psi_1^\dgr(\br)\psi_1^\dgr(\br') +
\sgm_1^*(\br',\br)\psi_1(\br')\psi_1(\br) -
$$
$$
- \rho_1(\br)\rho_1(\br') - |\rho_1(\br,\br')|^2 - |\sgm_1(\br,\br')|^2 \; .
$$
Therefore, in the HFB approximation, the condensate-function equation becomes 
$$
i\; \frac{\prt}{\prt t}\; \eta(\br) = \left ( -\; \frac{\nabla^2}{2m} +
U - \mu_0\right ) \eta(\br) +
$$
\be
\label{184}
+ \int \Phi(\br-\br') \left [ \rho(\br')\eta(\br) + \rho_1(\br,\br')
\eta(\br') + \sgm_1(\br,\br')\eta^*(\br') \right ]\; d\br' \; .
\ee
And the equation for the field operator of uncondensed atoms is
$$
i\; \frac{\prt}{\prt t}\;\psi_1(\br) = \left ( -\; \frac{\nabla^2}{2m} +
U - \mu_1 \right )\psi_1(\br) +
$$
\be
\label{185}
+ \int \Phi(\br-\br') \left [ \rho(\br')\psi_1(\br) + \rho(\br,\br')
\psi_1(\br') + \sgm(\br,\br')\psi_1^\dgr(\br') \right ]\; d\br' \; ,
\ee
with
$$
\rho(\br,\br')\equiv \eta(\br)\eta^*(\br') + \rho_1(\br,\br')\; ,
\qquad \sgm(\br,\br')\equiv \eta(\br)\eta(\br') + \sgm_1(\br,\br') \; .
$$
In the latter equation, the terms quadratic in $\psi_1(\br)$ are omitted.

For an equilibrium system in the HFB approximation, we have the equation 
$$
\mu_0\eta(\br) =  \left [ -\; \frac{\nabla^2}{2m} + U\right ] \eta(\br) +
$$
\be
\label{186}
+  \int \Phi(\br-\br') \left [ \rho(\br')\eta(\br) +
 \rho_1(\br,\br')\eta(\br') + \sgm_1(\br,\br')\eta^*(\br')\right ] \; d\br' \; .
\ee

In the case of a uniform system, when the condensate function reduces to a 
constant $\eta(\br) = \eta$, equation (\ref{186}) simplifies to the equality 
\be
\label{187}
\mu_0 = \rho\Phi_0 + \int \Phi(\br) \left [\rho_1(\br,0)
+\sgm_1(\br,0)\right ]\; d\br \; .
\ee

Let us find out the condensate-function equation for the system vacuum state. 
Recall \cite{Yukalov_1} that the condensate function $\eta({\bf r}, t)$ is the 
coherent field related to the coherent state $|\eta \rangle$ that is the vacuum state
in the Fock space ${\cal F}(\psi_1)$. The coherent state $|\eta \rangle$ is not an 
eigenstate of the Hamiltonian $H[\psi]$ defined on the Fock space ${\cal F}(\psi)$.  
But the coherent state $|\eta \rangle$ is an eigenstate and a ground state for the 
Hamiltonian $H[\eta,\psi_1]$ acting on the Fock space ${\cal F}(\psi_1)$, so that
$$
H[\eta,\psi_1] |\eta\rangle = H^{(0)} |\eta \rangle,
$$
with $H^{(0)}$ being the Hamiltonian (\ref{167}). 

When averaging over the coherent state $|\eta \rangle$, that is a vacuum state in 
the Fock space ${\cal F}(\psi_1)$, we have $\rho_1(\br,\br') = \sgm_1(\br,\br') = 0$, 
since
$$
\psi_1(\br)|0 \rangle_1 = \psi_1(\br)|\eta \rangle = 0 \; .
$$
and
$$
_1\langle 0|\psi_1^\dgr(\br')\psi_1(\br)|0 \rangle_1 = 0 \; .
$$
Thus, the equation for the vacuum-state condensate function is the equation 
for the coherent field:
\be
\label{188}
i\; \frac{\prt}{\prt t}\;\eta(\br,t) =
\left ( -\; \frac{\nabla^2}{2m} + U - \mu_0 \right ) \eta(\br,t)
+ \int \Phi(\br-\br')|\eta(\br',t)|^2 \eta(\br,t)\; d\br' \; .
\ee

In the dilute-gas approximation, when the interaction potential is modeled
by the delta function, we get 
\be
\label{189}
i\; \frac{\prt}{\prt t}\;\eta(\br,t) =\left ( -\; \frac{\nabla^2}{2m} +
U - \mu_0\right ) \eta(\br,t) + \Phi_0 |\eta(\br,t)|^2 \eta(\br,t) \; .
\ee
In equilibrium, this reduces to the eigenproblem
\be
\label{190}
\left [ -\; \frac{\nabla^2}{2m} +
U(\br) +\Phi_0|\eta(\br)|^2\right ]\eta(\br) = \mu_0 \eta(\br) \; .
\ee
The latter equation is the eigenproblem, generally, defining a whole 
spectrum of eigenvalues corresponding to different coherent modes. The 
eigenvalue $\mu_0$ corresponds to the lowest among the spectrum of such 
coherent modes. In all the cases, the normalization condition
$$
\int |\eta(\br)|^2 d\br = N_0
$$
is assumed. 

The mathematical structure of equations (\ref{188}) and (\ref{189}) 
corresponds to the general {\it Nonlinear Schr\"{o}dinger} (NLS) equations 
\cite{Faddeev_39}, while (\ref{190}) is the stationary NLS equation. Such 
equations are employed for modelling many problems, for instance, surface 
waves on a deep fluid \cite{Zakharov_40}, electromagnetic waves in fiber 
optics \cite{Kivshar_41,Kartashov_42}, and Bose-Einstein condensates 
\cite{Pethick_43,Pitaevskii_44,Lieb_45,Letokhov_46}. 

The most general form (\ref{188}) of the equation for Bose systems was, 
first, advanced by Bogolubov during his lectures in the Moscow State University 
and Kiev State University and published in 1949 in his famous 
{\it Lectures on Quantum Statistics} \cite{Bogolubov_47}. Since then, these 
lectures have been republished numerous times, e.g. in 
\cite{Bogolubov_16,Bogolubov_18,Bogolubov_19}. Bogolubov analyzed in detail 
the particular case of equilibrium uniform systems 
\cite{Bogolubov_8,Bogolubov_9,Bogolubov_16,Bogolubov_18,Bogolubov_19}  
as well as the general case of nonequilibrium systems exhibiting superfluid 
hydrodynamics \cite{Bogolubov_48}.

Equations (\ref{188}) and (\ref{189}) were investigated by Gross in a series 
of papers \cite{Gross_49,Gross_50,Gross_51,Gross_52,Gross_53} and different 
nonuniform solutions were found, including periodic and vortex solutions. 

Note that the Ginzburg-Landau equation \cite{Ginzburg_54,Landau_55} for 
superconductors has also the structure of the NLS equation and reduces to 
the above equations under zero vector potential. Abrikosov \cite{Abrikosov_56}
found vortex solutions for the Ginzburg-Landau equation. Therefore, such 
solutions should also exist for the NLS equations (\ref{188}) and (\ref{189}),
for which the vortex solutions were found as well
\cite{Gross_52,Gross_53,Ginzburg_57,Wu_58,Pitaevskii_59}. The role of vortices
in superfluid hydrodynamics was emphasized by Iordanskii \cite{Iordanskii_60}.
 
Comparing the equations for the condensate function (\ref{181}) and (\ref{188}),
we see that the vacuum state can be approximately achieved in a physical system,
when practically all atoms are condensed, which requires asymptotically low 
temperature and asymptotically weak interactions. 

In an equilibrium system at finite temperature, the condensate function 
$\eta(\br)$ realizes a minimum of the grand thermodynamic potential
$$
\Om = - T\ln{\rm Tr}_{ {\cal F}(\psi_1)} \; \exp\left\{
-\bt H[\eta,\; \psi_1]\right \} \; .
$$
The extremum condition
$$
\frac{\dlt\Om}{\dlt\eta^*(\br)} = 
\; \langle \frac{\dlt H[\eta,\;\psi_1]}{\dlt\eta^*(\br)} \rangle \; = \; 0
$$
is in agreement with the evolution equation (\ref{171}) for an equilibrium 
system, when the condensate function does not depend on time. Thermodynamic 
stability conditions guarantee that this extremum is a minimum. The so 
minimized grand potential $\Om$ characterizes an absolutely stable 
equilibrium system.

\subsection{Thermodynamic self-consistency}

Statistical ensemble is a pair of the Fock space describing the system 
and a statistical operator. The Fock space for a Bose-condensed system 
is ${\cal F}(\psi_1)$. The statistical operator for an equilibrium system 
can be found from the principle of minimal information 
\cite{Yukalov_1,Yukalov_61} that implies conditional maximization of 
entropy \cite{Gibbs_62,Gibbs_63}.

Statistical conditions for an equilibrium system are as follows. The 
statistical operator is normalized, such that 
$1 = \langle \hat 1 \rangle = {\rm Tr}\hat\rho$, where $\hat 1$ is the  
unity operator in the space $\cF(\psi_1)$. The internal energy is
$E = \langle \hat H \rangle = {\rm Tr}\hat\rho \hat H$. The number of 
condensed atoms is $N_0 = \langle \hat N_0 \rangle$, with 
$\hat N_0\equiv N_0 \hat 1$. The number of uncondensed atoms is
$N_1 = \langle \hat N_1 \rangle$. And the quantum-number conservation 
condition reads as $\langle \hat\Lbd \rangle = 0$. This defines the
information functional,
$$
I[\hat\rho] = {\rm Tr} \hat\rho \ln \hat\rho +\lbd_0 \left (
{\rm Tr} \hat\rho -1\right ) + \bt
\left ( {\rm Tr} \hat\rho\hat H - E \right ) - \bt \mu_0
\left ( {\rm Tr} \hat\rho \hat N_0 - N_0 \right ) -
$$
\be
\label{191}
- \bt \mu_1 \left ( {\rm Tr} \hat\rho \hat N_1 - N_1 \right ) -
\bt {\rm Tr} \hat\rho \hat\Lbd \; .
\ee

The minimization of the information functional yields the statistical operator
\be
\label{192}
\hat\rho = \frac{1}{Z} \;
\exp\left\{ -\bt H[\eta,\psi_1] \right \} \; ,
\ee
with the same grand Hamiltonian as in the evolution equations and with the 
partition function
$$
Z \equiv {\rm Tr}\exp \left \{ - \bt H[\eta,\psi_1] \right \} \; .
$$
Both dynamic and equilibrium properties are governed by the same grand Hamiltonian.

For an equilibrium system, the Bogolubov-Ginibre condition 
\be
\label{193}
\frac{\prt\Om}{\prt N_0} \; = 
\; \left \langle \frac{\prt H[\eta,\psi_1]}{\prt N_0} \right \rangle \; = \; 0
\ee
holds, since
$$
\frac{\prt\Om}{\prt N_0} =  \int \left [
\frac{\dlt\Om}{\dlt\eta(\br)} \; \frac{\prt\eta(\br)}{\prt N_0} +
\frac{\dlt\Om}{\dlt\eta^*(\br)} \;
\frac{\prt\eta^*(\br)}{\prt N_0} \right ] \; d\br
$$
and
$$
\frac{\dlt\Om}{\dlt\eta(\br)} = 0 \; .
$$

The free energy $F = \Om + \mu_0 N_0 + \mu_1 N_1$, under the total number 
of particles $N$ being fixed, can also be written as $F = \Om + \mu N$,
where $\mu$ is the system chemical potential. Comparison of these expressions
gives the system chemical potential 
\be
\label{194}
\mu=\mu_0 n_0 + \mu_1 n_1 \; .
\ee

Free energy and grand potential are connected through the differential 
relations
$$
dF = - SdT - PdV + \mu dN \; , \qquad
d\Om = - SdT - PdV - Nd\mu \; .
$$
It is easy to show that the thermodynamic and Gibbs entropies coincide,
$$
S=-\; \frac{\prt\Om}{\prt T} =
-{\rm Tr}\hat\rho \ln\hat\rho \; .
$$

For the particle variance, we have
\be
\label{195}
{\rm var}(\hat N) = T\; \frac{\prt N}{\prt\mu} \; .
\ee

This can be proved as follows. From the equations 
$$
N_0 = - \; \frac{\prt\Om}{\prt\mu_0} \; , \qquad
N_1 = -\; \frac{\prt\Om}{\prt\mu_1} \; , \qquad
N = -\; \frac{\prt\Om}{\prt\mu} = N_0 + N_1 
$$
we get
$$
\frac{\prt\Om}{\prt\mu} = \frac{\prt\Om}{\prt\mu_0} +
\frac{\prt\Om}{\prt\mu_1} \; .
$$
Differentiating again leads to the relation
$$
\frac{\prt^2\Om}{\prt\mu^2} = \frac{\prt^2\Om}{\prt\mu_0^2} +
\frac{\prt^2\Om}{\prt\mu_1^2} +
2\; \frac{\prt^2\Om}{\prt\mu_0\prt\mu_1} \; .
$$
By direct calculations, we find
$$
\frac{\prt^2\Om}{\prt\mu_0^2} = -\bt{\rm var}(\hat N_0) \; , \qquad
\frac{\prt^2\Om}{\prt\mu_1^2} = -\bt{\rm var}(\hat N_1) \; , \qquad
\frac{\prt^2\Om}{\prt\mu_0\prt\mu_1} = -
\bt{\rm cov}(\hat N_0,\hat N_1) \; .
$$
Hence,
$$
\frac{\prt N}{\prt\mu} = -\; \frac{\prt^2\Om}{\prt\mu^2} =
\bt{\rm var}(\hat N) \; ,
$$
which gives equation (\ref{195}).

In this way, it is straightforward to show that all thermodynamic relations
are valid \cite{Yukalov_24,Yukalov_30,Yukalov_38}.

\subsection{Nonuniform matter}

The HFB approximation can be used for the general case of a nonuniform system.
Defining the normal and anomalous averages
\be
\label{196}
\rho_1(\br,\br')\equiv\; \langle \psi_1^\dgr(\br')\psi_1(\br) \rangle \; , \qquad
\sgm_1(\br,\br')\equiv \; \langle \psi_1(\br')\psi_1(\br) \rangle \; ,
\ee
we remember that the field operators depend on time, $\psi_1(\br) = \psi_1(\br,t)$, 
while time is not explicitly shown. So that, in general, the consideration of 
the present section is applicable to nonequilibrium systems as well. 

We use the notation
$$
\rho(\br,\br')\equiv \eta(\br)\eta^*(\br') + \rho_1(\br,\br') \; , \qquad
\sgm(\br,\br')\equiv \eta(\br)\eta(\br') + \sgm_1(\br,\br') \; ,
$$
\be
\label{197}
\rho_0(\br) =|\eta(\br)|^2 \; , \qquad \rho_1(\br)=\rho_1(\br,\br)\; ,
\qquad \rho(\br) =\rho_0(\br) + \rho_1(\br) \; ,
\ee
and keep in mind that $\langle \psi_1(\br) \rangle = 0$. 

If the Hamiltonian would contain linear in $\psi_1(\br)$ terms, then 
$ \langle \psi_1(\br) \rangle$ would not be zero. Therefore the Lagrange 
multiplier $\lbd(\br)$ is chosen so that to kill the linear in 
$\psi_1(\br)$ terms. Then in the HFB approximation, we have $H^{(3)} = 0$,
implying that all matrix elements of this term are zero.

In the HFB approximation, Hamiltonian (\ref{166}) becomes 
$$
H_{HFB} = E_{HFB} + \int \psi_1^\dgr(\br) \left ( -\;
\frac{\nabla^2}{2m} + U - \mu_1\right )\psi_1(\br) \; d\br +
$$
$$
+ \int \Phi(\br-\br') \left [ \rho(\br')\psi_1^\dgr(\br)\psi_1(\br) +
\rho(\br',\br)\psi_1^\dgr(\br')\psi_1(\br) + \frac{1}{2}\;
\sgm(\br,\br')\psi_1^\dgr(\br')\psi_1^\dgr(\br) + \right.
$$
\be
\label{198}
+ \left.
\frac{1}{2}\sgm^*(\br,\br')\psi_1(\br')\psi_1(\br)\right ]\; d\br d\br' \; ,
\ee
where
\be
\label{199}
E_{HFB} = H^{(0)} - \; \frac{1}{2}\; \int \Phi(\br-\br') \left [
\rho_1(\br)\rho_1(\br') + |\rho_1(\br,\br')|^2 + |\sgm_1(\br,\br')|^2
\right ]\; d\br d\br' \; .
\ee

For a nonuniform system, the Bogolubov canonical transformations read as
\be
\label{200}
\psi_1(\br) = \sum_k \left [ u_k(\br) b_k + v_k^*(\br) b_k^\dgr
\right ]  \; , \qquad
\psi_1^\dgr(\br) = \sum_k \left [ u_k^*(\br) b_k^\dgr + v_k(\br)
b_k\right ] \; , 
\ee
where $b_k = b_k(t)$ and $k$ is a set of quantum numbers. The inverse 
transformations are
\be
\label{201}
b_k = \int \left [ u_k^*(\br)\psi_1(\br) - v_k^*(\br)\psi_1^\dgr(\br)
\right ]\; d\br \; , \qquad
b_k^\dgr = \int \left [ u_k(\br)\psi_1^\dgr(\br) - v_k(\br)\psi_1(\br)
\right ] \; d\br \; .
\ee
The requirement that $b_k$ and $b_k^\dgr$ be Bose operators yields 
the equations
$$
\sum_k \left [ u_k(\br) v_k^*(\br') - v_k^*(\br) u_k(\br')
\right ] = 0 \; , \qquad
\sum_k \left [ u_k(\br) u_k^*(\br') - v_k^*(\br) v_k(\br')
\right ] = \dlt(\br-\br') \; ,
$$
$$
\int \left [ u_k(\br) v_p(\br) - v_k(\br) u_p(\br)
\right ] \; d\br = 0 \; , \qquad
\int \left [ u_k^*(\br) u_p(\br) - v_k^*(\br) v_p(\br)
\right ]\; d\br = \dlt_{kp} \; .
$$

We introduce the quantities
\be
\label{202}
\om(\br,\br') \equiv \left [ -\; \frac{\nabla^2}{2m} + U(\br) - \mu_1
+ \int \Phi(\br-\br')\rho(\br')\; d\br'\right ] \dlt(\br-\br')
+ \Phi(\br-\br')\rho(\br,\br') 
\ee
and
\be
\label{203}
\Dlt(\br,\br') \equiv \Phi(\br-\br')\sgm(\br,\br') \; .
\ee

Requiring that the resulting Hamiltonian be diagonal imposes the
Bogolubov diagonalization equations 
$$
\int \left [ \om(\br,\br') u_k(\br') +\Dlt(\br,\br') v_k(\br')
\right ] \; d\br' = \ep_k u_k(\br) \; ,
$$
\be
\label{204}
\int \left [ \om^*(\br,\br') v_k(\br') +\Dlt^*(\br,\br') u_k(\br')
\right ]\; d\br' = -\ep_k v_k(\br) \; .
\ee

The same equations can be obtained by substituting the transformations 
for $\psi_1(\br)$ and $\psi_1^\dgr(\br)$ into their equations of motion 
within the $H_{HFB}$ approximation, that is the linearized equation 
for $\psi_1(\br)$.

Employing the Bogolubov equations (\ref{204}) transforms Hamiltonian 
(\ref{198}) into the diagonal Hamiltonian
\be
\label{205}
H_B= E_0 +\sum_k \ep_k b_k^\dgr b_k \; , \qquad
E_0 = E_{HFB} - \sum_k \ep_k \int |v_k(\br)|^2 d\br \; .
\ee
The operators $b_k$ represent collective excitations called {\it bogolons}. 

The Hamiltonian $H_B$ is defined on the space ${\cal F}(b_k)$, with the 
averages $\langle \ldots \rangle$ implying the averaging in the space 
${\cal F}(b_k)$ with $H_B$. This leads to the bogolon distribution
\be
\label{206}
\pi_k \equiv \; \langle b_k^\dgr b_k \rangle \; = \left ( e^{\bt\ep_k} -1 \right )^{-1} \; , 
\ee
while $\langle b_k b_p \rangle \; = \; 0$. The single-particle density matrix of 
uncondensed atoms reads as
\be
\label{207}
\rho_1(\br,\br') = \sum_k \left [ \pi_k u_k(\br) u_k^*(\br') +
(1+\pi_k) v_k^*(\br) v_k(\br') \right ] \; ,
\ee
the related atomic density being
\be
\label{208}
\rho_1(\br)  =\sum_k \left [ \pi_k |u_k(\br)|^2 + (1+\pi_k)
|v_k(\br)|^2\right ] \; .
\ee
The anomalous density matrix
\be
\label{209}
\sgm_1(\br,\br') = \sum_k \left [ \pi_k u_k(\br) v_k^*(\br') +
(1+\pi_k) v_k^*(\br) u_k(\br')\right ] 
\ee
gives the anomalous average
\be
\label{210}
\sgm_1(\br) \equiv \sgm_1(\br,\br)  =\sum_k (1+2\pi_k) u_k(\br)
v_k^*(\br) \; .
\ee
The concrete nature of the quantum numbers $k$ and the explicit expressions
for the coefficient functions $u_k$ and $v_k$ in the Bogolubov equations 
(\ref{204}) depend on the form of the external potential $U({\bf r})$.

\subsection{Uniform system}

When there is no external potential, $U(\br) = 0$, a large system is uniform.
Then one expands the field operators in plane waves,
$$
\psi_1(\br) = \sum_{k\neq 0} a_k \vp_k(\br) \; , \qquad
\vp_k(\br) = \frac{1}{\sqrt{V}}\; e^{i\bk\cdot\br} \; .
$$
The condensate function does not depend on spatial variables and, in 
equilibrium, can be chosen to be real, such that
$\eta(\br) = \eta =\eta^*= \sqrt{\rho_0}$.
The normal and anomalous single-particle density matrices can be written as
the expansions
$$
\rho_1(\br,\br') = \sum_{k\neq 0} n_k \vp_k(\br) \vp_k^*(\br') \; ,
\qquad n_k \equiv \; \langle a_k^\dgr a_k \rangle \; ,
$$
\be
\label{211}
\sgm_1(\br,\br') = \sum_{k\neq 0} \sgm_k \vp_k(\br) \vp_k^*(\br')\; ,
\qquad \sgm_k \equiv \; \langle a_k a_{-k} \rangle \; .
\ee
And the density matrices (\ref{197}) become
$$
\rho(\br,\br')=\rho_0 + \rho_1(\br,\br') \; , \qquad
\sgm(\br,\br')=\rho_0 + \sgm_1(\br,\br') \; , 
$$
with the total density $\rho(\br) = \rho = \rho_0 + \rho_1$. The interaction 
potential is assumed to enjoy the Fourier transform  
$$
\Phi_k =\int \Phi(\br) e^{-ik\cdot\br} \; d\br \; .
$$

The grand Hamiltonian (\ref{166}) is the sum of five terms, with the 
zero-order term
\be
\label{212}
H^{(0)} = \left ( \frac{1}{2}\; \rho_0\Phi_0 - \mu_0\right ) N_0 \; .
\ee
The first-order term is exactly zero, $H^{(1)} = 0$, because of the 
orthogonality of the plane waves with $k \neq 0$ and $k = 0$. The 
second-order term is
\be
\label{213}
H^{(2)} = \sum_{k\neq 0} \left [ \frac{k^2}{2m} + \rho_0 ( \Phi_0
+\Phi_k ) - \mu_1 \right ] a_k^\dgr a_k + \frac{1}{2} \sum_{k\neq 0}
\rho_0 \Phi_k\left ( a_k^\dgr a_{-k}^\dgr + a_{-k} a_k\right ) \; .
\ee
The third-order term reads as
\be
\label{214}
H^{(3)} = \sqrt{\frac{\rho_0}{V}}\; {\sum_{p,q}}' \;
\Phi_p\left ( a_q^\dgr a_{q-p} a_p + a_p^\dgr a_{q-p}^\dgr a_q
\right ) \; ,
\ee
where the prime implies that $\bp \neq 0$, ${\bf q} \neq 0$, and 
$\bp-{\bf q} \neq 0$. The fourth-order term takes the form
\be
\label{215}
H^{(4)} = \frac{1}{2V} \; \sum_k {\sum_{p,q}}' \;
\Phi_k a_p^\dgr a_q^\dgr a_{k+p} a_{q-k} \; ,
\ee
in which $\bp \neq 0$, ${\bf q} \neq 0$, $\bk + \bp \neq 0$, and 
$\bk-{\bf q} \neq 0$. 

In the HFB approximation, we have
$$
a_k^\dgr a_p a_q =  a_k^\dgr \langle a_p a_q \rangle + 
\langle a_k^\dgr \rangle a_p a_q + a_p \langle a_k^\dgr a_q \rangle + 
\langle a_p \rangle a_k^\dgr a_q +
$$
$$
+ a_q \langle a_k^\dgr a_p \rangle + \langle a_q \rangle a_k^\dgr a_p \; 
- \langle a_k^\dgr \rangle \langle a_p a_q \rangle -
\langle a_p \rangle \langle a_k^\dgr a_q \rangle - 
\langle a_q \rangle \langle a_k^\dgr a_p \rangle \; .
$$
This gives $H^{(3)} = 0$ in the HFB approximation, since 
$\langle a_k \rangle = 0$ for $k \neq 0$.

Employing the HFB approximation to the fourth-order term, we use the quantum
number conservation condition,
$$
\langle a_k^\dgr a_p \rangle \; = \; \dlt_{kp} \langle a_k^\dgr a_k \rangle \; , \qquad 
\langle a_k a_p \rangle \; = \; \dlt_{-kp} \langle a_k a_{-k} \rangle \; .
$$ 
Then we get
$$
H^{(4)} = \sum_{k\neq 0} \rho_1 \Phi_0\left ( a_k^\dgr a_k -\;
\frac{1}{2}\; n_k \right ) +
$$
\be
\label{216}
+ \frac{1}{V} \; \sum_{k,p\neq 0} \Phi_k \left [ n_{k+p} a_p^\dgr a_p
+ \frac{1}{2}\left ( \sgm_{k+p} a_p^\dgr a_{-p}^\dgr +
\sgm_{k+p}^* a_{-p} a_p \right ) - \; \frac{1}{2}\left (
n_{k+p} n_p + \sgm_{k+p}\sgm_p^* \right ) \right ] \; .
\ee

Let us introduce the notations
\be
\label{217}
\om_k \equiv \frac{k^2}{2m} + \rho \Phi_0 + \rho_0 \Phi_k +
\frac{1}{V} \; \sum_{p\neq 0} n_p \Phi_{k+p} - \mu_1 
\ee
and
\be
\label{218}
\Dlt_k \equiv \rho_0 \Phi_k + \frac{1}{V} \; \sum_{p\neq 0} \sgm_p \Phi_{k+p} \; .
\ee
These are related to the general nonuniform case by the expansions
$$
\om(\br,\br') = \sum_k \om_k \vp_k(\br)\vp_k^*(\br') \; , \qquad
\om_k = \int \vp_k^*(\br)\om(\br,\br')\vp_k(\br') \; d\br d\br' \; ,
$$
$$
\Dlt(\br,\br') = \sum_k \Dlt_k \vp_k(\br)\vp_k^*(\br') \; , \qquad
\Dlt_k = \int \vp_k^*(\br)\Dlt(\br,\br')\vp_k(\br') \; d\br d\br' \; .
$$

Thus the grand Hamiltonian of a uniform system in the HFB approximation reduces to
\be
\label{219}
H_{HFB} = E_{HFB} + \sum_{k\neq 0} \om_k a_k^\dgr a_k +
\frac{1}{2} \sum_{k\neq 0} \Dlt_k \left ( a_k^\dgr a_{-k}^\dgr +
a_{-k} a_k \right ) \; ,
\ee
with
\be
\label{220}
E_{HFB} = H^{(0)}  -\; \frac{1}{2}\; \rho_1 \Phi_0 N_1  -\;
\frac{1}{2V} \sum_{k,p\neq 0} \Phi_{k+p}\left ( n_k n_p +
\sgm_k \sgm_p\right ) \; .
\ee

Substituting to equations (\ref{204}) the functions 
$u_k(\br) = u_k \vp_k(\br)$ and $v_k(\br) = v_k \vp_k(\br)$ yields the 
Bogolubov diagonalization equations
\be
\label{221}
(\om_k -\ep_k) u_k + \Dlt_k v_k = 0 \; , \qquad
\Dlt_k u_k + (\om_k +\ep_k) v_k =0 \; ,
\ee
where
\be
\label{222}
\ep_k =\sqrt{\om_k^2 -\Dlt_k^2}
\ee
is the spectrum of collective excitations. 

The condition of the condensate existence
\be
\label{223}
\lim_{k\ra 0}\ep_k = 0 \; , \qquad \ep_k \geq 0
\ee
defines 
\be
\label{224}
\mu_1=\rho\Phi_0 + \frac{1}{V}\; \sum_{p\neq 0} \Phi_p (n_p - \sgm_p) \; .
\ee
The condensate chemical potential (\ref{187}) becomes
\be
\label{225}
\mu_0=\rho \Phi_0 + \frac{1}{V} \; \sum_{p\neq 0} (n_p + \sgm_p)\Phi_p \; .
\ee
As is evident, $\mu_0$ and $\mu_1$ are different, 
$$
\mu_0 -\mu_1 = \frac{2}{V} \; \sum_{p\neq 0} \sgm_p \Phi_p \; .
$$
In the dilute-gas approximation, we have
\be
\label{226}
\mu_0=(\rho+\rho_1+\sgm_1)\Phi_0 \; , \qquad \mu_1=(\rho+\rho_1-\sgm_1)\Phi_0 \; . 
\ee

Hamiltonian (\ref{219}) is diagonalized by the Bogolubov canonical 
transformation $a_k = u_k b_k + v_{-k}^* b_{-k}^\dgr$ similarly to this
procedure in the Bogolubov approximation, but with different $\om_k$ and 
$\Dlt_k$ defined in equations (\ref{217}) and (\ref{218}). For the coefficient 
functions, we now have
$$
u_k^2 = \frac{\sqrt{\ep_k^2+\Dlt_k^2}+\ep_k}{2\ep_k} =
\frac{\om_k+\ep_k}{2\ep_k} \; , \qquad
v_k^2 = \frac{\sqrt{\ep_k^2+\Dlt_k^2}-\ep_k}{2\ep_k} =
\frac{\om_k-\ep_k}{2\ep_k} \; ,
$$
$$
u_k^2 - v_k^2 = 1\; , \qquad u_k v_k = -\; \frac{\Dlt_k}{2\ep_k} \; ,
\qquad u_k^2+v_k^2 = \frac{\sqrt{\ep_k^2+\Dlt_k^2}}{\ep_k} =
\frac{\om_k}{\ep_k} \; .
$$

The normal and anomalous averages, in the momentum representation, are
\be
\label{227}
n_k = \frac{\sqrt{\ep_k^2+\Dlt_k^2}}{2\ep_k} \; {\rm coth}
\left ( \frac{\ep_k}{2T}\right ) - \; \frac{1}{2} \; , \qquad
\sgm_k = -\; \frac{\Dlt_k}{2\ep_k}  \; {\rm coth}
\left ( \frac{\ep_k}{2T}\right ) \; .
\ee
At small excitation energy, such that $\ep_k \ll \Dlt_k$ and $\ep_k \ll T$,
these averages behave as
$$
n_k \simeq \frac{T\Dlt_k}{\ep_k^2} + \frac{\Dlt_k}{12T} +
\frac{T}{2\Dlt_k} \; - \; \frac{1}{2} + \left (\frac{\Dlt_k}{3T} \; -
\; \frac{T}{\Dlt_k} \; - \; \frac{\Dlt_k^3}{90T^3}\right )
\frac{\ep_k^2}{8\Dlt_k^2} \; ,
$$
$$
\sgm_k \simeq -\; \frac{T\Dlt_k}{\ep_k^2} \; -\; \frac{\Dlt_k}{12T} +
\frac{\Dlt_k\ep_k^2}{720T^3} \; .
$$
While at high excitation energy, when $\ep_k \gg \Dlt_k$ and $\ep_k \gg T$,
their asymptotic forms are
$$
n_k \simeq \left ( \frac{\Dlt_k}{2\ep_k} \right )^2 -
\left ( \frac{\Dlt_k}{2\ep_k}\right )^4 + e^{-\bt\ep_k} \; , \qquad
\sgm_k \simeq - \; \frac{\Dlt_k}{2\ep_k}\left (1 + 2e^{-\bt\ep_k}
\right ) \; .
$$
Comparing these expressions, we see that the anomalous average $\sigma_k$ is
of order or larger than the normal average $n_k$, hence $\sigma_k$ can never 
be neglected \cite{Yukalov_26}. 

Diagonalizing Hamiltonian (\ref{219}), we come to the Bogolubov Hamiltonian
\be
\label{228}
H_B = E_B +\sum_{k\neq 0} \ep_k b_k^\dgr b_k \; , 
\ee
in which
$$
E_B = E_{HFB} - \;\frac{1}{2}\; \sum_{k\neq 0} (\om_k -\ep_k) \; .
$$
Here $\om_k$ is expression (\ref{217}) that can be written as
\be
\label{229}
\om_k = \frac{k^2}{2m} + \Dlt + \rho_0 (\Phi_k - \Phi_0) +
\frac{1}{V}\; \sum_{p\neq 0} n_p (\Phi_{k+p} - \Phi_p) \; ,
\ee
where
\be
\label{230}
\Dlt \equiv \lim_{k\ra 0} \Dlt_k \; = 
\rho_0 \Phi_0 + \frac{1}{V}\; \sum_{p\neq 0} \sgm_p \Phi_p \; .
\ee

In the long-wave limit $k \ra 0$, we have the following expansions:
$$
\Phi_{p+k} \simeq \Phi_p + \frac{1}{2}\; \Phi_p'' k^2 \; ,
\qquad \Phi_p'' \equiv \frac{\prt^2\Phi_p}{\prt p^2} \; ,
$$
$$
\om_k \simeq \Dlt + \frac{1}{2} \left ( \frac{1}{m} + \rho_0 \Phi_o''
+ \frac{1}{V} \; \sum_{p\neq 0} n_p \Phi_p'' \right ) k^2 \; ,
$$
$$
\Dlt_k \simeq \Dlt + \frac{1}{2} \left ( \rho_0 \Phi_o''
+ \frac{1}{V} \; \sum_{p\neq 0} \sgm_p \Phi_p'' \right ) k^2 \; .
$$
Using these expansions, it is straightforward to show that, in the long-wave 
limit $k\ra 0$, the spectrum of collective excitations (\ref{222}) is 
phonon-like,
\be
\label{231}
\ep_k \simeq ck \; , \qquad c \equiv \sqrt{\frac{\Dlt}{m^*} } \; ,
\ee
with the sound velocity $c$, in which 
$$
m^* \equiv \frac{m}{1 +\frac{m}{V} \sum_p (n_p-\sgm_p)\Phi_p''} 
$$
is effective mass. 

The HFB Hamiltonian (\ref{219}) as well as the Bogolubov Hamiltonian (\ref{228})
are quadratic in the field operators $a_k$ or $b_k$. Therefore, calculating 
observable quantities, we have to limit such calculations by the terms of 
second order in the field operators. Thus, calculating ${\rm var}(\hat N)$,
we have to keep only the terms of second order with respect to $a_k$. The
variance ${\rm var}(\hat N)$ is an important quantity characterizing the 
system stability. The variance can be represented as 
\be
\label{232}
{\rm var}(\hat N) = N \left\{ 1 + \rho \int [g(\br)-1]\; d\br \right \} \; ,
\ee
with the pair correlation function
\be
\label{233}
 g(\br) = 1 +\frac{2\rho_0}{\rho^2}\; {\rm Re}\; [\rho_1(\br,0) +
\sgm_1(\br,0)] 
\ee
in second order with respect to the field operators. Because of the equations 
$n_k^* = n_{-k} = n_k$ and $\sgm_k^* = \sgm_{-k} = \sgm_k$, the expressions 
$\rho_1(\br,0)$ and $\sgm_1(\br,0)$ are real and have the form
$$
\rho_1(\br,0)  =\frac{1}{V}\; \sum_{k\neq 0} n_k e^{i\bk\cdot\br} \; ,
\qquad \sgm_1(\br,0)  =\frac{1}{V}\; \sum_{k\neq 0} \sgm_k e^{i\bk\cdot\br} \; .
$$
This gives the pair correlation function
$$
g(\br) = 1 +\frac{2\rho_0}{\rho^2} \; \int (n_k +\sgm_k) \;
e^{k\bk\cdot\br} \; \frac{d\bk}{(2\pi)^3} \; .
$$
Since
$$
\int [g(\br)-1]\; d\br = \frac{2\rho_0}{\rho^2}\; \lim_{k\ra 0} (n_k + \sgm_k) \; ,
$$
we get
$$
\frac{{\rm var}(\hat N)}{N} = 1 + \frac{2\rho_0}{\rho}\; \lim_{k\ra 0}(n_k + \sgm_k) \; ,
$$
Under the condition $\rho_0/\rho\simeq 1$, we find
$$
\frac{{\rm var}(\hat N)}{N} = 1 + 2 \lim_{k\ra 0} (n_k + \sgm_k) \; , \qquad 
\lim_{k\ra 0} (n_k + \sgm_k) = \frac{1}{2}\left ( \frac{T}{\Dlt} \; - 1\right ) \; ,
$$
where
$$
\Dlt = \rho_0\Phi_0 + \int \sgm_k\Phi_k\;\frac{d\bk}{(2\pi)^3} \; = m^* c^2 \; .
$$
It is again worth stressing the importance of the anomalous average $\sgm_k$, 
without which ${\rm var}(\hat N)/N$ would be divergent. But calculating the 
variance correctly, we find the latter and, respectively, compressibility in 
the form 
\be
\label{234}
\frac{{\rm var}(\hat N)}{N} = \frac{T}{m^* c^2}\; , \qquad
\kappa_T = \frac{{\rm var}(\hat N)}{N\rho T} =\frac{1}{\rho m^* c^2} \; .
\ee
The particle fluctuations are thermodynamically normal for all $T < T_c$ and
the system is stable.

When temperature approaches the critical temperature, $T \ra T_c$, then
$\rho_0 \ra 0$, $\sgm_k \ra 0$ and $c \ra 0$. At the critical point, the 
compressibility $\kappa_T \ra \infty$, which is natural for a second-order
phase transition.

The equation for the superfluid fraction in the dilute-gas approximation
enjoys the same form  
$$
n_s = 1 -\; \frac{2Q}{3T} \; ,
$$
as in the Bogolubov approximation (\ref{134}), but with the new sound velocity 
$c$ defined in (\ref{231}). At the critical temperature $T_c$, the superfluid 
fraction $n_s \ra 0$ together with the condensate fraction $n_0$.   

The condensation temperature $T_c$ can be found from the expression for the 
average density, under the conditions $\rho_0 = 0$, $\rho_1 = \rho$, $\Dlt = 0$,
and $\ep_k = k^2/2m$. In the case of local interactions, the average density
at the critical temperature reads as
$$
\rho =\frac{1}{2}\; \int \left [ \coth\left ( \frac{k^2}{2mT_c}\right )
-1 \right ] \; \frac{d\bk}{(2\pi)^3} \; .
$$
Taking the integral
$$
\int_0^\infty \; (\coth x -1 )\sqrt{x}\; dx = \frac{1}{2}\;
\sqrt{\frac{\pi}{2}}\; \zeta\left (\frac{3}{2}\right ) \; ,
$$
we obtain
\be
\label{235}
T_c = \frac{2\pi}{m}\left [ \frac{\rho}{\zeta(3/2)}\right ]^{2/3} \; .
\ee
This is the same expression as for the ideal Bose gas, which is caused by
the use of the local delta-function interaction potential (\ref{102})
corresponding to the dilute-gas approximation.

\subsection{Nonlocal interactions}

If the interaction potential is not local, the critical temperature is 
different. To find out how the nonlocal interactions influence the 
transition temperature, let us consider the approach to $T_c$ from above.
For $T > T_c$, we can use the Hartree-Fock approximation, when the 
single-particle spectrum is
\be
\label{236}
\om_k = \frac{k^2}{2m} +\rho\Phi_0 + \frac{1}{V} \;
\sum_p n_p \Phi_{k+p} - \mu \; .
\ee
For the occurrence of Bose-Einstein condensation, it is necessary that
$$
\lim_{k\ra 0} \om_k = 0 \qquad (T=T_c+0) \; .
$$
This defines the chemical potential at the transition temperature
$$
\mu=\rho\Phi_0 + \frac{1}{V}\; \sum_p n_p \Phi_p \; ,
$$
which gives
\be
\label{237}
\om_k = \frac{k^2}{2m} + \frac{1}{V}\; \sum_p n_p (\Phi_{k+p}-\Phi_p) \; , 
\qquad (T = T_c) \; .
\ee

Calculating the sums over momentum, containing the distribution $n_k$, we 
can take into account that this distribution is maximal at zero $k$, hence     
the main contribution comes from the long-wave limit $k \ra 0$. Therefore 
it is admissible to invoke the approximation
\be
\label{238}
\sum_p n_p \Phi_{k+p} \cong \Phi_k \sum_p n_p \; .
\ee
Then spectrum (\ref{236}) reads as
$$
\om_k = \frac{k^2}{2m} +\rho(\Phi_0 + \Phi_k) - \mu \; .
$$
And the chemical potential at the transition point becomes 
$ \mu = 2\rho \Phi_0$, which yields
\be
\label{239}
\om_k = \frac{k^2}{2m} +\rho(\Phi_k - \Phi_0) \; .
\ee

To find the long-wave limit of the Fourier transform
$$
\Phi_k = \int \Phi(\br) e^{-i \bk \cdot\br} \; d\br
$$
for a potential depending on $r \equiv |\br|$, we substitute into the 
integral the expansion
$$
e^{-i\bk\cdot\br} \simeq 1 - i\bk \cdot \br - \; \frac{1}{2}(\bk\cdot\br)^2 \; .
$$
The integral of the linear term is zero. Hence
$$
\Phi_k \simeq \Phi_0 - \; \frac{1}{2}\; \int \Phi(r) (\bk\cdot\br)^2 \; d\br \; .
$$

Defining the effective interaction radius $r_{eff}$ by the equation  
\be
\label{240}
r_{eff}^2 \equiv \frac{\int r^2 \Phi(\br) \; d\br}{\int \Phi(\br) \; d\br} = \; 
\frac{4\pi}{\Phi_0} \int_0^\infty r^4 \Phi(\br) d\br \; ,
\ee
we obtain
$$
\Phi_k \simeq \left (1  -\; \frac{1}{12}\; k^2 r_{eff}^2\right )\Phi_0 \; .
$$

Thus for the long-wave limit of spectrum (\ref{239}), we have
\be
\label{241}
\om_k\simeq \frac{k^2}{2m^*} \; \; (k\ra 0) \; , 
\ee
with the effective mass
\be
\label{242}
m^*\equiv \frac{m}{1 - mr_{eff}^2 \rho \Phi_0/6} \; .           
\ee
As is seen, for repulsive interactions, the effective mass is larger than $m$. 

The critical temperature $T_c$ is found from the integral
$$
\rho =\int n_k\; \frac{d\bk}{(2\pi)^3} \; , \qquad
n_k = \left ( e^{\om_k/T_c} -1 \right )^{-1}\; , 
$$
which results in
\be
\label{243}
T_c \cong \frac{2\pi}{m^*} \left [ \frac{\rho}{\zeta(3/2)}\right ]^{2/3} \; .
\ee

To explicitly compare the transition temperature $T_c$ with the condensation 
temperature $T_0$ of ideal gas, one introduces the relative difference
\be
\label{244}
\frac{\Dlt T_c}{T_0} \equiv \frac{T_c}{T_0} \; - 1  \qquad 
\left ( T_0 \equiv \frac{2\pi}{m}\left [ \frac{\rho}{\zeta(3/2)}\right ]^{2/3} \right ) \; .
\ee
This gives
\be
\label{245}
\frac{\Dlt T_c}{T_0} = \frac{m}{m^*} - 1 \cong -\; \frac{1}{6}\; mr_{eff}^2 \rho \Phi_0 \ .
\ee

A nonlocal interaction potential leads to a lower critical temperature, as 
compared to the condensation temperature for a gas with local interactions, 
$T_c < T_0$. Notice that for a local potential, proportional to $\dlt(\br)$, 
the effective interaction radius $r_{eff} = 0$, as a result of which $m^* = m$
and $T_c$ reduces to $T_0$. 

As a nonlocal interaction potential, one often takes the Gaussian potential
$$
\Phi(\br) = A \exp\left ( - \; \frac{r^2}{2b^2}\right ) \; .
$$
Its Fourier transform is
$$
\Phi_k \equiv \int \Phi(\br) e^{-i\bk\cdot\br}\; d\br =
\frac{4\pi}{k} \; \int_0^\infty \; r \sin(kr) \Phi(r)\; d\br = 
\Phi_0 \exp \left ( \frac{-k^2 b^2}{2} \right ) \; .
$$
The Born-type normalization
$$
\Phi_0 \equiv \int \Phi(\br)\; d\br = 4\pi\; \frac{a_s}{m} \; , 
$$
yields
$$
\Phi_0 = (2\pi)^{3/2} A b^3 \; , 
\qquad A = \sqrt{\frac{2}{\pi} } \;\frac{a_s}{ m b^3} \; .
$$
Using the integral
$$
\int_0^\infty \; x^n e^{-qx^2}\; dx =
\frac{\Gm\left ( \frac{n+1}{2}\right )}{2q^{(n+1)/2}} \; ,
$$
for the effective interaction radius, defined in (\ref{240}), we get 
$r_{eff} = \sqrt{3} b$.

\subsection{Condensate and superfluid}

Let us consider in more detail the condensate and superfluid fractions
for a uniform system with local interactions, in the HFB approximation. 
For the condensate fraction we have
\be
\label{246}
n_0 = 1 - \; \frac{\rho_1}{\rho} \; , \qquad
\rho_1 = \int n_k \; \frac{d\bk}{(2\pi)^3} \; .
\ee
While the superfluid fraction is
\be
\label{247}
n_s = 1 - \; \frac{2Q}{3T}  \qquad
\left ( Q = \frac{{\rm var}(\hat\bP)}{2mN} \right ) \; .
\ee

For an equilibrium system,
$$
\langle \hat\bP \rangle \; = \; 0 \; , \qquad Q = \frac{\langle \hat\bP^2 \rangle}{2mN} \; .
$$
And for a uniform system,
$$
\langle \hat\bP^2 \rangle \; = \; \sum_{kp} (\bk\cdot\bp)
\langle \hat n_k \hat n_p \rangle \qquad \left ( \hat n_k \equiv a_k^\dgr a_k \right ) \; .
$$
In the Hartree-Fock-Bogolubov approximation,
$$
\langle \hat n_k \hat n_p \rangle \; = \; n_k n_p + \dlt_{kp} n_k(1 + n_k) -
\dlt_{-kp} \sgm_k^2 \; .
$$
Therefore
$$
Q = \frac{1}{\rho} \; \int \; \frac{k^2}{2m} \left ( n_k +
n_k^2 - \sgm_k^2\right ) \frac{d\bk}{(2\pi)^3} \; .
$$
Using the equality ${\rm cosh}x - 1 = 2{\rm sinh}^2 (x/2)$, we have 
$$
n_k + n_k^2 -\sgm_k^2 =
\frac{1}{4{\rm sinh}^2(\bt\ep_k/2)} \; .
$$
Then the dissipated heat becomes
\be
\label{248}
Q = \frac{1}{(4\pi)^2m\rho} \; \int_0^\infty \;
\frac{k^4dk}{{\rm sinh}^2(\bt\ep_k/2)} \; .
\ee

In the case of local interactions,
$$
\om_k = \frac{k^2}{2m} + mc^2 \; , \qquad \Dlt_k=mc^2 \; ,
$$
and the spectrum of collective excitations takes the form
\be
\label{249}
\qquad \ep_k =\sqrt{(ck)^2 + \left ( \frac{k^2}{2m}\right )^2 } \; .
\ee
The sound velocity $c$ is defined by the equation
\be
\label{250}
mc^2 = (\rho_0 + \sgm_1)\Phi_0 \; ,
\ee
with the anomalous average
$$
\sgm_1 = \int \sgm_k \; \frac{d\bk}{(2\pi)^3} \; .
$$

The density of uncondensed atoms can be reduced to the integral
\be
\label{251}
\rho_1 = \frac{(mc)^3}{3\pi^2} \left \{ 1 +
\frac{3}{2\sqrt{2}} \; \int_0^\infty
\left ( \sqrt{1+x^2} -1\right )^{1/2} \left [ {\rm coth}
\left ( \frac{mc^2}{2T}\; x\right ) -1 \right ] \; dx
\right \} \; .
\ee
And the anomalous average writes as
\be
\label{252}
\sgm_1 = - \int \frac{mc^2}{2\ep_k}\; {\rm coth}\left (
\frac{\ep_k}{2T}\right ) \; \frac{d\bk}{(2\pi)^3} \; .
\ee

This integral diverges for any finite $c$ and requires to be regularized.
In any type of regularization, the following properties must hold for $\sgm_1$. 
The anomalous average has to be zero for the ideal Bose gas. Hence there 
should exist the {\it ideal-gas condition}: 
\be
\label{253}
\sgm_1\ra 0 \qquad (\Phi_0 \ra 0 ) \; .
\ee
The anomalous average is nonzero together with $\rho_0$, since both of them
are caused by the gauge symmetry breaking. When the gauge symmetry is restored,
$\sgm_1$ must be zero, which is the {\it symmetry-restoration condition} 
\be
\label{254}
\sgm_1\ra 0 \qquad (\rho_0\ra 0) \; .
\ee

For temperatures outside of the critical region, when $T \ll T_c$, the following
regularization can be done. Notice that at zero temperature, $\sigma_1$ reduces to
the form
\be
\label{255}
\sgm_{eff} \equiv - mc^2 \int \frac{1}{2\ep_k} \; \frac{d\bk}{(2\pi)^3} \; .
\ee
We separate in integral (\ref{252}) the zero-temperature form by adding and 
subtracting one, which gives
\be
\label{256}
\sgm_1 = \sgm_{eff} - \int \frac{mc^2}{2\ep_k}
\left [ {\rm coth}\left ( \frac{\ep_k}{2T}\right ) -1\right ]
\frac{d\bk}{(2\pi)^3} \; .
\ee
 
The integral in $\sgm_{eff}$ is ultraviolet divergent, so that a regularization 
is needed. The regularization resorting to nonlocal interactions involves
complicated equations. The cutoff regularization essentially depends on the 
chosen cutoff momentum, whose choice is ambiguous. We resort to the dimensional 
regularization, that is asymptotically exact in the region of weak interactions,
and then we accomplish an analytic continuation to finite interactions.
Dimensional regularization yields 
$$
\int \frac{1}{2\ep_k} \; \frac{d\bk}{(2\pi)^3} = - \;
\frac{m^2c_{eff}}{\pi^2} \; ,
$$
where $c_{eff}$ is to be defined.

It is convenient to introduce the dimensionless anomalous averages
\be
\label{257}
\sigma_0 \equiv \frac{\sigma_{eff}}{\rho} \; , \qquad
\sigma \equiv \frac{\sigma_1}{\rho}
\ee
and the dimensionless Bogolubov sound velocity
\be
\label{258}
s_B \equiv \frac{mc_B}{\rho^{1/3}} \; , \qquad c_B \equiv \sqrt{\frac{\rho}{m} \Phi_0} \; .
\ee
Then the anomalous average $\sigma_0$ can be written in the form
\be
\label{259}
\sgm_0 = \frac{s_B^2 mc_{eff}}{\pi^2 \rho^{1/3}}  \;  \qquad (\Phi_0 \ra 0) \; .
\ee
Equation (\ref{250}) for the sound velocity can be written as 
\be
\label{260}
c = c_B \sqrt{n_0 + \sigma} \; .
\ee

In order to analytically continue the anomalous average (\ref{259}) from
weak to finite interactions, we combine it with expression (\ref{260}) in
the form of the iterative equation
\be
\label{261}
\sigma_0^{(n+1)} = \frac{s_B^3}{\pi^2}\sqrt{n_0 + \sigma_0^{(n)}} \; .
\ee
This shows that $c_{eff}$ is defined by expression (\ref{260}) with the 
appropriate iteration for $\sigma$. Starting the iteration from the zero 
approximation $\sigma_0 = 0$, we get the first approximation 
$\sigma_0^{(1)} = (s_B^3/\pi^2) \sqrt{n_0}$. The second approximation 
reads as
\be
\label{262}
\sigma_0^{(2)} = \frac{s_B^3}{\pi^2} \left ( n_0 + 
\frac{s_B^3}{\pi^2}\sqrt{n_0} \right )^{1/2} \; .
\ee
This form satisfies the ideal-gas and symmetry-restoration conditions.

In this way, at finite temperatures outside of the critical region, the 
anomalous average (\ref{256}) can be transformed to  
\be
\label{263}
\sgm_1 = \sgm_{eff} \; - \; \frac{(mc)^3}{2\sqrt{2}\pi^2} \;
\int_0^\infty \; \frac{(\sqrt{1+x^2}-1)^{1/2}}{\sqrt{1+x^2}}
\left [ {\rm coth}\left ( \frac{mc^2}{2T}\; x\right ) - 1
\right ] dx \; .
\ee

The dissipated heat (\ref{248}), at any temperature, reduces to
\be
\label{264}
Q = \frac{(mc)^5}{\sqrt{2}(2\pi)^2m\rho} \;
\int_0^\infty \; \frac{(\sqrt{1+x^2}-1)^{3/2} x\; dx}
{\sqrt{1+x^2}\;{\rm sinh}^2(mc^2x/2T)} \; .
\ee

From the above expressions, we can find the behavior of all fractions below
the transition temperature, where $T \ll T_c$. For instance, at low 
temperatures, such that $T/mc^2 \ll 1$, we have the density of uncondensed
atoms
$$
\rho_1 \simeq \frac{(mc)^3}{3\pi^2} + \frac{(mc)^3}{12}
\left ( \frac{T}{mc^2}\right )^2 \; , 
$$
the anomalous average
$$
\sgm_1 \simeq \sgm_{eff} \; - \; \frac{(mc)^3}{12}
\left ( \frac{T}{mc^2}\right )^2 \; ,
$$
and the dissipated heat
$$
Q \simeq \frac{\pi^2(mc)^5}{15m\rho} \left (
\frac{T}{mc^2}\right )^5 \; .
$$
This yields the condensate fraction
\be
\label{265}
n_0 \simeq 1 \; - \; \frac{(mc)^3}{3\pi^2\rho} \; - \;
\frac{(mc)^3}{12\rho} \left ( \frac{T}{mc^2}\right )^2 \; ,
\ee
and the superfluid fraction
\be
\label{266}
n_s \simeq 1 \; - \; \frac{2\pi^2(mc)^3}{45\rho} \left (
\frac{T}{mc^2}\right )^4 \; .
\ee
Numerical investigation \cite{Yukalov_64} shows that the condensate 
fraction as a function of the interaction strength is in good agreement
with Monte Carlo simulations \cite{Rossi_65}. 

When temperature tends to the critical $T_c$, it is necessary to take 
a different form of the anomalous average (\ref{252}), instead of the
noncritical expression (\ref{256}). Returning to the initial expression
(\ref{252}) of the anomalous average, we rewrite it as 
$$
\sigma_1 =  - \; \frac{(mc)^3}{2\sqrt{2}\pi^2} \;
\int_0^\infty \; \frac{(\sqrt{1+x^2}-1)^{1/2}}{\sqrt{1+x^2}}
{\rm coth} \left ( \frac{mc^2}{2T}\; x\right ) dx \; .
$$
From here, taking into account that $c \ra 0$ when $T \ra T_c$, and using
the integral
$$
\int_0^\infty \;
\frac{(\sqrt{1+x^2}-1)^{1/2}}{x\sqrt{1+x^2}} \; dx =
\frac{\pi}{\sqrt{2}} \; ,
$$  
we come to the expression
\be
\label{267}
\sgm_1 \simeq  - \; \frac{m^2cT}{2\pi} \; , \qquad (T \ra T_c) \; .
\ee
 
There is a temptation to consider the critical behavior in the vicinity
of $T_c$ by expanding the corresponding expressions in powers of
$mc^2/T_c \ll 1$. Then, in the zero order, when $c = 0$, we come to the 
same behavior as in the case of the ideal Bose gas. However, in higher 
orders, we confront divergent integrals. Hence, such an expansion is not
appropriate. Numerical solution proves \cite{Yukalov_66} that the 
condensate and superfluid fractions continuously tend to zero at $T_c$,
which implies the second order phase transition. 
   
It is important to stress that the described above self-consistent theory
is unique, providing, in the mean-field approximation \cite{Yukalov_64}, 
good agreement of the condensate fraction with Monte Carlo simulations 
\cite{Rossi_65} for arbitrarily strong interactions and correctly 
characterizing the Bose-Einstein condensation as a second-order phase 
transition occurring together with the superfluid transition \cite{Yukalov_66}.

\subsection{Gas parameter}

For numerical analysis, it is convenient to employ dimensionless quantities.
For this purpose, let us introduce the gas parameter
\be
\label{268}
\gm\equiv\rho^{1/3}a_s=\frac{a_s}{a} \; ,
\ee
characterizing the interaction strength, the dimensionless sound velocity
\be
\label{269}
s\equiv\frac{mc}{\rho^{1/3}} \; ,
\ee
and the dimensionless temperature
\be
\label{270}
t\equiv\frac{mT}{\rho^{2/3}}\; .
\ee

Note that for the ideal-gas critical temperature $T_c$, given by
expression (\ref{235}),        
$$
t_c \equiv \frac{mT_c}{\rho^{2/3}} =
\frac{2\pi}{[\zeta(3/2)]^{2/3}} = 3.312498 \; .
$$

The condensate fraction now reads as 
\be
\label{271}
n_0 = 1 \; - \; \frac{s^3}{3\pi^2}\left \{ 1 +
\frac{3}{2\sqrt{2}} \; \int_0^\infty
\left ( \sqrt{1+x^2}-1\right )^{1/2} \left [ {\rm coth}\left (
\frac{s^2 x}{2t}\right ) - 1 \right ] \; dx \right \} \; .
\ee
And the superfluid fraction is
\be
\label{272}
n_s = 1 \; - \; \frac{s^5}{6\sqrt{2}\pi^2t} \; \int
\frac{x\left ( \sqrt{1+x^2}-1\right )^{3/2}\; dx}
{\sqrt{1+x^2}\;{\rm sinh}^2(s^2x/2t)} \; .
\ee

We use the dimensionless anomalous density $\sgm \equiv \sgm_1/\rho$ and
the local interaction potential with the strength $\Phi_0 = 4\pi a_s/ m$.
Then we find
\be
\label{273}
\sgm = \sigma_0 \; - \; \frac{s^3}{2\sqrt{2}\pi^2}\; \int_0^\infty \;
\frac{\left (\sqrt{1+x^2}-1\right )^{1/2}}{\sqrt{1+x^2}}
\left [ {\rm coth}\left ( \frac{s^2x}{2t} \right ) - 1
\right ] dx \; .
\ee
Formula (\ref{250}) gives for the dimensionless sound velocity the equation
\be
\label{274}
s^2=4\pi\gm(n_0+\sgm) \; .
\ee

Let us consider the case of zero temperature. At zero $T = 0$, the above 
equations simplify to
\be
\label{275}
n_0 = 1  -\; \frac{s^3}{3\pi^2} \; , \qquad n_s = 1 \; , \qquad
\sgm = \sgm_0 \; .
\ee
Here $\sgm_0$ is defined by formula (\ref{262}). For weak interactions, 
when $\gm \ra 0$, we get 
$$
n_0 \simeq 1 - \; \frac{8}{3\sqrt{\pi}}\; \gm^{3/2} - \;
\frac{64}{3\pi}\; \gm^3 - \; \frac{256}{9\pi^{3/2}} \; \gm^{9/2} \; ,
$$
$$
s\simeq \sqrt{4\pi \gm} + \frac{16}{3}\; \gm^2 -
\frac{64}{9\sqrt{\pi}}\; \gm^{7/2} - \;
\frac{4480}{27\pi}\; \gm^5 \; ,
$$
$$
\sgm \simeq \frac{8}{\sqrt{\pi}}\; \gm^{3/2} +
\frac{64}{3\pi} \; \gm^3 - \; \frac{1408}{9\pi^{3/2}} \;
\gm^{9/2} \; .
$$

If the interactions are strong, so that $\gm \ra \infty$, then 
$$
n_0 \simeq 4 \times 10^{-5} \frac{1}{\gamma^{13}} \; , \qquad
s \simeq (3\pi^2)^{1/3} - \; \left ( \frac{\pi}{3} \right )^{2/3} n_0 \; ,
$$
$$
\sgm \simeq \frac{(9\pi)^{1/3}}{4}\; \gm^{-1} - n_0 \; . 
$$
These strong-interaction asymptotic expressions should be understood as
corresponding to a metastable uniform system, since under such strong 
interactions the system would crystallize.   

Recall that in the Bogolubov approximation, the condensate fraction is given 
by the form
$$
n_B = 1 \; - \; \frac{8}{3\sqrt{\pi}}\; \gm^{3/2} 
$$
and the sound velocity is $s_B = 2 \sqrt{\pi \gm n_B}$. The condensate depletion
becomes total, with $n_B = 0$ and $s_B = 0$ at the gas parameter
$$
\gm_B \equiv \frac{(9\pi)^{1/3}}{4} = 0.761618 \; .
$$

For superfluid $^4$He, the helium atoms can be represented by hard spheres 
of diameter $a_s = 2.203$ \AA. At saturated vapour pressure, $\rho a_s^3 = 0.21$, 
which corresponds to $\gm_{He} = 0.594$. This gives the condensate fraction
about $3\%$. Experiments suggest for the condensate fraction the values in the 
range between $2\%$ and $10\%$. The most recent data \cite{Diallo_67} give 
the zero-temperature condensate fraction around $7\%$. 

The ground-state energy is the internal energy 
$E \equiv \langle H \rangle + \mu N$ at zero temperature $(T = 0)$. Then
\be
\label{276}
\langle H \rangle = E_B = E_{HFB} + N \int \; \frac{\ep_k-\om_k}{2\rho} \;
\frac{d\bk}{(2\pi)^3} \; .
\ee
The last integral is divergent, but can be regularized by dimensional 
regularization,
$$
\int \; \frac{\ep_k-\om_k}{2\rho}\; \frac{d\bk}{(2\pi)^3} =
\frac{8(mc)^5}{15\pi^2m\rho} \; .
$$
Defining the dimensionless ground-state energy
\be
\label{277}
E_0 \equiv \frac{2mE}{\rho^{2/3}N} \; , 
\ee
we obtain
\be
\label{278}
E_0 = 4\pi\gm\left ( 1 + n_1^2 - 2\sgm n_1 - \sgm^2 +
\frac{4s^5}{15\pi^3\gm} \right ) \; .
\ee
For weak interactions, when $\gm \ll 1$, the ground-state energy becomes
\be
\label{279}
E_0 \simeq 4\pi\gm + \frac{512}{15}\; \sqrt{\pi}\; \gm^{5/2} +
\frac{512}{9}\; \gm^4 \; .
\ee
First two terms here exactly coincide with the Lee-Huang-Yang formula 
\cite{Lee_68,Lee_69}. 

As has been mentioned above, the strong-interaction expansion does not
have much sense, since at about $\gamma \approx 0.65$ the system 
crystallizes \cite{Rossi_65}. The formula for finite $\gamma$ can be 
obtained from expansion (\ref{279}) by means of the self-similar factor 
approximants \cite{Yukalov_70,Gluzman_71,Yukalov_72}, which give 
\be 
\label{280}
E_0 = 4\pi \gamma \left ( 1 + 2.934 \gamma^{3/2} \right )^{1.641} \; .
\ee
This expression reproduces the Lee-Huang-Yang energy at small $\gamma \ll 1$
and is in good agreement with the Monte Carlo simulations \cite{Rossi_65}
up to $\gamma \approx 0.6$.

\subsection{Conservation of particles}

Spontaneous gauge symmetry breaking is the necessary and sufficient condition 
for Bose-Einstein condensation \cite{Yukalov_10,Lieb_22,Yukalov_24}. But if 
gauge symmetry is broken, then the number of atoms, generally, is not conserved
in the process of calculations. However, there is no contradiction with the 
fact that the total number of atoms in the system can be fixed. This is because
when one requires that the number of particles be prescribed, this concerns the 
observable quantity, that is, the average number of particles. Actually, any 
conditions imposed on the system are senseful only with respect to observable
quantities that are defined as statistical averages. 
 
Recall that in the grand canonical ensemble, the number of particles is not 
fixed, independently whether gauge symmetry is broken or not. Moreover, when 
employing the field-operator representation, the number of particles is not
conserved under the action of field operators 
\cite{Yukalov_1,Schweber_73,Berezin_74,Yukalov_75}. But any such a 
nonconservation on the operator level does not contradict the conservation 
of the number of particles as an observable quantity.  

Let us show in what sense the number of particles is conserved, when gauge 
symmetry is broken by means of infinitesimal sources and the observable 
quantities are defined through quasiaverages. 
 
Usually, the operator of the number of particles 
$\hat N = \int \psi^\dgr(\br) \psi(\br) d\br$ commutes with the Hamiltonian,
$[\hat N, H] = 0$. The field operator can be represented as an expansion over
a basis of natural orbitals $\{\vp_k(\br)\}$, where it is possible to separate 
a term corresponding to Bose condensate:
$$
\psi(\br)=\psi_0(\br)+\psi_1(\br) \; , \qquad \psi_0(\br) = a_0\vp_0(\br) \; ,
\qquad \psi_1(\br) =\sum_{k\neq 0} a_k\vp_k(\br) \; .
$$
Here $k$ is a multi-index labelling quantum states. 

The gauge symmetry is broken by adding to the Hamiltonian a symmetry-breaking 
term, for instance, by defining the Hamiltonian
$$
H_\ep \equiv H -\ep \; \sqrt{\rho_0} \int_V \; \left [
\psi_0^\dgr(\br) + \psi_0(\br) \right ] d\br \; .
$$
The operator 
$$
\hat N_0 \equiv \int \psi_0^\dgr(\br) \psi_0(\br)\; d\br 
$$
is the number-of-particle operator of condensed atoms. Because of the 
commutation relations
$$
[\psi_0(\br),\psi_0^\dgr(\br')] = \vp_0(\br) \vp_0^*(\br') \; ,  
$$
$$
[\psi_0(\br),\hat N_0]=\psi_0(\br)\; , \qquad
[\psi_0^\dgr(\br),\hat N_0] = -\psi_0^\dgr(\br) \; ,
$$
we have 
$$
i\; \frac{d\hat N}{dt} = [ \hat N,H_\ep] \; = \; \ep\; \sqrt{\rho_0} \int_V \; \left [
\psi_0(\br) - \psi_0^\dgr(\br) \right ] \; d\br \; .
$$
This tells us that $\hat N$ is not an integral of motion.

However, the observable quantity is the average number of particles, for which
we get
$$
\frac{d}{dt} \langle \hat N \rangle_\ep \; = \; i\ep\; \sqrt{\rho_0} \;
\int_V \langle \psi_0^\dgr(\br) - \psi_0(\br) \rangle_\ep\; d\br \; .
$$
Quasiaverages in the Bogolubov theory \cite{Bogolubov_18,Bogolubov_19} are 
introduced as the averages under the Hamiltonian, containing the symmetry-breaking 
source, for which then the thermodynamic limit $N\ra\infty$ is taken, after which 
the source is removed, $\ep \ra 0$. The order of the limits cannot be interchanged.
Taking into account that in the Bogolubov theory 
$\langle \psi_0(\br) \rangle_\ep \simeq \eta(\br)$, we find
$$
\frac{d}{dt} \langle \hat N \rangle_\ep \; = \; i\ep\; \sqrt{\rho_0}
\int_V \; [\eta^*(\br)-\eta(\br)] \; d\br \; .
$$
Considering the reduced quantities, with respect to $N\equiv N(0)$,
we come to the equation
$$
\frac{d}{dt}\; \frac{ \langle \hat N \rangle_\ep}{N} =
i\ep\;\frac{\sqrt{\rho_0}}{N} \;
\int_V \; [\eta^*(\br)-\eta(\br)] \; d\br \; .
$$

If the system is uniform, so that 
$\eta(\br) \simeq \sqrt{\rho_0}$ for $N \ra \infty$, then
$$
\lim_{N\ra\infty}\; \frac{d}{dt}\; \frac{<\hat N>_\ep}{N} = 0 \; .
$$
Hence, the average number of atoms is conserved. 

In the general case of any kind of a system whether nonuniform or uniform, 
equilibrium or nonequilibrium, the Cauchy-Schwarz inequality is valid:
$$
\left | \int_V \; \eta(\br)\; d\br \right | \leq \sqrt{N_0 V} \; .
$$
Therefore
$$
\left | \frac{d}{dt} \langle \hat N \rangle_\ep \right | \leq 2\ep N_0 \; .
$$
Since the condensate fraction is $n_0 \equiv \lim_{N\ra\infty} \; N_0/N$,
we see that
$$
\lim_{N\ra\infty} \left | \frac{d}{dt}\;
\frac{\langle \hat N \rangle_\ep}{N} \right | \leq 2\ep n_0 \; .
$$
From here it follows that
$$
\lim_{\ep\ra 0}\; \lim_{N\ra\infty} \left | \frac{d}{dt}\;
\frac{\langle \hat N \rangle_\ep}{N} \right | = 0 \; .
$$
That is, the average number of particles is a well defined quantity and, 
being an observable, it is conserved on average. 

The same conclusion, that the observable number of particles is conserved,
being a well defined fixed quantity, can be derived in the case where the 
gauge symmetry is broken by means of the Bogolubov shift. Then the operator 
of the total number of particles
$$
\hat N[\eta,\psi_1] \equiv
\int \hat\psi^\dgr(\br) \hat\psi(\br)\; d\br
$$
commutes with the Hamiltonian $H[\eta,\psi_1]$. The symmetry is broken not
by adding infinitesimal terms, but in the Fock space $\cF(\psi_1)$ that now
is not gauge symmetric.

In the case of the Bogolubov shift, the operator of condensed atoms $\hat N_0$
becomes a nonoperator quantity $N_0$. This implies that the condensate 
fluctuations are negligible, since
$$
\lim_{N\ra\infty} \; \frac{{\rm var}(\hat N_0)}{N} = 0 \; .
$$
The observable number of atoms remains well defined fixed quantity 
$N = \langle {\hat N} \rangle$.

\section{Trapped atoms}

The trapping of atoms is described by external confining potentials. In the
presence of such potentials, the system becomes nonuniform, which induces
novel properties of Bose systems and requires to resort to different methods
of description. 

\subsection{Trapping potentials}
 
An atom consists of a nucleus and electrons, which produce a magnetic dipole 
moment $\vec\mu$ or an electric dipole moment $\bf d$. So, even neutral atoms 
can be manipulated by external magnetic or electric fields to be confined inside
a trap. There exist magnetic traps, optical traps, and their combinations. It is
not our aim to describe here the details of experimental atomic trapping, which 
requires separate consideration and can be found in other publications, e.g., in
the book by Letokhov \cite{Letokhov_46}. But we only briefly mention the physical
origin of the trapping fields. 
 
As is clear from their name, {\it magnetic traps} are based on the action of 
external magnetic fields on the atomic magnetic moments. An atom possesses a 
nuclear spin $\bf I$, with the largest eigenvalue $I$, and an electron spin $\bS$,
with the largest eigenvalue $S$. There also exists an electron angular momentum 
$\bf L$, with the largest eigenvalue $L$. The total electron momentum $J$ is the 
sum $\bJ = {\bf L} + \bS$. And the total atomic momentum (total atomic spin) $F$
is ${\bf F} = \bJ + {\bf I}$. The atomic magnetic dipole moment is
$\vec \mu = \mu_B g_f {\bf F}$, where $\mu_B = |e| / 2 m_e$ is the Bohr magneton
and $g_F$ is atomic Land\'e factor
$$
g_F = g_J \; \frac{F(F+1) + J(J+1) -I(I+1)}{2F(F+1)} \; , \qquad
g_J = 1 + \frac{J(J+1) + S(S+1) -L(L+1)}{2J(J+1)} \; .
$$
An external magnetic field $\bB(\br,t)$ creates the magnetic potential
$$
U_\mu(\br,t)= -\vec\mu\cdot\bB(\br,t) 
$$
acting on the atom. The required configuration of the total magnetic field
is formed by magnetic coils. 

In {\it optical traps}, an electric laser field $\bE(\br,t)$ creates an electric 
potential
$$
U_d(\br,t) = -\bd(\br,t)\cdot{\bf E}(\br,t) \; .
$$
The electric dipole moment is composed of an internal dipole moment $\bd_0$
and an induced dipole moment $\bd_{ind}(\br,t) = \al{\bf E}(\br,t)$, where
$\al$ is the dipole polarizability. Thus, $\bd(\br,t) = \bd_0 + \bd_{ind}(\br,t)$.

The characteristic variation frequencies of external electric and magnetic
fields are much larger than the typical frequencies of the atomic motion.
Therefore an atom feels an effective potential corresponding to temporal 
and quantum-mechanical averaging. The atom lifetime in a trap is much longer 
than the times of field variations. Thus, the effective trapping potential, 
acting on an atom, is  
$$
U(\br) = \lim_{\tau\ra\infty} \; \frac{1}{\tau}\; \int_0^\tau
\langle U_\mu(\br,t) +U_d(\br,t) \rangle dt \; .
$$
In magneto-optical traps, both electric and magnetic fields are employed 
for creating a trapping potential.

\subsection{Semiclassical approximation}

The simplest approximation, allowing for the treatment of atomic systems
in external trapping potentials, is the semiclassical, or quasiclassical, 
approximation. For an arbitrary external potential $U(\br)$, atomic states 
are described by the Schr\"{o}dinger equation
$$
\left [ -\; \frac{\nabla^2}{2m} + U(\br)\right ]\vp_k(\br) =
\ep_k \vp_k(\br) \; .
$$
Here $U(\br)$ is a confining potential that, in general, can be a sum of an
external potential and a self-consistent potential including atomic 
interactions. 

It is assumed that the confining potential is slowly varying in space. This
means that the mean interparticle distance $a$ is much smaller than the 
characteristic trap size (characteristic length of potential variation). It 
is also supposed that temperature is sufficiently high, such that the thermal 
length $\lbd_T = \sqrt{2\pi/mT}$ is much shorter than the characteristic 
lengths of atomic motion $l_0$,
\be
\label{281}
\frac{a}{l_0} \ll 1 \; , \qquad \frac{\lbd_T}{l_0} \ll 1 \; .
\ee
In other words, this assumes that particle characteristics change in space 
much faster than the trapping potential. The second of conditions (\ref{281})
can be written as 
\be
\label{282}
\frac{\om_0}{T} \ll 1 \; , \qquad \om_0 \equiv \frac{1}{ml_0^2} \; , 
\ee
telling that temperature is to be high as compared to the characteristic trap 
frequency $\om_0$. Fortunately, for trapped atoms, these conditions are almost 
always valid down to sufficiently low temperatures.

The existence of two scales, fast and slow, suggests to represent the solution 
to the Schr\"{o}dinger equation in the form of a product
$$
\vp_k(\br) =  u_k(\br) e^{i\bk\cdot\br} \; ,
$$
in which the function $u_k(\br)$ is slow, while the exponential is fast. Then 
$\nabla \vp_k(\br) \approx i \bk \vp_k(\br)$, so that the Schr\"{o}dinger equation
becomes
$$
\left [ \frac{k^2}{2m} + U(\br)\right ] u_k(\br) \cong
\ep_k(\br) u_k(\br) \; ,
$$
with the quasiclassical energy spectrum 
\be
\label{283}
\ep_k(\br) \equiv \frac{k^2}{2m} + U(\br) \; .
\ee

For large systems, summation over momenta is usually replaced by the 
integration. To understand how the similar procedure can be done in the
semiclassical approximation, let us recall this replacement for a 
uniform system in $d=3$. Then the replacement can be represented as
$$
\sum_k \longrightarrow V \int \; \frac{d\bk}{(2\pi)^3} =
\int \; \frac{d\bk}{(2\pi)^3} \; d\br \; \qquad 
\left ( V \equiv \int d\br \right ) \; .
$$
The same can be done for a $d$ - dimensional space, with the spatial
variables $\br = \{ r_\al \}$ and momenta $\bk = \{ k_\al \}$ where 
$\al = 1,2,\ldots,d$. Then the replacement is
\be
\label{284}
\sum_k \longrightarrow \int \; \frac{d\bk}{(2\pi)^d} \; d\br \; .
\ee
This replacement is what we need for the semiclassical approximation. 

When the integrand depends only on the momentum modulus $k$, then the 
integrals simplify. Thus, in the {\it one-dimensional} case,
$$
\int \; \frac{d\bk}{2\pi} = \frac{1}{\pi} \; \int_0^\infty dk \qquad (d=1) \; .
$$
And in the {\it $d$-dimensional} case, using the spherical coordinates, we
have
$$
d\bk \longrightarrow \; \frac{2\pi^{d/2}}{\Gm(d/2)} \; k^{d-1}\; dk
\qquad (k\equiv|\bk|) \; .
$$
Then 
\be
\label{285}
\int \; \frac{d\bk}{(2\pi)^d} = \frac{2}{(4\pi)^{d/2}\Gm(d/2)} \;
\int_0^\infty k^{d-1}\; dk 
\ee
for any $d\geq 1$. 

The quasiclassical momentum $k(\ep,\br)$ is defined as the solution 
of the equation $\ep_k(\br) = \ep$, giving $k = k(\ep,\br)$. In view of 
the Schr\"{o}dinger equation, we get
\be
\label{286}
k(\ep,\br) = \sqrt{2m [\ep - U(\br)]} \; .
\ee

The quantity under the square root is to be positive (non-negative), 
which defines the volume $\mathbb{V}_\ep$ available for quasiclassical 
motion,
\be
\label{287}
\mathbb{V}_\ep \equiv \{ \br|\; U(\br)\leq \ep\} \; .
\ee
Using $dk = (\prt k(\ep,\br)/\prt \ep) \; d\ep$, we get
$$
\int \frac{d\bk}{(2\pi)^d} \; d\br = \frac{2}{(4\pi)^{d/2}\Gm(d/2)} \;
\int k^{d-1}(\ep,\br) \; \frac{\prt k(\ep,\br)}{\prt\ep} \; d\ep d\br \; .
$$
The {\it density of states} is defined as
\be
\label{288}
\rho(\ep) \equiv \frac{2}{(4\pi)^{d/2}\Gm(d/2)} \;
\int_{\mathbb{V}_\ep} \; k^{d-1}(\ep,\br) \frac{\prt k(\ep,\br)}{\prt\ep}
\; d\br \; .
\ee
Taking into account the replacement
$$
\int \frac{d\bk}{(2\pi)^d} \; d\br \longrightarrow \int_0^\infty
\rho(\ep)\; d\ep \; ,
$$
and the derivative $\prt k(\ep,\br)/\prt\ep = m/k(\ep,\br)$, we obtain the
density of states
\be
\label{289}
\rho(\ep) = \frac{(2m)^{d/2}}{(4\pi)^{d/2}\Gm(d/2)} \;
\int_{\mathbb{V}_\ep} \; [\ep - U(\br) ]^{d/2-1}\; d\br \; ,
\ee
for any dimensionality $d \geq 1$. 

In particular, for the {\it one-dimensional} case, with the available
volume 
$$
\mathbb{V}_\ep =\{ x| \; -x_\ep \leq x \leq x_\ep\} \; ,
$$ 
where $x_\ep$ is found form the equality $U(x_\ep) = \ep$, the density of 
states reads as
$$
\rho(\ep) = \frac{\sqrt{2m}}{2\pi} \; \int_{-x_\ep}^{x_{\ep}} \;
\frac{dx}{\sqrt{\ep-U(x)}} \qquad (d=1) \; .
$$

An important situation is when the system is confined in a box of volume 
$V$, with no potential inside, $U(\br)\ra 0$. Then, in one dimension $(d = 1)$, 
the turning points are given by $x_\ep = L/2$, with $L$ being the system length.
The density of states becomes
$$
\rho(\ep) = \frac{\sqrt{2m}}{2\pi} \; L \ep^{-1/2} \qquad (d=1) \; .
$$
In the general {\it $d$-dimensional} case, the density of states for a system 
in a box is 
\be
\label{290}
\rho(\ep) = \frac{(2m)^{d/2}}{(4\pi)^{d/2}\Gm(d/2)} \; V \ep^{d/2-1}
\qquad (d\geq 1) \; .
\ee
Here the available volume coincides with the box volume 
$V = \int_{\mathbb{V}_\ep} d\br$.

\subsection{Power-law potentials}

A very general situation is when atoms are trapped by power-law potentials
of the form
\be
\label{291}
U(\br) = \sum_{\al=1}^d \; \frac{\om_\al}{2}\left |
\frac{r_\al}{l_\al} \right |^{n_\al} \; \qquad
\left ( l_\al \equiv \frac{1}{\sqrt{m\om_\al}} \right ) \; ,
\ee
with $l_\al$ being the effective trap length in the $\al$ - direction. 
The parameters $\om_\al$ and $n_\al$ are positive. 

As an example, consider a {\it one-dimensional trap} $(d = 1)$, with 
the potential 
$$
U(x) = \frac{\om_0}{2}\left | \frac{x}{l_0}\right |^n \; .
$$
Then
$$
U(x_\ep)= \ep \; , \qquad x_\ep = \left ( \frac{2\ep}{\om_0}
\right )^{1/n}\; l_0 \; ,
$$
and the available region corresponds to $-x_\ep \leq x \leq x_\ep$. The
density of states takes the form
$$
\rho(\ep) = \frac{\ep^{s-1}}{\pi n} \left ( \frac{2}{\om_0}\right )^s \;
\int_0^1 \; \frac{y^{-1+1/n}}{\sqrt{1-y}} \; dy \;  \qquad (d=1) \; ,
$$
where the notation
$$
s \equiv \frac{1}{2} + \frac{1}{n}
$$
is used, having the meaning of an effective {\it confining dimension}. 

To reorganize the above integral, recall the definition of the beta function, 
or Euler integral of the first kind,
$$
B(\mu,\nu) \equiv \int_0^1 x^{\mu-1} (1-x)^{\nu-1} \; dx \; ,
$$
where ${\rm Re}\; \mu > 0$ and ${\rm Re} \; \nu > 0$. The beta function enjoys
the property
$$
B(\mu,\nu) =\frac{\Gm(\mu)\Gm(\nu)}{\Gm(\mu+\nu)} = B(\nu,\mu) \; . 
$$
Using this, we find the density of states
$$
\rho(\ep) = \frac{\ep^{s-1}}{\gm_1\Gm(s)} \qquad (d=1) \; ,
$$
in which
$$
\gm_1  \equiv \frac{\sqrt{\pi}}{\Gm(1+1/n)} \left (
\frac{\om_0}{2}\right )^s \; .
$$

In the general case of $d \geq 1$, we have
$$
\rho(\ep) = \frac{2^s\ep^{s-1}}{\pi^{d/2}\Gm(d/2)\prod_{\al=1}^d n_\al
\om_\al^{1/2+1/n_\al}} \; \int \left ( 1  - \sum_{\al=1}^d y_\al
\right )^{-1+d/2} \; \prod_{\al=1}^d y_\al^{-1+1/n_\al}\; dy_\al \; .
$$
Here, the confining dimension is
\be
\label{292}
s\equiv \frac{d}{2} + \sum_{\al=1}^d \; \frac{1}{n_\al} \; .
\ee
For ${\rm Re} \; \mu > 0$ and ${\rm Re} \; \nu > 0$, and any $d\geq 1$, one has
$$
\int_0^u \; x^{\mu-1} (u-x)^{\nu-1} \; dx =  u^{\mu+\nu-1}\; B(\mu,\nu) \; ,
$$
$$
\prod_{\al=1}^d \; B\left (\frac{1}{n_\al} , \frac{d}{2}
+\sum_{i=1}^{\al-1}\; \frac{1}{n_i}\right ) = \frac{\Gm(d/2)}{\Gm(s)}\;
\prod_{\al=1}^d \; \Gm\left (\frac{1}{n_\al}\right ) \; .
$$
In that way, we find the density of states 
\be
\label{293}
\rho(\ep) = \frac{\ep^{s-1}}{\gm_d\Gm(s)} \;  \qquad (d\geq 1) \; ,
\ee
where
\be
\label{294}
\gm_d \equiv \frac{\pi^{d/2}}{2^s}\; \prod_{\al=1}^d \;
\frac{\om_\al^{1/2+1/n_\al}}{\Gm(1+1/n_\al)} \;  \qquad (d\geq 1) \; .
\ee
For a one-dimensional system $(d = 1)$, we have $\gm_1$ defined above. 

The case of a system in a box of volume $V$ is recovered, when
$$
n_\al \longrightarrow \; \infty \; , \qquad \prod_{\al=1}^d 2l_\al
\longrightarrow \; V \; , \qquad s  \longrightarrow \; \frac{d}{2}\; ,
\qquad l_0  \longrightarrow \; \frac{L}{2} \; , \qquad V=L^d \; .
$$

Here and in what follows we use the notations
$$
l_0 \equiv \left ( \prod_{\al=1}^d \; l_\al \right )^{1/d} \; , \qquad
\om_0 \equiv \left (\prod_{\al=1}^d \; \om_\al\right )^{1/d} \; .
$$

A very often met potential describes {\it harmonic traps}, when $n_\al = 2$.
Then $s = d$ and $\gm_d = \om_0^d$ for all $d \geq 1$. The density of states
becomes
\be
\label{295}
\rho(\ep)  =\frac{\ep^{d-1}}{\Gm(d)\om_0^d} \qquad 
(n_\alpha =2, \;  s=d\geq 1) \; .
\ee

In that way, it is straightforward to calculate any required sum over 
momenta by means of the replacement
$$
\sum_k \longrightarrow \int_0^\infty \rho(\ep)\; d\ep \; .           
$$

\subsection{Condensation in traps}

To illustrate the possibility of Bose-Einstein condensation inside a trap,
let us consider ideal trapped Bose gas. An important quantity for such 
a gas is the Bose-Einstein integral function (\ref{48}). The lower limit 
in this integral, for a large uniform system, is zero, corresponding to 
zero minimal energy. Bose-Einstein condensation in power-law potentials
with the integral function (\ref{48}) has been studied in Refs. 
\cite{Bagnato_76,Bagnato_77,Yan_78,Courteille_79}.  

However, when atoms are confined in a trap, the lower energy is not zero, 
but a finite positive quantity $\varepsilon_{min}$. Therefore the lower 
limit of integral (\ref{48}) has to be $u_{min} \equiv \beta \varepsilon_{min}$. 
Thus for trapped atoms, it has been suggested \cite{Yukalov_80} to introduce 
the generalized integral
\be
\label{296}
g_n(z) \equiv \frac{1}{\Gm(n)} \; \int_{u_{min}}^\infty
\frac{zu^{n-1}}{e^u-z}\; du \; . 
\ee
As is shown below, this generalization is principally important for 
trapped atoms. 

The total number of atoms of ideal Bose gas writes as
$$
N = N_0 + \int n_k\; \frac{d\bk}{(2\pi)^d}\; d\br \; , 
$$
with $n_k = \left [ e^{\bt(\ep_k-\mu)} -1 \right ]^{-1}$. In the 
generalized semiclassical approximation \cite{Yukalov_80}, this takes 
the form 
$$
\int  n_k({\bf r}) \; \frac{d\bk}{(2\pi)^d}\; d\br = \int_{\varepsilon_{min}}^\infty \;
n(\ep)\rho(\ep)\; d\ep \; , 
$$
with the distribution $n(\ep) = \left [ e^{\bt(\ep-\mu)}-1 \right ]^{-1}$. 
Changing the variable to $u = \bt \ep$ gives
$$
N = N_0 + T^s \int_{u_{min}}^\infty \; \frac{z\rho(u)}{e^u-z}\; du \; ,
$$
where $z\equiv e^{\bt\mu}$ is fugacity and $u_{min} \equiv \beta \varepsilon_{min}$.
For the power-law potential, this becomes
\be
\label{297}
N = N_0 + \frac{T^s}{\gm_d}\; g_s(z) \; .
\ee

At the point of condensation, $\mu\ra 0$ and $z\ra 1$, while $N_0 = 0$, which
yields the condensation temperature
\be
\label{298}
T_c =\left [ \frac{\gm_d N}{g_s(1)}\right ]^{1/s} \; .
\ee

Bose-Einstein condensation of ideal gas in power-law potentials has been studied
\cite{Bagnato_76,Bagnato_77,Yan_78,Courteille_79} using the semiclassical 
approximation employing the standard definition of the Bose-Einstein integral 
function (\ref{48}), with zero lower limit. Expression (\ref{48}) for $g_s(1)$ 
is finite for $s > 1$ and diverges for $s \leq 1$, which would imply that 
condensation can occur only for $s > 1$. In particular, condensation could not 
happen in a one-dimensional harmonic trap \cite{Bagnato_76,Bagnato_77}. But the 
situation is different if the more correct generalized form (\ref{296}) is 
employed. The difference arises in those cases where the standard form (\ref{48}) 
diverges, that is, for $s \leq 1$. When it is finite, that is, for $s > 1$, the 
lower limit can be replaced by zero. In this way, we need to attentively consider 
what happens for $s \leq 1$. 

Let us study one-dimensional Bose gas in a {\it harmonic trap}, when $n = 2$
and $d = 1$, hence $s = 1$. Taking into account that the trap houses a finite,
although large, number of atoms $N \gg 1$, we need to evaluate the integral 
$$
g_1(1) = \int_{u_{min}}^\infty \; \frac{du}{e^u-1} \qquad
(u_{min} \ll 1) \; .
$$
For small $u_{min}$, we have $g_1(1) \simeq - \ln u_{min}$. In the considered 
case of ideal gas, its spectrum is $\varepsilon = k^2/2m$. Thence 
$u_{min} = \bt k^2_{min}/2m$. The minimal wave vector is limited by the inverse
effective trap size defined by the expressions  
$$
l_0\equiv \left ( \prod_{\al=1}^d l_\al\right )^{1/d} \; , \qquad
\om_0 \equiv \left ( \prod_{\al=1}^d \om_\al\right )^{1/d} = \frac{1}{ml_0^2} \; . 
$$
Therefore the lower limit in the integral $g_s(1)$ is given by the formula
\be
\label{299}
u_{min} = \frac{\om_0}{2T} \; \qquad \left ( k_{min} =\frac{1}{l_0} \right ) \; .
\ee
This yields
$$
g_1(1) \simeq \ln\left ( \frac{2T}{\om_0}\right ) \qquad \left (
\frac{\om_0}{T} \ll 1 \right ) \; .
$$

For the case $n = 2$ and $s = 1$, parameter (\ref{294}) is $\gm_1 = \om_0$. Then
expression (\ref{298}) gives the equation  
\be
\label{300}
T_c = \frac{N\om_0}{\ln(2T_c/\om_0)} \qquad (d=1, \; n=2, \; s=1) \; .
\ee
This equation exactly coincides with that derived in purely quantum-mechanical 
calculation \cite{Ketterle_81}. We can rewrite equation (\ref{300}) in the 
equivalent form as
$$
T_c = \frac{N\om_0}{\ln(2N) - \ln\ln(2T_c/\om_0)} \; . 
$$
Using the inequality
$$
\frac{T_c}{\om_0} \ll e^{2N} \quad (N \gg 1) \; ,
$$
we obtain the condensation temperature
\be
\label{301}
T_c = \frac{N\om_0}{\ln(2N)} \qquad (d=1, n=2, s=1) \; .
\ee
To illustrate the correctness of the inequality used above, we notice that
$$
\frac{T_c}{\om_0} = \frac{N}{\ln(2N)} \ll e^{2N} \qquad (N\gg 1) \; .
$$
The validity of the semiclassical approximation is preserved, since
$$
\frac{\om_0}{T_c} = \frac{\ln(2N)}{N} \ll 1 \qquad (N\gg 1) \; .
$$
The existence of the finite condensation temperature (\ref{301}) suggests
that Bose condensation could occur in a one-dimensional harmonic trap.

And what happens in a one-dimensional trap for powers higher than harmonic? 
Then, for $d = 1$ and $n > 2$, one has $1/2 \leq s < 1$. Integral (\ref{296})
is estimated as
$$
g_s(1) \simeq \frac{1}{\Gm(s)} \; \int_{u_{min}}^\infty \;
\frac{du}{u^{2-s}} \qquad (0 < s < 1) \; .
$$
This leads to 
$$
g_s(1) \simeq \frac{u_{min}^{s-1}}{(1-s)\Gm(s)} = 
\frac{(2T/\om_0)^{1-s}}{(1-s)\Gm(s)} \; .
$$
As a result, we obtain the condensation temperature
\be
\label{302}
T_c  =\frac{\sqrt{\pi}(1-s)\Gm(s)}{2\Gm(1+1/n)}\; N\om_0 \qquad (0 < s < 1) \; .
\ee
This formula tells us that, it seems, Bose-Einstein condensation could 
occur in any one-dimensional $(d = 1)$ trap for all $n$. 

The behavior of the condensate fraction follows from equation (\ref{297})
that can be represented as
$$
\frac{N_0}{N} + \frac{g_s(z)}{g_s(1)}\left ( \frac{T}{T_c}\right )^s = 1 \; .
$$
As soon as $T < T_c$, one has $\mu \ra 0$ and $z \ra 1$. Therefore the 
condensate fraction depends on temperature as
\be
\label{303}
\frac{N_0}{N} =  1 - \left ( \frac{T}{T_c} \right )^s \qquad (T < T_c) \; .
\ee

Let us specify the condensation temperature for harmonic traps, when 
$n_\alpha = 2$, $s = d$, and
$$
\gamma_d = \omega_0^d  \qquad (n_\alpha = 2 \; , s=d) \; .
$$
For a one-dimensional harmonic trap, the condensation temperature is 
given by expression (\ref{301}). For {\it harmonic traps} of higher 
dimensionality, we have
\be
\label{304}
T_c = \om_0\left [ \frac{N}{\zeta(d)}\right ]^{1/d} \qquad (d\geq 2, n_\alpha = 2, s=d) \; .
\ee
For the dimensionality two and three, we should substitute here
$\zeta(2) = \pi^2/6$ and $\zeta(3) = 1.202057$, respectively.  

Above the transition temperature, $T > T_c$, there is no condensate, $N_0 = 0$,
the fugacity $ z < 1$, and the chemical potential $\mu$ is defined from the 
equation
\be
\label{305}
g_s(z) T^s = N\gm_d \qquad (T > T_c) \; .
\ee

For $T \gg T_c$, the fugacity is small, $z \ll 1$, and we can use 
the estimate $g_s(z) \simeq z$. Then
$$
\mu\simeq -T \ln\left ( \frac{T^s}{N\gm_d}\right ) \qquad (T\gg T_c) \; .
$$
Employing the relation $N \gm_d = g_s(1) T_c^s$ transforms the chemical 
potential to 
$$
\mu\simeq -T \ln\left [ \frac{1}{g_s(1)} \left ( \frac{T}{T_c}
\right )^s \right ] \qquad (T\gg T_c) \; .
$$

In the case of harmonic traps, the high-temperature chemical potential 
reads as
$$
\mu\simeq -T \ln\left [ \frac{1}{N}\left ( \frac{T}{\om_0}
\right )^d \right ] \qquad (T \gg T_c, n_\alpha =2) \; .
$$

In this way, it looks like Bose-Einstein condensation of ideal gas can 
occur in low-dimensional traps, contrary to the case of uniform Bose gas, 
for which the condensation in one-dimensional space $(d = 1)$ is impossible 
at all, the condensation temperature for two-dimensional gas is $T_c = 0$, 
and finite-temperature condensation can exist only for $d > 2$. However,
before concluding on the possibility of condensation, it is worth 
remembering that the uniform ideal Bose gas also formally displays a transition
temperature, but, actually, it does not exist below this temperature,
being unstable. A formal occurrence in theory of a transition temperature 
does not guarantee that the condensation can be real, if below this 
temperature the system is unstable. Recall that the problem of stability
requires the investigation of susceptibilities in thermodynamic limit.

\subsection{Thermodynamic limit}

For finite systems, there are no first- or second order phase transitions, 
but only crossovers can exist \cite{Yukalov_1}. First- or second order 
transitions require the consideration of thermodynamic limit. But how 
thermodynamic limit can be defined for trapped atoms?   

The physical meaning of thermodynamic limit for any statistical system, 
including trapped atoms, can be formulated as follows \cite{Yukalov_80}. 
This limit assumes that the number of atoms tends to infinity, 
$N \ra \infty$. At the same time, extensive observable quantities $A_N$, 
that are statistical averages of self-adjoint operators, increase
proportionally to $N$. Hence, the ratio of extensive observables to the
number of atoms tends to a constant,
\be
\label{306}
\lim_{N\ra\infty}\; \frac{A_N}{N} \ra \; const \; .
\ee

Taking, as an observable quantity, internal energy $E$, it is required 
that $E \propto N$ for all $T$, under $N \ra \infty$. In other words, 
thermodynamic limit implies that
$$
\lim_{N\ra\infty}\; \frac{E}{N} \ra \; const \; ,
$$
where a finite constant is assumed. Thus, thermodynamic limit can be defined 
by the conditions
\be
\label{307} 
N \ra \; \infty \; , \qquad E \ra \; \infty \; , \qquad
\frac{E}{N} \ra\; const \; .
\ee

In the case of power-low confining potentials,
$$
E = \int_0^\infty \;\ep n(\ep)\rho(\ep)\; d\ep \; , \qquad
\frac{E}{N} = \frac{sg_{s+1}(z)}{N\gm_d}\; T^{s+1} \; .
$$
Hence thermodynamic limit can be presented in the form
$$
N \ra\infty \; , \qquad \gm_d \ra 0 \; , \qquad N\gm_d\ra const \; .
$$
Since $N \gm_d = g_s(1) T_c^s$, thermodynamic limit means that
$g_s(1) T_c^s \ra const$, when $N \ra \infty$. 

To illustrate these conditions in more detail, let us consider the case
of {\it equipower traps}, for which $n_\al = n$. The characteristic 
frequencies $\om_\al$ can be different, so that the system can be anisotropic. 
The confining dimension is
\be
\label{308}
s =\left ( \frac{1}{2} + \frac{1}{n}\right ) d \; ,
\ee
being always larger than one-half, $s \geq 1/2$. For an equipower potential 
in $d$-dimensional space, parameter (\ref{294}) is  
\be
\label{309}
\gm_d = \frac{\pi^{d/2}}{\Gm^d(1+1/n)} \left ( \frac{\om_0}{2}\right )^s  
\qquad (d \geq 1) \; . 
\ee
As is seen, $\gm_d \propto \om_0^s$ for any $d \geq 1$. Therefore 
thermodynamic limit for such potentials takes the form
\be
\label{310}
N\ra\infty, \quad  \om_0\ra 0, \quad  N\om_0^s\ra const \; .
\ee
This shows that $\om_0 \propto N^{-1/s}\ra 0$, as $N \ra \infty$. Consequently, 
$l_0 \propto \om_0^{-1/2} \propto N^{1/2s}$. 

These properties confirm the validity of the semiclassical approximation, since,
for $N \gg 1$, one finds
$$
\frac{\om_0}{T_c} \; \propto \; \frac{1}{N} \quad (s<1) \; , \qquad
\frac{\om_0}{T_c} \; \propto \; \frac{\ln N}{N} \quad (s = 1) \; , \qquad
\frac{\om_0}{T_c} \; \propto \; \frac{1}{N^{1/s}} \quad (s >1) \; .
$$

The behavior of the critical temperatures, under $N \ra \infty$, is as follows:
$$
T_c \; \ra \; const \quad (s > 1) \; , \qquad
T_c \; \propto \; \frac{1}{\ln N} \ra \; 0 \quad (s = 1) \; , \qquad
T_c \; \propto \; N^{1-1/s} \; \ra \; 0 \quad (s < 1) \; .
$$
For {\it harmonic traps}, we get
$$
T_c \; \propto \; \frac{1}{\ln N}\; \ra \; 0 \quad (d = 1) \; ,
\qquad T_c \; \ra \;  const \quad (d\geq 2) \; .
$$

Thus, in thermodynamic limit, finite critical temperatures exist only 
for $ s > 1$.

\subsection{Specific heat}

Specific heat is one of the susceptibilities characterizing the system 
stability with respect to thermal energy fluctuations \cite{Yukalov_1}. 
For a system with a fixed number of atoms $N$, the specific heat is 
defined as
\be
\label{311}
C_N = \frac{{\rm var}(\hat H)}{NT^2} =\frac{1}{N} \;
\frac{\prt E}{\prt T} \; .
\ee
The number of particles is kept fixed, since the volume is not defined
for a system in a confining potential. Keeping in mind power-law potentials,
we get
$$
C_N = \frac{sT^s}{N\gm_d} \left [ (s+1) g_{s+1}(z) +
g_s(z) \; \frac{\prt\ln z}{\prt\ln T} \right ] \; .
$$

Above the transition temperature, $T > T_c$, we have $\mu < 0$, hence $z < 1$, 
and $s \geq 1/2$. The chemical potential $\mu = \mu(T)$ is a function of 
temperature above $T_c$. Differentiating equation (\ref{305}) over $\mu$ gives 
$$
\frac{\prt\ln z}{\prt\ln T} = - s\; \frac{g_s(z)}{g_{s-1}(z)}
\qquad ( T > T_c) \; .
$$
This yields the specific heat
\be
\label{312}
C_N = s(s+1) \; \frac{g_{s+1}(z)}{g_s(z)} \; - s^2\;
\frac{g_s(z)}{g_{s-1}(z)} \qquad ( T > T_c) \; .
\ee
Note that $\lim_{T\ra\infty} \; C_N = s$. 

Taking into account that for $z < 1$ all integrals $g_s(z)$ are finite, 
because $s \geq 1/2$, we see that $C_N$ is finite for all $N$. That is,
thermal fluctuations are thermodynamically normal everywhere above $T_c$. 

Below the condensation temperature, $T \leq T_c$, we have $\mu = 0$, 
hence $z = 1$. Then
$$
C_N = \frac{s(s+1)}{N\gm_d} \; g_{s+1}(1) T^s \qquad (T < T_c) \; .
$$
The other form of the specific heat for $T < T_c$ is
\be
\label{313}
C_N = s(s+1) \; \frac{g_{s+1}(1)}{g_s(1)} \left (
\frac{T}{T_c}\right )^s  \qquad (T < T_c) \; .
\ee
As we see, $C_N$ is finite for all $N$, including $N \ra \infty$, since 
$s \geq 1/2$. The finiteness of positive $C_N$ means that thermal 
fluctuations of energy are thermodynamically normal. Therefore the system 
is thermally stable.

At the condensation point $T_c$, the specific heat experiences a 
discontinuity that is characterized by a finite value of the difference 
$$
\Dlt C_N \equiv C_N(T_c+0) - C_N(T_c-0) \; , 
$$
for which we get
\be
\label{314}
\Dlt C_N = - s^2 \; \frac{g_s(1)}{g_{s-1}(1)} \; . 
\ee
For finite systems, the specific heat is always discontinuous, with the 
following jumps, depending on the confining dimension:
$$
\Dlt C_N =  - s^2 \; \frac{\zeta(s)}{\zeta(s-1)} \qquad (s > 2) \; ,
$$
$$
\Dlt C_N =  - \; \frac{2\pi^2}{3\ln(2T_c/\om_0)} \qquad (s=2) \; ,
$$
$$
\Dlt C_N =  - s^2 \zeta(s) (2-s)\Gm(s-1) \left (
\frac{\om_0}{2T_c}\right )^{2-s} \qquad ( 1< s< 2) \; ,
$$
$$
\Dlt C_N =  - \; \frac{\om_0}{2T_c} \; \ln\left ( \frac{2T_c}{\om_0}
\right ) \qquad (s=1) \; ,
$$
$$
\Dlt C_N =  - \; \frac{s^2(2-s)\Gm(s-1)\om_0}{2(1-s)\Gm(s) T_c}
\qquad ( s< 1) \; .
$$
Here, for $s = 1$, we use the relation
$$
g_0(1) = \lim_{z\ra 1} \; z\; \frac{\prt}{\prt z}\; g_1(z) \; , \qquad
g_0(1) \simeq \frac{2T}{\om_0} \; .
$$

The jump is negative, $\Dlt C_N < 0$, for $s\geq 1$ and positive, 
$\Dlt C_N > 0$, for $s < 1$, since $\Gm(s-1) < 0$ for $1/2 \leq s < 1$. 
Thus, for finite systems, $C_N$ is always discontinuous:
$$
C_N(T_c-0) > C_N(T_c+0) \qquad (s\geq 1) \; ,
$$
$$
C_N(T_c-0) < C_N(T_c+0) \qquad (s < 1) \; .
$$

The values of the jumps in thermodynamic limit, when $N \ra \infty$,
tend to zero for $s \leq 2$, but tend to a finite value for $s > 2$. 
When $s > 2 $, then $T_c \propto const$ and
$$
\Dlt C_N \;  \propto\; const < 0  \qquad ( s > 2 ) \; .
$$
For $s = 2$, we have $T_c \propto const$ and $\om_0 \propto N^{-1/2}$,
which gives
$$
\Dlt C_N \; \propto \; -\; \frac{1}{\ln N}\; \ra \; -0
\qquad (s=2) \; .
$$
When $1 < s < 2$, then $T_c \propto const$ and $\om_0 \propto N^{-1/s}$,
hence
$$
\Dlt C_N \;  \propto\; - N^{1-2/s} \; \ra \; -0 \qquad (1<s<2) \; .
$$
For $s = 1$, one has $T_c \propto 1/\ln N$ and $\om_0\propto 1/N$, from where
$$
\Dlt C_N \; \propto \; - \; \frac{(\ln N)^2}{N} \; \ra \; -0 \qquad (s=1) \; .
$$
Finally, if $s < 1$, then $T_c \propto N^{1-1/s}$ and $\om_0 \propto N^{-1/s}$,
which results in
$$
\Dlt C_N \; \propto \; \frac{1}{N} \; \ra \; +0 \qquad (s<1)\; .
$$

In the case of harmonic traps, $\Dlt C_N$ for $d = 1$ is the same as the value
for $s = 1$ above and $\Dlt C_N$ for $d = 2$ is the same as for $s = 2$. And
for a three-dimensional harmonic trap, we find
\be
\label{315}
\Dlt C_N = -\; \frac{54}{\pi^2}\; \zeta(3) \qquad (d=3) \; .
\ee

\subsection{Particle fluctuations}

Particle fluctuations are characterized by the number-of-particles dispersion
that define the isothermal compressibility
\be
\label{316}
\kappa_T = \frac{{\rm var}(\hat N)}{\rho T N} =\frac{1}{\rho^2}
\left ( \frac{\prt\rho}{\prt\mu}\right ) \; ,
\ee
where $\rho$ is average density of atoms in a trap and
$$
{\rm var}(\hat N) \equiv \; \langle \hat N^2 \rangle - \langle \hat N \rangle^2 \; =
T\; \frac{\prt N}{\prt\mu} = \frac{TN}{\rho} \left ( \frac{\prt\rho}{\prt\mu} \right ) \; .
$$

If Bose condensation is possible, then the operator of the total number 
of atoms is the sum $\hat N = \hat N_0 + \hat N_1$. Employing the Bogolubov 
shift fixes the number-of-particle operator of condensed atoms as a 
nonoperator quantity, $\hat N_0 = N_0$. Thence ${\rm var}(N_0) = 0$. As a 
result, ${\rm var}(\hat N) = {\rm var}(\hat N_1)$. So we need to calculate
the dispersion for the number-of-particle operator of uncondensed atoms 
$$
{\rm var}(\hat N) = {\rm var}(\hat N_1) = T \; \frac{\prt N_1}{\prt\mu} \; .
$$
For power-law potentials, 
$$
N_1 = \frac{T^s}{\gm_d}\; g_s(z) \; , 
$$
giving
$$
{\rm var}(\hat N_1) = \frac{T^s}{\gm_d}\; g_{s-1}(z) \; , 
$$
which also can be written as
\be
\label{317}
{\rm var}(\hat N_1) = N\; \frac{g_{s-1}(z)}{g_s(1)} \left (
\frac{T}{T_c}\right )^s \; .
\ee

Above the critical temperature, $T > T_c$, where $z < 1$, the integral
$g_n(z)$ is finite for $n > 0$ and $z < 1$. The integral $g_{s-1}(z)$ 
is finite for $s > 1$. For $s = 0$, we have
$$
g_0(z) = \frac{z}{1-z} \qquad (n=0, \; z<1) \; ,
$$
which also is finite for $z < 1$. Thus we come to the conclusion that
\be
\label{318}
{\rm var}(\hat N_1) \; \propto \; N \qquad (s\geq 1,\; T > T_c) \; ,
\ee
which implies thermodynamically normal fluctuations, with $\kappa_T$ 
being finite for $N \ra \infty$. Hence the system is stable. 

However for $s < 1$, we confront the integral
$$
g_n(z) \simeq \frac{z}{(1-z)|n|\Gm(n)} \left (
\frac{2T}{\om_0}\right )^{|n|}  \qquad (n\leq 0, \;  z < 1) \; ,
$$
which leads to the variance
$$
{\rm var}(\hat N_1) = \frac{zT}{(1-z)(1-s)\Gm(s-1)\gm_d}
\left (\frac{2}{\om_0} \right )^{1-s}  \qquad (s < 1, \; z < 1) \; .
$$
By using the relation
$$
\Gm(-x)\Gm(x) = - \; \frac{\pi}{x\sin(\pi x)} \; ,
$$
we find that $\Gm(s-1) < 0$ for $s < 1$. Therefore the compressibility
\be
\label{319}
\kappa_T \; \propto \; -N^{-1+1/s} \; \ra \; -\infty \qquad
( s < 1, \; T > T_c)
\ee
is divergent and negative, hence the system is unstable. 

In that way, above $T_c$, the system can be stable only for $s \geq 1$.
In particular, a one-dimensional trapped gas with $n > 2$ is unstable,  
which means that it cannot form a stable equilibrium system.

In the case of {\it harmonic traps}, when $s = d$, the ideal gas is always
stable above $T_c$, for any $d \geq 1$.

Below the condensation temperature, $T < T_c$, when $z \ra 1$, we obtain 
$$
{\rm var}(\hat N_1) = \frac{T^s}{\gm_d}\; g_{s-1}(1) \; .
$$
From here, we find
$$
{\rm var}(\hat N_1) = N\; \frac{\zeta(s-1)}{\zeta(s)} \left (
\frac{T}{T_c} \right )^s \qquad ( s > 2 )\; ,
$$
$$
{\rm var}(\hat N_1) = \frac{N}{\zeta(2)} \left ( \frac{T}{T_c} \right )^2 \;
\ln \left ( \frac{2T}{\om_0} \right ) \qquad (s = 2) \; ,
$$
$$
{\rm var}(\hat N_1) = \frac{N}{(2-s)\zeta(s)\Gm(s-1)} \left ( \frac{2T}{\om_0}
\right )^{2-s} \left ( \frac{T}{T_c} \right )^2 \qquad ( 1 < s < 2) \; ,
$$
$$
{\rm var}(\hat N_1) = 2\left ( \frac{T}{\om_0}\right ) ^2 \qquad (s=1)\; .
$$
And for $s < 1$, we get the same dispersion as for $s < 2$. However, for 
$s < 1$, the variance ${\rm var}(\hat N_1)$ becomes negative, since 
$\Gm(s-1) < 0$ for $1/2 < s < 1$. 

In thermodynamic limit, we find
$$
{\rm var}(\hat N_1) \; \propto \; N \quad (s > 2 )\; , \qquad
{\rm var}(\hat N_1) \; \propto \; N\ln N \quad (s = 2 )\; ,
$$
$$
{\rm var}(\hat N_1) \; \propto \; N^{2/s} \quad ( 1< s < 2 )\; ,
$$
$$
{\rm var}(\hat N_1) \; \propto \; N^2 \quad (s = 1 )\; , \qquad
{\rm var}(\hat N_1) \; \propto \; - N^{2/s} \quad (s < 1 ) \; .
$$
This leads to the following behavior of compressibility:
$$
\kappa_T  \; \propto \; const \quad (s  > 2) \; , \qquad
\kappa_T  \; \propto \; \ln N \quad (s  = 2) \; , \qquad
\kappa_T  \; \propto \; N^{-1+2/s} \quad ( 1< s < 2) \; ,
$$
$$
\kappa_T  \; \propto \; N \quad ( s = 1) \; , \qquad
\kappa_T  \; \propto \; -N^{-1+2/s} \quad ( s < 1) \; .
$$

Such an asymptotic behavior shows that the ideal gas, trapped in a
power-law potential becomes unstable below the formal condensation
temperature for all $s \leq 2$. The condensed system can be stable 
only for $s > 2$, which gives the {\it stability condition}
\be
\label{320}
\frac{d}{2} + \sum_{\al=1}^d \; \frac{1}{n_\al} > 2 \; .
\ee
While thermodynamically anomalous density fluctuations and divergent 
compressibility mean that equilibrium cannot be reached in a trapped 
system with $s \leq 2$.
 
In the particular case of {\it harmonic traps}, when $n_\alpha = 2$, 
$s = d$, and $\gm_d = \om_0^d$, we obtain
$$
{\rm var}(\hat N_1) = g_{d-1}(1) \left ( \frac{T}{\om_0} \right )^d \; .
$$
This results in
$$
{\rm var}(\hat N_1) = 2\left ( \frac{T}{\om_0} \right )^2 \qquad (d = 1) \; ,
$$
$$
{\rm var}(\hat N_1) = \left ( \frac{T}{\om_0} \right )^2 \ln
\left ( \frac{2T}{\om_0} \right )^2 \qquad (d = 2) \; ,
$$
$$
{\rm var}(\hat N_1) = N \; \frac{\pi^2}{6\zeta(3)} \left ( \frac{T}{T_c}
\right )^3 \qquad (d = 3) \; .
$$

In thermodynamic limit, we find
$$
{\rm var}(\hat N_1) \; \propto \; N^2 \quad (d =1) \; , \qquad
{\rm var}(\hat N_1) \; \propto \; N \ln N \quad (d =2) \; , \qquad
{\rm var}(\hat N_1) \; \propto \; N \quad (d =3) \; .
$$
Respectively, the behavior of the compressibility is as follows,
$$
\kappa_T \; \propto \; N \quad (d =1) \; , \qquad
\kappa_T \; \propto \; \ln N \quad (d = 2) \; , \qquad
\kappa_T \; \propto \; const \quad (d = 3) \; .
$$
This tells us that a harmonically trapped ideal Bose gas below $T_c$ is 
stable only for $d = 3$. In one-dimensional and two-dimensional harmonic 
traps, density fluctuations are rather strong, destroying equilibrium 
Bose-condensed gas. But for the three-dimensional harmonic potential,
fluctuations are thermodynamically normal \cite{Politzer_82}.

\subsection{Quantum limit}

In order to show that the generalized semiclassical approximation 
\cite{Yukalov_80} leads to the same conclusions as the purely 
quantum treatment of trapped gases, let us consider Bose gas in a 
one-dimensional harmonic trap, using quantum mechanical description
\cite{Ketterle_81}. This case is of interest, since the condensation
temperature does not exist, if the standard semiclassical approximation
is involved \cite{Bagnato_77}, while it exists when the generalized 
semiclassical approximation is employed \cite{Yukalov_80}. 

The case of interest corresponds to $d = 1$ and $n = 2$. The 
quantum-mechanical spectrum is 
$$
\ep_j = \left ( j +\frac{1}{2}\right ) \om \qquad (j=0,1,2,\ldots) \; ,
$$
where $\om = \om_0$. The distribution of atoms over the energy levels
$n_j = \left [ e^{\bt(\ep_j-\mu)} - 1 \right ]^{-1}$  can be written as 
$$
n_j = \frac{z}{e^{j\bt\om} - z} \; , \qquad 
z \equiv \exp\left\{ \bt\left ( \mu -\; \frac{1}{2}\; \om\right ) \right \} \; .
$$
The total  number of atoms
$$
N =\sum_{j=0}^\infty n_j = N_0 + N_1
$$
can be separated into two terms describing the ground-state population and 
the population of excited states, 
$$
N_0 = \frac{z}{1-z} \; , \qquad N_1  =\sum_{j=1}^\infty \;
\frac{z}{e^{j\bt\om}-z} \; .
$$

Condensation occurs when $\mu \ra \om/2$ and $z \ra 1$. Replacing the 
summation over $j$ by integration, we get
$$
N_1 \cong \int_{1/2}^\infty \; \frac{zdj}{e^{j\bt\om}-z} \; .
$$
For $\bt \om \ll 1$, it has been checked numerically \cite{Ketterle_81} 
that a good approximation of the above integral is
$$
N_1 \cong - \; \frac{1}{\bt\om} \; \ln \left (1 - z e^{-\bt\om/2}
\right ) \; .
$$
For $N_1 = N$, this gives the equation for the critical temperature $T_c$,  
$$
N = \frac{T_c}{\om} \; \ln \left ( \frac{2T_c}{\om} \right ) \; ,
$$
which coincides with the equation (\ref{300}) in the generalized semiclassical 
approximation. 

Studying particle fluctuations, we use the derivative 
$$
\frac{\prt N_1}{\prt\mu}  =\frac{z/\om}{e^{\bt\om/2}-z} \; ,
$$
which, under $\bt \om \ll 1$ and $z \ra 1$, yields
$$
{\rm var}(\hat N_1) = 2\left ( \frac{T}{\om}\right )^2 \qquad (T \leq T_c) \; .
$$
Again, this variance exactly coincides with the corresponding expression in 
the generalized semiclassical approximation.

In this way, in the region of applicability of the generalized semiclassical 
approximation, the latter is equivalent to quantum calculations, at the same 
time being essentially simpler.

\subsection{Density distributions}

The density of atoms is the sum $\rho(\br) = \rho_0(\br) + \rho_1(\br)$ of
the density of condensed atoms $\rho_0(\br) = |\eta(\br)|^2$ and of the density
$\rho_1(\br)$ of uncondensed atoms. In the case of ideal gas in a harmonic trap,
the condensate function is the ground-state wave function of a $d$-dimensional 
harmonic oscillator
$$
\eta(\br) = \sqrt{N_0}\left ( \frac{m\om_0}{\pi}\right )^{d/4} \;
\exp\left ( -\; \frac{m}{2}\; \sum_{\al=1}^d \om_\al r_\al^2\right ) \; .
$$
That is, the real-space density of condensed atoms is
$$
\rho_0(\br) = N_0 \left ( \frac{m\om_0}{\pi}\right )^{d/2} \;
\exp\left ( - m \sum_{\al=1}^d \om_\al r_\al^2\right ) \; . 
$$
Recall that
$$
\int \rho_0(\br)d\br =  N_0 \; , \qquad
\om_0 \equiv \left ( \prod_{\al=1}^d \om_\al\right )^{1/d} \; .
$$

At the trap center $\br = 0$, we have 
$$
\rho_0(0) = N_0 \left ( \frac{m\om_0}{\pi} \right )^{d/2} \; .
$$
When $T$ diminishes, $N_0$ increases. In thermodynamic limit, when
$N \ra \infty$, so that $N_0 \propto N$ and $\om_0 \propto N^{-1/d}$,
the central density is
\be
\label{321}
\rho_0(0) \; \propto \; \sqrt{N} \qquad (d\geq 1) \; .
\ee

The distribution of {\it uncondensed atoms} in the semiclassical approximation, 
valid for arbitrary trapping potentials, reads as
$$
n_k(\br) = \frac{1}{\exp\{ \bt[\ep_k(\br)-\mu]\}-1} \; .
$$
In the case of ideal gas, the spectrum is
$$
\ep_k(\br) = \frac{k^2}{2m} + U(\br) \; . 
$$
The real-space distribution of uncondensed atoms writes as
$$
\rho_1(\br) = \int n_k(\br)\; \frac{d\bk}{(2\pi)^d} \; = \;
\frac{1}{\lbd_T^d}\; g_{d/2}(z(\br)) \quad (d\geq 1) \; ,
$$
where
$$
z(\br) \equiv \exp\left \{ \bt [\mu-U(\br) ]\right \} \; ,
\qquad \lbd_T \equiv \sqrt{\frac{2\pi}{mT}} \; .
$$
The normalization conditions are
$$
N_1 = \int n_k(\br) \;  \frac{d\bk}{(2\pi)^d} \; d\br \; = \;
\int \rho_1(\br)\; d\br \; .
$$

Below the condensation temperature, $T \leq T_c$, we can set $\mu = U(0) = 0$.
Then the density of uncondensed atoms at the center of the trap $\br = 0$ 
is given by the expression
$$
\rho_1(0) = \frac{1}{\lbd_T^d} \; g_{d/2}(1) \qquad ( d\geq 1) \; ,
$$
in which, according to the generalized semiclassical approximation 
\cite{Yukalov_80}, 
$$
g_{1/2}(1) \cong 2 \; \sqrt{ \frac{2T}{\pi\om_0} }\; , \qquad
g_1(1) \cong \ln \left (  \frac{2T}{\pi\om_0}\right ) \; .
$$

Thus, for arbitrary trapping potentials, the central density of uncondensed 
atoms is
$$
\rho_1(0) = \frac{2}{\lbd_T} \; \sqrt{ \frac{2T}{\pi\om_0} } \qquad
(d=1) \; ,
$$
$$
\rho_1(0) = \frac{1}{\lbd_T^2} \; \ln \left ( \frac{2T}{\om_0}
\right ) \qquad (d=2) \; ,
$$
\be
\label{322}
\rho_1(0) = \frac{1}{\lbd_T^3} \; \zeta\left (\frac{3}{2}\right )
\qquad (d= 3) \; .
\ee

When $T\ra 0$, then $\rho_1(0) \ra 0$. But at any finite temperature $T > 0$,
when $N \ra \infty$ and $\om_0\propto N^{-1/s}$, then the central density
of uncondensed atoms behaves as
\be
\label{323}
\rho_1(0) \; \propto \; N^{1/2s} \quad (d=1) \; , \qquad
\rho_1(0) \; \propto \; \ln N  \quad (d=2) \; , \qquad
\rho_1(0) \; \propto \; const \quad (d=3) \; .
\ee
This shows that, for dimensionality $d = 1, 2$ and large $N$, uncondensed 
atoms would collapse to the trap center.  

Comparing the central densities of uncondensed and condensed atoms for 
harmonic traps at large $N \ra \infty$, we get
$$
\frac{\rho_1(0)}{\rho_0(0)}\; \ra \; const \quad (d=1)\; ; \qquad
\frac{\rho_1(0)}{\rho_0(0)}\; \ra \; 0 \quad (d\geq 2)\; .
$$
As we see, for one-dimensional traps, the condensate density at the trap 
center cannot be distinguished from the density of uncondensed atoms. This
is connected with the instability of the condensed gas in one-dimensional 
harmonic traps, caused by anomalous particle fluctuations.  

Let us consider the momentum distribution of uncondensed atoms 
$$
n_k \equiv \int n_k(\br) \; d\br
$$
for ideal gas in a harmonic trap. Then we find
$$
n_k = \left ( \frac{2\pi l_0^2}{\lbd_T}\right )^d
g_{d/2}(z_k) \qquad (d \geq 1) \; ,
$$
where
$$
z_k \equiv \exp\left \{ \bt\left ( \mu -\; \frac{k^2}{2m} \right )
\right \} \; .
$$
Below the condensation temperature of ideal gas in a harmonic trap 
of arbitrary anisotropy, $\mu = 0$ for $T \leq T_c$. 

Let us consider the zero-momentum value of $n_k$ at $\bk = 0$ denoted as
$$
\tilde\rho_1(0) \equiv \lim_{k \ra 0} n_k \; .
$$
For different dimensionalities, we find 
$$
\tilde\rho_1(0) =  4 l_0 \left ( \frac{T}{\om_0} \right ) \qquad
(d = 1) \; ,
$$
$$
\tilde\rho_1(0) = 2\pi l_0^2 \; \frac{T}{\om_0}\; \ln\left (
\frac{2T}{\om_0}\right ) \qquad (d = 2) \; ,
$$
\be
\label{324}
\tilde\rho_1(0) = \zeta\left ( \frac{3}{2}\right ) l_0^3 \left ( 2\pi\;
\frac{T}{\om_0}\right )^{3/2} \qquad (d = 3) \; .
\ee
When $T \ra 0$, the zero-momentum density $\tilde \rho_1(0) \ra 0$.

However, for any finite $T$, in thermodynamic limit $N \ra \infty$, 
when $\om_0 \propto N^{-1/d}$, we obtain
\be
\label{325}
\tilde\rho_1(0) \; \propto \; N^{3/2} \quad (d=1) \; , \qquad
\tilde\rho_1(0) \; \propto \; N \ln N \quad (d=2) \; , \qquad
\tilde\rho_1(0) \; \propto \; N \quad (d=3) \; .
\ee
This demonstrates that the zero-momentum density of uncondensed atoms
is anomalously large for dimensionality $d = 1,2$. 

The momentum distribution of condensed atoms is $N_0 \delta_{k0}$. 
Comparing $\tilde \rho_1(0)$, for harmonic traps, with the zero-momentum 
density of condensed atoms $N_0 \propto N$, in thermodynamic limit 
$N \ra \infty$, we have
$$
\frac{\tilde\rho_1(0)}{N_0} \; \propto \; N^{1/2}  \quad (d=1) \; ; \qquad 
\frac{\tilde\rho_1(0)}{N_0} \; \propto \; \ln N \quad (d = 2) \; ; \qquad 
\frac{\tilde\rho_1(0)}{N_0} \; \propto \; const \quad (d = 3) \; .
$$
We see that the zero-momentum density of uncondensed atoms, in low-dimensional
harmonic traps of dimensionality $d = 1,2$, is much larger than the 
zero-momentum density of condensed atoms. Such an anomalous situation is 
connected with the instability of ideal condensed gas in these traps.

\subsection{Finite-size corrections}

The generalized semiclassical approximation \cite{Yukalov_80} makes it 
possible to explicitly estimate corrections caused by the finite number 
of atoms in a trap. Recall that in this approximation the Bose-Einstein 
integral (\ref{48}) is replaced by the generalized integral 
$$
g_n(z) = \; \frac{1}{\Gm(n)} \;
\int_{u_{min}}^\infty \; \frac{zu^{n-1}}{e^u-z} \; du \; , 
$$
with the nonzero lower limit
$$
u_{min} =\frac{\bt k_{min}^2}{2m} = \frac{\om_0}{2T} \qquad 
\left ( k_{min} = \frac{1}{l_0} \right ) \; .
$$
This integral can be splitted in two parts as 
$$
\int_{u_{min}}^\infty = \int_0^\infty  - \int_0^{u_{min}} \; .
$$
Then it follows
$$
\frac{1}{\Gm(n)} \; \int_{u_{min}}^\infty \; \frac{zu^{n-1}}{e^u-z}\;
du = g_n^{(0)}(z) - \; \frac{1}{\Gm(n)}\; \int_0^{u_{min}} \;
\frac{zu^{n-1}}{e^u-z} \; du \; ,
$$
where $g_n^{(0)}(z)$ is given by integral (\ref{48}) with zero lower limit.

Taking into account that $u_{min}\ll 1$ and setting in the second term
$z \approx 1$, we get
$$
\int_0^{u_{min}} \; \frac{u^{n-1}}{e^u-1}\; du \simeq
\frac{u^{n-1}_{min}}{n-1} \qquad (n > 1) \; .
$$
Therefore, we have
$$
g_n(z) =  g_n^{(0)}(z) - \; \frac{u_{min}^{n-1}}{(n-1)\Gm(n)} \; .
$$

Let us consider power-law potentials. Then, taking into consideration 
that $g_s^{(0)}(1) = \zeta(s)$ for $s > 1$, so that 
$g_s(1) \ra \zeta(s)(1-\phi)$, were we define the coefficient 
of finite-size corrections
$$
\phi \equiv \frac{u_{min}^{s-1}}{(s-1)\Gm(s)\zeta(s)} \; ,
$$
we find the condensation temperature
\be
\label{326}
T_c^s = \frac{\gm_d N}{\zeta(s)(1-\phi)} \; .
\ee

This temperature should be compared with the condensation temperature 
$$
T_0 \equiv \left [ \frac{\gm_d N}{\zeta_s(1)}\right ]^{1/s} 
$$
without finite-size corrections. Their ratio is
$$
\left ( \frac{T_c}{T_0} \right )^s \simeq 1 + \phi \qquad ( \phi \ll 1) \; .
$$
For small $\phi$, we have
$$
 \frac{T_c}{T_0} = \frac{1}{s} \; \phi \qquad ( \phi \ll 1) \; .
$$
Introducing the relative temperature shift
$$
\frac{\Dlt T_c}{T_0} \equiv \frac{T_c - T_0}{T_0} \; , 
$$
we obtain
\be
\label{327}
\frac{\Dlt T_c}{T_0} = \frac{1}{s(s-1)\Gm(s)\zeta(s)} \left (
\frac{\om_0}{2T_0}\right )^{s-1} \qquad (s > 1) \; .
\ee
For large $N$, one has $\om_0 \propto N^{-1/s}$, because of which
$$
\frac{\Dlt T_c}{T_0} \; \propto \; \frac{1}{N^{1-1/s}} \qquad (s > 1) \; .
$$
For power-law potentials, with a confining dimension $s > 1$, the shift
of the condensation temperature is positive.  

In the case of {\it harmonic traps}, the temperature shift, caused by 
finite-size corrections, is   
\be
\label{328}
\frac{\Dlt T_c}{T_0} =
\frac{1}{d(d-1)\Gm(d)2^{d-1}[\zeta(d)]^{1/d}N^{1-1/d}} \qquad (s=d>1) \; .
\ee
In particular, for a two-dimensional harmonic trap, the shift is
\be
\label{329}
\frac{\Dlt T_c}{T_0} = \frac{1}{4\sqrt{\zeta(2)}\;N^{1/2}} \qquad (d=2) \; ,
\ee
while for a three-dimensional trap, 
\be
\label{330}
\frac{\Dlt T_c}{T_0} = \frac{1}{48[\zeta(3)]^{1/3}\;N^{2/3}} \qquad (d=3) \; .
\ee

Thus, for harmonic traps, finite-size corrections shift the transition 
temperature upward.

\subsection{Quantum corrections}

In the approximations above, it has been assumed that at the condensation 
point the chemical potential of ideal Bose gas becomes zero. However, 
strictly speaking, for finite quantum systems, the chemical potential 
becomes not zero, but tends to the lowest energy level in the trap, 
$\mu \longrightarrow \ep_0 \equiv \min_k \ep_k$. The energy $\ep_0$ is 
the quantum lowest energy of the spectrum related to the Schr\"odinger
equation. For example, for a one-dimensional harmonic oscillator, 
$\ep_0 = \om_0/2$. For a $d$-dimensional harmonic oscillator
$$
\ep_0 = \frac{1}{2}\; \sum_{\al=1}^d \om_\al \; = \frac{d}{2} \; \overline \om \qquad
\left ( \overline\om \equiv \frac{1}{d} \; \sum_{\al=1}^d \om_\al \right ) \; . 
$$

In the general case of {\it power-law potentials}, the fraction of uncondensed
atoms is
$$
n_1 = \frac{N_1}{N} =\frac{g_s(z)}{g_s(1)} \left ( \frac{T}{T_0}\right )^s \; ,
\qquad N_1 =\frac{T^s}{\gm_d}\; g_s(z) \qquad \left ( z\equiv e^{\bt\mu}\right ) \; .
$$
When temperature tends to the condensation temperature, $T \ra T_c$, then 
$\mu \ra \ep_0$, with this lowest energy being much smaller than $T_c$, because
of which
$$
g_s(z)\simeq g_s(1)+g_{s-1}(1)\bt\ep_0 \qquad \left ( \frac{\ep_0}{T_c} \ll 1 \right ) \; .
$$

Introducing the coefficient of quantum corrections
$$
\phi_q \equiv \frac{g_{s-1}(1)\ep_0}{g_s(1)T_0} =
\frac{g_{s-1}(1)\ep_0}{(\gm_d N)^{1/s}[g_s(1)]^{1-1/s}} \; ,
$$
we get the fraction of uncondensed atoms
\be
\label{331}
n_1  =\left ( \frac{T}{T_0} \right )^s + \phi_q \left (
\frac{T}{T_0}\right )^{s-1} \; ,
\ee
in which $T_0$ is the condensation temperature without quantum corrections. 
Respectively, the fraction of condensed atoms $n_0 \equiv 1-n_1$ reads as
\be
\label{332}
n_0 = 1  -\left ( \frac{T}{T_0} \right )^s - \phi_q \left (
\frac{T}{T_0}\right )^{s-1} \; .
\ee

At the condensation temperature, $T = T_c$, we have $n_0 = 0$ and $n_1 = 1$.
Then formula (\ref{332}) leads to the equation  
$$
t_c^s +\phi_q t_c^{s-1}-1 = 0  \qquad 
\left ( t_c \equiv \frac{T_c}{T_0} \right ) 
$$
for the condensation temperature. The quantum-correction coefficient is 
assumed to be small, $|\phi_q| \ll 1$. For the relative shift of the 
condensation temperature we get 
$$
\frac{\Dlt T_c}{T_0} = \frac{T_c}{T_0} \; - 1 \;   = -\; \frac{1}{s}\; \phi_q \; .
$$

Recall that in the generalized semiclassical approximation \cite{Yukalov_80},
we have the following Bose integrals
$$
g_s(1) = \zeta(s) \quad (s > 1 ) \; , \qquad
g_s(1) \cong \ln \left (\frac{2T}{\om_0}\right ) \quad (s = 1 ) \; ,
$$
$$
g_s(1) \cong \frac{(2T/\om_0)^{1-s}}{(1-s)\Gm(s)} \quad ( 0\neq s < 1) \; ,
\qquad g_0(1) \cong \frac{2T}{\om_0} \quad (s=0) \; .
$$
Using these expressions and the property $\Gm(s) = (s-1) \Gm(s-1)$, we obtain
the condensation-temperature shifts, caused by quantum corrections,
$$
\frac{\Dlt T_c}{T_0} = -\; \frac{\zeta(s-1)\ep_0}{s\zeta(s)T_0}
\qquad ( s > 2) \; ,
$$
$$
\frac{\Dlt T_c}{T_0} = -\; \frac{\ep_0}{2\zeta(2)T_0}\; \ln \left(
\frac{2T_0}{\om_0}\right ) \qquad (s=2) \; ,
$$
$$
\frac{\Dlt T_c}{T_0} = -\; \frac{(s-1)\ep_0}{s(2-s)\Gm(s)\zeta(s)T_0}
\left ( \frac{2T_0}{\om_0}\right )^{2-s} \qquad ( 1 < s < 2) \; ,
$$
$$
\frac{\Dlt T_c}{T_0} = -\; \frac{2\ep_0}{\om_0\ln(2N)} \quad (s=1) \; ,
$$
\be
\label{333}
\frac{\Dlt T_c}{T_0} =\frac{2(1-s)^2\ep_0}{s(2-s)\om_0} \quad (s < 1) \; .
\ee
Notice that for all confining dimensions $s \geq 1$ the shifts are negative, 
and only for $s < 1$ the relative shift becomes positive.

Since $\ep_0$ tends to zero when the number of atoms tends to infinity, 
the quantum corrections at the same time are the finite-size corrections.
Assuming that $\ep_0 \propto \om_0$ and using the relation 
$\om_0 \propto N^{-1/s}$ valid for power-law potentials, we come to the 
behavior
$$
\frac{\Dlt T_c}{T_0} \; \propto \; - \; \frac{1}{N^{1/s}} \quad
( s > 2)\; , 
$$
$$
\frac{\Dlt T_c}{T_0} \; \propto \; - \; \frac{\ln N}{\sqrt{N}} \quad
( s = 2) \; ,
$$
$$
\frac{\Dlt T_c}{T_0} \; \propto \; - \; \frac{1}{N^{1-1/s}} \quad
( 1< s <2)\; ,
$$
$$
\frac{\Dlt T_c}{T_0} \; \propto \; - \; \frac{1}{\ln N} \quad
( s = 1)\; , 
$$
\be
\label{334}
\frac{\Dlt T_c}{T_0} \; \propto \; const \quad (s < 1) \; .
\ee
In the last case, where $s < 1$, quantum corrections cannot be treated 
as small. But the systems with $s < 1$ are strongly unstable having
negative compressibility.

For {\it harmonic traps}, when $\gm_d = \om_0^d$ and $s = d$, we find 
the shifts
$$
\frac{\Dlt T_c}{T_0} = -\; \frac{1}{2\ln N} \qquad ( d=1) \; ,
$$
$$
\frac{\Dlt T_c}{T_0} = -\;
\frac{\overline\om}{4\sqrt{\zeta(2)}\;\om_0\sqrt{N}}\; \ln \left [
\frac{4N}{\zeta(2)}\right ] \qquad (d=2) \; ,
$$
\be
\label{335}
\frac{\Dlt T_c}{T_0} = -\;
\frac{\zeta(2)\overline\om}{2[\zeta(3)]^{2/3}\om_0 N^{1/3}} \qquad
(d=3) \; .
\ee
Quantum corrections reduce the condensation temperature. 

The influence of quantum corrections has been studied for a three-dimensional
harmonic trap by Grossmann and Holthaus \cite{Grossmann_83,Grossmann_84}. 
For a spherical harmonic trap, for which $\overline \om = \om_0$, they 
found the coefficient exactly coinciding with the coefficient in the last 
of shifts (\ref{335}) that equals    
$$
\frac{\zeta(2)}{2[\zeta(3)]^{2/3}}= 0.727504 \; .
$$
For anisotropic traps, $\overline \om > \om_0$, and the temperature shift 
is larger than for the isotropic case. 

For increasing number of atoms $N \ra \infty$ in a harmonic trap, the 
behavior of the shifts is as follows:
$$
\frac{\Dlt T_c}{T_0} \; \propto \; -\; \frac{1}{\ln N} \quad (d=1) \; ,
\qquad \frac{\Dlt T_c}{T_0} \; \propto \; -\; \frac{\ln N}{\sqrt{N}}
\quad (d=2) \; , \qquad
\frac{\Dlt T_c}{T_0}\; \propto\; -\; \frac{1}{N^{1/3}} \quad (d=3) \; .
$$
For harmonic traps, quantum corrections are larger than finite-size
corrections considered in the previous section. Also the quantum 
corrections to the condensation temperature are of opposite sign, as 
compared to the finite-size corrections. But the quantum corrections are
more important.

\subsection{Finite box}

For a finite system, boundary conditions can play role. Thus, let us consider
ideal Bose gas in a finite box of volume $V = L^d$. When using periodic 
boundary conditions, the wave vector is quantized so that 
$k_\al = (2 \pi/L) n_\al$, with $n_\al = 0,1,2,\ldots$. Thence the minimal 
wave vector is $k_\al \ra 0$ and the minimal energy is $\ep_0 \ra 0$. This 
corresponds to a uniform system for which there are neither finite-size nor 
quantum corrections. 

But for atoms, confined in a finite box, more appropriate is the 
{\it zero boundary condition}
$$
\psi_n(\br)=0 \qquad (r_\al=0,L) 
$$
for a wave function labelled by the multi-index $n \equiv \{ n_\al \}$, 
with $n_\al = 1,2,3,\ldots$. The wave function is
$$
\psi_n(\br) = \left ( \frac{2}{L} \right )^{d/2} \; \prod_{\al=1}^d
\sin\left ( \frac{\pi n_\al}{L}\; r_\al\right ) \qquad (\br=\{ r_\al\}) \; . 
$$
It enjoys the symmetry property $|\psi_n(-\br)| = |\psi_n(\br)|$. The energy 
spectrum reads  
$$
\ep_n = \frac{\pi^2}{2mL^2}\; \sum_{\al=1}^d n_\al^2 \; , \qquad
\ep_0 = \frac{\pi^2 d}{2mL^2} \; .
$$
Because of the wave-function symmetry under the inversion $\br \ra -\br$, 
it is sufficient to consider only positive coordinates.

Deriving the density of states, we have to keep in mind that not all parts 
of the box volume are available for the particle motion. Thus a $d$-dimensional 
cube has $2d$ planes on which the wave function is zero, hence these planes
are not available for motion. Therefore from the total density of states in 
the box, we need to subtract $\rho_{sur}(\ep)$ corresponding to the density of 
states on the surface $2d (L/2)^{d-1}$. In that way, the density of states is
\be
\label{336}
\rho(\ep) = \frac{(2m)^{d/2} L^d}{(4\pi)^{d/2}\Gm(d/2)}\; e^{(d-2)/2} - \;
\frac{(2m)^{(d-1)/2}\; 2d(L/2)^{d-1}}{(4\pi)^{(d-1)/2} \; \Gm(d/2-1/2)} \;
\ep^{(d-3)/2} \; .
\ee
For a three-dimensional box, this gives
\be
\label{337}
\rho(\ep) = \frac{(2m)^{3/2}L^3}{(4\pi)^{3/2}\Gm(3/2)} \; \ep^{1/2} \; -
\; \frac{3mL^2}{4\pi} \qquad (d=3) \; .
\ee

The total number of atoms is
$$
N = N_0 + \int_{\ep_0}^\infty \; n(\ep)\rho(\ep)\; d\ep \; ,
$$ 
where
$$
n(\ep) = \frac{z}{e^{\bt\ep}-z} \qquad \left ( z = e^{\bt\mu}\right ) \; .
$$
Note that, in agreement with the generalized semiclassical approximation
\cite{Yukalov_80}, we always limit the integration over energy by the 
minimal energy. Dividing the equation for the total number of atoms by $N$, 
we have the equation 
$$
1 = n_0 + \frac{1}{\rho\lbd_T^3\Gm(3/2)} \; \int_{u_0}^\infty\;
\frac{z\sqrt{u}}{e^u-z} \; du \; - \; \frac{3}{2\rho\lbd_T^2L} \;
\int_{u_0}^\infty \; \frac{zdu}{e^u-z} \; ,
$$
where $u_0 \equiv \bt \ep_0 \ll 1$. 

For temperature approaching $T_c$, we have $\mu \ra \ep_0$. Taking this 
into account makes it straightforward to evaluate the integrals
$$
\frac{1}{\Gm(3/2)} \; \int_{u_0}^\infty \; \frac{z\sqrt{u}}{e^u-z}\; du
\simeq g_{3/2}(z) - 4 \; \sqrt{\frac{u_0}{\pi}} \; ,
$$
$$
\int_{u_0}^\infty \; \frac{zdu}{e^u-z} = -\ln\left ( 1 - ze^{-u_0}
\right ) \simeq -\ln u_0 \; .
$$
Then for the fraction of condensed particles near $T_c$, we find
$$
n_0 =  1 -\; \frac{1}{\rho\lbd_T^3}\left [ g_{3/2}(z) -
4 \; \sqrt{\frac{u_0}{\pi}} + \frac{3\lbd_T}{2L}\; \ln u_0 \right ] \; .
$$
Since $z \simeq 1 + u_0$, as $T \ra T_c$, we have
$$
g_{3/2}(z) \simeq g_{3/2}(1)+ {1/2}(1) u_0 \; , \qquad
g_{1/2}(1) \simeq \frac{2}{\sqrt{\pi u_0}} \; ,
$$
which gives
$$
g_{3/2}(z) \simeq \zeta\left (\frac{3}{2}\right ) +
2\;\sqrt{\frac{u_0}{\pi}} \; .
$$
At $T_c$ the condensate fraction $n_0 \ra 0$. Then the equation for the 
condensate fraction leads to the equation
$$
\zeta\left (\frac{3}{2}\right ) - 2\; \sqrt{\frac{u_0}{\pi}} +
\frac{3\lbd_T}{2L}\; \ln u_0 = \rho \lbd_T^3 \; .
$$
From here, introducing the notation
$$
\lbd_0 \equiv \sqrt{\frac{2\pi}{mT_0}} =
\left [ \frac{\zeta(3/2)}{\rho}\right ]^{1/3} \; , \qquad 
T_0 = \frac{2\pi}{m}\left [ \frac{\rho}{\zeta(3/2)} \right ]^{2/3} \; , 
$$
we get
$$
\left ( \frac{T_c}{T_0}\right )^{3/2} =  1  - \;
\frac{2\sqrt{u_0}}{\sqrt{\pi}\;\zeta(3/2)} + \frac{3\lbd_0}{2\zeta(3/2)L}\;
\ln u_0 \; .
$$
Using the smallness of $u_0 \ll 1$, we derive from the previous equation the 
shift of the condensation temperature, caused by the finiteness of the box, 
$$
\frac{\Dlt T_c}{T_0} = \frac{4\sqrt{u_0}}{3\sqrt{\pi}\;\zeta(3/2)} \; -
\; \frac{\lbd_0}{\zeta(3/2)L}\; \ln u_0 \; .
$$
In view of the equality
$$
u_0 = \frac{\ep_0}{T_0} = \frac{3\pi}{4} \left (\frac{\lbd_0}{L}
\right )^2 \; , 
$$
we find the relative temperature shift
$$
\frac{\Dlt T_c}{T_0} =\frac{2\lbd_0}{\zeta(3/2)L} \left [
\frac{1}{\sqrt{3}} \; - \; \ln\left ( \frac{\sqrt{3\pi}\;\lbd_0}{2L}
\right ) \right ] \; .
$$
Since
$$
\frac{\lbd_0}{L} = \left [ \frac{\zeta(3/2)}{N} \right ]^{1/3} = \frac{1.37725}{N^{1/3}} \; ,
$$
we come to the temperature shift
$$
\frac{\Dlt T_c}{T_0} = \frac{2}{3[\zeta(3/2)]^{2/3}N^{1/3}} \;
\left [ \ln N + \sqrt{3} \; - \; \frac{3}{2}\; \ln\left (\frac{3\pi}{4}
\right ) - \ln \zeta\left( \frac{3}{2}\right ) \right ] \; .
$$
Simplifying the latter, we obtain
\be
\label{338}
\frac{\Dlt T_c}{T_0} = \frac{0.351467}{N^{1/3}} \left ( \ln N -0.513781 \right ) \; .
\ee
For $N \geq 2$, the expression in the brackets is positive. The second term 
in the brackets can be neglected for $N \gg 1$. Therefore
\be
\label{339}
\frac{\Dlt T_c}{T_0} = 0.351467 \; \frac{\ln N}{N^{1/3}} \; .
\ee
This expression is in agreement with the shift found by Grossmann and Holthaus
\cite{Grossmann_85}.

As has been found in section 2.6, the ideal uniform Bose-condensed gas is 
unstable, exhibiting anomalous particle fluctuations and infinite 
compressibility. Particle fluctuations of the ideal gas in a box are also 
anomalous. By calculating ${\rm var}(\hat N) = T \prt N_1/\prt\mu$ for the
three-dimensional space $d = 3$, we use the expression
$$
N_1 = \frac{N}{\rho\lbd_T^3}\left [ \zeta\left (\frac{3}{2}\right ) -
2\sqrt{\frac{u_0}{\pi}} + \frac{3\lbd_T}{2L}\; \ln u_0 \right ] \; .
$$
Taking into account that $u_0 = \bt \mu$ and that at the critical point, 
$\mu \ra \ep_0\ = 3 \pi^2/2mL^2$, we find the derivative
$$
\frac{\prt N_1}{\prt\mu} =\left ( 1 -\; \frac{1}{\sqrt{3}}\right )
\frac{m^2T}{2\pi^3}\; V^{4/3} \; .
$$
Thus we find that particle fluctuations of atoms in a box are anomalous:
\be
\label{340}
{\rm var}(\hat N) = \frac{0.423}{2\pi^3}\; (mT)^2 \; V^{4/3} \; .
\ee
Therefore the ideal Bose-condensed gas in a box is as unstable as the 
uniform ideal Bose-condensed gas.

\subsection{Interaction corrections}

The condensation temperature of a trapped Bose gas depends on atomic 
interactions. The spectrum of an interacting system, above the condensation 
temperature $T_c$, in the semiclassical Hartree-Fock approximation, reads  
$$
\ep_k(\br) = \frac{k^2}{2m} + U(\br) + 2\Phi_0\rho(\br) \; ,
$$
where we keep in mind local interactions (\ref{102}) with the interaction
strength $\Phi_0 = 4 \pi a_s/ m$. The semiclassical atomic distribution 
can be written in the form
$$
n_k(\br) = \frac{z_{int}(\br)}{\exp(\bt k^2/2m)-z_{int}(\br)} \; ,
$$
with the notation
$$
z_{int}(\br) \equiv z(\br) \exp\left \{ -2\bt \Phi_0\rho(\br)\right \}\; ,
\qquad z(\br) = \exp\{ -\bt [\mu-U(\br)] \} \; .
$$

The atomic spatial density is
$$
\rho(\br) = \int n_k(\br) \; \frac{d\bk}{(2\pi)^d} \; = 
\frac{1}{\lbd_T^d}\; g_{d/2}(z_{int}(\br)) \; .
$$

Here we take into account only interaction corrections, caused by weak
interactions, and neglect quantum corrections considered in section 4.11. 
Therefore $\mu \ra 2 \Phi_0 \rho(0)$, when $T \ra T_c$.  

Interactions are assumed to be weak, such that 
\be
\label{341}
\left | \frac{\Phi_0\rho(0)}{T_c}\right | \ll 1 
\ee
for all $N$. Then we can employ the approximation  
$$
g_{d/2}(z_{int}(\br)) \simeq g_{d/2}(z(\br)) -  2\bt \Phi_0 \rho(\br)
g_{d/2-1} (z(\br)) \; .
$$
As a result, the normalization condition for the total number of atoms can 
be represented in the form
$$
N = \int \rho(\br)\; d\br = \frac{T^s}{\gm_d} \; g_s(z) - \; \frac{2\Phi_0}{T\lbd_T^d} \;
\int \rho(\br) g_{d/2-1} (z(\br))\; d\br \; ,
$$
with $z = e^{\bt \mu}$. At the transition point, where $\mu = 2 \Phi_0 \rho(0)$, 
we can write
$$
g_s(z) \simeq g_s(1) + g_{s-1}(1) 2\bt \Phi_0 \rho(0) \; , \qquad
\rho(0) = \frac{1}{\lbd_T^d}\; g_{d/2}(1) \; ,
$$
and the spatial density in the integrand is
$$
\rho(\br) = \frac{1}{\lbd_T^d} \; g_{d/2} \left ( e^{-\bt U}\right ) \; ,
$$
with $U = U(\br)$,

In order that $g_{s-1}(1)$ be limited at large $N$, the confining dimension 
should be $s > 2$. This is in agreement with the fact that the Bose-condensed 
state of trapped atoms is unstable when the confining dimension $s$ is smaller 
than $2$, as is found in section 4.7. The stability condition (\ref{320}) 
requires that $s > 2$, since for $s$ smaller than or equal to $2$, particle 
fluctuations are thermodynamically anomalous. In the present case, 
interactions yield small corrections for all $N$ only when $s > 2$ and $d > 2$.

For power-law trapping potentials, the condensation temperature $T_0$ of 
ideal gas is defined by the relation $\gm_g N = T_0^s g_s(1)$. Our aim is 
to find out how this temperature changes when interactions are switched on. 

For what follows, we introduce the {\it interaction-correction coefficient}    
\be
\label{342}
\phi_{int} \equiv \frac{2\Phi_0}{g_s(1)T_0\lbd_0^d} \left [
g_{d/2}(1) g_{s-1}(1) - S_d(s) \right ] \; ,
\ee
in which the notation
$$
S_d(s) \equiv \frac{\gm_d}{\lbd_T^d T^s} \; \int g_{d/2} \left (
e^{-\bt U} \right ) g_{d/2-1}\left ( e^{-\bt U}\right )\; d\br
$$
is used and 
$$
\lbd_0 \equiv \sqrt{\frac{2\pi}{mT_0}} \; .
$$

The above normalization condition for the total number of atoms, at $T = T_c$,
can be written in the form 
$$
t_c^s + \phi_{int} t_c^{s-1} -1 = 0 \qquad \left ( t_c \equiv \frac{T_c}{T_0} \right ) \; .
$$
By assumption, the interactions are weak, such that the interaction-correction
coefficient is small, $|\phi_{int}| \ll 1$. Then for the relative 
condensation-temperature shift, we have
\be
\label{343}
\frac{\Dlt T_c}{T_0} = - \; \frac{1}{s} \; \phi_{int} \; .
\ee

Evaluating the interaction-correction coefficient (\ref{342}) we keep 
in mind the power-law confining potential
$$
U(\br) = \sum_{\al=1}^d \; \frac{\om_\al}{2}\left | \frac{r_\al}{l_\al}
\right |^{n_\al} \qquad \left ( l_\al \equiv \frac{1}{\sqrt{m\om_\al}}
\right ) \; ,
$$
we take the integral
$$
\int_0^\infty \; \exp\left ( -x^\nu \right ) \; dx = \frac{1}{\nu} \;
\Gm\left ( \frac{1}{\nu} \right ) \qquad ({\rm Re} \; \nu > 0) \; ,
$$
and use the representations
$$
S_d(s) = \sum_{m,n=1}^\infty \; \frac{n}{(mn)^{d/2}(m+n)^{s-d/2}} \; , \qquad
g_s(z) = \sum_{j=1}^\infty \; \frac{z^j}{j^s} \; . 
$$
Since we deal with the case $d > 2$ and $s > 2$, we take $d = 3$. 

The relative condensation-temperature shift (\ref{343}) becomes
\be
\label{344}
\frac{\Dlt T_c}{T_0} = - c_3(s) \; \frac{a_s}{\lbd_0} \; ,
\ee
where $a_s$ is scattering length and 
$$
c_3(s) \equiv \frac{4}{s\zeta(s)} \left [ \zeta\left ( \frac{3}{2}
\right ) \zeta(s-1) - S_3(s) \right ] \; . 
$$

Taking into account that
$$
\zeta(s-1) \simeq \frac{1}{s-2} \quad ( s\ra 2) \; , \qquad
\zeta(s-1) < 0 \quad ( s < 2)
$$
we find that $S_3(s)$ logarithmically diverges as $s \ra 2$, and
$$
c_3(s) \simeq \frac{12\zeta(3/2)}{\pi^2(s-2)} \ra \infty \qquad (s\ra 2) \; .
$$
This is in agreement with the fact that trapped Bose-condensed system with
$s < 2$ is unstable.  

Let us consider a three-dimensional harmonic trap, when $s = d = 3$. Then
$$ 
S_3(3)=1.2084708 \; , \quad c_3(3)=3.426032 \; .
$$
And the shift of the condensation temperature $T_c$ of weakly interacting 
Bose gas, with respect to that of the ideal gas, using the notations
$$
\lbd_0 \equiv \sqrt{\frac{2\pi}{mT_0}} \; , \quad 
T_0 =\om_0 \left [ \frac{N}{\zeta(3)} \right ]^{1/3} \; ,
$$
becomes
\be
\label{345}
\frac{\Dlt T_c}{T_0}= - 3.426032 \; \frac{a_s}{\lambda_0} \; ,
\ee
in agreement with the result of Pitaevskii and Stringari \cite{Pitaevskii_44}. 

We may study the interaction corrections for increasing confining dimension 
$(s \ra \infty)$, when $\zeta(s) \ra 1$ and $S_d(s) \ra 0$. Then 
$$
c_3(s) \simeq \frac{4}{3}\; \zeta\left ( \frac{3}{2} \right ) \quad (s \ra \infty) \; .
$$
The limit $s \ra \infty$ implies that $n_\al \ra 0$, if $d$ is fixed. Taking 
into account that
$$
\gm_d \simeq \frac{\pi^{d/2}}{\Gm(s/d)} \left ( \frac{\om_0}{2}\right )^s \; ,
$$
and
$$
\Gm(z) \simeq \sqrt{\frac{2\pi}{z}}\; z^z e^{-z} \qquad ( |z|\gg 1) \; ,
$$
we have
$$
T_0 \simeq \frac{\om_0 N^{1/s} e}{2s} \; , \qquad 
\lbd_0 \simeq \sqrt{\frac{4\pi s}{m\om_0 N^{1/s}e}} \; .
$$
Since $\om_0 \propto N^{-1/s}$, then $\om_0 N^{1/s}$ is finite. In this 
way, we come to the conclusion that 
$$
\frac{\Dlt T_c}{T_0} \; \propto \; -\; \frac{1}{\sqrt{s}}\;
\ra \; 0 \qquad (s\ra \infty) \; .
$$
The temperature shift tends to zero because the limit $n_\al \ra 0$, under 
fixed $d$, effectively removes the trapping potential, making the system
uniform. And for a uniform system, the condensation temperature of Bose 
gas, in mean-field approximation, coincides with that of ideal Bose gas.

\subsection{Thomas-Fermi approximation}

In the general case of interacting atoms at finite temperature, trapped
in a confining potential $U(\br)$, we have to employ the techniques of
section 3.10 treating nonuniform matter.

For dilute gas, one accepts local interactions, with the delta-function 
potential $\Phi(\br) = \Phi_0 \dlt(\br)$. Resorting to the 
Hartree-Fock-Bogolubov approximation for a nonuniform matter, as in 
section 3.10, for the local potential, we have
\be
\label{346}
\om(\br,\br') = \om(\br)\dlt(\br-\br') \; , \qquad
\om(\br) = -\; \frac{\nabla^2}{2m} + U(\br) + 2\Phi_0 \rho(\br) - \mu_1 \; ,
\ee
and 
\be
\label{347}
\Dlt(\br,\br') =\Dlt(\br)\dlt(\br-\br') \; , \qquad
\Dlt(\br) = \Phi_0\left [ \rho_0(\br) +\sgm_1(\br) \right ] \; .
\ee
Here the local densities are
$$
\rho(\br)=\rho_0(\br) + \rho_1(\br) \; , \qquad 
\rho_0(\br) = |\eta(\br)|^2 \; .
$$
The Bogolubov equations (\ref{204}) become
\be
\label{348}
\om(\br) u_k(\br) +\Dlt(\br)v_k(\br) = \ep_k u_k(\br) \; ,
\qquad
\om^*(\br) v_k(\br) +\Dlt^*(\br)u_k(\br) = -\ep_k v_k(\br) \; .
\ee
Since $\om(\br)$ is real, $\Dlt(\br)$ also can be treated as real. 
Additional conditions for the coefficient functions $u_k(\br)$ and 
$v_k(\br)$ are as in section 3.10. 

The condensate-function equation for an equilibrium system (\ref{186})
reduces to   
\be
\label{349}
\left [ -\; \frac{\nabla^2}{2m} + U(\br)\right ] \eta(\br) +
\Phi_0 \left [ |\eta(\br)|^2 \eta(\br) + 2 \rho_1(\br)\eta(\br) +
\sgm_1(\br) \eta^*(\br) \right ] = \mu_0 \eta(\br) \; .
\ee
For an equilibrium system, the functions $\Dlt(\br)$, $\sgm_1(\br)$,
and $\eta(\br)$ can be considered to be real.

The condensate chemical potential $\mu_0$ can be found from the normalization
condition
$$
N_0 = \int |\eta(\br)|^2 \; d\br \; .
$$
The density of uncondensed atoms $\rho_1(\br)$ and the anomalous average
$\sgm_1(\br)$ are defined in section 3.10. The condition of condensate 
existence
$$
\min_k \ep_k = 0  \qquad \left ( \ep_k \geq 0 \right ) 
$$
defines $\mu_1$. The number of condensed atoms is connected with the total 
number of atoms by the relation
$$
N_0 = N - N_1 \; , \qquad N_1 = \int \rho_1(\br)\; d\br \; .
$$

If the interaction term in the condensate-function equation (\ref{349})
is much larger than the kinetic-energy term, the latter can be omitted,
which is named the Thomas-Fermi approximation. Then we get the condensate 
density
\be 
\label{350}
\rho_0(\br) = \frac{\mu_0 - U(\br)}{\Phi_0} - 2 \rho_1(\br) -\sgm_1(\br) \; .
\ee
This expression is valid for the Fermi region, where the density $\rho_0(\br)$
is non-negative. And outside the Fermi region, one sets $\rho_0(\br) = 0$.

\subsection{Local-density approximation}

The local-density approximation is similar to the semiclassical approximation. 
It is applicable for slowly varying in space external potentials $U(\br)$, 
that is, for weakly nonuniform matter. The meaning of the slow spatial 
variation is defined below. 

Considering the Hartree-Fock-Bogolubov approximation for nonuniform matter 
in section 3.10, one can look for the solutions of the Bogolubov equations 
in the form
$$
u_k(\br) = u(\bk,\br)\vp_k(\br) \; , \qquad
v_k(\br) = v(\bk,\br) \vp_k(\br) \; ,
$$
with the plane wave $\vp_k(\br) \equiv e^{i \bk \cdot \br}/\sqrt{V}$. 
The latter is assumed to be a fast varying function of the spatial variable
$\br$, as compared to the slowly varying functions $u(\bk,\br)$ and 
$v(\bk,\br)$. 

Upon differentiating the coefficient functions, one has
$$
\nabla u_k(\br) =
\vp_k(\br)\left (\nabla +i\bk\right ) u(\bk,\br) \; , \qquad
\nabla v_k(\br) = \vp_k(\br)\left (\nabla +i\bk\right ) v(\bk,\br) \; .
$$
The coefficient functions are slowly varying in $\br$ in the case of 
weak system nonuniformity, so that 
$$
\left | \nabla u(\bk,\br) \right | \ll
k \left | u(\bk,\br)\right | \; , \qquad
\left | \nabla v(\bk,\br) \right | \ll
k \left | v(\bk,\br)\right | \; .
$$

The trapping potential $U(\br)$ and density $\rho(\br)$ are slowly varying
functions of $\br$, such that the dependence on the spatial variable in 
all quantities comes from these functions. Then it is possible to use the
expressions appearing for uniform systems, with replacing the average density
$\rho$ by the local density $\rho(\br)$. To this end, for a dilute gas with 
local interactions, $\om_k$ changes to 
\be
\label{351}
\om(\bk,\br) = \frac{k^2}{2m} + U(\br) + 2\Phi_0 \rho(\br) - \mu_1 \; ,
\ee
$\Dlt_k$ changes to 
\be
\label{352}
\Dlt(\bk,\br) = \Phi_0 \left [ \rho_0(\br) + \sgm_1(\br)\right ] \; ,
\ee
and $\ep_k$ changes to $\ep(\bk,\br)$. As a result, the Bogolubov equations
read as
$$
\left [ \om(\bk,\br) -\ep(\bk,\br)\right ] u(\bk,\br) + \Dlt(\bk,\br)
v(\bk,\br) = 0 \; ,
$$
\be
\label{353}
\Dlt^*(\bk,\br) u(\bk,\br) + \left [ \om^*(\bk,\br) + \ep(\bk,\br)\right ]
v(\bk,\br) = 0\; .
\ee

With $\om(\bk,\br)$ being real, one can treat $\sgm_1(\br)$ as real, hence 
$\Dlt(\bk,\br)$ is also real. Therefore one can look for real solutions 
$u(\bk,\br)$ and $v(\bk,\br)$. Similarly to the case of a uniform system,
we have
$$
u^2(\bk,\br)-v^2(\bk,\br)=1 \; .
$$
The local spectrum of collective excitations is
$$
\ep(\bk,\br) =\sqrt{\om^2(\bk,\br) -\Dlt^2(\bk,\br)} \; .
$$
And, analogously to the uniform case, we find
$$
u^2(\bk,\br) = \frac{\om(\bk,\br)+\ep(\bk,\br)}{2\ep(\bk,\br)} \; ,
\qquad
v^2(\bk,\br) = \frac{\om(\bk,\br)-\ep(\bk,\br)}{2\ep(\bk,\br)} \; .
$$
Quasiparticle excitation distribution 
$\pi_k \equiv \langle b_k^\dgr b_k \rangle$ changes to 
$$
\pi(\bk,\br) =\left \{ \exp\left [ \bt\ep(\bk,\br)\right ] -1
\right \}^{-1} \; .
$$

The condition of condensate existence, in the local-density interpretation, is
\be
\label{354}
\lim_{k\ra 0} \ep(\bk,\br) = 0 \; , \qquad \ep(\bk,\br)\geq 0 \; .
\ee
Then $\mu_1$ becomes 
\be
\label{355}
\mu_1(\br) = U(\br) + \left [ \rho_0(\br) + 2\rho_1(\br) -
\sgm_1(\br) \right ] \Phi_0 \; .
\ee
The local sound velocity $c(\br)$ is given by the equation
\be
\label{356}
mc^2(\br) = \left [ \rho_0(\br) +\sgm_1(\br)\right ] \Phi_0 \; .
\ee
As a result, we get
\be
\label{357}
\om(\bk,\br) = \frac{k^2}{2m} + mc^2(\br)\; , \qquad
\Dlt(\bk,\br) = mc^2(\br) \; .
\ee
The spectrum of collective excitations takes the local form 
\be
\label{358}
\ep(\bk,\br) =\sqrt{c^2(\br) k^2 + \left ( \frac{k^2}{2m}
\right )^2 } \; .
\ee

Following this way, we replace $n_k$ by $n(\bk,\br)$ and $\sgm_k$ by 
$\sgm(\bk,\br)$, defined by the expressions
$$
n(\bk,\br) = \frac{\om(\bk,\br)}{2\ep(\bk,\br)}\; {\rm coth} \left [
\frac{\ep(\bk,\br)}{2T}\right ] \; - \; \frac{1}{2} \; ,
$$
\be
\label{359}
\sgm(\bk,\br) =-\; \frac{mc^2(\br)}{2\ep(\bk,\br)}\; {\rm coth} \left [
\frac{\ep(\bk,\br)}{2T}\right ] \; .
\ee
The local density of uncondensed atoms and the anomalous average are 
given by the integrals
\be
\label{360}
\rho_1(\br) = \int n(\bk,\br) \; \frac{d\bk}{(2\pi)^3} \; , \qquad
\sgm_1(\br) = \int \sgm(\bk,\br) \; \frac{d\bk}{(2\pi)^3} \; .
\ee

Thus, almost all main equations, derived for uniform systems, can be 
straightforwardly transformed for weakly nonuniform systems, employing 
the local-density approximation.

\section{Coherent states}

The phenomenon of coherence possesses two sides, state coherence and  
transition coherence. The state coherence characterizes correlations 
between static properties of the considered objects, while the transition 
coherence describes correlated dynamical processes \cite{Yukalov_86}. 
Generally, these two sides are interconnected.

\subsection{Coherence characteristics}

The appearance of Bose-Einstein condensate implies the appearance of coherent
properties in the system. One can say that Bose-Einstein condensate is the 
coherent fraction of the considered system. In that sense, the condensate 
fraction $n_0 = N_0/N$ plays the role of coherence measure for the system 
as a whole. 

Generally, coherent phenomena in space and time can be described by means 
of correlation functions. The so-called {\it first-order coherence} is 
characterized by the first-order correlation function
\be
\label{361}
C(\br,t,\br',t') \equiv
\frac{\rho(\br,t,\br',t')}{\sqrt{\rho(\br,t)\rho(\br',t')}} \; ,
\ee
in which
$$
\rho(\br,t,\br',t') \equiv\; \langle \psi^\dgr(\br',t')\psi(\br,t) \rangle \; ,
\qquad \rho(\br,t) \equiv \; \langle \psi^\dgr(\br,t)\psi(\br,t) \rangle \; .
$$
By means of this correlation function, one can describe the spatial features
of coherence. 

For an isotropic system, one can define the coherence radius $r_{coh}$ by 
the expression
\be
\label{362}
r_{coh}^2 \equiv \frac{\int r^2 \; |C(\br,t,0,0)|^2 \; d\br}{\int |C(\br,t,0,0)|^2 \; d\br} \; ,
\ee
with the integration over the system volume. Generally speaking, this radius 
depends on time. 

For a cylindrically symmetric system, with the symmetry axis $z$, it is possible 
to introduce two characteristic lengths, the transverse coherence radius $r_{coh}$, 
given by the relation
$$
r_{coh}^2 \equiv \frac{\int r_\perp^2 \; |C(\br,t,0,0)|^2 \; d\br}{\int |C(\br,t,0,0)|^2 \; d\br} \; ,
$$
where $r_\perp \equiv \sqrt{x^2 + y^2}$, and the coherence length $l_{coh}$,
defined by the formula
$$
l_{coh}^2 \equiv \frac{\int z^2 \; |C(\br,t,0,0)|^2 \; d\br}{\int |C(\br,t,0,0)|^2 \; d\br} \; .
$$

For nonequilibrium systems, one can introduce the coherence time $t_{coh}$ 
by the equation
\be
\label{363}
t_{coh}^2 \equiv \frac{\int t^2 |C(\br,t,0,0)|^2 \; dt}{\int |C(\br,t,0,0)|^2 dt} \; ,
\ee
with the integration over time from $t = 0$ to $t = t_{obs}$, where $t_{obs}$
is the observation time. The latter, in particular, can tend to infinity. Generally,
the coherence time depends on spatial location.   

Depending on the coherence radius (in the case of isotropic systems), it 
is possible to distinguish three cases of first-order coherence:
\begin{eqnarray}
\begin{array}{ll}
r_{coh}\lesssim a & ({\rm spatial \; chaos}) \; , \\ \nonumber
a\ll r_{coh} \ll L & ({\rm local \; coherence}) \; , \\
r_{coh}\sim L & ({\rm global \; coherence}) \; .
\end{array}
\end{eqnarray}
Here $a$ is the mean interparticle distance and $L$ is the system linear size. 

In the case of a Bose-condensed system, accomplishing the Bogolubov shift 
$\psi(\br,t) \longrightarrow \eta(\br,t) + \psi_1(\br,t)$, we have
$$
\rho(\br,t,\br',t')=\eta(\br,t)\eta^*(\br',t')+\rho_1(\br,t,\br',t') \; .
$$
Then, in equilibrium, the coherence radius is of the system length, 
$r_{coh} \sim L$, and the coherence time is of order of the observation time,
$t_{coh} \sim t_{obs}$, provided that $\eta(\br,t) \not \equiv 0$.

The {\it second-order coherence} is characterized by the pair correlation 
function
\be
\label{364}
g(\br,\br') = \frac{\rho_2(\br,\br',\br,\br')}{\rho(\br)\rho(\br')} \; ,
\ee
in which
$$
\rho_2(\br,\br',\br,\br')= \langle \psi^\dgr(\br')\psi^\dgr(\br)\psi(\br)
\psi(\br') \rangle
$$
is the second-order density matrix. For compactness, time $t$ is not 
explicitly written. 

Assuming that the interaction potential is integrable, it is possible to 
distinguish two opposite limiting situations. One case corresponds to a purely 
coherent system, when
$$
\rho(\br) \; \ra \; |\eta(\br)|^2 = \rho_0(\br) \; , \qquad
\rho_2(\br,\br',\br,\br') \; \ra \; \rho_0(\br) \rho_0(\br') \; .
$$
Then $g(\br,\br') \ra 1$ exhibiting global coherence. 

Let us stress that the assumption of the interaction-potential integrability
is crucial, since if it is not integrable, then $g(\br,\br) = 0$. 

The opposite situation is when there are no correlations in the system, 
so that the Hartree-Fock approximation is valid, 
$$
\rho_2(\br,\br',t,t') \; \ra \; \rho_1(\br) \rho_1(\br') +
|\rho_1(\br,\br')|^2 \; ,
$$
with only noncondensed particles being present, $\rho_1(\br) = \rho(\br)$.
Then
$$
g(\br,\br')  \; \ra \; 1 +
\frac{|\rho_1(\br,\br')|^2}{\rho_1(\br)\rho_1(\br')} \; .
$$
And $g(\br,\br) \ra 2$, since $\rho_1(\br,\br) = \rho_1(\br)$. 

Summarizing, these opposite cases of second-order coherence are 
characterized by the conditions
\begin{eqnarray}
\label{365}
g(\br,\br) =\left\{ \begin{array}{ll}
1\; , & {\rm coherence} \\ 
2\; , & {\rm chaos} \; .
\end{array} \right. 
\end{eqnarray}

For example, the Hartree-Fock-Bogolubov approximation for a uniform 
Bose-condensed system gives
$$
g(\br,\br) =  1 + \frac{2\rho_0}{\rho^2} (\rho_1 +\sgm_1) \; .
$$
Then there is global second-order coherence, if $\rho = \rho_0$ and 
$\rho_1 = \sgm_1 = 0$, since $g(\br,\br) = 1$. But, generally, there 
occurs partial second-order coherence, when $1 < g(\br,\br) < 2$. 

We stress it again that an integrable interaction potential is assumed.
Otherwise, $g(\br,\br)$ would always be zero.

\subsection{Interference effects}

In the presence of several coherent parts of the whole system, there arise
{\it interference effects} between these coherent parts. Thus, let the 
condensate function be composed of two parts, called coherent modes, so 
that
$$
\eta(\br,t)=\eta_1(\br,t)+\eta_2(\br,t) \; .
$$
The {\it interference density}
$$
\rho_{int}(\br,t) \equiv |\eta(\br,t)|^2 - \sum_n |\eta_n(\br,t)|^2
$$
describes interference fringes arising as a result of the mode interference.
As is clear,
$$
\rho_{int}(\br,t)=2{\rm Re}\; \eta_1^*(\br,t)\eta_2(\br,t) \; .
$$
Assume that 
$$
\eta_n(\br,t)=\eta_n(t)e^{i\bk_n\cdot\br} \; ,
$$
with real $\eta_n^*(t) = \eta_n(t)$. Such expressions can arise in the
process of expansion of two Bose condensates from different traps or from one
trap separated in two parts. Then
$$
\rho_{int}(\br,t)=2\rho_{12}(t)\cos(\bk_{12}\cdot\br) \; , 
$$
where $\rho_{12}(t) \equiv \eta_1(t) \eta_2(t)$ and $\bk_{12} \equiv \bk_1 - \bk_2$.

The other characteristic, describing interference effects, is the
{\it interference current}
$$
\bj_{int}(\br,t) \equiv \bj(\br,t) - \sum_n \bj_n(\br,t) \; ,
$$
in which the total current $\bj(\br,t)$ and partial currents $\bj_n(\br,t)$,
respectively, are
$$
\bj(\br,t) \equiv -\; \frac{i}{2m} \left [\eta^*(\br,t) \;
\nabla\eta(\br,t) - \eta(\br,t)\; \nabla\eta^*(\br,t)\right ] \; ,
$$
$$
\bj_n(\br,t) \equiv -\; \frac{i}{2m} \left [\eta_n^*(\br,t) \;
\nabla\eta_n(\br,t) - \eta_n(\br,t)\; \nabla\eta_n^*(\br,t)
\right ] \; .
$$
The interference current writes
$$
\bj_{int}(\br,t)= 2{\rm Re}\;\bj_{12}(\br,t) \; ,
$$
where
$$
\bj_{12}(\br,t) \equiv  -\; \frac{i}{2m} \left [\eta_1^*(\br,t)\;
\nabla\eta_2(\br,t) - \eta_2(\br,t)\; \nabla\eta_1^*(\br,t)
\right ] \; .
$$

If the coherent modes depend on time as
$$
\eta_n(\br,t) = \eta_n(\br)e^{-i\ep_n t} \; ,
$$
with real $\eta_n^*(\br) = \eta_n(\br)$, the interference current takes 
the form
$$
\bj_{int}(\br,t)= 2\bj_{12}(\br)\sin(\om_{12}t) \; ,
$$
in which $\om_{12} \equiv \ep_1 - \ep_2$ and
$$
\bj_{12}(\br) \equiv \frac{1}{2m} \left [ \eta_1(\br)\;
\nabla\eta_2(\br) - \eta_2(\br)\; \nabla\eta_1(\br)\right ] \; . 
$$

The interference current $\bj_{int}(\br,t)$ is called the
{\it Josephson current}. Such currents appear, e.g., when Bose-Einstein 
condensate is separated into the wells of a double-well trap, or 
when several coherent modes are excited in a single trap
 \cite{Yukalov_23,Yukalov_24}.

\subsection{Coherent field}

As has been explained above, a Bose-condensed system is described in the
Fock space $\mathcal{F}(\psi_1)$ generated by the field operators $\psi_1$. 
The vacuum state of this space is the state $|\eta \rangle$, such that
$\psi_1 |\eta \rangle = 0$. At the same time, we know that a coherent state
is an eigenstate of a field operator \cite{Yukalov_1,Klauder_87}. In the 
present case, the coherent state is the eigenstate of the field operator 
$\psi = \eta + \psi_1$. Here $\eta$ is the condensate function, while $\psi_1$ 
is the field operator of uncondensed atoms. The condensate-function equation, 
as defined in (\ref{171}), is    
$$
i\; \frac{\prt}{\prt t}\; \eta(\br,t) =
\left \langle \frac{\dlt H[\eta,\psi_1]}{\dlt\eta^*(\br,t)} \right \rangle \; . 
$$ 
If we assume that all the system is in the vacuum state and average over 
this state $|\eta \rangle$, then we obtain the coherent-field equation (\ref{188})
or, for a dilute gas, equation (\ref{189}). Thus, the coherent-field equation  
is a particular case of the equation for the condensate wave function, when all 
the system is assumed to be in the vacuum state.    

The assumption that the whole system is in its vacuum state is equivalent to
the supposition that all particles are in condensate, so that $N_0 \approx N$.
The physical situation, when the system is practically completely condensed, 
corresponds to the case of asymptotically weakly interacting gas at very low 
temperature. The related conditions are:
\begin{eqnarray}
\begin{array}{ll}
Weak\; interaction: & \rho|a_s|^3\ll 1 \; . \\ \nonumber
Low \; temperature: & \rho\lbd_T^3\gg 1 \; .
\end{array}
\end{eqnarray}
Here $\lbd_T\equiv\sqrt{2\pi/mT}$. 

A dilute gas interacts through the local potential $\Phi(\br) = \Phi_0\dlt(\br)$,
where $\Phi_0 =  4\pi a_s/m$, with $a_s$ being scattering length. In the case 
of repulsion, $a_s > 0$, and for attraction, $a_s < 0$. 

If the whole system is in a coherent state, being completely Bose-condensed,
it is described by the Hamiltonian
\be
\label{366}
H^{(0)} = \int \eta^*(\br) \left ( - \; \frac{\nabla^2}{2m} + U -
\mu_0 \right ) \eta(\br)\; d\br + \frac{1}{2}\; \Phi_0 \;
\int |\eta(\br)|^4 \; d\br \; .
\ee
And the coherent-field is defined by the equation 
\be
\label{367}
i\; \frac{\prt}{\prt t}\; \eta(\br,t) =
\frac{\dlt H^{(0)}}{\dlt\eta^*(\br,t)} \; , \qquad
\left ( \int |\eta(\br,t)|^2 \; d\br = N_0 \right ) \; .
\ee
This leads to the Nonlinear Schr\"odinger (NLS) equation
\be
\label{368}
 i\; \frac{\prt}{\prt t}\; \eta(\br,t) = H[\eta] \; \eta(\br,t) \; ,
\ee
with the Nonlinear Schr\"odinger Hamiltonian
\be
\label{369}
H[\eta] \equiv -\; \frac{\nabla^2}{2m} + U(\br,t) + \Phi_0|\eta|^2
- \mu_0 \; .  
\ee
In equilibrium, $\eta(\br,t) = \eta(\br)$, and one gets the eigenproblem
\be
\label{370}
-\; \frac{\nabla^2}{2m} + U(\br) + \Phi_0|\eta(\br)|^2 = \mu_0 \eta(\br) 
\ee
in the form of a stationary NLS equation.  

In this way, the correct meaning of the NLS equation for a Bose system is 
that this equation describes the coherent field corresponding to the vacuum 
state of a system that is assumed to be completely Bose-condensed. This 
equation is widely used for Bose systems with very weak interactions at 
zero temperature.

\subsection{Hydrodynamic equations}

The NLS equation is often employed for describing weakly interacting 
superfluids at zero temperature, since coherent systems are superfluid. 
If all the system is assumed to be coherent, it is superfluid as a whole. 
This follows from the definition of the superfluid fraction 
$$
n_s =  1 - \; \frac{\bt}{3m N}\; {\rm var}(\hat\bP) \; , \qquad
{\rm var}(\hat\bP) \equiv \; \langle \hat\bP^2 \rangle - \langle \hat\bP \rangle^2 \; .
$$
If the whole system is coherent, then its operator of momentum is
$$
\hat\bP = \int \eta^*(\br)\; (-i\nabla)\eta(\br)\; d\br \; .
$$
From this, we have ${\rm var}(\hat\bP) = 0$. Therefore $n_s = 1$. Hence the 
total system is superfluid.

One employs the Madelung representation
\be
\label{371}
\eta(\br,t) =\sqrt{\rho_0(\br,t)}\; e^{iS(\br,t)} \; , \qquad
\rho_0(\br,t) =|\eta(\br,t)|^2 \; .
\ee
Introducing the density of current
\be
\label{372}
\bj_0(\br,t) = -\;\frac{i}{2m} \left (\eta^*\; \nabla\eta -
\eta\; \nabla\eta^*\right ) \; =\rho_0(\br,t)\bv(\br,t) 
\ee
and the velocity field
\be
\label{373}
\bv(\br,t) \equiv \frac{1}{m}\; \nabla S(\br,t) \; ,
\ee
and substituting them into the NLS equation, one derives the continuity 
equation and the velocity-field equation
$$
\frac{\prt\rho_0}{\prt t} + \nabla\cdot\bj_0 = 0 \; ,
$$
\be
\label{374}
m\; \frac{\prt\bv}{\prt t} + \nabla\left ( U + \rho_0\Phi_0 +
\frac{mv^2}{2} + U_q \right )  = 0 \; ,
\ee
with the notation for the quantum potential
$$
U_q \equiv - \; \frac{\nabla^2\sqrt{\rho}}{2m\sqrt{\rho_0}} \; .
$$

If there is no strong nonuniformity in the system, so that there are no 
sharp gradients, one neglects the quantum potential $U_q$, retaining only
first-order derivatives. This results in the hydrodynamic equations for 
a superfluid
\be
\label{375}
\frac{\prt\rho_0}{\prt t} + \nabla\cdot(\rho_0\bv) = 0 \; , \qquad
m\; \frac{\prt\bv}{\prt t} + \nabla \left ( \mu_{eff} + \frac{mv^2}{2}
\right )  = 0 \; ,
\ee
with the effective chemical potential 
$$
\mu_{eff}(\br,t) \equiv U(\br,t) +\rho_0(\br,t)\Phi_0 \; .
$$

Recall that these hydrodynamic equations are valid only at zero temperature
and very weak interactions, since uncondensed atoms are not taken into account.

\subsection{Thomas-Fermi limit}

In section 4.14, the Thomas-Fermi approximation has been introduced for the 
condensate-function equation. Now we shall analyze this approximation for
the coherent-field equation (\ref{370}) that is a stationary NLS equation.     
  
Let us consider an isotropic harmonic trap, for which 
$\om_0 \equiv \om_x = \om_y = \om_z$, so that the trapping potential is
$$
U(\br) = \frac{m}{2}\; \om_0^2 r^2 \qquad (r\equiv |\br|) \; .
$$
In spherical coordinates $\br = \{ r,\vt,\vp\}$, we have
$$
\nabla^2 = \frac{1}{r^2}\; \frac{\prt}{\prt r}\left ( r^2\;
\frac{\prt}{\prt r}\right ) + \frac{1}{r^2\sin\vt} \;
\frac{\prt}{\prt\vt} \left (\sin\vt\; \frac{\prt}{\prt\vt}\right ) +
\frac{1}{r^2\sin^2\vt}\; \frac{\prt^2}{\prt\vp^2} \; .
$$

It is convenient to pass to dimensionless quantities by scaling the 
spatial coordinate in units of the oscillator length,
$$
r\; \longrightarrow \;  r/l_0 \; , \qquad \left ( l_0 \equiv
\frac{1}{\sqrt{m\om_0}} \right) \; ,
$$
and defining the dimensionless energy 
\be
\label{376}
E \equiv \frac{\mu_0}{\om_0} \; .
\ee
Keeping this scaling in mind, we treat below the variable $r$ as being   
dimensionless. To return back to the dimensional spatial variable, we have 
to make the substitution $r \longrightarrow r/l_0$. 

The ground-state solution to equation (\ref{370}) does not depend on the 
spherical angles, which allows us to look for a solution in the form
$$
\eta(\br) = \sqrt{\frac{N_0}{4\pi l_0^3}} \; \frac{\chi(r)}{r} \; ,
$$
respecting the normalization conditions
$$
\int |\eta(\br)|^2\; d\br =  N_0 \; , \qquad
\int_0^\infty |\chi(r)|^2\; dr =  1 \; .
$$

By introducing the dimensionless coupling parameter
\be
\label{377}
g \equiv \frac{\Phi_0 N_0}{\om_0 l_0^3} = 4\pi\; \frac{a_s}{l_0}\; N_0 \; ,
\ee
equation (\ref{370}) reduces to
\be
\label{378}
-\; \frac{1}{2}\; \frac{d^2\chi}{dr^2} + \frac{r^2}{2}\;\chi +
\frac{g}{4\pi r^2}\; \chi^3 = E\chi \; ,
\ee
with the boundary condition $\chi(0) = 0$.

From the above equation, it is straightforward to find the asymptotic behaviour
of the solution at short and large distance,
$$
\chi(r)\simeq c_1 r + c_3 r^3 + c_5 r^5 \qquad (r\ra 0) \; ,
$$
\be
\label{379}
\chi(r) \; \propto \; r\exp\left ( -\;\frac{r^2}{2}\right ) \qquad
(r\ra \infty) \; .
\ee

The Thomas-Fermi limit corresponds to the strong-coupling limit $g \ra \infty$,
when the interaction term is much larger then the kinetic-energy-term. Then 
one can neglect the kinetic term, coming to the equation 
$$
\frac{r^2}{2}\;\chi + \frac{g}{4\pi r^2}\; \chi^3 = E_{TF}\chi \; .
$$
This results of the Thomas-Fermi approximate solution
\be
\label{380}
\chi_{TF}(r) \simeq r\Theta(r_{TF}-r)\;
\sqrt{\frac{2\pi}{g}\left ( r_{TF}^2 - r^2\right )} \; ,
\ee
where the Thomas-Fermi radius is 
\be
\label{381}
r_{TF} \equiv \sqrt{2E_{TF}}
\ee
and $\Theta(x)$ is the unit-step function
\begin{eqnarray}
\Theta(x) = \left\{ \begin{array}{ll}
1\; , & x > 0 \\ \nonumber
0 \; , & x < 0 \end{array} \; .
\right. 
\end{eqnarray}

From the normalization integral, we get the expression for the energy
\be
\label{382}
E_{TF} = \frac{1}{2}\left ( \frac{15}{4\pi}\; g\right )^{2/5} =
0.536689\; g^{2/5} \; , 
\ee
that is valid for large $g \ra \infty$. 

Thus the condensate density becomes 
\be
\label{383}
\rho_0(\br) = |\eta(\br)|^2 = \frac{N_0}{2gl_0^3}\;
\Theta(r_{TF}-r) \left ( r_{TF}^2  - r^2\right ) \; , 
\ee
with the Thomas-Fermi radius
\be
\label{384}
r_{TF} =  \left ( \frac{15}{4\pi}\; g\right )^{1/5} \; \gg 1 \; .
\ee

In the center of the trap, the central density is
$$
\rho_0(0) = \left ( \frac{15}{4\pi} \right )^{2/5} \;
\frac{N_0}{2g^{3/5}l_0^3} \; .
$$

For the average kinetic energy, we have
$$
\overline K \equiv \frac{1}{N_0} \; \int \eta^*(\br) \left (
-\; \frac{\nabla^2}{2m}\right ) \eta(\br)\; d\br =
$$
$$
 -\; \frac{1}{2ml_0^2} \; \int_0^\infty \;
\chi\; \frac{d^2\chi}{dr^2}\; dr \; = \frac{\om_0}{2}  \; \int_0^\infty \;
\left | \frac{d\chi}{dr} \right |^2 \; dr \; .
$$
Using here the expression
$$
\left | \frac{d\chi}{dr} \right |^2 = \frac{2\pi}{g} \; r_{TF}^2
\left ( \frac{r_{TF}^2}{r_{TF}^2-r^2} \; - \;
\frac{4r^2}{r_{TF}^2} \right ) \; \Theta(r_{TF} - r) \; ,
$$
and taking the integral
$$
\int \frac{dx}{1-x^2} = \frac{1}{2}\; \ln\; \frac{1+x}{1-x} \; ,
$$
we see that $\overline K$ diverges.

In order to correct the result for the average kinetic energy, one considers 
a boundary layer of thickness $\dlt$, such that 
$\dlt/r_{TF} \ll 1$. Then the integral
$$
\int_0^{r_{TF}-\dlt} \; \left | \frac{d\chi}{dr} \right |^2 \; dr
\simeq \frac{\pi}{g} \; r_{TF}^3 \left ( \ln\; \frac{2r_{TF}}{\dlt}\;
- \; \frac{8}{3}\right )
$$
is finite. The second term here can be neglected for small $\dlt$, which
yields 
$$
\overline K \cong \frac{\pi\om_0}{2g}\;  r_{TF}^3 \ln\;
\frac{2r_{TF}}{\dlt} \; .
$$
Taking the boundary-layer width $\dlt \sim r_{TF}^{-1/3}$ and neglecting $\ln 2$,
as compared to $\ln(r_{TF}/\dlt)$, we find 
\be
\label{385}
\frac{\overline K}{\om_0} \cong \frac{2\pi}{3g} \; r_{TF}^3
\ln r_{TF} \; \cong \frac{5}{2r_{TF}^2}\; \ln r_{TF} \; ,
\ee
with the dimensionless coupling parameter
$$
g =\frac{4\pi}{15}\; r_{TF}^5 \; .
$$

The Thomas-Fermi approximation is connected to the effective thermodynamic 
limit, when $N \ra \infty$ and $\om_0 \ra 0$, so that $N \om_0^3\ra const$.
Really, in this limit $\om_0 \propto N^{-1/3}$ and $l_0 \propto N^{1/6}$. 
Since $N_0 \sim N$, we have $g \propto N^{5/6}$ and $\rho_0(0) \propto const$.
The Thomas-Fermi radius $r_{TF} \propto g^{1/5} \propto N^{1/6}$, hence 
$r_{TF} l_0 \propto N^{1/3}$. Therefore the energy 
$E_{TF} \propto g^{2/5} \propto N^{1/3}$. And the total energy per atom is 
finite, $E_{TF} \om_0 \; \propto \; const$. 

The average kinetic energy per atom is
$$
\frac{\overline K}{\om_0} \; \propto \; \frac{\ln g}{g^{5/3}} \;
\propto \; \frac{\ln N}{N^{1/3}} \; , 
$$
which gives
$$
\overline K \; \propto \; \frac{\ln N}{N^{2/3}} \; \ra \; 0
\qquad (N\ra\infty) \; .
$$

This behaviour shows that the Thomas-Fermi limit is the same as the 
strong-coupling limit, or thermodynamic limit.

\subsection{Optimized approximants}

If the coupling parameter $g$ is not necessarily large, which can occur 
for a small scattering length or a small number of atoms, then a more 
delicate procedure for evaluating the observable quantities, e.g., 
energy, is required. To find accurate approximations for observable 
quantities in the whole range of the coupling parameter 
$g \in [0,\infty)$, one can resort to {\it optimized perturbation theory}
\cite{Yukalov_88,Yukalov_89,Yukalov_90,Yukalov_91}. 

Suppose we are looking for a real function $E(g)$ of a real variable $g$.
For concreteness, we can keep in mind the energy as a function of the 
coupling parameter. But the theory is applicable to any other function
of a real variable. Complicated problems are rarely solved exactly. The 
standard situation is when the problem can be treated only by perturbation 
theory, yielding a sequence of perturbative approximants $E_i(g)$,
with the index $i = 0,1,2,\ldots$ enumerating the approximation order. 
Practically in all realistic cases, the sequence $\{ E_i(g) \}$ diverges
for any finite value of $g$. How then one could extract information from
a divergent sequence?

This can be done by resorting to optimized perturbation theory, whose main 
idea is the introduction of control functions $u_i = u_i(g)$ that would 
transform the divergent sequence into a convergent one. This can be 
realized by including a set of parameters $u_i$ either through initial
approximations or by means of a sequence transformation, so that, instead
of $E_i(g)$, we would get $E_i(g,u_i)$. The next step is the definition
of the control functions $u_i = u_i(g)$ leading to the optimized 
approximants 
$$
\tilde E_i(g) \equiv E_i(g,u_i(g)) \; .
$$
The role of the control functions $u_i(g)$ is to control the convergence 
of the optimized sequence $\{ \tilde E_i(g) \}$. 

Since, by their meaning, control functions have to produce convergence, they
should be related to a convergence criterion. Thus the Cauchy criterion 
of uniform convergence states that the sequence $\{ \tilde E_i(g) \}$ 
converges if and only if for each $\ep$, there exists $N_\ep$, such that
$$
\left |\tilde E_{i+j}(g) -\tilde E_i(g)\right | < \ep \; , \qquad
\left ( \forall i > N_\ep\; , \quad j \geq 1 \right ) \; .
$$ 

It is clear that the fastest convergence would occur when the control 
functions could provide the minimum
$$
\min_{u_i} \left |  E_{i+1}(g,u_{i+1}) - E_i(g,u_i) \right | \; .
$$
However, this expression contains two control functions. We need to 
simplify the Cauchy criterion to be able to get solvable equations 
defining optimal control functions. For this purpose, we represent the 
above minimum in the form  
$$
\min_{u_i} \left | E_{i+1}(g,u_i) - E_i(g,u_i) + ( u_{i+1} - u_i )\;
\frac{\prt}{\prt u_i}\; E_i(g,u_i) \right | \; ,
$$
taking into account the change of the approximation order plus the 
variation of the control function. Then we use the inequality
$$
 \left | E_{i+1} - E_i + ( u_{i+1} - u_i )\;
\frac{\prt}{\prt u_i}\; E_i \right | \leq |E_{i+1} - E_i| +
\left | (u_{i+1}-u_i)\; \frac{\prt}{\prt u_i}\; E_i \right | \; ,
$$
where $E_i = E_i(g,u_i)$. An approximate minimum of the right-hand side 
can be found from one of the following optimization conditions.
 
It is possible to set
$$
E_{i+1}(g,u_i)-E_i(g,u_i)=0 \; , \qquad u_i=u_i(g) \; .
$$
If this equation has either multiple or no solution, control functions 
can be found from the minimum 
$$
\min_{u_i} \left |  E_{i+1}(g,u_i) -  E_i(g,u_i) \right | \; .
$$

The other admissible optimization condition can be
$$
(u_{i+1}-u_i)\; \frac{\prt}{\prt u_i}\; E_i(g,u_i) = 0 \; .
$$
If here there are either multiple or there is no solution, control functions 
can be defined either setting $u_i = u_{i-1}$ or from the minimum
$$
\min_{u_i} \left | (u_{i+1}-u_i)\; \frac{\prt}{\prt u_i}\; E_i(g,u_i)
\right | \; .
$$

The simplest form of an optimization condition is
$$
\frac{\prt}{\prt u_i}\; E_i(g,u_i) = 0 \; .
$$ 
If this equation gives no solution, then it is possible to look for 
the minimum  
$$
\min_{u_i} \left | \frac{\prt}{\prt u_i}\; E_i(g,u_i) \right | \; .
$$
The existence of the latter minimum requires the condition    
$$
\frac{\prt^2}{\prt u_i^2}\; E_i(g,u_i) = 0 \; ,
$$
defining the control function as an inflection point. 

In practical applications, one can use any of the above conditions, 
since, strictly speaking, they are similar to each other. It is 
reasonable to try first the simplest of these conditions.

\subsection{Coherent modes}

To illustrate the optimized perturbation theory, let us find the eigenvalues 
of the stationary NLS equation, whose eigenfunctions are called coherent 
modes \cite{Courteille_79,Yukalov_92}. The eigenvalues describe collective 
energy levels of trapped atoms in the coherent state. The coherent modes can
also be termed topological coherent modes, since the related density 
distributions differ from each other by different numbers of zeros.    
 
We consider a cylindrical trap, with a transverse, or radial, frequency 
$\om_\perp \equiv \om_x = \om_y$ and a longitudinal, or axial, frequency 
$\om_z$. The trap anisotropy is characterized by the anisotropy parameter,
\be
\label{386}
\lbd\equiv \frac{\om_z}{\om_\perp} \; . 
\ee
It is natural to use the cylindrical coordinates
$$
\br=\{ r_\perp,\vp,r_z\} \; , \qquad
r_\perp=\sqrt{r_x^2+r_y^2} \; ,  \qquad \vp=\arctan \left (
\frac{r_y}{r_x}\right ) \; .
$$

The dimensionless coupling parameter is defined as
\be
\label{387}
g \equiv 4\pi\; \frac{a_s}{l_\perp}\; N_0 \qquad 
\left ( l_\perp \equiv \frac{1}{\sqrt{m\om_\perp}} \right ) \; . 
\ee
It is convenient to pass to the dimensionless spatial variables
$$
r \equiv \frac{r_\perp}{l_\perp} \; , \qquad z \equiv
\frac{r_z}{l_\perp} \; , 
$$
with $r \in [0,\infty)$, $\vp \in [0,2\pi)$, and $z \in (-\infty, +\infty)$. 
The dimensionless condensate function is introduced through the relation
$$
\eta(\br) =\sqrt{ \frac{N_0}{l_\perp^3}}\; \psi(r,\vp,z) \; ,
\qquad \int |\psi(r,\vp,z)|^2 \; rdr \; d\vp\; dz = 1 \; .
$$
Finally, the dimensionless energy is given by the expression
\be
\label{388}
E \equiv \frac{\ep}{\om_\perp} \; .
\ee

In these variables, the stationary NLS equation takes the form
\be
\label{389}
\hat H_{NLS}\psi = E\psi \; ,
\ee
with the nonlinear Schr\"odinger Hamiltonian
$$
\hat H_{NLS} \equiv -\; \frac{\nabla^2}{2} + \frac{1}{2}\left (
r^2 +\lbd^2 z^2\right ) + g|\psi|^2 \; ,
$$
in which
$$
\nabla^2 = \frac{\prt^2}{\prt r^2} + \frac{1}{r}\; \frac{\prt}{\prt r}
+ \frac{1}{r^2}\; \frac{\prt^2}{\prt\vp^2} + \frac{\prt^2}{\prt z^2} \; .
$$

The eigenproblem (\ref{389}) defines the spectrum of eigenvalues, the 
lowest of which corresponds to the chemical potential 
$\mu_0 = (\min E) \omega_\perp$.   

The control parameters $u$ and $v$, playing the role of effective radial 
and axial frequencies, are included in the initial approximation
of the Hamiltonian 
\be
\label{390}
\hat H_0 = -\; \frac{\nabla^2}{2} + \frac{1}{2}\left ( u^2 r^2 +
\lbd^2 v^2 z^2\right ) \; ,
\ee
whose eigenvalues are
\be
\label{391}
E^{(0)}_{nmk} = (2n+|m|+1)u +\left ( k+ \frac{1}{2}\right )\lbd v \; .
\ee
Here $n = 0,1,2,\ldots$ is radial quantum number, $m = 0,\pm 1,\pm 2,\ldots$
is azimuthal quantum number, and $k = 0,1,2,\ldots$ is axial quantum number. 
The eigenfunction, corresponding to the zero-order Hamiltonian, reads as
$$
\psi_{nmk}(r,\vp,z) = \left [ \frac{2n!\; u^{|m|+1}}{(n+|m|)!}
\right ]^{1/2} r^{|m|} \; \exp\left ( -\;\frac{u}{2}\; r^2\right ) \;
L_n^{|m|}\left ( ur^2\right ) \; \frac{e^{im\vp}}{\sqrt{2\pi}}\; \times
$$
\be
\label{392}
\frac{1}{\sqrt{2^k\; k!}} \left ( \frac{\lbd v}{\pi} \right )^{1/4}
\exp\left ( -\; \frac{\lbd}{2}\; vz^2\right ) \; H_k(\sqrt{\lbd v}\; z)\; .
\ee
The notations are used for the associated Laguerre polynomial
$$
L_n^m(x) = \frac{e^x x^{-m}}{n!}\; \frac{d^n}{dx^n}
\left ( e^{-x} x^{n+m}\right ) =  \sum_{k=0}^n (-1)^k
\frac{(n+m)!}{(n-k)! \; (m+k)!\; k!}\; x^k 
$$
and Hermite polynomial
$$
H_k(x) = (-1)^k e^{x^2} \frac{d^k}{dx^k}\; e^{-x^2} = n! \;
\sum_{k=0}^{[n/2]} (-1)^k\; \frac{(2x)^{n-2k}}{k!\;(n-2k)!} \; .
$$

The Rayleigh-Schr\"odinger perturbation theory gives the first-order
approximation 
\be
\label{393}
E_{nmk}^{(1)} = \left ( \psi_{nmk},\hat H_{NLS}\psi_{nmk}\right ) \; 
 = E_{nmk}^{(1)}(g,\lbd,u,v) \; ,
\ee
where
\be
\label{394}
p\equiv 2n+|m|+1 \; , \quad  q\equiv(2k+1)\lbd \; .
\ee

Accomplishing integration, we meet the integral
$$
I_{nmk} \equiv \frac{1}{u\sqrt{\lbd v}} \; \int |\psi_{nmk}(r,\vp,z)|^4
\; rdr\; d\vp \; dz \; =
$$
$$
\frac{2}{\pi^2}\left [ \frac{n!}{(n+|m|)!\; 2^k\; k!}\right ]^2
\int_0^\infty \; x^{2|m|}\; e^{-2x} \left [ L_n^{|m|}(x)\right ]^4 dx\;
\int_0^\infty \; e^{-2t^2}\; H_k^4(t)\; dt \; .
$$
Then we obtain
\be
\label{395}
E_{nmk}^{(1)} =\frac{p}{2}\left ( u + \frac{1}{u}\right ) +\frac{q}{4}
\left ( v + \frac{1}{v}\right ) + \frac{1}{2}\; u\sqrt{v}\; x \; ,
\ee
where we introduce the notation for the effective coupling parameter 
\be
\label{396}
x\equiv 2g\sqrt{\lbd} \; I_{nmk} \; .
\ee

The optimization conditions
\be
\label{397}
\frac{\prt}{\prt u}\; E_{nmk}^{(1)} = 0 \; , \qquad
\frac{\prt}{\prt v}\; E_{nmk}^{(1)} = 0 
\ee
result in the equations 
$$
1 -\; \frac{1}{u^2}  + \frac{\sqrt{v}\; x}{p}  = 0 \; , \qquad1 -\;
\frac{1}{v^2}  + \frac{ux}{q\sqrt{v}}  = 0 \; ,
$$
defining the control functions $u = u_{nmk}(g,\lbd)$ and $v = v_{nmk}(g,\lbd)$. 
From the optimization conditions, it follows that the control functions 
are such that $0 \leq u \leq 1$ and $0 \leq v \leq 1$ for $x \geq 0$, while
$u \geq 1$ and $v \geq 1$ for $x \leq 0$. The marginal case of an ideal gas 
is recovered in the limit 
$$
\lim_{x\ra 0} u = \lim_{x\ra 0} v = 1 \; .
$$
Thus we come to the optimized approximant
$$
\tilde E \equiv
\tilde E_{nmk}^{(1)}(g,\lbd,u_{nmk}(g,\lbd), v_{nmk}(g,\lbd)) \; =
$$
\be
\label{398}
\frac{p}{u} + \frac{q}{4}\left ( v + \frac{1}{v}\right ) \; .
\ee

This approximant describes well the energy levels of atoms in a trap for the 
whole region of the effective coupling parameter $x \in [0,\infty)$. In the 
limit of weak coupling $x \ra \pm 0$, the control functions are
$$
u\simeq 1 - \; \frac{1}{2p}\; x + \frac{p+3q}{8p^2q}\; x^2 - \;
\frac{3p^2+16pq+20q^2}{64p^3q^2}\; x^3 \; ,
$$
$$
v\simeq 1  -\; \frac{1}{2q}\; x + \frac{p+q}{4pq^2}\; x^2  - \;
\frac{7p^2+20pq+12q^2}{64p^2q^3}\; x^3 \; .
$$
In the limit $x \ra \pm 0$ they tend to one, approaching the case of 
noninteracting atoms in a harmonic trap, as it should be. 

Under strong coupling $x \ra +\infty$, the control functions behave as
$$
u\simeq \left ( \frac{p^3}{q}\right )^{1/5} x^{-2/5} +
\frac{q^2-3p^2}{5(pq^3)^{1/5}}\; x^{-6/5} +
\frac{3p^4-p^2q^2-q^4}{5pq}\; x^{-2} \; ,
$$
$$
v\simeq \left ( \frac{q^2}{p}\right )^{2/5} x^{-2/5} +
\frac{2(p^2-2q^2)}{5}\left ( \frac{q}{p^3}\right )^{2/5} \;
x^{-6/5} +
\frac{6q^4-4p^2q^2-p^4}{5p^2}\; x^{-2} \; .
$$
Note that there are no real solutions for $x \ra -\infty$, since strongly
attractive atomic interactions make the system unstable.

The weak-coupling expansion for the energy levels reads as 
$$
\tilde E \simeq a_0 + a_1 x + a_2 x^2 + a_3 x^3 \qquad \left ( x\ra\pm 0 \right) \; ,
$$
with the coefficients
$$
a_0 = p+\frac{q}{2}\; , \qquad a_1 =\frac{1}{2}\; ,
\qquad a_2 = -\; \frac{p+2q}{16pq} \; , \qquad
a_3 =\left ( \frac{p+2q}{8pq} \right )^2 \; .
$$

In the strong coupling limit $x \ra +\infty$, the energy becomes
$$
\tilde E \simeq b_0 x^{2/5} + b_1 x^{-2/5} + b_2 x^{-6/5} + b_3 x^{-2}\; ,
$$
where
$$
b_0 = \frac{5}{4}\left ( p^2q\right )^{1/5}\; , \qquad
b_1 = \frac{2p^2+q^2}{4(p^2q)^{1/5}} \; ,
$$
$$
b_2 = -\; \frac{3p^4-2p^2q^2+2q^4}{20(p^2q)^{3/5}} \; , \qquad
b_3 = \frac{2p^6-p^4q^2-2p^2q^4+2q^6}{20p^2q} \; .
$$

The shape of the atomic cloud in the trap is characterized by the 
mean-square radius and the mean-square length defined by the expressions
\be
\label{399}
r_0^2\equiv(\psi_{nmk},r^2\psi_{nmk}) \; , \qquad z_0^2\equiv(\psi_{nmk},z^2\psi_{nmk}) \; ,
\ee
for which we obtain
\be
\label{400}
r_0^2 = \frac{p}{u} \; , \qquad z_0^2 = \frac{2k+1}{2\lbd v} \; .
\ee

For weak coupling, it follows
$$
r_0^2 \simeq p\left ( 1 +\frac{x}{2p}\right )\; , \quad
z_0^2 \simeq \frac{2k+1}{2\lbd}\left [ 1 + \frac{x}{2(2k+1)\lbd}
\right ] \qquad (x\ra 0) \; .
$$
And when the coupling is strong, then
$$
r_0^2 \simeq \left (p^2 q\right )^{1/5}\; x^{2/5} \; , \quad z_0^2
\simeq \frac{(p^2q)^{1/5}}{2\lbd^2}\; x^{2/5} \qquad (x\ra+\infty) \; .
$$

The ratio of the mean-square radius to the mean-square length defines the  
aspect ratio
$$
\frac{r_0^2}{z_0^2} =\frac{2p\lbd v}{(2k+1)u} \; .
$$
In the weak and strong coupling limits, respectively, 
$$
\frac{r_0^2}{z_0^2} \simeq \frac{2p\lbd}{2k+1} \quad (x\ra\pm 0)\; ,
\qquad \frac{r_0^2}{z_0^2} \simeq 2\lbd^2 \quad (x\ra+\infty)\; .
$$

Of special interest is the ground-state energy, with $n = m = k = 0$, $p = 1$, 
and $q = \lbd$. At low temperature, atoms pile down to this lowest energy level.
For the ground state, we have
$$
x =\frac{2g\sqrt{\lbd}}{(2\pi)^{3/2}} \; , \qquad I_{000}=(2\pi)^{-3/2} \; .
$$
The ground-state energy follows from expression (\ref{398}),
$$
\tilde E = \frac{1}{u} + \frac{\lbd}{4}\left ( v + \frac{1}{v}\right ) \; ,
$$
with control functions given by the solutions to the optimization equations
$$
1 -\; \frac{1}{u^2}+\sqrt{v}\; x = 0 \; , \qquad
1 -\; \frac{1}{v^2} + \frac{ux}{\lbd\sqrt{v}} = 0 \; .
$$

In the weak-coupling limit $x \ra 0$, the energy behaves as  
$$
\tilde E \simeq a_0 + a_1 x + a_2 x^2 + a_3 x^3 \; ,
$$
with the coefficients
$$
a_0= 1+\frac{\lbd}{2}\; , \qquad a_1 = \frac{1}{2}\; ,
\qquad a_2 = -\;\frac{1+2\lbd}{16\lbd} \; , \qquad
a_3=\left ( \frac{1+2\lbd}{8\lbd} \right )^2 \; .
$$
In terms of the coupling parameter $g$, this is equivalent to
$$
\tilde E \simeq 1 + \frac{\lbd}{2} + \frac{g\sqrt{\lbd}}{(2\pi)^{3/2}}
\quad (g\ra 0) \; . 
$$

And in the strong-coupling limit, we find
$$
\tilde E \simeq 0.547538(g\lbd)^{2/5} \qquad (g\ra\infty) \; .
$$
This result is only $2\%$ different from the Thomas-Fermi limit.

Thus, optimized perturbation theory allows us to derive, for any quantity 
of interest, rather accurate approximations valid in the whole region
of parameters.

\subsection{Attractive interactions}

As is mentioned in the previous section, the energy levels of trapped atoms
may not exist at sufficiently strong coupling, when the coupling parameter 
(\ref{387}) is negative. A negative coupling parameter corresponds to 
attractive atomic interactions, having a negative scattering length $a_s < 0$, 
hence the coupling parameter $g < 0$, and the effective coupling parameter 
(\ref{396}) is also negative, $x < 0$. The eigenproblem (\ref{389}) does not
have real solutions for the ground-state energy, if $x < x_c < 0$. This means 
that there is a critical line $x_c = x_c(\lbd)$ separating the values of 
$x > x_c$, for which real eigenvalues exist, from the values $x < x_c$, where 
there are no real solutions. Respectively, there should exist a critical value 
$g_c(\lbd)$, giving the related critical line  
\be
\label{401}
g_c(\lbd) = \frac{(2\pi)^{3/2}}{2\sqrt{\lbd}}\; x_c(\lbd) \; .
\ee
According to definition (\ref{387}), the coupling parameter is connected with 
the number of atoms. Therefore, there exists a critical number of atoms $N_c$,
above which the system of trapped atoms becomes unstable, while it is stable
for $N < N_c$. The critical number $N_c$ depends on $\lbd$ and characteristic 
trap lengths $l_\perp$, $l_z$ or $l_0$, that is, on trap frequencies. The 
trap lengths and frequencies are connected through the relations  
$$
l_0 \equiv \frac{1}{\sqrt{m\om_0}} \; = \frac{l_\perp}{\lbd^{1/6}} = \lbd^{1/3} l_z \; ,
$$
$$
\om_0 \equiv(\om_x\om_y\om_z)^{1/3} \; =\lbd^{1/3}\om_\perp = \frac{\om_z}{\lbd^{2/3}} \; .
$$

Varying the trap lengths, it is possible to find the maximal critical number
of atoms, below which the system is stable. Depending on which of the trap lengths 
is assumed to be varied, it is possible to present the critical number $N_c$ 
in three forms:
$$
N_c = \al_\perp(\lbd) \; \frac{l_\perp}{a_s} = \al_z(\lbd) \;
\frac{l_z}{a_s} = \al_0(\lbd)\; \frac{l_0}{a_s} \; ,
$$
where
$$
\al_\perp(\lbd) \equiv \frac{g_c(\lbd)}{4\pi} = \sqrt{\frac{\pi}{2}}\;
\frac{x_c(\lbd)}{2\sqrt{\lbd}} \; ,
$$
$$
\al_z(\lbd) \equiv \frac{g_c(\lbd)}{4\pi}\; \sqrt{\lbd} =
\sqrt{\frac{\pi}{2}}\; \frac{x_c(\lbd)}{2} \; ,
$$
$$
\al_0(\lbd) \equiv \frac{g_c(\lbd)}{4\pi}\;\lbd^{1/6} =
\sqrt{\frac{\pi}{2}}\; \frac{x_c(\lbd)}{2\lbd^{1/3}} \; .
$$

The critical value $x_c$ can be found by resorting to optimized perturbation 
theory analyzed in the previous section \cite{Yukalov_92}. As an example, 
let us consider a spherical trap, where $\lbd = 1$ and $u = v$. Then the 
optimization equation for the control function $u$ reads  
$$
xu^{5/2} + u^2 -1 = 0 \; .
$$
This equation possesses real solutions only for $x > x_c$, where
$$
x_c = -\; \frac{4}{5^{5/4}} = -0.534992\; , 
$$
hence
$$
g_c =\frac{(2\pi)^{3/2}}{2}\; x_c = -4.21296 \; . 
$$
For $x < x_c$ there are no real solutions. 

In the general case, when $\lbd \neq 1$, it is necessary to study the general
form of the optimization equations presented in the previous section. This 
study can be done numerically. In order to obtain analytic expressions, we 
involve self-similar approximation theory \cite{Yukalov_70,Gluzman_71,Yukalov_72}, 
as is explained in Ref. \cite{Yukalov_92}. This yields the effective critical 
coupling
\be
\label{402}
x_c(\lbd) = -\; \frac{2\lbd}{1+2\lbd} \; , 
\ee
with $|x_c| \leq 1$, and the critical coupling parameter
\be
\label{403}
g_c(\lbd) = - (2\pi)^{3/2} \; \frac{\sqrt{\lbd}}{1+2\lbd} \; .
\ee

From these expressions, we have
$$
\al_\perp(\lbd)  = \sqrt{\frac{\pi}{2}}\left (
\frac{\sqrt{\lbd}}{1+2\lbd} \right ) \; , \qquad
\al_z(\lbd) = \sqrt{\frac{\pi}{2}}\left ( \frac{\lbd}{1+2\lbd} \right ) \; ,
$$
$$
\al_0(\lbd) = \sqrt{\frac{\pi}{2}}\left ( \frac{\lbd^{2/3}}{1+2\lbd} \right ) \; .
$$

The trap shape is characterized by the value of $\lambda$. The trap has a 
cigar shape, if $\lambda < 1$, a disk shape, when $\lambda > 1$, and it is
spherical, for $\lambda = 1$. In the limit of $\lambda \ra 0$ or 
$\lambda \ra \infty$, we get
\begin{eqnarray}
\al_\perp(\lbd) \simeq \left\{ \begin{array}{ll}
1.253\; \lbd^{1/2} & (\lbd\ra 0) \\ \nonumber
0.627\; \lbd^{-1/2} & (\lbd\ra\infty) \; ,
\end{array} \right.
\end{eqnarray}

\begin{eqnarray}
\al_z(\lbd) \simeq \left\{ \begin{array}{ll}
1.253\; \lbd & (\lbd\ra 0) \\ \nonumber
0.627 & (\lbd\ra\infty) \; ,
\end{array} \right.
\end{eqnarray}

\begin{eqnarray}
\al_0(\lbd) \simeq \left\{ \begin{array}{ll}
1.253\; \lbd^{2/3} & (\lbd\ra 0) \\ \nonumber
0.627\; \lbd^{-1/3} & (\lbd\ra\infty) \; .
\end{array} \right.
\end{eqnarray}

The maximal number of atoms, that can form a stable coherent system in 
a trap, depends on which of the trap lengths is fixed and which is varied. 
Fixing a particular trap length, we look for a maximum of the related 
$\alpha_j(\lambda)$, where $j = \perp, z, 0$. If the length $l_\perp$ 
is fixed, then 
$$
\max_\lbd \al_\perp(\lbd) =0.443 \qquad (\lbd=0.5) \; ,
$$
and the cigar-shape trap is optimal, with $\lbd = 0.5$.

When $l_z$ is fixed, then
$$
\max_\lbd \al_z(\lbd) =0.627 \qquad (\lbd\ra \infty) \; ,
$$
and the disk-shape trap is optimal, with $\lbd\ra\infty$.

Finally, when $l_0$ is fixed, then
$$
\max_\lbd \al_0(\lbd) =0.418 \qquad (\lbd=1) \; ,
$$
hence the spherical trap is optimal, when $\lbd = 1$. 

In the general situation, the critical number of atoms can be represented
\cite{Yukalov_92} in the form
\be
\label{404}
N_c = \sqrt{\frac{\pi}{2}}\;
\frac{l_xl_yl_z}{|a_s|(l_x^2+l_y^2+l_z^2)} \; .
\ee
This can be rewritten as 
\be
\label{405}
\frac{|a_s|}{l_0}\; N_c = \sqrt{\frac{\pi}{2}}\;
\frac{l_0^2}{l_x^2+l_y^2+l_z^2} \; ,
\ee
where $l_0\equiv (l_xl_yl_z)^{1/3}$. In terms of the trap frequencies, 
the latter reduces to
$$
\frac{|a_s|}{l_0}\; N_c = \sqrt{\frac{\pi}{2}}\; \frac{1}{\om_0}\left (
\frac{1}{\om_x} + \frac{1}{\om_y} + \frac{1}{\om_z}\right )^{-1} \; .
$$

The accuracy of the approximate analytical expressions (\ref{404}) and 
(\ref{405}) has been estimated for the case of a spherical trap and compared  
with the numerically found value of $N_c$. The difference of the analytical
and numerical values is only about $10\%$.

\subsection{Collective excitations}

The stationary NLS equation (\ref{389}) gives the spectrum of coherent modes
that are the stationary solutions corresponding to the coherent field.
Around each of the stationary solutions, there can exist small nonequilibrium
oscillations describing collective excitations.

To study collective excitations, arising around a coherent mode, it is 
necessary to consider the time-dependent NLS equation
\be
\label{406}
i\; \frac{\prt}{\prt t}\; \eta(\br,t) =  H[\eta(\br,t)]\; \eta(\br,t) \; ,
\ee
with the nonlinear Schr\"odinger Hamiltonian
$$
H[\eta] \equiv -\; \frac{\nabla^2}{2m} + U(\br) +\Phi_0|\eta|^2 -\ep \; .
$$
Again, we write here $\ep$, instead of $\mu_0$, keeping in mind that there
can exist the whole variety of coherent modes with the related energy 
spectrum. The lowest of these energies gives $\mu_0$.  

Elementary collective excitations are described by a linearized equation,
with respect to small deviations from a given coherent mode,
$$
\eta(\br,t)=\eta(\br)+\dlt\eta(\br,t) \; .
$$
Small deviations from a stationary solution $\eta(\br)$ correspond to small
oscillations of the coherent mode defined by the stationary NLS equation
$$
\left [ - \; \frac{\nabla^2}{2m} + U(\br) + \Phi_0|\eta(\br)|^2 \right ]
\eta(\br) = \ep\eta(\br) \; .
$$

Linearizing with respect to $\dlt\eta(\br,t)$, and representing the latter as
\be
\label{407}
\dlt\eta(\br,t) = u(\br) e^{-i\om t} + v^*(\br) e^{i\om t} \; ,
\ee
implies the linearization with respect to $u(\br)$ and $v(\br)$. This 
linearization results in the Bogolubov equations
$$
\left \{ \om - H[\eta(\br)] - \Phi_0|\eta(\br)|^2 \right \} u(\br) -
\Phi_0\eta^2(\br) v(\br) = 0 \; ,
$$
$$
\left\{ \om + H[\eta(\br)] + \Phi_0|\eta(\br)|^2\right \} v(\br) +
\Phi_0\left [ \eta^*(\br)\right ]^2 u(\br) = 0 \; .
$$
These equations are appropriate for describing small deviations from 
any coherent topological mode.

If one is interested in the excitations of the ground-state coherent field,
the latter can be taken real, $\eta(\br) = \eta^*(\br)$. For a slowly 
varying trapping potential $U(\br)$, one can use the local-density 
approximation, assuming that the condensate density 
$\rho_0(\br) \equiv |\eta(\br)|^2$ is also slowly varying. Considering 
a dilute gas and employing the Thomas-Fermi approximation, that is, neglecting 
the kinetic term in the stationary NLS equation, yields
$$
\ep\cong U(\br) + \rho_0(\br)\Phi_0 \; , 
$$
which results in the condensate density
\be
\label{408}
\rho_0(\br) \cong \frac{\ep-U(\br)}{\Phi_0} \qquad \left ( U(\br)\leq \ep \right ) \; .
\ee
The Bogolubov equations reduce to
$$
\left ( \om +\frac{\nabla^2}{2m} \; -\; \rho_0\Phi_0\right ) u(\br) -
\rho_0\Phi_0 v(\br) = 0 \; , 
$$
$$
\left ( \om -\; \frac{\nabla^2}{2m} + \rho_0\Phi_0\right ) v(\br) +
\rho_0\Phi_0 u(\br) = 0 \; ,
$$
where $\rho_0 = \rho_0(\br)$ is a slowly varying function of ${\bf r}$. 

Following the local-density approximation, one sets
$$
u(\br) = u_0 e^{i\bk\cdot\br} \; , \qquad
v(\br) = v_0 e^{i\bk\cdot\br} \; ,
$$
where $u_0 = u_0(\br)$ and $v_0 = v_0(\br)$ are slowly varying functions.
Then the Bogolubov equations transform into 
$$
\left ( \om -\; \frac{k^2}{2m} - \rho_0\Phi_0\right )u_0  -
\rho_0\Phi_0 v_0 = 0 \; , \qquad
\rho_0\Phi_0 u_0 + \left ( \om + \frac{k^2}{2m} +\rho_0\Phi_0\right )
v_0 = 0 \; .
$$
The existence of nontrivial solutions requires that the determinant 
of the system of equations be zero. This gives the local Bogolubov spectrum
\be
\label{409}
\ep(\bk,\br) = \sqrt{c^2(\br)k^2 +\left ( \frac{k^2}{2m}\right )^2} \; ,
\ee
with the local sound velocity
\be
\label{410}
c(\br) \equiv \sqrt{\frac{\rho_0(\br)}{m}\; \Phi_0} \; .
\ee

In the long-wave limit, the spectrum is of phonon type, 
$$
\ep(\bk,\br) \simeq c(\br)k \qquad  (k \ra 0) \; .
$$
And for large $k$, one gets the spectrum
$$
\ep(\bk,\br) \simeq \frac{k^2}{2m} \qquad (k\ra\infty)
$$
of the single-particle type. 

The obtained spectrum of collective excitations, for a purely coherent system,
looks similar to spectrum (\ref{358}) derived in the local-density approximation, 
but here the sound velocity has the Bogolubov form (\ref{410}), since in a purely
coherent system, there are no uncondensed atoms and the anomalous average is zero.

\subsection{Moving superfluid}

Recall that a totally coherent system is superfluid, with the superfluid fraction
$n_s = 1$. For a moving system, with a constant superfluid velocity 
$\bv = const$, the coherent field, in the laboratory frame, is denoted by 
$\eta_v(\br,t)$. The time-dependent NLS equation is
\be
\label{411}
i\; \frac{\prt}{\prt t}\; \eta_v(\br,t) = H[\eta_v(\br,t)]\eta_v(\br,t) \; .
\ee
This evolution equation is invariant under the Galilean transformation
\be
\label{412}
\eta_v(\br,t) =\eta(\br-\bv t,t)\exp\left\{ i\left ( m\bv\cdot\br -\;
\frac{mv^2}{2}\; t\right )\right \} \; .
\ee

To find the spectrum of collective excitations, we consider a small deviation
from the function $\eta(\br-\bv t)$, looking for a solution in the form
$$
\eta(\br-\bv t,t) =\eta(\br-\bv t) + u_v(\br)e^{-i\om t} +
v_v^*(\br) e^{i\om t} \; .
$$
Linearizing (\ref{411}) with respect to $u_v(\br)$ and $v_v(\br)$, we use the 
equality
$$
i \; \frac{\prt}{\prt t}\; \eta(\br-\bv t) = \bv \cdot \hat\bp \;
\eta(\br-\bv t) \; ,
$$
where $\hat \bp \equiv -i \nabla$ is the momentum operator. The linearization
results in the Bogolubov equations
$$
\left \{ \om -\bv \cdot \hat\bp - H[\eta(\br-\bv t)] - \Phi_0
|\eta(\br-\bv t)|^2\right \} u_v(\br) - \Phi_0 \eta^2(\br-\bv t)
v_v(\br) = 0 \; ,
$$
$$
\left \{ \om -\bv \cdot \hat\bp + H[\eta(\br-\bv t)] + \Phi_0
|\eta(\br-\bv t)|^2\right \} v_v(\br) + \Phi_0 \left [\eta^*(\br-\bv t)
\right ]^2  u_v(\br) = 0 \; .
$$

Considering excitations over the ground state, in the local-density 
approximation, we get the collective spectrum of a moving superfluid,
\be
\label{413}
\ep_v(\bk,\br - \bv t) = \bv \cdot \bk + \sqrt{c^2(\br - \bv t) k^2 +\left (\frac{k^2}{2m}
\right )^2 },
\ee
with the similar definition of the sound velocity $c(\br - \bv t)$, as 
for an immovable system, but with $\rho_0 = \rho_0(\br - \bv t)$. The spectrum
can be rewritten as
$$
\ep_v(\bk, \br - \bv t) = \ep(\bk, \br - \bv t) +\bv \cdot\bk \; ,
$$
with $\ep(\bk, \br - \bv t)$ being the spectrum (\ref{409}) in a system at rest. 

Since excitations increase the system energy, the spectrum must be non-negative 
for all $\bk$, so that $\ep_v(\bk, \br - \bv t) \geq 0$. Otherwise, the 
coherent mode would be unstable. In the long-wave limit, this implies that
$$
\ep_v(\bk, \br - \bv t) \simeq c(\br - \bv t) k +\bv \cdot\bk \qquad (k\ra 0) \; .
$$
There always exists a vector $\bk$, such that $\bv \cdot \bk = - v k$, where 
$v \equiv |\bv|$. Therefore, for the stability of motion, it is required that
$$
\min_k\; \frac{\ep_v(\bk, \br - \bv t)}{k} = c(\br - \bv t) - v > 0 \; .
$$
Hence the critical velocity of superfluid motion has to be smaller than 
the sound velocity,
\be
\label{414}
\min_k\; \frac{\ep(k, \br - \bv t)}{k} =  c(\br - \bv t) > v \; ,
\ee
which is equivalent to the local Landau criterion. The motion is dynamically
stable, provided that $v < c(\br - \bv t)$. The stability criterion can be 
represented as the stability condition for a moving wave packet:
\be
\label{415}
\rho_0(\br-\bv t) > \frac{mv^2}{\Phi_0} = \frac{(mv)^2}{4\pi a_s} \; .
\ee
A wave packet, moving faster than the sound velocity, becomes unstable.

\subsection{Coherent mixtures}

It is possible to create mixtures of several coherent components. To remain
coherent, the components have to consist of atoms with weak interactions  
between all components enumerated by $i = 1,2,\ldots$, so that
\be
\label{416}
\sqrt{\rho_i\rho_j}\; a_{ij}^3 \ll 1 \qquad \left ( \rho_i \equiv
\frac{N_i}{V} \right ) \; ,
\ee
where $\rho_i$ is the mean density of an $i$-th component, $N_i$ is 
the number of particles in a component, and $a_{ij}$ is a scattering length 
for scattering between an atom of the $i$-th component and an atom of the
 $j$-th component.

The system is dilute if all interactions are short-range, such that
\be
\label{417}
\sqrt{\rho_i\rho_j}\;r_{ij}^3\ll 1 \; ,
\ee
where $r_{ij}$ is an effective interaction radius of interactions between 
atoms of the $i$-th and $j$-th species. 

Temperature has to be low, satisfying the inequality
\be
\label{418}
\rho_i\lbd_i^3 \ll 1 \qquad \left ( \lbd_i \equiv \sqrt{\frac{2\pi}{m_iT}}
\right ) \; ,
\ee
$m_i$ being atomic mass in the $i$-th component. 

Under these conditions, the dilute-gas approximation for the interaction 
potential is admissible. There are three local interaction potentials
$$
\Phi_{ij}(\br) = \Phi_{ij}\dlt(\br) \; , \qquad
\Phi_{ij} \equiv \int \Phi_{ij}(\br)\; d\br \; , 
$$
\be
\label{419}
\Phi_{ij}=4\pi\; \frac{a_{ij}}{m_{ij}} \; , \qquad m_{ij}\equiv
\frac{2m_im_j}{m_i+m_j} \; ,
\ee
with $a_{ij} = a_{ji}$ and $m_{ij}$ being reduced mass. 

The Hamiltonian of a coherent mixture is
$$
H_{mix} = \sum_i \int \eta_i^*(\br,t) \left ( - \; \frac{\nabla^2}{2m_i} +
U_i  - \ep_i\right ) \eta_i(\br,t) \; d\br + 
$$
\be
\label{420}
\frac{1}{2}\; \sum_{ij}
\Phi_{ij} \; \int |\eta_i(\br,t)|^2 |\eta_j(\br,t)|^2 \; d\br \; ,
\ee
where $U_i = U_i(\br,t)$ are external trapping potentials. The quantities
$\ep_i$ play the role of the species chemical potentials, guaranteing the 
normalization conditions
$$
\int |\eta_i(\br,t)|^2 \; d\br = N_i \; .
$$
We use the notation $\ep_i$, instead of $\mu_i$ for generality, keeping 
in mind that the species can be not in their ground states, but in excited
states corresponding to higher coherent topological modes. When both 
components in the mixture are in their ground states, then  $\mu_i = \ep_i$.

The evolution equations for the species are
\be
\label{421}
i\; \frac{\prt}{\prt t}\; \eta_j(\br,t) =
\frac{\dlt H_{mix}}{\dlt\eta_j^*(\br,t)} \; = H_j[\eta(\br,t)]\; \eta_j(\br,t) \; ,
\ee
with the nonlinear Schr\"odinger Hamiltonian
$$
H_i[\eta] \equiv -\; \frac{\nabla^2}{2m_i} + U_i(\br) + \sum_j
\Phi_{ij} |\eta_j|^2 -\ep_i \; .
$$

In equilibrium, $\eta_j(\br,t) = \eta_j(\br)$, and we get the stationary NLS
equations
\be
\label{422}
\left [  -\; \frac{\nabla^2}{2m_i} + U_i(\br) + \sum_j
\Phi_{ij} |\eta_j(\br)|^2 \right ] \; \eta_i(\br) = \ep_i\eta_i(\br) \; .
\ee

However not all different species can be uniformly mixed. But this requires
the validity of special conditions. Otherwise the mixture can stratify 
\cite{Pethick_43}. Stratification implies spatial separation into independent 
species occupying different locations in the system volume. Each of the species 
occupies its own volume $V_i$, whose sum composes the whole volume 
$V = \sum_i V_i$. The Hamiltonian of the stratified species is the sum
\be
\label{423}
H_{sep} = \sum_i \int \eta_i^*(\br) \left ( -\; \frac{\nabla^2}{2m_i} +
U_i -\ep_i\right )\eta_i(\br)\; d\br + \frac{1}{2}\; \sum_i \Phi_{ii}
\int  |\eta_i(\br)|^4 \; d\br 
\ee
of the partial species Hamiltonians, without interactions between the 
components.

The mixture does not stratify, if mixture stability conditions are valid.
Thus {\it thermodynamic stability condition} requires that the mixture free
energy be smaller than the free energy of the stratified system. At zero 
temperature, this reduces to the stability condition for the internal energies  
\be
\label{424}
E_{mix} < E_{sep} \; ,
\ee
where
$$
E_{mix} = H_{mix} + \sum_i \ep_i N_i \; , \qquad
E_{sep} = H_{sep} + \sum_i \ep_i N_i \; .
$$

Assuming that the components are in their ground states, one can resort to
the local-density approximation, where the densities $\rho_i = |\eta_i(\br)|^2$
are slowly varying in space. Then in all equations, one can consider the 
densities as almost constant. 

The internal energy of the mixture can be written in the form
$$
E_{mix} = \frac{1}{2} \; \sum_{ij} \Phi_{ij}\; \frac{N_iN_j}{V} \; +
\; \sum_i U_i N_i \; ,
$$
where $U_i$ includes all single-particle energy terms. Separating here the sum 
into two terms, as
$$
\sum_{ij} \; \ra \; \sum_{i=j} + \sum_{i\neq j} \; ,
$$
we get
\be
\label{425}
E_{mix} = \frac{1}{2} \; \sum_i \Phi_{ii}\; \frac{N_i^2}{V} \;  +
\frac{1}{2} \; \sum_{i\neq j} \Phi_{ij}\; \frac{N_iN_j}{V} \;
+ \sum_i U_i N_i \; .
\ee
While the stratified system has the internal energy
\be
\label{426}
E_{sep} = \frac{1}{2} \; \sum_i \Phi_{ii}\; \frac{N_i^2}{V_i} \;  +
\; \sum_i U_i N_i \; .
\ee

The separated species are assumed to be in equilibrium with each other.
The condition of mechanical equilibrium requires that the pressures in 
each of the component be equal
\be
\label{427}
P_i = -\; \frac{\prt E_{sep}}{\prt V_i} = P_j \; = -\; \frac{\prt E_{sep}}{\prt V_j} \; .
\ee
The pressure in an $i$-th component is
$$
P_i = \frac{1}{2}\; \Phi_{ii} \left ( \frac{N_i}{V_i}\right )^2 \; .
$$
Thence condition (\ref{427}) yields
$$
\frac{P_i}{P_j} = \frac{\Phi_{ii}}{\Phi_{jj}} \left (
\frac{N_i V_j}{N_j V_i}\right )^2 = 1 \; , 
$$
from where it follows
$$
\frac{V_j}{V_i} = \frac{N_j}{N_i}\; \sqrt{\frac{\Phi_{jj}}{\Phi_{ii}} } \; .
$$
Using the equality
$$
\frac{V}{V_i} = 1 + \sum_{j(\neq i)} \; \frac{V_j}{V_i} \; ,
$$
we get
$$
\frac{V}{V_i} = 1 + \sum_{j(\neq i)} \; \frac{N_j}{N_i}\;
\sqrt{\frac{\Phi_{jj}}{\Phi_{ii}} } \; .
$$
The energy of a stratified system (\ref{426}) becomes
\be
\label{428}
E_{sep} = \frac{1}{2}\; \sum_i \Phi_{ii}\; \frac{N_i^2}{V} \;
+ \; \frac{1}{2}\; \sum_{i\neq j} \; \frac{N_iN_j}{V} \;
\sqrt{\Phi_{ii} \Phi_{jj}} + \sum_i U_i N_i \; .
\ee

In this way, the stability condition (\ref{424}) reduces to 
$$
\sum_{i\neq j} N_iN_j \left ( \Phi_{ij} - \sqrt{\Phi_{ii} \Phi_{jj}}
\right ) < 0 \; .
$$
Requiring that this condition be satisfied for any $N_i$ and $N_j$ 
leads to the conclusion that it is necessary and sufficient that condition 
\be
\label{429}
\Phi_{ij} < \sqrt{\Phi_{ii} \Phi_{jj}} \qquad ( i\neq j)
\ee 
be valid. 

Generalizing the mixture stability condition to nonzero temperatures, 
leads \cite{Yukalov_93} to the inequality 
\be
\label{430}
\Phi_{ij} < \sqrt{\Phi_{ii} \Phi_{jj}} + \frac{TV}{N_i N_j} \Delta S_{mix} \; ,
\ee
with $i\neq j$ and the entropy of mixing
$$
\Delta S_{mix} = -N_i \ln \frac{N_i}{N} - N_j \ln \frac{N_j}{N} \; .
$$
Hence finite temperature facilitates the mixture stability.

\subsection{Spectrum branching}

The stability of a mixture can also be investigated by studying the spectra 
of collective excitations \cite{Nepomnyashchy_94,Yukalov_95,Yukalov_96}. 
For this purpose, we consider the dynamical equations (\ref{421}), looking 
for small deviations from stationary solutions,
$$
\eta_j(\br,t) = \eta_j(\br) + \dlt\eta_j(\br,t) \; ,
$$
with
$$
\dlt\eta_j(\br,t) = u_j(\br)e^{-i\om t} + v_j^*(\br) e^{i\om t} \; .
$$
Linearizing with respect to $u_j(\br)$ and $v_j(\br)$, we obtain the equations
$$
(\om - H_j[\eta]) u_j(\br) -  \eta_j(\br) \sum_i \Phi_{ij} \left [
\eta_i^*(\br) u_i(\br) +\eta_i(\br) v_i(\br) \right ] = 0 \; ,
$$
$$
(\om + H_j[\eta]) v_j(\br) +  \eta^*_j(\br) \sum_i \Phi_{ij} \left [
\eta_i(\br) v_i(\br) + \eta_i^*(\br) u_i(\br) \right ] = 0 \; .
$$
Here the Hamiltonian $H_j[\eta]$ is expressed through stationary solutions 
$\eta_i(\br)$. In what follows, we consider $\eta_i(\br)$ as ground states.
Then in the local-density approximation, we have
$$
\left ( \om + \frac{\nabla^2}{2m_j}\right ) u_j(\br) - \sum_i
\sqrt{\rho_i\rho_j}\; \Phi_{ij} [u_i(\br) + v_i(\br)] = 0 \; ,
$$
$$
\left ( \om -\; \frac{\nabla^2}{2m_j}\right ) v_j(\br) + \sum_i
\sqrt{\rho_i\rho_j}\; \Phi_{ij} [v_i(\br) + u_i(\br)] = 0 \; .
$$
Substituting here
$$
u_j(\br) =u_j e^{i\bk\cdot\br} \; , \qquad
v_j(\br) =v_j e^{i\bk\cdot\br} \; ,
$$
we get an algebraic system of linear equations with respect to $u_j$ and $v_j$.

Note that in the local-density approximation, $u_j$ and $v_j$ are treated as 
slow functions of spatial variable $\br$ depending on the latter through the local
densities and the slowly varying external potentials. Respectively, the spectra 
of collective excitations depend on $\bk$ as well as on $\br$. In order to 
simplify notations, we omit below the dependence of the spectra on the spatial
variable $\br$. This dependence can be restored by taking into account the local
densities $\rho_i = |\eta_i(\br)|^2$. 

The existence of nontrivial solutions of a system of linear equations implies
their zero determinant. In what follows, we consider a binary mixture $(j = 1,2)$.
And the following notations will be used:
$$
\ep_j(\bk) \equiv
\sqrt{c_j^2 k^2+ \left ( \frac{k^2}{2m_j}\right )^2} \; ,
\qquad c_j \equiv \sqrt{\frac{\rho_j}{m_j}\; \Phi_{jj}} \; ,
$$
\be
\label{431}
\ep_{12}(\bk) \equiv c_{12} k \; , \qquad
c_{12}^2 \equiv \sqrt{\frac{\rho_1\rho_2}{m_1m_2}}\; \Phi_{12} \; .
\ee
The requirement of zero determinant results in the equation
\be
\label{432}
\left [\om^2 -\ep_1^2(\bk)\right ] \left [\om^2 -\ep_2^2(\bk)\right ]
= \ep_{12}^4(\bk) \; .
\ee
Solving the latter with respect to $\om$ yields the spectra
\be
\label{433}
\ep_{\pm}^2(\bk) = \frac{1}{2}\left \{ \ep_1^2(\bk) + \ep_2^2(\bk) \pm
\sqrt{[\ep_1^2(\bk)-\ep_2^2(\bk)]^2 + 4\ep_{12}^4(\bk)} \right \} \; .
\ee
The spectrum $\ep_+(\bk)$ describes the density wave that is common for 
both species. While the spectrum $\ep_-(\bk)$ corresponds to the 
oscillations of the components with respect to each other.

In the long-wave limit, both these spectra are of phonon type
\be
\label{434}
\ep_\pm(\bk) \simeq c_\pm k \qquad (k\ra 0) \; ,
\ee
with the sound velocities
\be
\label{435}
c_\pm^2 \equiv \frac{1}{2}\left [ c_1^2 + c_2^2 \pm
\sqrt{(c_1^2-c_2^2)^2 + 4c_{12}^4} \right ] \; .
\ee

But neither of $c_\pm$ is the hydrodynamic sound velocity
\be
\label{436}
s^2 \equiv \left ( \frac{\prt P}{\prt\rho_m}\right )_{TN} \; ,
\ee
where $\rho_m$ is the mass density $\rho_m \equiv \sum_i m_i\rho_i$. 
The pressure in the mixture is
$$
P_{mix} = -\; \frac{\prt E_{mix}}{\prt V} \; = 
\frac{1}{2}\; \sum_{ij} \Phi_{ij}\rho_i\rho_j \qquad
\left ( \rho_i \equiv \frac{N_i}{V}\right ) \; .
$$
Using the approximate relation $\prt \rho_i/\prt \rho_m \approx \rho_i/\rho_m$
gives
$$
s^2 \approx \sum_{ij} \Phi_{ij}\; \frac{\rho_i\rho_j}{\rho_m} \; .
$$
Thus, for a binary mixture, the hydrodynamic sound velocity is
\be
\label{437}
s^2 =
\frac{\Phi_{11}\rho_1^2+\Phi_{22}\rho_2^2+2\Phi_{12}\rho_1\rho_2}{m_1\rho_1
+m_2\rho_2} \; .
\ee

In the short-wave limit, the spectra are of the single-particle type,
$$
\ep_+(\bk) \simeq \frac{k^2}{2m_1}\; , \qquad \ep_-(\bk) \simeq
\frac{k^2}{2m_2} \qquad (k\ra\infty) \; .
$$

The mixed system is dynamically stable, provided that the spectra of 
collective excitations are positive, $\ep_\pm(\bk) > 0$, except maybe zero $k$.
Therefore, in view of equation (\ref{433}), it should be
$\ep_{12}^2(\bk) < \ep_1(\bk)\ep_2(\bk)$. In the long-wave limit, this gives
\be
\label{438}
c_{12}^2 < c_1 c_2 \qquad (k\ra 0) \; .
\ee
And using notations (\ref{431}), we see that the dynamic stability condition
(\ref{438}) for an immovable mixture is the same as the thermodynamic stability 
condition $\Phi_{12} < \sqrt{\Phi_{11}\Phi_{22}}$. 

When the components move with the velocities $\bv_1$ and $\bv_2$, respectively,
then, instead of equation (\ref{432}), we obtain
\be
\label{439}
\left [ (\om - \bv_1\cdot \bk)^2 - \ep_1^2(\bk)\right ] 
\left [ (\om - \bv_2\cdot \bk)^2 - \ep_2^2(\bk)\right ] = \ep_{12}^4(\bk) \; .
\ee

It is convenient to pass to the coordinate system describing the relative 
motion of the components, where $\bv_1 = -\bv_2 \equiv \bv$. Then the 
dynamical instability occurs when $\om \ra 0$ as $|\bv| \ra v_c$. 
When $\om \ra 0$, equation (\ref{439}) takes the form 
$$
\left [ y_k^2 k^2 - \ep_1^2(\bk)\right ] \left [ y_k^2 k^2 -
\ep_2^2(\bk)\right ]  = \ep_{12}^4 \; , 
$$
in which
$$
y_k \equiv \frac{\bv_c\cdot\bk}{k} = v_c\cos\vt \; .
$$
At the same time, the solution to the above equation with respect 
to $y_k^2$ is  
$$
y_k^2 = \frac{\ep_\pm^2(\bk)}{k^2} \; . 
$$
Therefore the critical velocity is defined as the minimal solution of 
the equation
$$
(v_c\cos\vt)^2 = \frac{\ep_\pm^2(\bk)}{k^2} \; , 
$$
understanding that $v_c$ is to be minimal for all $\bk$. Thus we need 
to maximize the left-hand side of the above equation, while minimizing its 
right-hand side. The maximum of the left-hand side corresponds to 
$\cos\vt = 1$. And, since $\ep_-(\bk) \leq \ep_+(\bk)$, we find the 
critical velocity
\be
\label{440}
v_c = \min_k\; \frac{\ep_-(\bk)}{k} =  c_- \; .
\ee
When $v = |\bv|$ reaches $v_c$ the system stratifies. The condition for 
the stability of moving components is $v < v_c = c_-$.  

The effect of the dynamical stratification cannot be treated by the 
thermodynamic stability condition, in which kinetic energy enters additively.

Note that in the laboratory frame, the velocities $\bv_1$ and $\bv_2$ can 
be directed arbitrarily. If they point in opposite directions, the appearing 
instability can be called counterflow instability. And if they point in one 
direction, the arising instability can be termed coflow instability. However,
in the relative system of coordinates, where $\bv_1 = -\bv_2 \equiv \bv$, 
their velocities are opposite to each other. Hence it is always possible
to call the resulting instability the {\it counterflow instability} or simply
{\it flow instability}.

\subsection{Quantum vortices}

The NLS equation
$$
i\; \frac{\prt}{\prt t}\; \eta(\br,t) = H[\eta]\eta(\br,t) \; , 
$$
with the NLS Hamiltonian
$$
H[\eta] = -\; \frac{\nabla^2}{2m} + U(\br) +\Phi_0|\eta|^2 - \ep \qquad 
\left ( \Phi_0 =  4\pi\; \frac{a_s}{m} \right ) \; ,
$$
describes various coherent topological modes. One of the most interesting modes
is the quantum vortex filament. 

To describe this mode, it is convenient to pass to cylindrical coordinates
$\br = \{ r_\perp, \vp, z \} = \{ \br_\perp, z \}$, where $\br_\perp = \{ r_\perp, \vp \}$.
Then
$$
\nabla^2 =\nabla^2_\perp + \frac{\prt^2}{\prt z^2} \; , \qquad
\nabla^2_\perp = \frac{\prt^2}{\prt r_\perp^2} + \frac{1}{r_\perp}\;
\frac{\prt}{\prt r_\perp} + \frac{1}{r_\perp^2}\; \frac{\prt^2}{\prt\vp^2} \; .
$$
The trapping harmonic potential reads as
$$
U(\br) = \frac{m}{2}\left ( \om^2_\perp r_\perp^2 + \om_z^2 z^2
\right ) \qquad 
\left ( l_\perp \equiv \frac{1}{\sqrt{m\om_\perp}} \; , \quad 
l_z \equiv \frac{1}{\sqrt{m\om_z}} \right ) \; .
$$

Let us consider a disk-shape trap, where $l_z \ll l_\perp$, hence 
$\om_\perp \ll \om_z$. We assume that there is no scattering on the 
trap boundary and $|a_s| \ll l_z$. We can look for the solution of the 
evolution equation by separating the variables,  
$$
\eta(\br,t) \cong \sqrt{N_0}\; \chi(r_\perp)e^{i\nu\vp} \psi(z) e^{-iE_zt} \; , 
$$
with the normalization conditions
$$
\int |\chi(r_\perp)|^2\; d\br_\perp = 1 \quad (d\br_\perp = 2\pi r_\perp dr_\perp) \; ,
$$
$$
\int |\psi(z)|^2 \; dz = 1 \; .
$$
The quantum number $\nu$ is called winding number, taking the values 
$\nu = 0,\pm 1,\pm 2,\ldots$. Keeping in mind a cylindrical trap, we denote
its radius and length by $R$ and $L$, respectively.

Assuming that the longitudinal part of the trapping potential is much larger 
than the interaction term, we get the longitudinal confinement characterized 
by the harmonic-oscillator equation
$$
\left ( -\; \frac{1}{2m}\; \frac{\prt^2}{\prt z^2} + \frac{m}{2}\;
\om_z^2 z^2 \right ) \psi(z) = E_z \psi(z) \; ,
$$
whose ground-state solution is
$$
\psi(z) = \frac{1}{\pi^{1/4}\sqrt{l_z}}\; \exp\left ( -\;
\frac{z^2}{2l_z^2}\right ) \; , \qquad E_z = \frac{1}{2}\; \om_z \; .
$$
Introducing the effective length $L_{eff}$ by the equation
$$
\frac{1}{L_{eff}} \equiv \int_{-L/2}^{L/2} \; |\psi(z)|^4 \; dz \; ,
$$
for the harmonic oscillator, we get $L_{eff} = \sqrt{2\pi} \; l_z$. In the 
uniform case, when $\psi(z) \ra 1/\sqrt{L}$, the effective length would be 
$L_{eff} \ra L$. 
 
Substituting $\eta(\br,t)$ into the evolution equation, multiplying the 
latter by $\psi^*(z)$, and integrating over $z$, we come to the effective 
radial equation.
\be
\label{441}
\left [ -\; \frac{1}{2m}\left ( \frac{d^2}{dr_\perp^2} +
\frac{1}{r_\perp} \; \frac{d}{dr_\perp} \right ) +
\frac{\nu^2}{2mr_\perp^2} + \frac{m}{2}\; \om_\perp^2 r_\perp^2 +
\frac{\Phi_0}{L_{eff}}\; N_0 |\chi(r_\perp)|^2 -\ep\right ]\chi(r_\perp)
= 0 \; .
\ee
As is seen, the quantity
$$
\frac{\Phi_0}{L_{eff}} = \sqrt{8\pi}\; \frac{a_s}{ml_z}
$$
plays the role of an effective interaction strength. Here we assume that 
the interactions are repulsive, such that $\Phi_0 > 0$.

For what follows, we need to introduce the {\it healing length}
\be
\label{442}
\xi \equiv \frac{1}{\sqrt{2m\ep}} \; , 
\ee
which depends on the eigenvalue $\ep$. And we define the dimensionless
real function
$$
f(r) \equiv \sqrt{\frac{\Phi_0N_0}{\ep L_{eff}}}\; \chi(r_\perp) \qquad 
\left ( r \equiv \frac{r_\perp}{\xi} \right ) \; .
$$
Then equation (\ref{441}) reduces to 
\be
\label{443}
\left ( \frac{d^2}{dr^2} + \frac{1}{r}\; \frac{d}{dr} \; - \;
\frac{\nu^2}{r^2}\; - \; \frac{\xi^4}{l_\perp^4}\; r^2 + 1\right )
f - f^3 = 0 \; ,
\ee
with the normalization condition
\be
\label{444}
\frac{\pi L_{eff}}{m\Phi_0 N_0}\; \int_0^{R/\xi} \; f^2(r)\; rdr = 1
\ee
defining $\ep$. The boundary conditions are
$$
f_\nu(r)\; \propto\; r^{|\nu|} \qquad (r\ra 0) \; ,
$$
$$
f_\nu(r)\; \propto\; r^\nu \exp\left ( -\; \frac{\xi^2}{2l_\perp^2}\; r^2
\right ) \qquad (r\ra\infty) \; .
$$

The healing length is assumed to be much shorter than the trap characteristic 
transverse length $l_\perp \sim R$, so that $\xi \ll l_\perp$ and $\xi \ll R$.
Therefore, for $1 \ll r \ll R/\xi$, it is possible to neglect the trapping term,
thus, obtaining the equation
\be
\label{445}
\left ( \frac{d^2}{dr^2} + \frac{1}{r}\; \frac{d}{dr} -\;
\frac{\nu^2}{r^2} + 1 \right ) f_\nu - f_\nu^3  = 0 \; .
\ee
This equation is invariant with respect to the sign change $f_\nu \ra -f_\nu$,
because of which any sign can be chosen. Below we choose positive $f(r)$. 
The boundary conditions can be presented in the form
\be
\label{446}
\lim_{r\ra 0} f_\nu(r) = 0 \; , \qquad \lim_{r\ra\infty} f_\nu(r) = 1 \; ,
\ee
keeping here in mind that $r \ra \infty$ implies $r \gg 1$.   

When there are no vortices, hence $\nu = 0$, then $f_0(r) = 1$. The 
normalization condition (\ref{444}) gives
\be
\label{447}
\ep_0 = \rho_0\Phi_0 \; , \qquad \rho_0 \equiv
\frac{N_0}{\pi R^2 L_{eff}} \; .
\ee
This yields the healing length 
\be
\label{448}
\xi_0 \equiv \frac{1}{\sqrt{2m\ep_0}}=\frac{1}{\sqrt{2m\rho_0\Phi_0}} \; 
= \frac{1}{\sqrt{2}\; mc} \; ,
\ee
where $c$ is the Bogolubov sound velocity
$$
c\equiv \sqrt{\frac{\rho_0}{m}\;\Phi_0} \; .
$$
This healing length can be employed as an approximation for the healing length
in the presence of a vortex.  

The vortex with the winding number $\nu = 1$ is called basic. For this
basic vortex there are the asymptotic expansions
$$
f_1(r) \; \propto \; r\left ( 1 + c_2 r^2 + c_4 r^4 + c_6 r^6\right )
\qquad (r\ra 0) \; ,
$$
\be
\label{449}
f_1(r) \simeq 1 -\; \frac{1}{2}\; r^{-2} -\; \frac{9}{8}\; r^{-4} -
\; \frac{161}{16}\; r^{-6} \qquad \left ( 1\ll r \ll \frac{R}{\xi}
\right )\; .
\ee

Note that the Thomas-Fermi approximation is not applicable for solving 
equation (\ref{445}), since neglecting the kinetic term would give
$$
f_{TF}(r) = \sqrt{1 -\;\frac{1}{r^2}} \qquad (r\leq 1) \; ,
$$
which is divergent at $r \ra 0$.

Approximate analytic solution of equation (\ref{445}), satisfying 
expansions (\ref{449}) can be found by employing self-similar approximation 
theory that gives \cite{Yukalov_97} the filament solution 
\be
\label{450}
f_1(r) \cong \frac{r/2}{\sqrt{1+r^2/4}} \; .
\ee
This is a rather good approximation, whose maximal residual for the 
related equation (\ref{445}) is of order $0.1$ at $r \sim 1$. It is possible 
to essentially improve the accuracy of solutions by using higher-order 
approximants of self-similar approximation theory. But here we limit 
ourselves by the simple expression (\ref{450}). Recall that this simple 
expression is valid, if boundary effects are negligible. 

The normalization condition (\ref{444}) gives
$$
\frac{R^2}{4\xi^2} = \frac{mR^2}{2}\; \ep_0 + \ln\left ( 1 +
\frac{R^2}{4\xi^2}\right ) \;  \qquad
\left ( \xi^2 = \frac{1}{2m\ep_1} \right ) \; .
$$
Since $\xi \ll R$, we can write
\be
\label{451}
\ep_1 = \ep_0 + \frac{4}{mR^2}\; \ln \left ( \frac{R}{2\xi}\right ) \; .
\ee

The vortex energy per particle is defined as 
\be
\label{452}
\ep_{vor}\equiv\ep_1-\ep_0 \; .
\ee
This energy can be treated as small, compared to $\ep_0$. Therefore it 
is possible to iterate $\ep_1$ with $\xi_0$, given by expression (\ref{448}),
confirming that the ratio
$$
\frac{\ep_{vor}}{\ep_0} =
\frac{8\xi_0^2}{R^2}\; \ln\left ( \frac{R}{2\xi_0}\right )
$$
is small for $\xi_0 \ll R$. 

Thus for the vortex energy per particle, we have
\be
\label{453}
\ep_{vor} = \frac{4}{mR^2}\; \ln\left ( \frac{R}{\xi_0}\right ) \; .
\ee
The total vortex energy is 
$$
N_0\ep_{vor}  = \frac{4\pi}{m}\; \rho_0 L_{eff}\ln\left (
\frac{R}{2\xi_0}\right ) \; .
$$

The energy, required for producing a vortex, can be provided by rotation
with the angular momentum $\hat{\bf L} \equiv -i \left [\br \times \nabla \right ]$.
In cylindrical coordinates, the $z$-th component of the angular momentum is
$\hat L_z = - i\;\prt/\prt \vp$. Rotation, with the angular velocity $\Om$ 
around the $z$-axis, is described by the operator of kinetic energy of rotation
$\Om \hat L_z$, which gives the kinetic energy $\nu \Om$. The critical rotation 
frequency $\Om_\nu$, necessary for creating a vortex with a winding number $\nu$,
is defined by the relation 
\be
\label{454}
\nu\Om_\nu=\ep_\nu-\ep_0 \; .
\ee
For the basic vortex this yields
\be
\label{455}
\Om_1 = \frac{4}{mR^2}\;\ln\left (\frac{R}{2\xi_0}\right ) \; .
\ee

The energy of a vortex with a winding number $\nu$ is proportional 
to $\nu^2$, while the orbital moment is proportional to $\nu$. Therefore 
the energy of several basic vortices is lower than the energy of one 
large vortex with the same angular momentum.

\subsection{Dark solitons}

In the previous section, a disk-shape trap was considered, resulting
in a quasi-two-dimensional radial equation. Now we shall consider an 
elongated pencil-shape trap leading to a quasi-one-dimensional equation. 

An elongated trap is characterized by the inequality $l_\perp \ll l_z$,
hence $\om_z \ll \om_\perp$. The scattering length is assumed to be such 
that $|a_s| \ll l_\perp$, so that there are no geometric effects caused by 
the boundary. The solution to the evolution equation is presented in
the variable-separated form
$$
\eta(\br,t) = \sqrt{N_0}\; \chi(r_\perp)\psi(z,t) e^{-iE_\perp t} \; ,
$$
with the same normalizations for $\chi(r_\perp)$ and $\psi(z,t)$ as in
the previous section. The transverse confinement is assumed to be strong,
such that the oscillator term prevails over the interaction one. Hence
the transverse equation is 
$$
\left ( -\; \frac{\nabla^2_\perp}{2m} + \frac{m}{2}\; \om_\perp^2
r_\perp^2 \right )\chi(r_\perp) = E_\perp \chi(r_\perp) \; .
$$
The solution for the transverse harmonic ground state is
$$
\chi(r_\perp) = \frac{1}{\sqrt{\pi}\; l_\perp}\; \exp\left ( -\;
\frac{r_\perp^2}{2l_\perp^2}\right ) \; .
$$

Let us define the effective area $A_{eff}$ by the equation
$$
\frac{1}{A_{eff}} \equiv 2\pi \int_0^R \; |\chi(r_\perp)|^4 \;
r_\perp dr_\perp \; .
$$
In the uniform case, when $\chi(r_\perp) \ra 1/\sqrt{\pi R^2}$, the effective 
area $A_{eff} \ra \pi R^2$. While for the harmonic potential, the effective area
becomes $A_{eff} = 2 \pi l_\perp^2$.

Substituting $\eta(\br,t)$ into the evolution equation, multiplying the latter
by $\eta^*(r_\perp)$, and integrating over $\br_\perp$ yields the effective
axial equation
\be
\label{456}
i\; \frac{\prt}{\prt t}\; \psi(z,t) = \left [ -\; \frac{1}{2m} \;
\frac{\prt^2}{\prt z^2} + \frac{m}{2}\; \om_z^2 z^2 +
\frac{\Phi_0}{A_{eff}} \; N_0 |\psi(z,t)|^2 - \ep \right ] \psi(z,t) \; .
\ee
Here the expression
$$
\frac{\Phi_0}{A_{eff}} = \frac{2a_s}{ml_\perp^2}  
$$
plays the role of an effective one-dimensional interaction strength. Again we 
assume that the interaction is repulsive, that is, $\Phi_0 > 0$.

If the trap is sufficiently long, such that $\xi \ll l_z \sim L$ and $\xi \ll L$,
then, one can neglect the boundary effects, omitting the harmonic term and 
introducing the variable
\be
\label{457}
x \equiv \frac{z-vt}{\xi} \qquad \left ( \xi= \frac{1}{\sqrt{2m\ep}}
\right ) \; .
\ee
It is convenient to pass to the dimensionless function
$$
f(x) \equiv \sqrt{\frac{\Phi_0 N_0}{\ep A_{eff}}}\; \psi(z,t) \; ,
$$
for which equation (\ref{456}) reduces to
\be
\label{458}
\frac{d^2f}{dx^2} + \left ( 1 -|f|^2\right ) f =  2i\gm \; \frac{df}{dx} \qquad 
\left ( \gm \equiv \sqrt{\frac{mv^2}{2\ep}} \right ) \; .
\ee
The equation is invariant with respect to the sign inversion $f \ra -f$. The 
boundary conditions can be written as 
\be
\label{459}
\lim_{x\ra-\infty}\; f(x) = -1\; , \qquad
\lim_{x\ra+\infty}\; f(x) = 1 \; .
\ee
And the normalization condition reads as
\be
\label{460}
\frac{\ep A_{eff}}{\Phi_0 N_0} \; \int_{-L/2}^{L/2} \;
\left |f\left (\frac{z-vt}{\xi}\right )\right |^2 \; dz = 1 \; ,
\ee
which defines $\ep$. 

The ground state has the form
\begin{eqnarray}
f_0(x)=\left\{ \begin{array}{rr}
-1\; , & x< 0 \\ \nonumber
1\; , & x > 0
\end{array} \right.
\end{eqnarray}
defining the eigenvalue 
\be
\label{461}
\ep_0=\rho_0\Phi_0 = mc^2 \qquad \left ( \rho_0 \equiv \frac{N_0}{A_{eff}L} \right ) \; , 
\ee
in which $c$ is the sound velocity, as in the previous section.

A nontrivial solution of equation (\ref{458}), satisfying the boundary 
conditions (\ref{459}), is the kink soliton
\be
\label{462}
f(x) = i\sqrt{2}\;\gm +\sqrt{1-2\gm^2}\;\tanh\left (
\frac{x}{\sqrt{2}}\; \sqrt{1-2\gm^2}\right ) \; .
\ee
The corresponding density
$$
|f(x)|^2 = 2\gm^2 + \left ( 1 - 2\gm^2\right )\tanh^2 \left (
\frac{x}{\sqrt{2}}\; \sqrt{1-2\gm^2}\right )
$$
is minimal at $x = 0$, where 
$$
\min_x |f(x)|^2 = 2\gm^2 \; ,
$$
because of which this solution is named dark soliton. 

The normalization condition (\ref{460}) gives
\be
\label{463}
\ep=\ep_0 + \frac{2c}{L}\;\sqrt{1-2\gm^2} \; .
\ee
For large $L \gg \xi$, the second term is much smaller than $\ep_0$. 

The soliton energy per particle is
\be
\label{464}
\ep_{sol}\equiv\ep-\ep_0 \; , 
\ee
which is small, as compared to $\ep_0$, since
$$
\frac{\ep_{sol}}{\ep_0} \; \propto \; \frac{\xi}{L} \ll 1
\qquad (\xi \ll L) \; .
$$
Then the expression for the parameter $\gamma$ can be iterated with $\ep_0$,
resulting in
$$
\gm\cong \sqrt{\frac{mv^2}{2\ep_0}} = \frac{v}{\sqrt{2}\; c} \; .
$$
The healing length is
$$
\xi \cong \frac{1}{\sqrt{2m\ep_0}} = \frac{1}{\sqrt{2}\; mc} \; .
$$

In this way, for the soliton energy, we find
\be
\label{465}
\ep_{sol} = \frac{2c}{L} \; \sqrt{1 -\; \frac{v^2}{c^2}} \; .
\ee
When $v$ increases, then $\ep_{sol}$ diminishes. The acceleration of the 
soliton velocity to $v \ra c$ yields $\ep_{sol} \ra 0$.

\subsection{Bright solitons}

Let us consider the quasi-one-dimensional evolution equation (\ref{456}) 
in the stationary case. And assume that in the longitudinal direction the 
trapping potential is essentially smaller than the interaction term. Then
equation (\ref{456}) reduces to the equation
\be
\label{466}
\left [ -\; \frac{1}{2m} \; \frac{\prt^2}{\prt z^2} +
\frac{\Phi_0}{A_{eff}} \; N_0 |\psi(z)|^2 - \ep\right ] \psi(z) = 0\; .
\ee
Suppose that interactions are attractive, $\Phi_0 < 0$. Impose the boundary 
conditions
$$
\lim_{z\ra\pm\infty}\; \psi(z) = 0 \; .
$$
For $\Phi_0 < 0$, these conditions are satisfied only for negative energy,
$\ep = -|\ep|$. Thus we come to the equation 
\be
\label{467}
\left (\frac{1}{2m} \; \frac{d^2}{dz^2} + \frac{|\Phi_0|}{A_{eff}}\;
N_0|\psi|^2 - |\ep|\right ) \psi = 0 \; .
\ee
The dimensionless spatial variable is defined as 
\be
\label{468}
x\equiv \frac{z}{\xi} \; \qquad \left ( \xi \equiv \frac{1}{\sqrt{2|m\ep|}} \right ) \; .
\ee
And the dimensionless function is
$$
f(x) \equiv \sqrt {\left |\frac{\Phi_0}{\ep}\right |\frac{N_0}{A_{eff}}} \;
\psi(z) \; .
$$
Then equation (\ref{467}) reduces to 
\be
\label{469}
\frac{d^2 f}{dx^2} + \left (|f|^2 - 1\right ) f = 0 \; , 
\ee
with the boundary condition
\be
\label{470}
\lim_{x\ra\pm\infty} \; f(x) = 0 \; .
\ee
The equation is invariant with respect to the replacement $f \ra -f$. 
Therefore it is possible to take $f$ as a positive or a negative solution. 
Below we define it as being positive. The normalization condition reads as
\be
\label{471}
\left | \frac{\ep}{\Phi_0}\right | \frac{A_{eff}}{N_0} \;
\int_{-L/2}^{L/2} \; \left | f\left ( \frac{z}{\xi}\right )\right |^2 \;
dz  = 1 \; .
\ee

There is no uniform ground state satisfying the boundary conditions (\ref{470}).
The soliton solution is
\be
\label{472}
f(x) = \frac{\sqrt{2}}{\cosh x} \; .
\ee
The normalization condition gives
\be
\label{473}
\ep = -\; \frac{m}{8}\left (\rho_0 \Phi_0 L\right )^2 \; .
\ee
This defines the healing length
$$
\xi =\frac{2}{\rho_0|m\Phi_0|L} \; .
$$
The soliton is named bright, since it is of bell shape, with a maximum 
at $x = 0$.

The other situation, when a bright soliton arises, is the case of a gap 
soliton, when the effective mass is negative, $m = -|m|$, while the 
interactions are repulsive, $\Phi_0 > 0$. In that case, instead of equation
(\ref{467}), we get the equation
\be
\label{474}
\left ( \frac{1}{2|m|}\; \frac{d^2}{dz^2} + \frac{\Phi_0}{A_{eff}}\;
N_0|\psi|^2 -\ep \right )\psi = 0 \; .
\ee
This is equivalent to the previous case, but with positive energy $\ep > 0$,
for which the normalization condition gives
$$
\ep =\frac{|m|}{8}\; \left ( \rho_0\Phi_0 L\right )^2 \; ,
$$
with the same healing length.

The evolution equation for the condensate wave function is invariant
under the Galilean transformation. Therefore it is straightforward to 
generalizes the stationary solution to the nonstationary situation,
obtaining a bright soliton moving with velocity $v$, 
\be
\label{475}
\psi_v(z,t) = \psi(z-vt)\exp\left\{ i\left ( mvz - \;
\frac{mv^2}{2}\; t\right )\right \} \; .
\ee
In all soliton solutions, it is admissible to make the shift $z \ra z-z_0$, 
with $z_0$ being the initial soliton location.

\section{Conclusion}

This Tutorial is the continuation of the first part \cite{Yukalov_1}, where
a general mathematical foundation of quantum statistics has been given for
both Bose-Einstein and Fermi-Dirac statistics. In the present part, the 
properties of Bose systems are treated, with an emphasis on physics of 
systems with Bose-Einstein condensate. Specifics, related to trapped atoms,
are emphasized. 

Since this is a Tutorial, the explanations are sufficiently detailed in order 
that the reader could follow the main steps of calculations. Of course, it is 
impossible to treat all so numerous topics related to Bose systems. Here only 
some of the topics are considered, which to the understanding of the author, 
are of principal interest for describing the basic properties of Bose-condensed 
systems. Many important problems are not touched here. For instance, those 
considering nonequilibrium Bose-condensed systems, including such a fascinated
topic as quantum turbulence \cite{Tsubota_98, Nemirovskii_99}. The choice, to 
some extent, can be subjective. The selection of the material is done on the 
basis of lectures the author has given in the Free University of Berlin, 
Germany, and S\~ao Paulo State University, Brazil. 
     
The methods described here for treating cold trapped atoms can be employed for 
considering other finite quantum systems \cite{Birman_100}. 

For a Tutorial, there has been no intention to give an extensive list of 
references, as it would be necessary for a review article. Only those papers 
are cited that are of historical interest or directly used. 

\vskip 2mm

{\bf Acknowledgments} 

The author is grateful for discussions to V.S. Bagnato, R. Graham, H. Kleinert,
and E.P. Yukalova. Financial support from the RFBR (grant $\#$ 14-02-00723) is
appreciated.

\end{document}